\documentclass{jfm}
\usepackage{rotating,graphicx}
\usepackage{graphicx}
\usepackage{epstopdf, epsfig}

\usepackage{mathtools}
\usepackage{amsmath}
\usepackage{amssymb}
\usepackage{booktabs}
\usepackage{array}
\usepackage{amssymb}   % for \checkmark
\usepackage{lscape}

\usepackage{lineno}
\usepackage[utf8]{inputenc}
\usepackage[T1]{fontenc}
\usepackage{mathptmx}
\usepackage{soul}
\usepackage{microtype}
\usepackage{placeins}
\usepackage{booktabs} % To thicken table lines
\usepackage{bm} % in preamble
\usepackage{tabularx}
\usepackage{graphicx}% Include figure files
\usepackage{caption}
\usepackage{hyperref}
\usepackage{subcaption}
\usepackage{algpseudocode}
\usepackage{algorithm}

\usepackage{amsmath}
\usepackage{gensymb}
\let\Psi\varPsi
\usepackage{placeins}
\usepackage{listings}
\usepackage[dvipsnames]{xcolor}

\usepackage{placeins} % To sue floatBarrier

\usepackage[toc,page]{appendix}

\usepackage{cleveref}
\newcommand\myshade{85}
\colorlet{mylinkcolor}{violet}
\colorlet{mycitecolor}{Aquamarine}
\colorlet{myurlcolor}{Aquamarine}
\usepackage{xcolor}

\hypersetup{
  linkcolor  = mylinkcolor!\myshade!black,
  citecolor  = mycitecolor!\myshade!black,
  urlcolor   = myurlcolor!\myshade!black,
  colorlinks = true,
}

\definecolor{codegreen}{rgb}{0,0.6,0}
\definecolor{codegray}{rgb}{0.5,0.5,0.5}
\definecolor{codepurple}{rgb}{0.58,0,0.82}
\definecolor{backcolour}{rgb}{0.95,0.95,0.92}

\lstdefinestyle{mystyle}{
    backgroundcolor=\color{backcolour},   
    commentstyle=\color{codegreen},
    keywordstyle=\color{magenta},
    numberstyle=\tiny\color{codegray},
    stringstyle=\color{codepurple},
    basicstyle=\ttfamily\footnotesize,
    breakatwhitespace=false,         
    breaklines=true,                 
    captionpos=b,                    
    keepspaces=true,                 
    numbers=left,                    
    numbersep=5pt,                  
    showspaces=false,                
    showstringspaces=false,
    showtabs=false,                  
    tabsize=2
}

\newcommand\R{\mbox{\text{Re}}}

\newcommand\Ka{\mbox{\text{Ka}}}

\newcommand\K{\mbox{\text{K}}}
\newcommand\E{\mbox{\text{E}}}
\newcommand\D{\mbox{\text{D}}}
\newcommand\M{\mbox{\text{Ma}}}
\newcommand{\Ell}{\mathcal{L}}

\newcommand\Ct{\mbox{\text{Ct}}}
\newcommand\Vr{\mbox{\text{Vr}}}

\usepackage{comment}

\shorttitle{Integral modelling of weakly evaporating 3D liquid film with variable substrate heating}
\shortauthor{F. Pino}

\title{Integral modelling of weakly evaporating 3D liquid film with variable substrate heating}

\author{Fabio Pino\aff{1}
  \corresp{\email{fp448@cam.ac.uk}} }

\affiliation{\aff{1}Department of Applied Mathematics and Theoretical Physics (DAMTP), University of Cambridge, Wilberforce Rd, Cambridge CB3 0WA}

\begin{document}
\maketitle

\begin{abstract}
Analysing the dynamics of phase-changing liquid films is essential for enhancing the performance of thermal management systems. Still, direct simulation of the full governing equations is computationally expensive. To circumvent this limitation, I derived a weighted-integral boundary-layer (WIBL) model under long-wave assumptions, weak evaporation, and strong surface tension, also accounting for variable substrate heating. In the linear regime, the WIBL reproduces the growth rates and cut-off wavenumbers of unstable modes with significantly higher accuracy than commonly used Benney-type models for $\R<40$, and with accuracy comparable to that of the Orr-Sommerfeld equations. The linear analysis further reveals a threshold separating streamwise- and spanwise-dominated instabilities in hanging films, arising from the competition between the Kapitza and Rayleigh-Taylor mechanisms; the WIBL accurately predicts this threshold for small $\R$ and inclination angles. In the nonlinear regime, with space- and time-varying substrate heating, the WIBL model captures the evolution of the free-surface thickness and temperature with an accuracy of approximately 6\% relative to the original Navier-Stokes equations. Three-dimensional simulations show that a condensing film undergoes dry-out due to Kapitza instability, whereas unsteady substrate heating promotes spanwise momentum spreading, modifies wave dynamics, and prevents dry-out. The WIBL model provides a good level of accuracy at a low computational cost, enabling extensive parametric studies, nonlinear stability analyses, and the design of optimal substrate-heating control strategies.
\end{abstract}


% --------- Introduction --------- 
\section{Introduction}
Phase-changing liquid films are used for transferring intense heat loads in refrigeration systems \citep{fernandez2014refrigerant} and thermal-management devices such as vapour chambers \citep{bulut2019review}, grooved heat pipes \citep{bertossi2009modeling}, and pulsating heat pipes \citep{zhang2023physics}. In vapour chambers specifically, heat generated by localised 3D electronic components must be efficiently spread over a larger condenser surface to match the dimensions of the heat sink \citep{singh2025numerical}. This multidirectional heat-spreading process is accompanied by capillary-driven liquid transport from the condenser towards the evaporator, leading to non-negligible inertial and extensional effects within the liquid film. Accurately accounting for these effects is crucial for predicting film evolution, interfacial stability, and dry-out phenomena. However, the analysis of phase-changing film flows and the development of control strategies to enhance heat transfer are hindered by the high computational cost of the full governing equations \citep{patankar2018validated} and the limited validity range of existing reduced-order models \citep{nikolayev2021physical,mohamed2020linear}. To address these limitations. I developed a novel three-dimensional (3D) integral model that captures the dominant linear and nonlinear dynamics over a broad range of Reynolds numbers while reducing the computational cost by several orders of magnitude.

The dynamics of phase-changing liquid films are commonly described by long-wave models, valid for film aspect ratios $\varepsilon \ll 1$ \citep[see reviews by][]{oron1997long, craster2009dynamics, chattopadhyay2019review}. These models are classified by their range of validity in $\R$ \citep[Section~4.9]{kalliadasis2011falling}. At low $\R$, the film dynamics is described by a single Benney–type (BE) equation derived by expanding the dependent variables in powers of $\varepsilon$ \citep{benney1966long, burelbach1988nonlinear}. Using a modified BE equation, \citet{burelbach1988nonlinear} studied evaporating films over horizontal plates with particular attention to dry-out dynamics. For inclined films, \citet{joo1991long} showed that nonlinear interactions among the Kapitza (H), Marangoni (M), and evaporative (E) modes govern wave breaking, secondary structure formation, and dry-out, with strong sensitivity to initial conditions. In 3D, the mode-coupled H–M and H–E modes generate spanwise-fingering patterns that cannot arise from either mode alone \citep{joo1996mechanism,joo1993two}.

For larger $\R$, \citet{mohamed2020linear} showed that the linearised Benney equation overpredicts maximum growth rates and underpredicts cutoff wavenumbers compared with the linearised Navier–Stokes solution, particularly for $\R \gtrsim 10$. Integral-equation models derived via Galerkin projection \citep{ruyer2002further} are better suited for moderate $\R$ regimes. Representative examples include the Integral Boundary-Layer (IBL) model \citep{aktershev2013nonlinear}, employing a parabolic trial function, and the Weighted Integral Boundary-Layer (WIBL) model, using polynomial trial and test functions \citep{craster2009dynamics,kalliadasis2011falling}. To reduce discrepancies with the full Navier–Stokes solution, \citet{scheid2008interaction} developed a simplified second-order WIBL formulation retaining $O(\varepsilon^2)$ viscous corrections while neglecting second-order inertial terms. Although successful in various liquid-film problems \citep{samanta2014shear,scheid2016critical}, such models have not yet been used for phase-changing films.

Developing a simplified second-order WIBL model for phase-changing films is crucial for designing control strategies that enhance heat transfer via free-surface waves. For non-evaporating films, experiments \citep{frisk1972enhancement} and simulations \citep{goren1968mass} show that surface waves can increase heat transfer by up to 80\% relative to flat-film conditions \citep{miyara1999numerical}. More recent studies \citep{collignon2023heat, yoshimura1996enhancement} indicate that both periodic and solitary waves substantially enhance heat and mass transfer. In condensing films, \citet{aktershev2013nonlinear} found that specific wave structures increase heat transfer through the combined effects of localised thinning and large recirculation zones.

Achieving such wave states requires control strategies that strongly influence the film dynamics. In practical situations, such as sealed heat pipes \citep{zhang2021liquid, pagliarini2023pulsating}, the free surface is inaccessible, and boundary modifications for mass injection are infeasible. Modulating the substrate temperature thus emerges as the most practical control approach, motivating the development of non-isothermal reduced-order models. \citet{cellier2020new} introduced a novel two-dimensional second-order thermal model known as \textit{$\theta$ model}, using thermal relaxation modes as trial functions. This model agrees well with the full energy equation even at large Biot and Péclet numbers (up to $O(100)$). Yet such approaches have not been extended to variable-substrate heating and phase-changing films.

In this work, I develop a model for a liquid film flowing over an incline, in contact with its pure vapour phase, which is assumed to be passive and at saturation pressure and temperature (Section~\ref{sec:prob_description}). A 3D simplified second-order WIBL model is coupled to a modified \textit{$\theta$ model} accounting for varying substrate temperatures (Section~\ref{sec:methodology}). Assuming weakly evaporating conditions, the interfacial mass flux is taken as a linear function of the free-surface temperature, with the proportionality coefficient derived from the Linearised Boltzmann-equation Moment Method \citep[§2.6.4]{aursand2019comparison}. \cite{sultan2005evaporation} and, more recently, \cite{mohamed2024stability} developed a more general mass-flux closure accounting for both kinetic- and diffusion-driven phase change. The diffusion-limited regime is typically encountered in evaporating films or droplets surrounded by air, where the gas phase becomes nearly saturated with vapour near the free surface. In vapour chambers, however, the intense convective transport in the vapour phase prevents significant accumulation of vapour near the interface. Therefore, gas-phase diffusion does not constitute the dominant resistance to phase change. For this reason, and consistently with common practice in the vapour-chamber literature, a kinetic model was adopted for the mass-flux closure \citep{patankar2018validated}.

The capabilities of the present reduced-order modelling framework for non-isothermal liquid films are summarised in Table~\ref{tab:model_comparison}, which compares its features with those of existing models available in the literature.

The simplified model is validated against the full governing equations. In the linear regime, predictions are compared with linearised 3D Navier-Stokes equations for both evaporating and condensing films. In the nonlinear regime, film evolution under spatially and temporally varying substrate temperatures is compared with the solution of the Navier-Stokes equation in \textsc{COMSOL} Multiphysics. Finally, the simplified 3D model is used to investigate the effects of substrate-temperature modulation, showing that it can prevent dry-out in evaporating films and modify free-surface wave structures, thereby providing a foundation for control strategies.

\begin{table}
\centering
\renewcommand{\arraystretch}{1.25}
\setlength{\tabcolsep}{6pt}

\begin{tabular}{lcccccc} & Model & Dim. & \begin{tabular}[c]{@{}c@{}}Var. \\ Temp. \end{tabular}  & \begin{tabular}[c]{@{}c@{}}Substrate. \\ Temp. Var.\end{tabular} & \begin{tabular}[c]{@{}c@{}}Phase \\ Change \end{tabular} & Order \\
\citet{burelbach1988nonlinear} & Benney & 2D & -- & -- &weakly evap. & $O(\varepsilon)$ \\
\citet{joo1991long} & Benney & 3D & -- & -- &weakly evap. & $O(\varepsilon)$ \\
\citet{scheid2002nonlinear}& Benney & 2D & Linear & \checkmark &-- & $O(\varepsilon)$\\ 
\citet{ruyer2005thermocapillary} & WIBL & 2D & WIBL & -- & -- &  $O(\varepsilon^2)$ \\
\citet{scheid2008interaction}& WIBL & 3D & WIBL & -- & -- &  $O(\varepsilon^2)$ \\
\citet{thompson2019robust} & WIBL  & 2D & Projection & \checkmark & -- &  $O(\varepsilon)$ \\
\citet{cellier2020new} & WIBL & 2D & $\theta$ model & -- &  -- & $O(\varepsilon^2)$ \\
\textbf{This work} & WIBL & 3D & $\theta$ model & \checkmark & weakly evap. &  $O(\varepsilon^2)$ \\
\end{tabular}
\caption{Comparison of representative long-wave reduced-order models for non-isothermal thin liquid films, highlighting their dimensionality, treatment of the temperature field, inclusion of spatial and temporal substrate-temperature variations ($\checkmark$), accounting for phase change, and long-wave asymptotic order.}
\label{tab:model_comparison}
\end{table}

% ---------- Problem Description ---------- 
\section{Problem Description}
\label{sec:prob_description}
\begin{figure}
    \centering
    \includegraphics[width=0.8\textwidth]{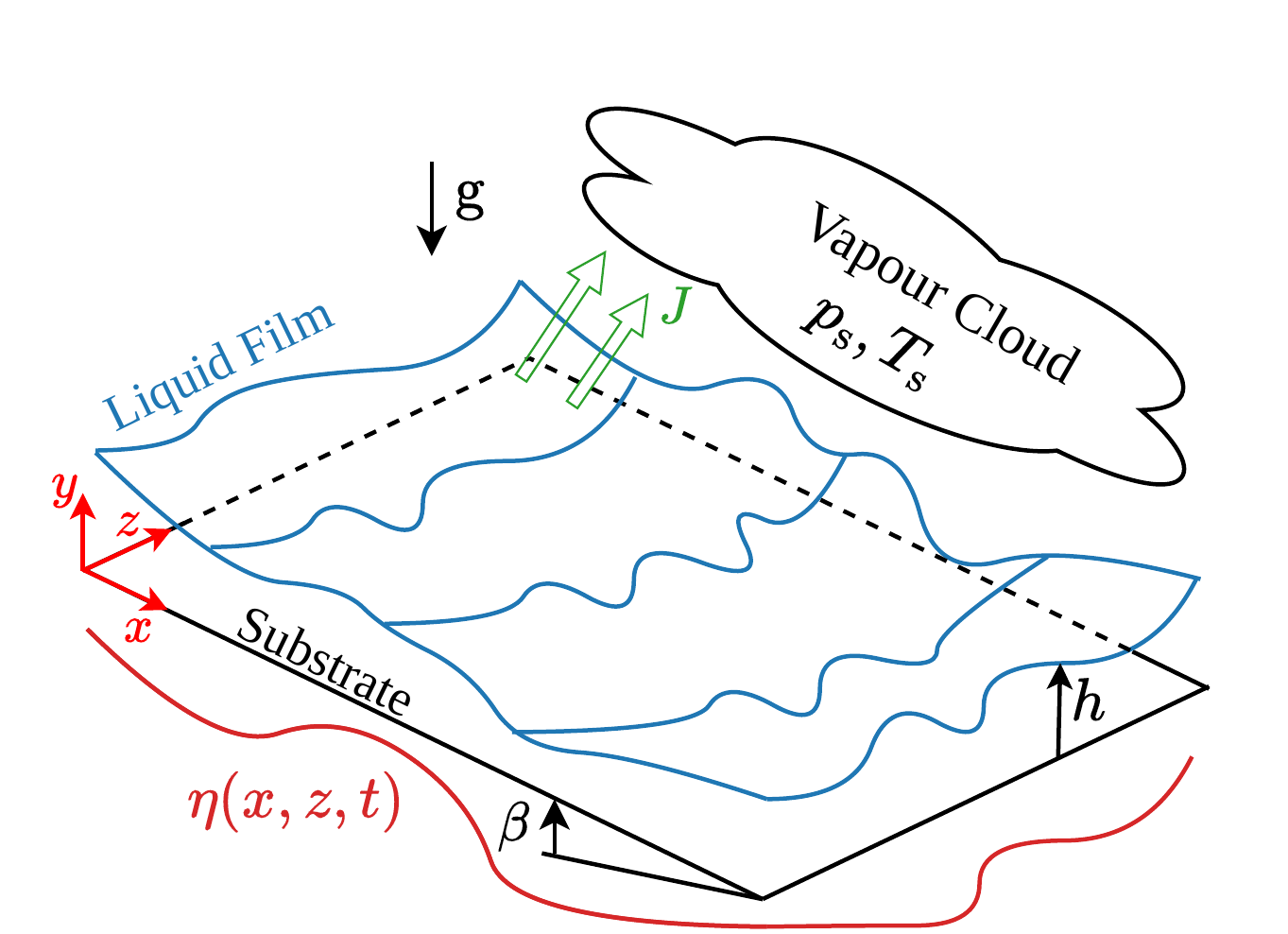}
    \caption{Schematic representation of a 3D liquid film flowing under the effect of gravity $g$ over a substrate inclined at an angle $\beta$ to the horizontal, with an imposed temperature distribution $\eta(x,z,t)$. The phase-changing film is bounded above by its pure vapour phase at saturation pressure $p_s$ and temperature $T_s$, with an interfacial mass flux $J$ across the free surface.}
    \label{fig:scheme}
\end{figure}

Figure~\ref{fig:scheme} shows the 3D evaporating/condensing liquid film flowing over a flat plate inclined at an angle $\beta$ to the horizontal. The liquid is in contact with its pure vapour phase, which is assumed mechanically and thermodynamically passive and held at the saturation pressure $p_s$ and temperature $T_s$ (as in \citep{Burelbach1988}). The liquid has density $\rho$, dynamic viscosity $\mu$, thermal conductivity $\lambda$, surface tension $\sigma$, latent heat $L_v$ and thermal capacity at constant pressure $C_p$, while the vapour density is $\rho_v$.

A Cartesian reference frame $\mathcal{O}(x,y,z)$ is defined on the substrate with the $x$-axis streamwise, the $y$-axis normal to the substrate pointing toward the free surface, and the $z$-axis spanwise. At the free surface, a local orthonormal reference frame $\mathcal{O}(\bm{n},\bm{t}_x,\bm{t}_z)$ is defined, where $\bm{n}$ is the outward unit normal and $\bm{t}_x$, $\bm{t}_z$ are the tangential unit vectors along the streamwise and spanwise directions, respectively. These vectors are defined as:
\begin{equation}
\label{eq:local_ref_frame}
    \bm{n} = \frac{(-\partial_x h, 1, -\partial_z h)^T}{n}, \qquad\qquad
    \bm{t}_x = \frac{(1, \partial_x h, 0)^T}{t_x}, \qquad\qquad
    \bm{t}_z = \frac{(0, \partial_z h, 1)^T}{t_z},
\end{equation}
where $n$, $t_x$ and $t_z$ read:
\begin{equation}
    n = (1 + (\partial_x h)^2 + (\partial_z h)^2)^{1/2},\qquad t_x = (1+(\partial_x h)^2)^{1/2},\qquad t_z = (1+(\partial_z h)^2)^{1/2}.
\end{equation}

In the remaining, I use the shorthand notation $\partial_x h = \partial h/\partial x$.

The film flows down the inclined plate under the effect of the gravitational acceleration vector $\bm{g}=(g\sin{(\beta)},-g\cos{(\beta)},0)^{T}$. The solid substrate is held at a prescribed temperature $\eta(x,z,t)$, while the liquid is characterised by its free-surface height $h(x,y,t)$, velocity field $\bm{u}=(u,v,w)^{\mathrm{T}}$, pressure $p(x,y,z,t)$ and temperature $T(x,y,z,t)$ fields with the free-surface temperature $T_{fs}(x,z,t)$, streamwise $q_x(x,z,t)$ and spanwise $q_z(x,z,t)$ flow rates defined as:
\begin{equation}
\label{eq:def_flow_rates}
    T_{fs}(x,z,t)=T|_{y=h},\qquad q_x(x,z,t) = \int_{0}^{h}\, u(x,y,z,t)\, dy, \qquad q_z(x,z,t) = \int_{0}^{h}\, w(x,y,z,t)\, dy.
\end{equation}

The difference between the liquid temperature at the free surface and the vapour saturation temperature induces both a change in surface tension and a mass flux $J(x,z,t)$ due to phase change. Following common practice for thermocapillary flows and previous studies of evaporating films \citep{palmer1976hydrodynamic}, I model these relations as linear functions of $T_{fs}$, which read:
\begin{equation}
\label{eq:lin_approx}
    \sigma\bigl(T_{fs}\bigr)=\sigma_0-\frac{d\sigma}{dT}\Big|_{T_s}\bigl(T_{fs}-T_s\bigr), 
    \qquad\qquad\qquad\qquad
    J=\zeta\bigl(T_{fs}-T_s\bigr),
\end{equation}
where $\sigma_0$ is the surface tension at the saturation temperature $T_s$, and $\zeta$ is the interfacial mass-transfer coefficient. This coefficient is defined as \citep[§2.6.4]{aursand2019comparison}:
\begin{equation}
\label{eq:lin_approx_1}
    \zeta \;=\; (1.64\,\alpha)\,\rho^v \, \frac{L_v}{T_s^{3/2}}\,
    \sqrt{\frac{M_w}{2\pi R_g}},
\end{equation}
where $\alpha$ is the accommodation coefficient and $M_w$ is the molecular weight of the liquid, and $R_g$ is the universal gas constant. 

The linear mass flux relation is defined as the linearised Boltzmann Equation Moment Method (BEMM) model, which accounts for the mass, momentum, and energy balance in the Knudsen layer between the free surface and the vapour bulk. This simplified relation is valid in the limit of weak evaporation, corresponding to small driving forces $\Delta Z$. The driving force is the difference between the saturation pressure associated with the free-surface temperature $p_s(T_{fs})$ of the liquid and the vapour pressure $p_s$. By linearising the saturation line $p_s(T)$, $\Delta Z$ is approximated as a function of $T_{fs}$, reading:
\begin{equation}
\label{eq:driving_force}
    \Delta Z \;=\; \frac{L_v M_w}{R_g T_s}\,\frac{T_{fs}-T_s}{T_s}.
\end{equation}
Following the linear assumption \eqref{eq:lin_approx}, the linear approximation for the mass flux at the free-surface is valid for $ \Delta Z\ll 1$.

% --------- Scaling Quantities --------- 
\subsection{Scaling Quantities and Nondimensional Groups}
\label{subsec:scaling}
The scaling quantities are based on the characteristic viscous scales and the mean film thickness $d_0$ at the initial time $t=0$ \citep{burelbach1988nonlinear,williams1982nonlinear,mohamed2020linear}, which corresponds to the \textit{Nusselt flat film thickness} in non-volatile conditions \citep[Section 2.2]{kalliadasis2011falling}. The scaling quantities read: 
\begin{equation}
\label{eq:scaling}
    h_{ref} = d_0, \qquad u_{ref} = \frac{\nu}{d_0}, \qquad p_{ref} = \frac{\rho\nu^2}{d_0^2},\qquad J_{ref} = \frac{\lambda\Delta T_{ref}}{d_0 L_v},\qquad t_{ref} = \frac{d_0^2}{\nu},
\end{equation}
where $\Delta T_{ref}$ is the reference temperature difference.

Based on these scaling quantities, the dependent and independent variables are scaled accordingly as follows:
\begin{equation}
\label{eq:scaling_original}
    (x,y,z)\rightarrow (\hat{x},\hat{y},\hat{z})\,h_{ref}, \qquad t\rightarrow\hat{t}\,t_{ref},\qquad (u,v,w)\rightarrow (\hat{w},\hat{v},\hat{w})\,u_{ref}, \qquad p\rightarrow \hat{p}\,p_{ref},
\end{equation}
with the substrate temperature profile $\eta$, the temperature field $T$ and the free-surface temperature $T_{fs}$ scaled as:
\begin{equation}
\label{eq:scaling_original_1}
\eta\rightarrow T_s + \hat{\eta}\,\Delta T_{ref},\qquad T\rightarrow T_s + \hat{\theta}\,\Delta T_{ref},\qquad T_{fs}\rightarrow T_s + \hat{\theta}_s\,\Delta T_{ref},
\end{equation}
where $\hat{\bullet}$ denotes the nondimensional variables. 

For the derivation of the reduced order models in Subsection \ref{subsec:ROM_der}, slow spatial $\hat{X}$ and $\hat{Z}$ and time $\hat{T}$ scales are introduced. In addition, to guarantee the consistency of the continuity equation, the wall-normal velocity $v$ is assumed to be a small quantity. The slow scales read: 
\begin{equation}
\label{eq:ROM_stretching}
     \hat{X} = \varepsilon\,\hat{x},\qquad\qquad\qquad \hat{Z} = \varepsilon\,\hat{z},\qquad\qquad\qquad \hat{T} = \varepsilon\, \hat{t},\qquad\qquad \hat{v} = \varepsilon\, \hat{V},
\end{equation}
where $\varepsilon\ll 1$ and $\hat{\bullet}$ accounts for $O(1)$ quantities.

\begin{table}
\centering  
\renewcommand{\arraystretch}{1.2}
\begin{tabular}{c@{\hspace{0.6cm}}c@{\hspace{0.6cm}}c@{\hspace{0.6cm}}c@{\hspace{0.6cm}}}
Nondimensional Number & Definition & Physical meaning\\
\toprule
Reynolds ($\R)$  & $(d_0^3\,g\,\sin{(\beta)})/\nu^2$ & Inertial over viscous stresses\\
Inclination ($\Ct)$ & $(d_0^3\,g\,\cos{\beta})/\nu^2$ &  Measure hydrostatic pressure\\
Surface Tension ($\Gamma$) & $(\sigma_0 d_0)/(3\rho\nu^2)$ & Surface tension over inertia  \\
Prandtl ($\Pr)$  & $(\nu\rho C_p)/\lambda$ & momentum diffusivity over thermal diffusivity\\
Marangoni ($\M)$  & $(\gamma\Delta T_{ref}d_0)/(2\mu\kappa)$ & surface tension gradient over inertia \\
Evaporation ($\E$) & $J_{ref}/(\rho u_{ref})$ & Ratio viscous to evaporation time scales  \\
Vapour recoil ($\Vr$) & $\E^2/\D$ & Vapour recoil force  \\
kinetic energy ($\Pi$)  & $\E^2/(\D^2\Ell)$ & Measure kinetic energy flux at the free surface \\
non-equilibrium ($\K$) & $J_{ref}/(\zeta \Delta T_{ref})$ &  Degree thermodynamic non equilibrium free-surface
\end{tabular}
\caption{Nondimensional numbers with their definition and physical meaning}
\label{tab:adim_num}
\end{table}

The nondimensional groups arising from the scalings \eqref{eq:scaling_original} and \eqref{eq:scaling_original_1} are classified into three main categories: hydrodynamic, thermal, and phase change. The hydrodynamic nondimensional numbers are the Reynolds ($\R$) and the inclination number ($\Ct$) and the non-dimensional surface tension number ($\Gamma$) which are defined as:
\begin{equation}
\label{eq:adim_num_00}
    \R =\frac{d_0^3\,g\,\sin{\beta}}{\nu^2},\qquad\qquad\Ct = \frac{d_0^3\,g\,\cos{\beta}}{\nu^2} = \R\,\cot\beta,\qquad\qquad \Gamma = \frac{\sigma_0\,d_0}{3\rho\nu^2}.
\end{equation}

This definition of $\R$ corresponds to three times the Reynolds number based on the Nusselt scaling, $\R_N$ \citep[Section~2.2]{kalliadasis2011falling}. Furthermore, replacing $d_0$ with the capillary length $l_v$ in $\Gamma$ yields the classical expression of the Kapitza number $\Ka$.

The relevant thermal nondimensional numbers are the Prandtl ($\Pr$) number, the Marangoni ($\M$) number, and the Marangoni–Reynolds ($\R_{M}$) number, which quantifies the ratio of inertial to viscous effects in thermocapillary-driven motion. These nondimensional numbers are defined as:
\begin{equation}
\label{eq:adim_thermal_nums}
     \Pr = \frac{\nu\rho C_p}{\lambda},\quad\quad \M = \frac{(d\sigma/dT)\,d_0\,\rho\,C_p\,\Delta T_{ref}}{2\mu\lambda},\quad\quad  \R_M = \frac{\M}{\Pr} = \frac{\rho U_{tc} d_0}{2 \mu},
\end{equation}
where $U_{tc}$ is the characteristic thermocapillary velocity scale reading: 
\begin{equation}
    U_{tc} = \frac{\gamma\Delta T_{ref}}{\mu}.
\end{equation}

The nondimensional group associated to the phase change are the evaporation number number ($\E$), accounting for the ratio between characteristic viscous $t_{ref}$ over evaporative $t_E$ time scales and the vapour recoil number ($\Vr$) quantifying the strength of the vapour recoil effect and the kinetic energy ($\Pi$) number accounting for the intensity of the kinetic energy lost at the free-surface due to the phase change. These numbers read: 
\begin{equation}
    \E = \frac{t_{ref}} {t_E} =\frac{\lambda\Delta T_{ref}}{L_v\,\rho\,\nu},\qquad\qquad \Vr = \frac{2\E^2\,\rho}{3\rho_v},\qquad\qquad \Pi = \frac{\E^2\,\rho^2}{2\rho_v^2}\frac{\nu^2}{d_0^2\,L}.
\end{equation}

The non-equilibrium number $\K$ accounts for the thermodynamic equilibrium at the free surface, and is defined as:
\begin{equation}
    \K = \frac{J_{ref}}{\zeta \Delta T_{ref}} = \Big(\frac{\lambda\,T_s^{3/2}}{r(\alpha)\,d_0\,\rho_{v}\,L^2}\Big)\sqrt{\frac{2\pi\,R_g}{M_w}}.
\end{equation}
A value of $\K=0$ corresponds to the quasi-equilibrium limit, in which the free-surface temperature $T_{fs}$ is equal to the saturation temperature $T_s$, which in nondimensional terms corresponds to $\hat{\theta}_s=0$. In the non-volatile limit ($\E \rightarrow 0$), the mass flux at the free surface $\hat{J}$ is zero and $\K$ reduces to the inverse of the Biot number, which quantifies the heat transfer from the liquid to the vapour phase.

The reader is referred to Table \ref{tab:adim_num} for an overview of the nondimensional numbers, including their definitions and physical meanings.

% ------------- Nondimensional Governing Equations
\subsection{Nondimensional governing equations}
\label{subsec:gov_eqs_adim}
The phase-changing liquid film is described by the nondimensional mass, momentum and energy conservation equations, which read:
\begin{subequations}
\label{eq:nondim_gov_eq}
\begin{equation}
\label{eq:cont_adim}
    \nabla\cdot\hat{\bm{u}} = 0
\end{equation}
\begin{equation}
\label{eq:gov_momentum}
\partial_t\hat{\bm{u}} + \hat{\bm{u}}\cdot\nabla\hat{\bm{u}} = -\nabla\hat{p} + \nabla^2\hat{\bm{u}} + \hat{\bm{g}}  
\end{equation}
\begin{equation}
\label{eq:energy_gov_eq}
Pr(\partial_t\hat{\theta} + \hat{\bm{u}}\cdot\nabla\hat{\theta}) = \nabla^2\hat{\theta}  
\end{equation}
\end{subequations}
where the nondimensional gravitational acceleration vector $\hat{\bm{g}}$ is given by:
\begin{equation}
    \hat{\bm{g}} = (\R,-\Ct,0) ^T.
\end{equation}
The kinematic boundary condition at the substrate ($\hat{y}=0$) read:
\begin{equation}
\label{eq:bc_noslip}
     \hat{\bm{u}} = \bm{0},\qquad\qquad\qquad\qquad\hat{\theta} = \hat{\eta}.
\end{equation}

At the free-surface ($\hat{y}=\hat{h}$), the kinematic boundary condition enforcing the film continuity reads:
\begin{equation}
\label{eq:kin_cond_adim}
-\E\,\hat{J}\,n = \partial_t\hat{h} + \hat{u}\,\partial_x\hat{h} - v + \hat{w}\,\partial_z\hat{h},
\end{equation}
the thermal boundary, which expresses the heat exchanged with the vapour phase via latent heat and kinetic energy, reads:
\begin{equation}
\label{eq:bc_thermal}
    \hat{J} + \Pi\hat{J^3} = - \nabla\hat{\theta}\cdot\bm{\hat{n}},
\end{equation}
where $\hat{J}$ is expressed as a function of $\hat{\theta}_s$ via the nondimensional version of the constitutive relation \eqref{eq:lin_approx}, reading:
\begin{equation}
\label{eq:J_cond_FS_adim}
    \hat{J} = \K^{-1}\,\hat{\theta}_s.
\end{equation}

The normal and tangential stress boundary conditions read:
\begin{subequations}
\label{eq:bc_dyn}
\begin{equation}
\label{eq:bc_dyn_p}
    \frac{3\Vr\hat{J}^2}{2} - \hat{p} + (\mathsfbi{\hat{P}}\cdot\bm{\hat{n}})\cdot\bm{\hat{n}} = -\Big(3\Gamma - \frac{\M}{\Pr}\hat{\theta}_s\Big)\nabla_s\cdot\bm{\hat{n}},
    \end{equation}
\begin{equation}
\label{eq:bc_dyn_sh_0}
(\mathsfbi{\hat{P}}\cdot\bm{\hat{n}})\cdot\bm{\hat{t}}_x = \frac{2\M}{\Pr}\partial_x\hat{\theta}_s,\, \qquad\qquad\qquad\qquad (\mathsfbi{\hat{P}}\cdot\bm{\hat{n}})\cdot\bm{\hat{t}}_z = \frac{2\M}{\Pr}\partial_z\hat{\theta}_s,\,
\end{equation}
\end{subequations}
where $\mathsfbi{\hat{P}}=\nabla\bm{\hat{u}}+\nabla\bm{\hat{u}}^T$ is the deviatoric stress tensor and $\nabla_s=(\mathsfbi{I} - \bm{n}\otimes\bm{n})\cdot\nabla$ is the \textit{surface gradient operator}, with $\otimes$ representing the dyadic product. For simplicity, the set of governing equations with the corresponding boundary conditions is referred to as the \textit{full model equations} in contrast to the \textit{reduced order model equations}, which are presented in subsection \ref{subsec:ROM_der}.

By neglecting phase-change effects ($\E = \Vr = \Pi = 0$) and adopting the Nusselt scaling, the above governing equations and boundary conditions reduce to those describing a non-isothermal falling liquid film \citep[Section~2.1]{kalliadasis2011falling}.

% ------------- Flat film solutions ------------- 
\subsection{Weakly evaporating/condensing flat state solutions}
\label{sec:flat_film}
Considering weak evaporating/condensing conditions ($\E\ll1$), with a constant and uniform substrate temperature $\hat{\eta}$, the full model equations admit a slowly-varying ($\partial_t = O(\E)$) and flat ($\partial_x \sim 0$ and $\partial_z \sim 0$) film solution. Adapting the derivation described by \citep{burelbach1988nonlinear}, this solution is given by:
\begin{subequations}
\label{eqs:basic_state_equations}
\begin{equation}
    \overline{H}(\hat{t}) = -K + \sqrt{(\K+1)^2-2\,\E\,\hat{\eta}\,\hat{t}},\qquad\overline{J} = \frac{\hat{\eta}}{\overline{H}(\hat{t})+\K},\qquad \overline{U} = \R\,\overline{H}\,(\hat{t})^2\Big(\bar{y} - \frac{\bar{y}^2}{2}\Big),
\end{equation}
\begin{equation}
    \overline{V} = 0, \quad\qquad \overline{W} = 0,\quad\qquad\overline{P} = \Ct\,\overline{H}\,(\hat{t})(1-\bar{y}) + \frac{3}{2}\Vr\bar{J}^2,\quad\qquad\overline{\Theta} = \hat{\eta}-\overline{J}\,\bar{y}\,\overline{H},
\end{equation}
\end{subequations}
where $\bar{y}=\hat{y}/\hat{h}$.

In terms of streamwise $\overline{Q}_x$ and spanwise $\overline{Q}_z$ flow rates and free-surface temperature $\overline{\Theta}_s$, the solution \eqref{eqs:basic_state_equations} reads: 
\begin{equation}
\label{eq:ss_rel_q_theta}
    \overline{Q}_x = \int_0^{\overline{H}}\,\overline{U}\,d\hat{y} = \R\frac{\overline{H}^3}{3},\qquad\qquad \overline{Q}_z = \int_0^{\overline{H}}\,\overline{U}\,d\hat{y} =0,\qquad\qquad \overline{\Theta}_s = \overline{\Theta}|_{\hat{y}=\overline{H}} = \frac{\hat{\eta}\K}{\overline{H}+\K}.
\end{equation}

Depending on the value of $\hat{\eta}$, the film is classified as either evaporating ($\hat{\eta}>0$), when the substrate temperature exceeds the vapour saturation temperature, or condensing ($\hat{\eta}<0$), when the substrate temperature is below the vapour saturation temperature.

In evaporating conditions, the film thins uniformly and disappears at the dry out $\hat{t}_d$ given by:
\begin{equation}
\label{eq:dry_out_time}
  \hat{t}_d = \frac{1 + 2\,\K}{2\,\E\,\hat{\eta}}.
\end{equation}

For $\hat{\eta}=1$, the solution recovers the result of \citet{mohamed2020linear} for an evaporating film on an inclined substrate. Furthermore, assuming a horizontal plate ($\Ct = 0$), it reduces to the classical solution for an evaporating film originally derived by \citet{burelbach1988nonlinear}.

When $\E = 0$ (the no‐evaporation or quasi‐equilibrium limit), the solution \eqref{eqs:basic_state_equations}  reduces to the classic Nusselt flat‐film solution for falling films considered in \citet[Section 2.2]{kalliadasis2011falling} and by \cite{kelly1986instability} for layers with thermocapillarity.

% ------------- Methodology ------------- 
\section{Methodology}
\label{sec:methodology}
In this section, I present the Benney-like single-equation model (derived in \cite{joo1991long}) and describe the derivation of the WIBL reduced-order model with temperature variations (Subsection~\ref{subsec:ROM_der}). I also derive the equations governing the dynamics of small-amplitude waves (Subsection~\ref{subsec:linear_dyn}) by linearising the WIBL and the full governing equations about the weakly evaporating flat-state solution.

% ------------- Derivation reduced order models ------------- 
\subsection{Derivation reduced order model}
\label{subsec:ROM_der}
Both the Benney-like and the WIBL reduced-order models rest on the assumptions of weak evaporation $\E$, strong surface tension $\Gamma$, and small variations of kinetic energy $\Pi$ at the free surface, reading:
\begin{equation}
\label{eq:ass_ROMS_nondimnum}
    \E = \varepsilon\,\overline{\E}, \qquad\qquad 
    \Gamma = \varepsilon^{-3}\,\overline{\Gamma}, \qquad\qquad 
    \Pi = \varepsilon^2\,\overline{\Pi}.
\end{equation}
where $\overline{\bullet}$ denotes $O(1)$ quantities and $\varepsilon\ll 1$ is the \textit{liquid film parameter}, representing the ratio of the characteristic film thickness to the film length. This quantity also accounts for the slow variation of the film in both space and time, as expressed in \eqref{eq:ROM_stretching}.

The starting point for the derivation of the simplified models is the set of second-order boundary-layer equations consistent up to $O(\varepsilon^2)$ (reported in  Appendix~\ref{appx:boundary_layer_equations}). These equations are obtained by rescaling the nondimensional governing equations and boundary conditions from Subsection~\ref{subsec:gov_eqs_adim} onto the slow temporal and spatial scales defined in~\eqref{eq:ROM_stretching}, applying the assumptions in~\eqref{eq:ass_ROMS_nondimnum}, and retaining terms up to $O(\varepsilon^{2})$.

Based on this set of equations, the integral continuity equation is obtained by integrating the continuity equation \eqref{eq:cont_adim} across the film thickness and making use of the kinematic condition at the free surface \eqref{eq:kin_cond_adim}, which yields:
\begin{equation}
\label{eq:cont_eq_ROMS}
    \partial_T \hat{h} + \partial_X \hat{q}_x + \partial_Z \hat{q}_z = -\overline{\E}\,\hat{J}.
\end{equation}

The main difference between the Benney-like and WIBL approaches lies in the representation of the flow rates, $\hat{q}_x$ and $\hat{q}_z$ and the free-surface temperature $\hat{\theta}_s$. In the Benney-like model, these quantities are enslaved to the film thickness, $\hat{h}$, through an explicit analytical relation, whereas in the WIBL framework, $\hat{q}_x$, $\hat{q}_z$ and $\hat{\theta}_s$ are free to evolve independent from $\hat{h}$ via dedicated integral equations.

% -------- Benney like equation
\subsubsection{Single film-thickness evolution equation}
\label{subsec:sigle_equation_model}
The Benney-like model, derived in 2D by \citet{joo1991long} and then extended in 3D by \citet{joo1993two}, neglects terms of order $O(\varepsilon)$ or higher in the boundary layer equation. The starting point of its derivation consists in expanding the dependent variables in $\varepsilon$, reading: 
\begin{equation}
\label{eq:exp_Benney}
    \hat{u}\approx\hat{u}_0 + \varepsilon\,\hat{u}_1,\quad\, \hat{V}\approx\hat{V}_0 + \varepsilon\,\hat{V}_1,\quad\,  \hat{w}\approx\hat{w}_0 + \varepsilon\,\hat{w}_1,\quad\, \hat{p}\approx\hat{p}_0 + \varepsilon\,\hat{p}_1,\quad\, \hat{\theta}\approx\hat{\theta}_0.
\end{equation}

Injecting \eqref{eq:exp_Benney} in the boundary layer equations, solving the equations at successive orders in $\varepsilon$, computing the flow rate using expressions \eqref{eq:def_flow_rates}, and then substituting the result along with the approximated mass flux $\hat{J}$ into \eqref{eq:cont_eq_ROMS}, one obtains the single evolution equation for $\hat{h}$, which in 3D reads:
\begin{align}
\label{eq:Benney_compact}
\hat{h}_T 
&+ \frac{\bar{\E}}{\hat{h}+\K}
+ \R\,\hat{h}^2\hat{h}_X
+ \frac{2\R^2}{15}\partial_X(\hat{h}^6 \hat{h}_X)
+ \frac{5\bar{\E}\R}{24}\,\partial_X\!\left(\frac{\hat{h}^4}{\hat{h}+\K}\right) \nonumber\\
&+ \nabla\!\cdot\!\Bigg\{\Bigg[\left(\frac{\hat{h}}{\hat{h}+\K}\right)^2
\left[\frac{\K\M}{\Pr} 
+ \Vr\left(\frac{\hat{h}}{\hat{h}+\K}\right) \right]
- \frac{\Ct}{3}\hat{h}^3
\Bigg]\nabla\hat{h} \\
&+ \bar{\Gamma}\,\hat{h}^3 \nabla(\nabla^2\hat{h})\Bigg\}
+ \bar{\E}\Pr \left(\frac{\hat{h}}{\hat{h}+\K}\right)^3
\left[
\frac{\bar{\E}}{3(\hat{h}+\K)}
+ \frac{\R}{120}(7\hat{h}-15\K)\hat{h}\,\hat{h}_X
\right]
= 0. \nonumber
\end{align}
Neglecting terms associated with liquid-film phase change ($\overline{\E} = \Vr = \overline{\Pi}=0$) and adopting Nusselt scaling, the above equation reduces to the standard equation for a thermocapillary liquid film \citep{scheid2005validity}.

% ------- Derivation WIBL model
\subsubsection{Second order simplified Weighted Integral Boundary Layer (WIBL) equations}
\label{subsec:WIBL_2nd_order}
To derive the WIBL model, I assume that the film’s dynamics remain close to the quasi-steady solution introduced in Subsection~\ref{sec:flat_film}, with only small corrections of order $O(\varepsilon)$ for the inertial term and $O(\varepsilon^2)$ for the other terms. This assumption implies that the flow preserves coherent wall-normal structure, determined by the viscous–gravity balance in the momentum equations and by the diffusion-dominated behaviour of the energy equation. The assumptions and parameter regimes under which the present WIBL model is valid are summarised in Table~\ref{tab:validity}.
\begin{table}
\centering
\renewcommand{\arraystretch}{1.25}
\setlength{\tabcolsep}{7pt}

\begin{tabular}{p{4cm} p{9cm}}
\textbf{Assumption / regime} & \textbf{Validity range and justification} \\

Long-wave approximation 
& Film aspect ratio $\varepsilon \ll 1$. Breaks down when short-wave modes dominate or strong capillary ridges form. \\

Weak-to-moderate inertia 
& $\R \lesssim 20$-$30$. Accuracy decreases for larger $\R$ where $O(\varepsilon^2)$ inertial corrections and short-wave instabilities become significant. \\

Strong surface tension 
& $\Gamma = O(\varepsilon^{-3})$, surface tension terms appearing at $O(1)$. Alternatively, as I consider $O(\varepsilon^2)$ terms, this assumption can be relaxed to $\Gamma=O(\varepsilon^{-2}$).\\

Linearised phase-change mass flux 
& Valid for small interfacial temperature jumps  
$\Delta Z\ll 1$ (\eqref{eq:driving_force}). Corresponds to linearised kinetic boundary conditions and diffusion-limited vapour transport.\\

Moderate evaporation/condensation intensity 
& Evaporation number $\E = O(\varepsilon)$ or smaller.  
For large $\E$, interfacial temperature variations induce short-scale dynamics not captured by long-wave theory. \\

Variable density and viscosity 
& Viscosity and density are considered constant. This assumption is valid as long as the temperature variation across the film is not significant and the non-equilibrium number is small $\K\ll 1$ \\

Moderate substrate temperature variation 
& Variations on scales $O(1/\varepsilon)$ or slower. Sharp heating fronts generate small-scale structures that the model does not capture. \\

No vapour dynamics
& Vapour is assumed to be mechanically and thermally passive, maintained at its saturation pressure and temperature. This assumption is valid for sufficiently rapid vapour equilibration and a large vapour thermal capacity relative to that of the liquid film.
\end{tabular}
\caption{Modelling assumptions and range of validity of the present WIBL-$\theta$ model for evaporating and condensing films.}
\label{tab:validity}
\end{table}

Within this framework, the WIBL and $\theta$ models are obtained by computing the $O(\varepsilon)$ corrections analytically and subsequently applying gauge conditions to derive the evolution equations for $\hat{q}_x$, $\hat{q}_z$, and $\hat{\theta}_s$.

The velocity $\hat{u}$, $\hat{V}$ and $\hat{w}$ and the temperature $\hat{\theta}$ are decomposed into the sum of its weakly evaporating steady-state profile $\overline{U},\overline{\Theta}$, plus terms representing the deviation of the flow rate and temperature distributions from the base state $\hat u^{(0)}$, $\hat w^{(0)}$ and $\hat{\theta}^{(0)}$ at $O(1)$, and small corrections $\hat u^{(1)}$, $\hat w^{(1)}$ and $\hat \theta^{(1)}$ at $O(\varepsilon)$. The resulting approximate expressions for the field quantities are given by:
\begin{equation}
\label{eq:vel_decomp}
\hat{u} \approx \overline{U} + \hat{u}^{(0)} + \hat u^{(1)}, \qquad\qquad
\hat w \approx \overline U + \hat w^{(0)} + \hat w^{(1)}, \qquad\qquad
\hat \theta \approx \overline{\Theta} + \hat \theta^{(0)} + \hat \theta^{(1)},
\end{equation}
Although the steady-state solution has only a streamwise velocity component, the spanwise velocity $\hat{w}$ is assumed to represent a perturbation of the streamwise base profile, following standard practice in liquid film reduced-order modelling.

The wall normal velocity $\hat{V}$ is calculated by integrating the continuity equation \eqref{eq:bl_cont} with only $O(1)$ terms and considering the non-penetrability boundary condition at the boundary \eqref{eq:bc_bl_equations_wall}, which gives:
\begin{equation}
\label{eq:V_bl}
    \hat{V} = -\int\partial_X(\overline{U} + \hat{u}^{(0)}) + \partial_Z(\overline{U} + \hat{w}^{(0)})\;d\hat{y}.
\end{equation}

The quantities $\hat{u}^{(0)}$, $\hat{w}^{(0)}$, and $\hat{\theta}^{(0)}$ are written as the product of a function of $\bar{y}$, which characterises the wall-normal structure of the flow, and amplitude functions $a_u^{(0)}(\hat{X},\hat{Z},\hat{T})$, $a_w^{(0)}(\hat{X},\hat{Z},\hat{T})$, and $a_\theta^{(0)}(\hat{X},\hat{Z},\hat{T})$, corresponding respectively to the streamwise velocity, the spanwise velocity, and the free-surface temperature. 
The components $\hat{u}^{(0)}$, $\hat{w}^{(0)}$, and $\hat{\theta}^{(0)}$ therefore read:
\begin{equation}
\label{eq:decom_0}
\hat u^{(0)} = a_u^{(0)}\!\left(\bar{y} - \frac{\bar{y}^2}{2}\right), 
\qquad\qquad
\hat w^{(0)} = a_w^{(0)}\!\left(\bar{y} - \frac{\bar{y}^2}{2}\right),
\qquad\qquad
\hat \theta^{(0)} = a_\theta^{(0)}\,v_1(\bar{y}),
\end{equation}
where the wall-normal velocity profile structure is equal to the one in the weakly evaporating film solution in \eqref{eqs:basic_state_equations} and $v_1(\bar{y})$ represents a linear combination of the first two relaxation thermal eigenmodes with $v_1(\bar{y}=1)=1$. The derivation of this function will be detailed later.

For convenience in the physical interpretation of the results, the values of the amplitude functions are expressed in terms of $\hat{h}$,  $\hat{q}_x$, $\hat{q}_z$ and $\hat{\theta}_s$. Replacing the decompositions \eqref{eq:vel_decomp} up to $O(1)$ with \eqref{eq:decom_0} in the definitions  \eqref{eq:def_flow_rates} gives:
\begin{equation}
\label{eq:def_flow_rates_model}
\int_0^{\hat h} \bigl(\overline U + \hat u^{(0)}\bigr)\,d\hat y \;=\; \hat q_x, 
\qquad\qquad
\int_0^{\hat h} \bigl(\overline U + \hat w^{(0)}\bigr)\,d\hat y \;=\; \hat q_z,
\qquad\qquad
\overline\Theta\bigl(\hat h\bigr) + \hat\theta\bigl(\hat h\bigr)^{(0)} = \hat\theta_s.
\end{equation}

Solving \eqref{eq:def_flow_rates_model} gives the values of the amplitude functions, reading:
\begin{equation}
\label{eq:amp_decom}
a_u^{(0)} = \frac{3(\hat q_x - \overline Q_x)}{\hat h}, \qquad\qquad
a_w^{(0)} = \frac{3(\hat q_z - \overline Q_z)}{\hat h}, \qquad\qquad
a_\theta^{(0)} = \hat\theta_s - \overline\Theta\bigl(\hat h\bigr).
\end{equation}

The wall-normal temperature structure $v_1$ is obtained as a linear combination of the two thermal relaxation eigenmodes associated with the eigenvalues with the largest real parts. These modes are determined by solving the linearised temperature equation without convective effects, then approximating their shapes using polynomial functions.

The linearised temperature equation is obtained assuming that the liquid film is in equilibrium under weak evaporation, with a flat-film thickness $\overline{H}$, and considering a small perturbation of the temperature profile, $\tilde{\theta} \ll \overline{\Theta}$. The solution of the perturbed energy equation~\eqref{eq:energy_gov_eq} without convective effects is then sought in the form of normal modes in the long-wave limit (i.e.\ $\partial_x \rightarrow 0$ and $\partial_{xx} \rightarrow 0$), which yields:
\begin{equation}
\label{eq:ansatz_thermal_modes}
    \tilde{\theta} = \hat{\tau}(\bar{y})\,exp(\omega \hat{t}),
\end{equation}
where $\hat{\tau}(\bar{y})$ is the complex amplitude and $\omega$ is the complex pulsation.

Injecting \eqref{eq:ansatz_thermal_modes} into the temperature equation for the perturbation temperature and considering the boundary conditions \eqref{eq:kin_cond_adim} and \eqref{eq:bc_thermal}, gives the following eigenvalue problem with eigenfunction $\hat{\tau}(\bar{y})$ and eigenvalue $\omega$, which reads:
\begin{equation}
\label{eq:eigen_prob_thermal_modes}
    \Pr\overline{H}^2\lambda\tilde{\tau} = \partial_{yy}\tilde{\tau},\qquad\qquad \hat{\tau}|_0 = 0, \qquad\qquad (\partial_{y}\hat{\tau})|_{\bar{y}=1} + \tilde{J}\;\overline{H} = 0,
\end{equation}
where the wall-normal derivatives are expressed in terms of $\bar{y}$ and $\tilde{J}$ is assumed proportional to the variation of temperature at the free surface $\tilde{J} = \tilde{\theta}|_{\overline{y}=1}\K^{-1}$.

The solution of the eigenvalue problem \eqref{eq:eigen_prob_thermal_modes} reads:
\begin{equation}
    \hat{\tau} = \sin(l\,\bar{y}),\qquad\qquad\qquad\qquad \omega=-\frac{l^2}{\Pr\overline{H}^2},
\end{equation}
where $l$ is a solution of the following equation:
\begin{equation}
\label{eq:l_thermal_eigen}
    \cot{(l)}\,l + \overline{H}/K = 0.
\end{equation}

All the eigenvalues $\omega$ are negative real numbers corresponding to the relaxation modes from the self-similar temperature profile. Considering asymptotic condition $\hat{h}/\K\rightarrow 0$ and $\hat{h}/\K\rightarrow \infty$ the solution of \eqref{eq:l_thermal_eigen} reads:
\begin{equation}
\label{eq:thermal_asym}
l_n = n\pi \quad {\rm for} \quad \overline{H}/\K\rightarrow 0,\qquad \text{and}\qquad l_n = \frac{\pi}{2} + \pi n \quad {\rm for} \quad \overline{H}/\K\rightarrow\infty,
\end{equation}
with $n\in\mathbb{N}^+$.
\begin{figure}
    \begin{subfigure}[b]{0.33\textwidth}
        \centering
        \includegraphics[width=\textwidth]{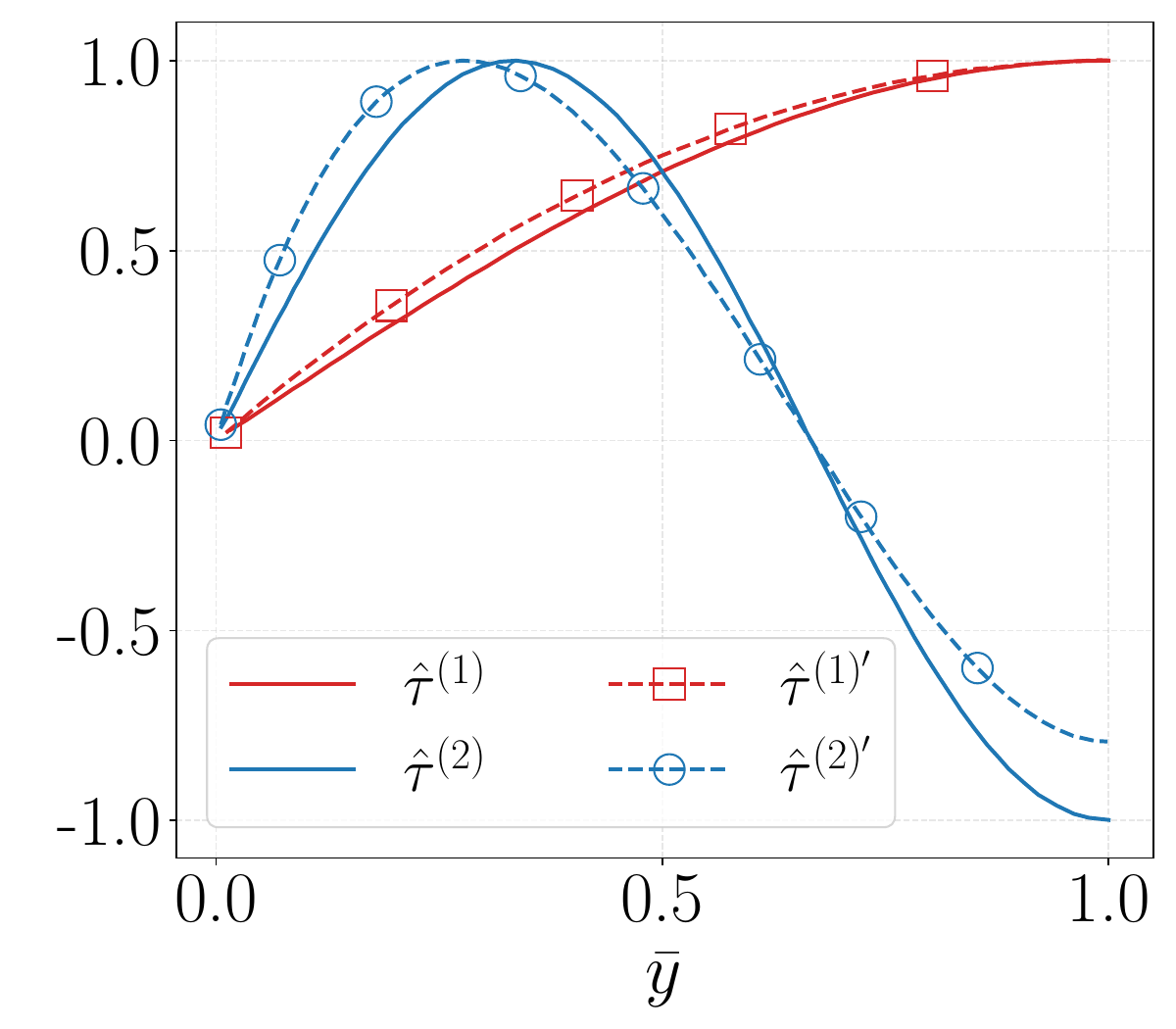}
        \caption{}
    \end{subfigure}
    \begin{subfigure}[b]{0.33\textwidth}
        \centering
        \includegraphics[width=\textwidth]{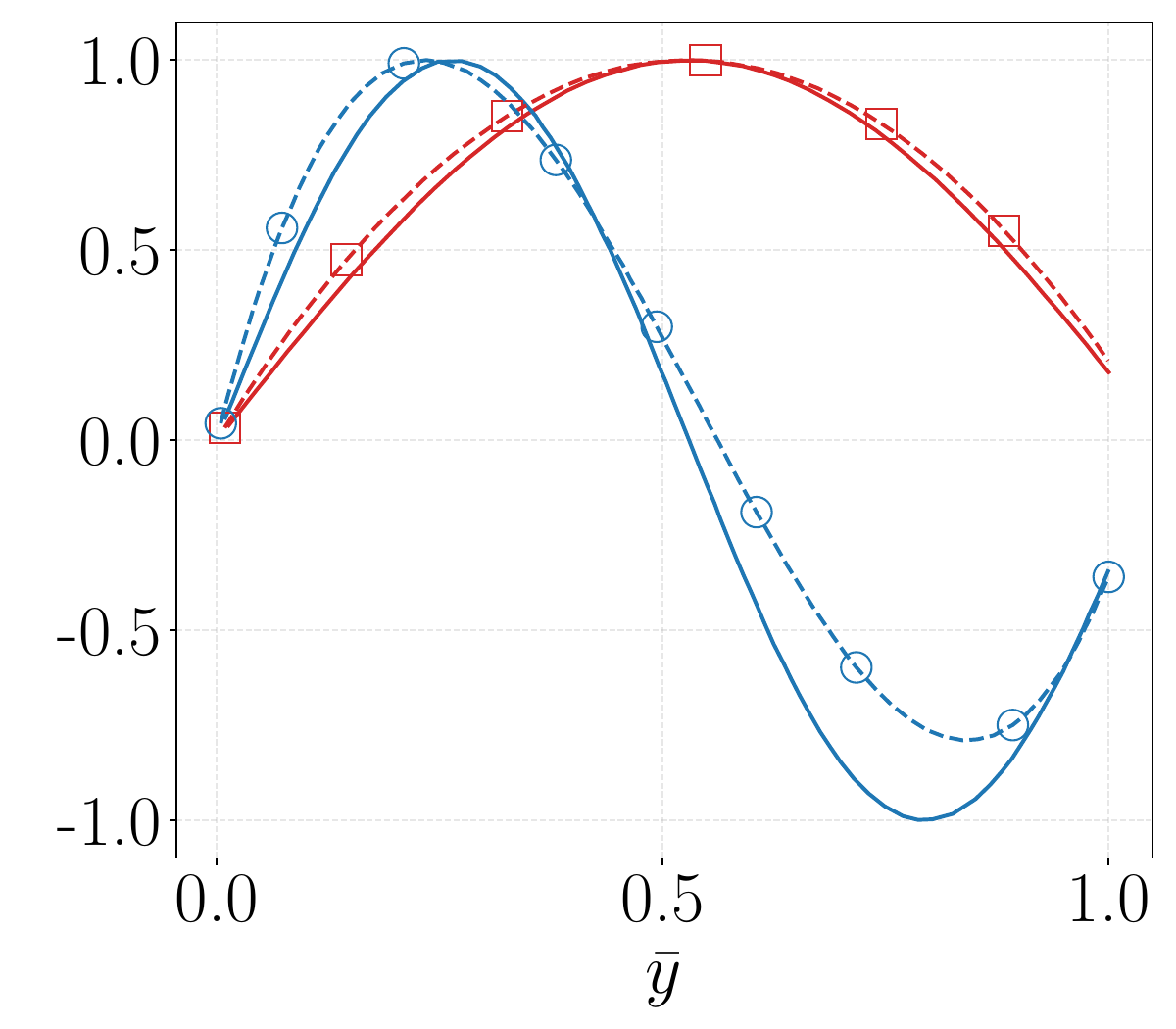}
        \caption{}
    \end{subfigure}
    \begin{subfigure}[b]{0.33\textwidth}
        \centering
        \includegraphics[width=\textwidth]{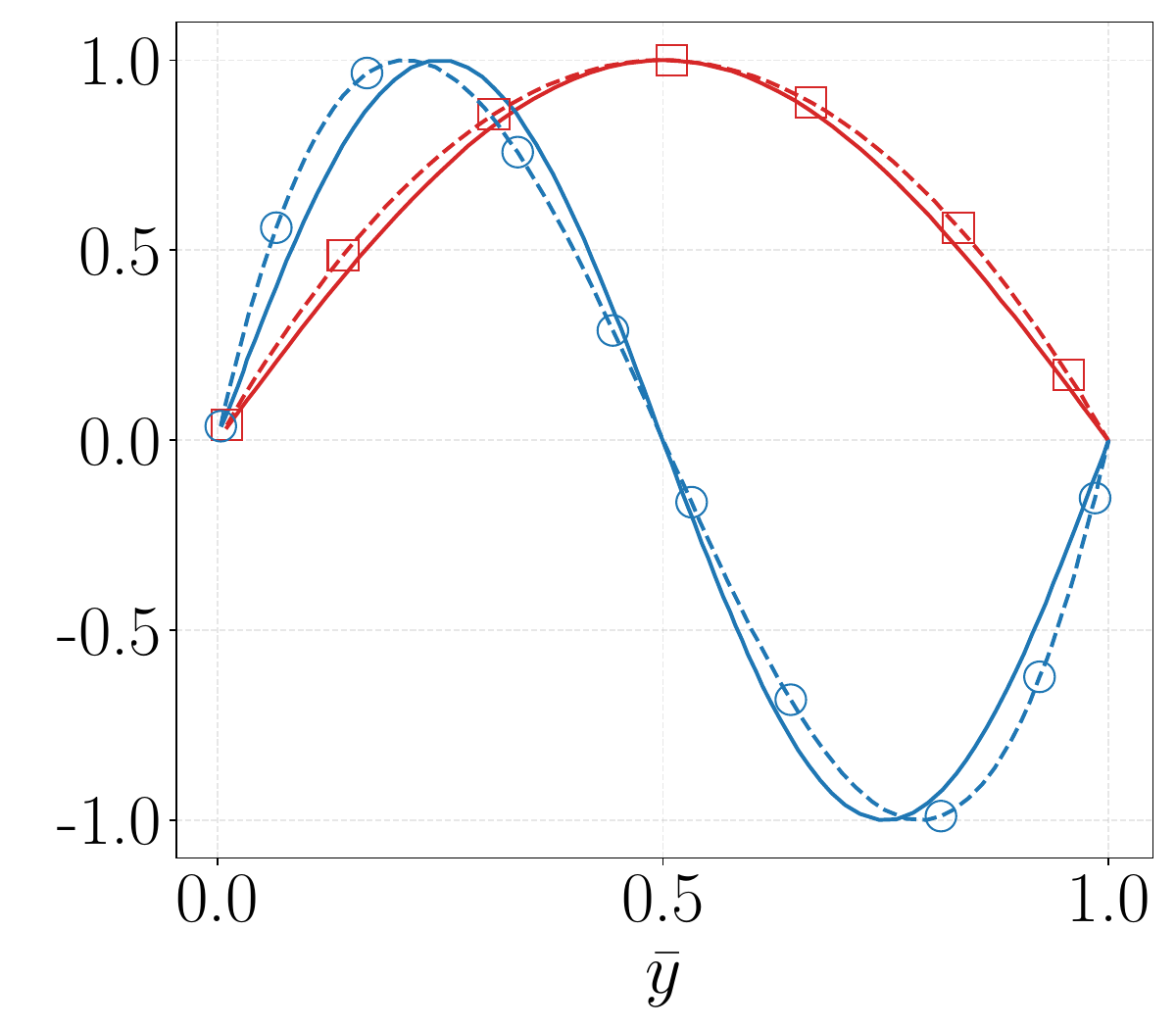}
        \caption{}
    \end{subfigure}
    \caption{Long-wave eigenfunctions $\nu_{k=0}^1$ (red contentious line) and $\nu_{k=0}^2$ (blue contentious line) and their polynomial approximations  $\hat{\nu}_1$ (red dashed line with squares) and $\hat{\nu}_2$ (blue dashed line with circles) for $\overline{H}/\K$ equal to (a) $10^{-3}$, (b) $16$ and (c) $10^{5}$.}
    \label{fig:thermal_modes_comp}
\end{figure}

At large times, only the relaxation eigenmodes with the largest real eigenvalues affect the temperature evolution. Thus only the first two eigenmodes, $\hat{\tau}^{(1)}$ and $\hat{\tau}^{(2)}$, associated with $n = 1$ and $n = 2$, are considered. For convenience in the construction of the integral model, their values are expressed through approximating functions, $\hat{\tau}^{(1)'}$ and $\hat{\tau}^{(2)'}$, respectively, which are based on low-order polynomials with rational fractions. Following the derivation in \citep{cellier2020gradient}, the approximated first two eigenfunctions read:
\begin{subequations}
\label{eq:approximated_eigenfunctions}
\begin{equation}
    \hat{\tau}^{(1)'} = \overline{y}(2-\overline{y}) + \frac{\overline{H}}{\K}\overline{y}(1-\overline{y}),\quad\quad
    \hat{\tau}^{(2)'} = -12\bar{y}\Big(\frac{2}{3}-\bar{y}\Big)\Big(\frac{5}{4}-\bar{y}\Big) + 2\frac{\overline{H}}{\K}\bar{y}(1-\bar{y})\Big(\bar{y}-\frac{1}{2}\Big).
\end{equation}
\end{subequations}

Figure~\ref{fig:thermal_modes_comp} shows a comparison between the first two long-wave eigenfunctions, $\nu_{k=0}^1$ (red continuous line) and $\nu_{k=0}^2$ (blue continuous line), computed numerically, and their corresponding polynomial approximations, $\hat{\nu}_1$ (red dashed line with squares) and $\hat{\nu}_2$ (blue dashed line with circles), for $\overline{H}/\K$ equal to (a) $10^{-3}$, (b) $16$, and (c) $10^{5}$. The eigenfunctions are scaled by their maximum absolute value. As shown, the polynomial approximations reproduce the behaviour of the numerical eigenfunctions with reasonable accuracy, particularly in the regime of large $\overline{H}/\K$.

The structure of the temperature profile in the wall-normal direction, $v_1(\bar{y})$, introduced in \eqref{eq:amp_decom}, is expressed as a linear combination of the approximated eigenfunctions \eqref{eq:approximated_eigenfunctions}. The coefficients of this linear combination are determined by imposing the following boundary conditions:
\begin{equation}
    \tilde{\nu}_1\big|_{1} = 1, 
    \qquad\qquad 
    \partial_{yy}\tilde{\nu}_1\big|_{1} = 0.
\end{equation}
The resulting expression for $v_1(\bar{y})$ reads:
\begin{equation}
    \hat{\nu}_1 = \bar{y}\Big[3 - 3\bar{y} + \bar{y}^2 
    + \frac{\overline{H}}{\K}\big(2 - 3\bar{y} + \bar{y}^2\big)\Big].
\end{equation}

Now that I have defined the wall-normal structures and the amplitude functions of the first-order terms for the velocity and temperature fields \eqref{eq:vel_decomp}, I proceed to compute the first-order corrections $\hat u^{(1)}, \hat w^{(1)}, \hat \theta^{(1)}$ and then derive the integral equations.

The values of the correction terms $\hat{u}^{(1)}$, $\hat{w}^{(1)}$, and $\hat{\theta}^{(1)}$ are obtained by solving the system \eqref{eq:first_order_system} together with the boundary conditions \eqref{eq:bc_wall_WIBL_der} and \eqref{eq:first_order_gradients}, which for brevity have been reported in Appendix \ref{appx_subsec_gov_eqs_oesp}.

The evolution equations for $\hat{q}_x$, $\hat{q}_z$ and $\hat{\theta}_s$, are obtained by imposing the following conditions:
\begin{equation}
\label{eq:gauge_cond}
\int_{0}^{\hat{h}} \hat{u}^{(1)} \, d\hat{y} = 0, 
\qquad\qquad
\int_{0}^{\hat{h}} \hat{w}^{(1)} \, d\hat{y} = 0,
\qquad\qquad
\hat{\theta}^{(1)}\big|_{\hat{y}=\hat{h}} = 0,
\end{equation}
which are consistent with \eqref{eq:def_flow_rates_model} and can be viewed as gauge conditions, defining one among an infinite number of possible velocity decompositions \citep{samanta2011falling}. By imposing conditions \eqref{eq:gauge_cond}, I obtain a set of nonlinear hyperbolic partial differential equations, which, for conciseness, are presented below in a compact form:
\begin{subequations}
\label{eqs:WIBL_equations}
\begin{align}
    \varepsilon\,\partial_T \hat{q}_x\,=&\, F_{x,el}^{(0)} +
        F_{x,p}^{(0)}  + F_{x,g}^{(0)} + \varepsilon\,\Big(F_{x,in}^{(1)} + F_{x,shfs}^{(1)} + F_{x,p}^{(1)}\Big) + \label{eq:int_qx}\\ &\quad+ \varepsilon^2\,\Big(F_{x,shb}^{(2)} + F_{x,el}^{(2)} + F_{x,shfs}^{(2)} + F_{x,p}^{(2)}\Big),\nonumber\\
    \varepsilon\,\partial_T \hat{q}_z\,=&\, F_{z,el}^{(0)} +
        F_{z,p}^{(0)} + \varepsilon\,\Big(F_{z,in}^{(1)} + F_{z,shfs}^{(1)} + F_{z,p}^{(1)}\Big) +\label{eq:int_qz}\\&\quad +\varepsilon^2\,\Big(F_{z,shb}^{(2)} + F_{z,el}^{(2)} + F_{z,shfs}^{(2)} + F_{z,p}^{(2)}\Big),\nonumber\\
    \varepsilon\,\partial_T \hat{\theta}_s\,=&\, 
        F_{\theta,diff}^{(0)}  + \varepsilon\,F_{\theta,conv}^{(1)} + \varepsilon^2\,\Big(F_{\theta,diff}^{(2)} + F_{\theta,fs}^{(2)}\Big),\label{eq:int_temp}
\end{align}
\end{subequations}
where $\hat{J}$ is linked to $\hat{\theta}_s$ by the constitutive relation \eqref{eq:J_cond_FS_adim} and the expressions of the individual terms, along with their physical interpretations, 
are provided in Appendix \ref{sec:des_terms_WIBL}.

The final set of equations for the WIBL model, valid up to $O(\varepsilon)$ in the inertial terms and up to $O(\varepsilon^2)$ in the remaining terms with substrate temperature variations in space and time $\hat{\eta}(\hat{X},\hat{T})$, consists of the continuity equation \eqref{eq:cont_eq_ROMS}, together with the evolution equations \eqref{eqs:WIBL_equations}, complemented by the constitutive relation \eqref{eq:J_cond_FS_adim}. 

The non-volatile limit corresponds to local thermodynamic equilibrium at the liquid-vapour interface, for which the vapour pressure equals the saturation pressure at the free surface ($p_v=p_{s}(T_{fs})$) and no net mass flux occurs across the interface ($\hat{J}=0$). Nevertheless, small temperature variations may still exist along the free surface without substantially altering the local equilibrium conditions. These temperature gradients induce variations in surface tension, thereby generating thermocapillary stresses. The more restrictive isothermal limit corresponds to a uniform temperature field in both the liquid and vapour phases, including at the free surface. As a result, neither phase-change effects ($\overline{\E}=\Vr=\overline{\Pi}=0$) nor thermocapillary stresses $(\M=0)$ can arise in this case.

For the non-volatile case and under Nusselt (isothermal) scaling, the continuity and flow rate equations reduces to the thermocapillary falling-film formulation of \citet[Chapter~8]{kalliadasis2011falling}, while, assuming $\hat{\eta}$ constant, the thermal equation recovers the two-dimensional $\theta$ model derived by \citet{cellier2020new}, with $\K^{-1}=\Bi$.

The derived WIBL model is consist for $\varepsilon\rightarrow 0$, as expanding the variables $\hat{q}_x$, $\hat{q}_z$ and $\hat{\theta}_s$ in $\varepsilon$ and solving at successive orders yields the relations used in the construction of the single-equation models in Subsection \ref{subsec:sigle_equation_model}.

% ------------- Methodology ------------- 
\subsection{Linear Dynamics}
\label{subsec:linear_stab_analysis}
In this subsection, I present the linearised governing equations used to investigate linear stability characteristics of the base state solution introduced in Subsection~\ref{sec:flat_film}.

To derive these equations, the dependent variable of the Navier-Stokes equations and of the WIBL model are decomposed into their base-state components \eqref{eqs:basic_state_equations} and \eqref{eq:ss_rel_q_theta}, and a small perturbation, reading:
\begin{equation}
\label{eq:linear_decomposition_0}
\begin{gathered}
    \hat{u} = \overline{U}(\hat{y}) + \iota\Tilde{u}(x,y,t),\qquad \hat{v} = \overline{V}(\hat{y}) + \iota\Tilde{v}(x,y,t),\qquad \hat{w} = \overline{W}(\hat{y}) +\iota\Tilde{w}(x,y,t),\\
    \hat{\theta} = \overline{\Theta}(\hat{y}) + \iota\Tilde{\theta}(x,y,t), \,\quad \hat{p} = \overline{P}(\hat{y}) + \iota\Tilde{p}(x,y,t),\,\quad
    \hat{J} = \overline{J} + \iota\Tilde{J}(x,t), \,\quad
    \hat{h} = \overline{H} + \iota\Tilde{h}(x,t),\\
    \hat{\theta}_s = \overline{\Theta}_s + \iota\Tilde{\theta}_s,\qquad \hat{q}_x = \overline{Q}_x + \iota\Tilde{q}_x,\qquad \hat{q}_z = \overline{Q}_z + \iota\Tilde{q}_z
\end{gathered}
\end{equation}
where $\iota\ll 1$ is a small number and $\tilde{\bullet}$ represent the perturbation quantities at $O(1)$.

Substituting the decomposition \eqref{eq:linear_decomposition_0} into the full-model equations  
(presented in Subsection~\ref{subsec:gov_eqs_adim}) and into the WIBL equations  
\eqref{eq:cont_eq_ROMS}, \eqref{eqs:WIBL_equations}, and \eqref{eq:J_cond_FS_adim},  
and retaining terms up to $O(\iota)$ gives the linearised system governing the evolution of perturbation quantities. Following \citet{joo1991long} and \citet{mohamed2020linear}, the term proportional to $\E$ in the free-surface kinematic condition and in the linearised integral continuity equation is omitted, as it represents the ratio between the initial perturbation amplitude and the evolving base-state film thickness, and therefore does not affect the temporal growth or decay of the perturbations.

The linearised full-model equations are further simplified by eliminating the streamwise and spanwise velocity components.  
For clarity, the complete derivation of these equations is provided in the supplementary material. 

The solution of the linearised equations is sought in the form of normal modes, reading:
\begin{equation}
\label{eq:normal_mode_sol}
\begin{pmatrix}\tilde{v}\\\hat{\theta}\\\hat{h}\\\hat{J}\end{pmatrix} = \Re\begin{pmatrix}\phi(\hat{y})\\\hat{\tau}(\hat{y})\\\xi \\\mathcal{J}\end{pmatrix}\rm{exp}(i(\bm{k}\cdot (\hat{x},\hat{z})^T-\omega\,\hat{t})),\;\begin{pmatrix}\tilde{q}_x\\\tilde{q}_z\\\tilde{\theta}_s\end{pmatrix} = \Re\begin{pmatrix}q_x^{\prime}\\q_z^{\prime}\\\theta_s^{\prime}\end{pmatrix}\rm{exp}(i(\bm{k}\cdot (\hat{x},\hat{z})^T-\omega\,\hat{t}))
\end{equation}
where $\phi(\hat{y})$, $\tau(\hat{y})$, $\xi$, $\mathcal{J}$, $q_x^{\prime}$, $q_z^{\prime}$ and $\theta_s^{\prime}$ are the complex amplitudes, $\bm{k} = (k_x,k_z)^T$ is the wavenumber vector, with $k_x$ and $k_z$ the real streamwise and spanwise components, $\omega=\omega_r + i \omega_i$ is the complex pulsation with $\omega_i$ representing the perturbation growth rate. The disturbance propagate in the direction defined by the angle $\varphi = \tan^{-1}(k_z/k_x)$ with respect to the $\hat{x}-$axis with phase velocity $c=\omega_r/\lVert \bm{k} \rVert$. For convenience in the long-wave solution, the cosine of the angle $\varphi$ is expressed by the variable $\hat{k}_{x1}$, which is defined as:
\begin{equation}
    \hat{k}_{x1} = \cos{(\varphi)} = k_x/\lVert \bm{k} \rVert.
\end{equation}

Substituting \eqref{eq:normal_mode_sol} into the linearised equations yields the \textit{dispersion relation} linking $\bm{k}$ to $\omega$. 

For the reduced-order models, the dispersion relation is given by a nonlinear algebraic equation, which is solved numerically using the \textit{Python} routine \textit{scipy.optimize.root}. For the linearised full model equations, the dispersion relation is provided by the solution of a generalised eigenvalue problem, where the amplitude functions are the eigenfunctions and $\omega$ is the eigenvalue. This eigenvalue problem, known as \textit{Orr-Sommerfeld} eigenvalue problem, reads:
\begin{subequations}
\label{eq:OS_main_eqs}
    \begin{equation}
    \label{eq:OS_main_eqs_1}
        (D^2-\lVert \bm{k} \rVert^2)^2\phi + i[(\omega - k_x\overline{U})(D^2-\lVert \bm{k} \rVert^2)+D^2\overline{U}k_x] \phi = 0,
      \end{equation}
    \begin{equation}
    \label{eq:OS_main_eqs_2}
        (D^2 - \lVert \bm{k} \rVert^2)\tau - \Pr[\phi D\overline{\Theta} -i (\omega -k_x\overline{U})\tau]= 0,
    \end{equation}
\end{subequations}

and with boundary conditions at $\hat{y}=0$:
\begin{equation}
\label{eq:OS_main_bc0}
    \phi = 0, \qquad\qquad\qquad \tau= 0
\end{equation}
with boundary conditions at the free-surface ($\hat{y}=\hat{h}$), reading    :
\begin{subequations}
\label{eq:OS_main_bc1}
    \begin{equation}
    \label{eq:OS_main_bc1_1}
        \xi = \frac{\phi|_{\hat{h}}}{i(k_x\overline{U}|_{\overline{H}}-\omega)}
    \end{equation}
    \begin{equation}
    \label{eq:OS_main_bc1_2}
    \begin{gathered}
        \Big[(D^2-3 \lVert \bm{k} \rVert^2) + i(\omega -k_x\overline{U}|_{\overline{H}})\Big]D\phi +\\+ \lVert \bm{k} \rVert^2\Big[-3\mathcal{J}\,\overline{J}\,\Vr + \xi\Big(-\Ct + \lVert \bm{k} \rVert^2\Big(2\frac{\M}{\Pr}\overline{\Theta}|_{\overline{H}}-3\Gamma\Big)\Big)\Big] = 0
    \end{gathered}
    \end{equation}
    \begin{equation}
    \label{eq:OS_main_bc1_3}
        (D^2 +\lVert \bm{k} \rVert^2)\phi + \lVert \bm{k} \rVert^2\frac{2\M}{\Pr}(\xi D\Theta + \tau) - ik_x(D^2\overline{U})|_{\overline{H}}\,\xi = 0
    \end{equation}
    \begin{equation}
    \label{eq:OS_main_bc1_4}
        (3\overline{J}^2\Pi + 1)\mathcal{J} + D\hat{\tau}|_{\overline{H}} + \xi D^2\Theta|_{\overline{H}} = 0
    \end{equation}
    \begin{equation}
    \label{eq:OS_main_bc1_5}
        \mathcal{J} = \frac{\tau|_{\overline{H}}}{\K} + \frac{D\Theta|_{\overline{H}}}{\K}\xi,
    \end{equation}
\end{subequations}
where $D(\bullet)=\partial_{y}(\bullet)$ is the wall-normal differential operator.

In case of streamwise perturbations ($k_z=0$), the above generalised eigenvalue problem reduces to the one used by \citep{mohamed2020linear}. Furthermore, neglecting the phase change terms $\mathcal{J}=0$, the problem reduces to the one for a falling liquid film with thermocapillary effects, which is reduced to the Orr-Sommerfeld problem in \citet[section 3.2]{kalliadasis2011falling}.

Under long-wave conditions, $\lVert \boldsymbol{k} \rVert \ll 1$, the OS eigenvalue problem  
\eqref{eq:OS_main_eqs}, \eqref{eq:OS_main_bc0}, and \eqref{eq:OS_main_bc1} was also solved by expanding both the eigenfunctions and the eigenvalue $\omega$ in a power series of $\lVert \boldsymbol{k} \rVert$, truncated at second order. The expansion takes the form:
\begin{equation}
\label{eq:expansion_lw_3DOS}
    \phi \approx \phi_0 \lVert \bm{k} \rVert  + i \phi_1 \lVert \bm{k} \rVert^2 , \ \quad
    \tau \approx \tau_0 + \tau_1 \lVert \bm{k} \rVert,\quad
    \xi \approx \xi_0 + i\xi_1 \lVert \bm{k} \rVert^2, \ \quad
    \omega \approx \omega_0 \lVert \bm{k} \rVert + i \omega_1 \lVert \bm{k} \rVert^2.
\end{equation}
Moreover, a strong surface tension condition is assumed, which corresponds to taking the term associated with the surface tension number $\Gamma$ in the normal stress boundary condition \eqref{eq:OS_main_bc1_2} to be of order $O(1)$. This assumption can be expressed as
\begin{equation}
    \Gamma = \lVert \boldsymbol{k} \rVert^{-2} \, \Gamma_1.
\end{equation}

The approximate values of the eigenfunctions and eigenvalue are obtained by substituting the expansions \eqref{eq:expansion_lw_3DOS} into the Orr-Sommerfeld eigenvalue problem and solving the resulting equations at successive orders in $\lVert \boldsymbol{k} \rVert$. The growth rate and the phase speed obtained with this asymptotic expansion coincide with the solution of the dispersion relation obtained by \citep{joo1991long} with the Benny-type equation valid up to $O(\varepsilon)$. The resulting solutions are presented in subsubsection~\ref{subsec:linear_dyn}.

Regarding the numerical implementation of the OS problem, the eigenvalue problem \eqref{eq:OS_main_eqs}, \eqref{eq:OS_main_bc0}, and \eqref{eq:OS_main_bc1} was solved numerically using the Chebyshev-Tau spectral method \citep[Chapter~VII]{johnson1996chebyshev,lanczos1988applied} using the \textit{Python} routine \texttt{numpy.linalg.eig}. Following \citet{mcfadden1990elimination}, the presence of spurious eigenvalues is mitigated by zeroing the last two rows of the differentiation matrices associated with the second derivatives on the right-hand side of the discretised generalised eigenvalue problem. To ensure convergence, a mesh-independence study was performed by comparing the error in the most unstable eigenvalue across different values of $N$, using $N=80$ as the reference (see supplementary material). The results indicate that $N=20$ provides sufficient accuracy for the most unstable mode while maintaining low computational costs. To further minimise the occurrence of spurious eigenvalues, the solution of the Orr-Sommerfeld problem was repeated with $N=20$ and $N=50$, and only those eigenvalues whose relative difference in magnitude, measured using the Euclidean norm $\lVert \cdot \rVert$, remained below $0.1$ were retained, following the recommendation of \citet{gardner1989modified}.

To investigate the growth rate of three-dimensional waves, I formulated an optimisation problem to identify combinations of nondimensional parameters and flow conditions that yield the largest growth rate, $\omega_i$, for a perturbation with $k_z \neq 0$. This optimisation was carried out using the nonlocal algorithm \textit{differential evolution}, as implemented in the \texttt{scipy.optimize} library. In addition, I explored the $(\R, \beta)$ parameter space while keeping the remaining nondimensional groups fixed to determine the boundary separating cases in which the most unstable mode is purely streamwise from those in which it is fully three-dimensional. This threshold was obtained by performing a line search in $\R$ for each prescribed value of $\beta$, and repeating the procedure across successive values of $\beta$.

% ----- Nonlinear simulation setup -------
\subsection{Setup nonlinear simulations}
\label{subsec:nonlin_sim}

\begin{table} 
\centering \setlength{\tabcolsep}{6pt} 
\begin{tabularx}{\textwidth}{@{} *{9}{>{\centering\arraybackslash}X} @{}} \multicolumn{9}{c}{\textbf{Physical properties}} \\ \cmidrule(lr){1-9} $\rho$ & 995.6 & kg/m$^{3}$ & $\mu$ & 0.85 & mPa$\cdot$s & $R_g$ & 8.314 & J/(mol$\cdot$K) \\ $\rho_v$ & 0.03 & kg/m$^3$ & $\gamma_0$ & 71.97 & mN/m & $M_g$ & 0.018 & kg/mol \\ $\mathrm{d}\gamma/\mathrm{d}T$ & 0.15 & mN/(m$\cdot$K) & $L_v$ & 2.256 & MJ/kg & $C_p$ & 4.18 & kJ/(kg$\cdot$K) \\ $\alpha$ & 0.10 & - & $\lambda$ & 0.60 & W/(m$\cdot$K) & $\nu$ & 0.85 & mm$^{2}$/s
\end{tabularx}\\ \vspace{0.2cm} 
\begin{tabularx}{\textwidth}{@{} *{7}{>{\centering\arraybackslash}X} @{}} \multicolumn{3}{c}{\textbf{Reference quantities}} & \multicolumn{4}{c}{\textbf{Nondimensional groups}} \\ \cmidrule(lr){1-3}\cmidrule(lr){4-7} $h_{\rm ref}$ & 162.691 & $\mu$m & $\E$ & 0.01 & Re & 15 \\ $\Delta T_{\rm ref}$ & 31.96 & K & $\Ct$ & 56 & Vr & 2.21 \\ $x_{\rm ref}$ & 2850 & $\mu$m & Ma & $7.75\cdot10^{-4}$ & K & 0.04\\ $t_{\rm ref}$ & 31 & ms & Pr & 6 & $\Gamma$ & 5378\\ 
\end{tabularx} 
\caption{Physical properties, reference quantities, and nondimensional groups for the nonlinear simulation in 2D and 3D, considering a water liquid film over a flat plate inclined at $\beta=15^{\circ}$ with respect to the horizontal at $T=300\,$K.}
\label{tab:parameters_units} 
\end{table}
\begin{table}
\centering
\begin{tabular}{p{0.5cm} p{3cm} p{3cm} p{3cm}}
\multicolumn{1}{c}{Dim.} &
& &  \\

% 2D row
\multicolumn{1}{c}{2D} &
\multicolumn{3}{c}{$\hat{\eta}= 0.3 \sin\!\left(\frac{2\pi}{L_x}\hat{X} - \frac{2\pi}{10}\hat{T}\right)\!\left(1 - e^{-\hat{T}}\right)$} \\
\hline

 &
\multicolumn{1}{c}{Isothermal} &
\multicolumn{1}{c}{Constant Evaporation} &
\multicolumn{1}{c}{Constant Condensation} \\

% 2D row & 
&
\multicolumn{1}{c}{$\hat{\eta}=0$} &
\multicolumn{1}{c}{$\hat{\eta}=1$} &
\multicolumn{1}{c}{$\hat{\eta}=-1$} \\

% Empty row for 2D spacing
\multicolumn{1}{c}{} &
\multicolumn{1}{c}{} &
\multicolumn{1}{c}{} &
\multicolumn{1}{c}{} \\

% 3D rows
\multicolumn{1}{c}{3D} &
\multicolumn{2}{p{6cm}}{\centering Oscillating Temperature \\
$\hat{\eta} = 0.3 \cos\!\left(-\frac{2\pi}{10}\hat{T}\right)
\left(1 - e^{-\hat{T}}\right)$} &
\multicolumn{1}{p{6cm}}{\centering Variable Temperature \\
$\hat{\eta} = 0.3 \cos\!\left(\frac{2\pi}{L}\hat{X}
+ \frac{2\pi}{L}\hat{Z} - \frac{2\pi}{10}\hat{T}\right)
\left(1 - e^{-\hat{T}}\right)$} \\
\end{tabular}
\caption{Summary of substrate temperature functions used for the 2D and 3D nonlinear simulations.}
\label{tab:temperature_conditions}
\end{table}

In this subsection, I outline the setup used for the two- and three-dimensional nonlinear simulations of an evaporating liquid film subject to spatially varying substrate heating. The two-dimensional evaporating film is simulated by solving the WIBL model via a Fourier pseudo-spectral method (described in Appendix \ref{appx:Fourier_Spectral_method}) and the full governing equations implemented in \textsc{COMSOL}. In contrast, the three-dimensional simulations are performed exclusively with the WIBL model, as the computational cost of solving the full model equations is prohibitive with the available resources, as highlighted by \citet{dietze2014three} for a 3D falling film.

In both the 2D and 3D cases, I consider a film flowing over a substrate inclined at $\beta = 15^{\circ}$ with water as the working fluid, in equilibrium with its vapour at the saturation temperature, $T_\mathrm{sat} = 300~\mathrm{K}$. The liquid thermophysical properties are evaluated at $300~\mathrm{K}$ and treated as constant in both space and time. Weak evaporation conditions are assumed, corresponding to an evaporation number $\E = 0.01$. The values of these properties, along with the relevant nondimensional parameters, are summarised in Table~\ref{tab:parameters_units}.

Different substrate–heating configurations were investigated. In the two-dimensional simulation, spatio-temporal temperature variations were considered. For the three–dimensional simulations, five cases were examined: isothermal conditions, constant evaporation, constant condensation, oscillatory temperature, and harmonic temperature variation in both space and time. In the space-time variable–heating case, the harmonic function follows an exponential ramp to avoid abrupt transients at startup. Table~\ref{tab:temperature_conditions} summarises the substrate–temperature conditions used in each configuration.

For the variable–heating simulations, the amplitude of the substrate–temperature fluctuations is set to 0.3 with zero mean, to achieve intense heating while ensuring the constraint on the free–surface temperature given by \eqref{eq:driving_force}. This constraint determines the validity limit of the linearised mass–flux approximation \eqref{eq:lin_approx} with \eqref{eq:lin_approx_1}. By substituting the definition of the nondimensional free–surface temperature from \eqref{eq:scaling_original_1} into \eqref{eq:driving_force}, the condition that must be satisfied by the free–surface temperature becomes:
\begin{equation}
\hat{\theta}s \ll \frac{R_g,T_s^2}{L_v,M_w,\Delta T{\mathrm{ref}}} = 0.56,
\end{equation}
which remains above the maximum temperature the substrate reaches during the simulation.

For the two-dimensional simulations, a domain of nondimensional length $L = 60$ is considered, with periodic boundary conditions. The simulation is run for a nondimensional time $T = 100$, starting from an initially flat liquid film, $\hat{h}(\hat{X},0) = 1$, and an isothermal liquid at the saturation temperature, $\hat{\theta}(\hat{X},\hat{y}) = 0$. The velocity, pressure, and flow rate fields are initialised according to the base-state relations given in \eqref{eqs:basic_state_equations} and \eqref{eq:ss_rel_q_theta}.

The solutions obtained from the WIBL model, expressed in terms of the liquid-film thickness $\hat{h}_{\mathrm{WIBL}}$ and the free-surface temperature $\hat{\theta}_{s,\mathrm{WIBL}}$, are compared against the corresponding full-model solutions, denoted by $\hat{h}_{\mathrm{full}}$ and $\hat{\theta}_{s,\mathrm{full}}$. The overall percentage errors for the liquid-film thickness, $E_{h,\mathrm{rel}}(\hat{T})$, and for the free-surface temperature, $E_{\theta_s,\mathrm{rel}}(\hat{T})$, are defined as:
\begin{equation}
\label{eq:error_WIBL_COMSOL}
E_{h,\mathrm{rel}}(\hat{T}) = \frac{
\sqrt{\int_0^L \!\big\| \hat{h}_{\mathrm{WIBL}} - \hat{h}_{\mathrm{full}} \big\|_2 \, \mathrm{d}\hat{X}}
}{\sqrt{\int_0^L \!\big\| \hat{h}_{\mathrm{full}} \big\|_2 \, \mathrm{d}\hat{X}}},\quad E_{\theta_s,\mathrm{rel}}(\hat{T}) = \frac{\sqrt{\int_0^L \!\big\| \hat{\theta}_{s,\mathrm{WIBL}} - \hat{\theta}_{s,\mathrm{full}} \big\|_2 \, \mathrm{d}\hat{X}}
}{\sqrt{\int_0^L \!\big\| \hat{\theta}_{s,\mathrm{full}} \big\|_2 \, \mathrm{d}\hat{X}}},
\end{equation}

where $\| \bullet \|_2$ denotes the $L_2$ norm of the variable evaluated at the grid points, with the COMSOL solution remapped over the grid used to solve the WIBL model.

For the 3D simulations, the domain is a square region defined as $\Omega = \{(\hat{X},\hat{Z}) \in \mathbb{R}^2 \, | \, 0 < \hat{X} < L,\, 0 < \hat{Z} < L\}$ with $L = 60$, and periodic boundary conditions are applied in both directions. The final simulation time is $\hat{T} = 100$, consistent with the 2D runs.

To trigger the development of nonlinear waves, the initial condition consists of a small localised perturbation of the film thickness defined as:
\begin{equation}
    \hat{h}(\hat{X},\hat{Z},0) = 1 + 0.1 \exp\!\left(-\frac{(\hat{X} - L/4)^2 + (\hat{Z} - L/2)^2}{10}\right),
\end{equation}
a uniform substrate temperature equal to the vapour saturation temperature ($\hat{\eta} = 0$), and a flow rate defined by the steady-state relation \eqref{eq:ss_rel_q_theta}.

The results obtained with the different substrate heating profiles are compared in terms of the time evolution of the minimum film thickness at each time step $\hat{T}_k$, $\hat{h}_{\min}$, and the spatially averaged mass flux at the free surface, which read:
\begin{equation}
\label{eq:int_variable_comp_3D}
    \hat{h}_{min}(\hat{T}_k) = min(\hat{h}(\hat{X},\hat{T}_k),\qquad\qquad\langle \hat{J} \rangle_{\Omega}(\hat{T}_k) = \frac{1}{|\Omega|} \int_{\Omega} \frac{\hat{\theta}_s(\hat{X},\hat{Z},\hat{T}_k)}{\K} \, \mathrm{d}\hat{X} \, \mathrm{d}\hat{Z}.
\end{equation}

Regarding the numerical implementation, the WIBL equations are discretised in space using a Fourier pseudo-spectral method \citep{dutykh2016brief} on an equispaced grid of 200 nodes for both the two- and three-dimensional cases (see Appendix~\ref{appx:Fourier_Spectral_method} for details). Time integration is carried out using the adaptive Runge–Kutta method of order 5(4) available in \texttt{scipy.integrate} library.

The full-model equations are solved in \textsc{COMSOL Multiphysics} using an Arbitrary Lagrangian–Eulerian (ALE) moving-mesh formulation, in which the interface displacement in the $y$-direction satisfies the kinematic boundary condition at the free surface \eqref{eq:kin_cond_adim}. A mesh convergence study (reported in the supplementary material) shows that a mesh with 16574 elements provides sufficient accuracy. To adequately resolve the thermal and momentum boundary layers that develop near both the substrate and the free surface, the mesh contains ten layers of quadrilateral elements with a stretching factor of 1.5. Time integration is performed using the implicit Backwards Differentiation Formula (BDF) solver with automatic step-size control and a relative tolerance of $10^{-4}$ for residual error control.

% ------------------------------------
%               Results               
% ------------------------------------
\section{Results}
In this section, I present the results for the linear (Subsection~\ref{subsec:linear_dyn}) and nonlinear (Subsection~\ref{subsec:res_nonlinear_sims}) dynamics of the 3D evaporating and condensing liquid film, with particular emphasis on comparing the WIBL prediction with that of the full model equations. 

% ---- Linear dynamics ---- 
\subsection{Linear Stability Analysis}
\label{subsec:linear_dyn}
This subsection presents a three-dimensional linear stability analysis of the weakly evaporating/condensing flat-film solution introduced in Subsection~\ref{sec:flat_film}. The substrate temperature $\hat{\eta}$ is assumed uniform and treated as a control parameter, with $\hat{\eta}=1$ corresponding to evaporation and $\hat{\eta}=-1$ to condensation. Neutral stability curves, growth rates, and phase speeds are analysed as functions of the wavenumber and key nondimensional parameters. Predictions from the WIBL model are compared with numerical solutions of the Orr-Sommerfeld eigenvalue problem and with long-wave asymptotic results, which provide analytical estimates near the instability threshold and explicit expressions for the critical parameters. The long-wave solution also serves as a benchmark for assessing the accuracy of the WIBL model, as its predicted growth rates and phase speeds are consistent with those of the two-dimensional Benney equation \citep{joo1991long}.

For the comparison between the WIBL, Orr-Sommerfeld (OS), and long-wave (LW) predictions, the following parameters are fixed: $\E=0.2$, $\Pi=0$, $\Gamma=1000$, $\Pr=7$, and $\K=0.01$. The inclination angle $\beta$, Reynolds number $\R$, vapour-recoil number $\Vr$, Marangoni number $\M$, mean film thickness $\overline{H}$, and nondimensional time $\hat{t}$ are specified on a case-by-case basis. Unless otherwise stated, all results are reported using the scaling \eqref{eq:scaling_original}. WIBL predictions are shown by red dashed lines with squares, long-wave asymptotic results by blue dash-dotted lines with triangles, and Orr-Sommerfeld solutions by solid black lines. For reference, the hydrodynamic mode alone is indicated by dashed grey lines.

In what follows, I first analyse the long-wave asymptotic solution, then compare WIBL and Orr–Sommerfeld predictions for streamwise and spanwise perturbations, and finally identify the transition between streamwise- and spanwise-dominated instabilities in the $\R$–$\beta$ parameter space.

In the long-wave limit $\lVert \boldsymbol{k} \rVert \ll 1$, the Orr–Sommerfeld eigenvalue problem \eqref{eq:OS_main_eqs}–\eqref{eq:OS_main_bc1} is solved via an asymptotic expansion in $\lVert \boldsymbol{k} \rVert$ \eqref{eq:expansion_lw_3DOS}. Substituting this expansion into the governing equations and solving at leading order yields:
\begin{equation}
    \phi_0(\hat{y}) = \hat{y}^2, \qquad\qquad \tau_0(\hat{y}) = \frac{2i\,\hat{\eta}\,\hat{y}}{\hat{k}_{x1} \R\,(\overline{H}+\K)^2}, \qquad\qquad \omega_0 = \R\,\overline{H}^2\,\hat{k}_{x1},
\end{equation}

and at solving at first order yields:
\begin{subequations}
\begin{equation}
\begin{aligned}
    \phi_1(\hat{y}) =& \frac{\hat{y}^3\left((h+\K)^3 \left(60\,\Gamma_1+20\, \Ct+\overline{H}\,\hat{k}_{x1}^2\,\R^2\hat{y}\,(\hat{y}-5 \overline{H})\right)-60\,\hat{\eta}^2 \Vr\right)}{60\,\hat{k}_{x1}\,\R\,(\overline{H}+\K)^3}, 
\end{aligned}
\end{equation}
\begin{equation}
\begin{aligned}
\tau_1(\hat{y}) =& \frac{\hat{y}}{60\,\hat{k}_{x1}^2\,\Pr\,\R^2(\overline{H}+\K)^5} \Big[ 
-5\overline{H}^4\,\Pr\,\big(72\,\Gamma_1 + 24\,\Ct - \hat{k}_{x1}^2\,\R^2\,(\K^3\,(6\Pr+8)\\& + 12\,\K\,\Pr\,\hat{y}^2 - 3\,\Pr\,\hat{y}^3)\big) + \overline{H}^3\,\Pr\,\big(-360\,\Ct\,\K + \hat{k}_{x1}^2\,\Pr\,\R^2\,(20\,\K^4 + 60\,\K^2\,\hat{y}^2\\& - 50\,\K\,\hat{y}^3 + 3\,\hat{y}^4) -1080\,\Gamma_1\,\K\big) + \overline{H}^2\,\K\,\Pr\,\big(-360\,\Ct\,\K + \hat{k}_{x1}^2\,\Pr\,\R^2\hat{y}^2\,(20\,\K^2\\& - 60\,\K\,\hat{y} + 9\,\hat{y}^2) - 1080\,\Gamma_1\,\K\big) + 3\,\overline{H}\,\big(-10\,\K^3\,\Pr\,(12\,\Gamma_1 + 4\,\Ct + \hat{k}_{x1}^2\,\Pr\,\R^2\,\hat{y}^3)\\& + 3\,\K^2\,\hat{k}_{x1}^2\,\Pr^2\,\R^2\,\hat{y}^4 + 80\,\K\,\M + 120\,\Pr\,\Vr\big) - 8\,\overline{H}^7\,\hat{k}_{x1}^2\,(\Pr -5)\,\Pr\,\R^2\\& + 2\,\overline{H}^6\,\K\,\hat{k}_{x1}^2 \,\Pr(60-13 \Pr)\,\R^2 + 4\,\overline{H}^5\,\hat{k}_{x1}^2\,\Pr\,\R^2\,(5\,\Pr\,\bar{y}^2 - 2\,\K^2\,(\Pr-15))\\& + \K^2\,\big(\K\,\hat{k}_{x1}^2\,\Pr^2\,\R^2\,\hat{y}^3\,(3\hat{y} - 5\K) + 240\,\M\big)
\Big],
\end{aligned}
\end{equation}
\begin{equation}
\label{eq:growth_rate_o1_lw_OS}
    \omega_1 = -\Gamma_1\,\overline{H}^3 - \frac{\Ct}{3}\,\overline{H}^3 + \frac{2}{15}\,\overline{H}^6\,\hat{k}_{x1}^2\, \R^2
    +\frac{\hat{\eta}^2 \,\overline{H}^3\,\Vr}{(\overline{H}+\K)^3}+\frac{\hat{\eta}\,\overline{H}^2\,\K\,\M}{\Pr\,(\overline{H}+\K)^2}.
\end{equation}
\end{subequations}

The eigenvalues $\omega = \omega_0\lVert \boldsymbol{k}\rVert + i\omega_1\lVert \boldsymbol{k}\rVert^2$ coincide with those obtained by solving the algebraic dispersion relation associated with the 3D Benney equation, as reported in \citet{joo1993two}. In addition, considering the nondimensional film thickness $ \overline{H} = 1$, evaporating conditions ($\hat{\eta} = 1$), and purely streamwise perturbations ($\hat{k}_{x1}=1,\hat{k}_z=0$), the leading-order solution coincides with that found by \citet{mohamed2020linear}. 

    For the isothermal case, $\omega_0$ is real, corresponding to a travelling wave with neither growth nor decay, known as \textit{Goldstone mode} \citep[Section 3.5]{kalliadasis2011falling}. The phase speed associated with this neutral wave, $c_0 = \omega_0$, is proportional to $\R\,\overline{H}^{2}$ and depends on the orientation of the wavenumber vector $\hat{k}_{x1}$. As a consequence, thinner films or perturbations more closely aligned with the 2D streamwise direction ($\hat{k}_{x1}=1$) propagate faster than 3D perturbations with a component in the spanwise direction ($\hat{k}_{x1}\neq 0$). The normal velocity eigenfunction, $\phi_0$, depends solely on $\hat{y}^2$, while the temperature eigenfunction, $\hat{\tau}$, is linear in $\hat{y}$ and inversely proportional to the orientation of $\bm{k}$, $\R$, and $(\overline{H} + \K)^2$. Considering spanwise perturbations ($k_{x1} \to 0$), stronger convective effects (larger $\R$) or thinner films (smaller $\overline{H}$) lead to an increased amplitude of the temperature variation. As $\tau_0$ is imaginary, it is shifted by $\pi/2$ relative to the film profile $\xi$, with temperature maxima coinciding with troughs and temperature minima coinciding with crests. A similar behaviour occurs in the condensation case ($\hat{\eta}=-1$), with a phase shift of $-\pi/2$.

Considering the first-order solution, $\omega_1$, this is a purely imaginary quantity representing the temporal growth rate of the perturbation. In its expression, the hydrodynamic contribution, associated with $\Ct$, and the inertial contribution, associated with $\R$, are independent of $\hat{\eta}$ and therefore unaffected by the phase-change regime of the film. The vapour recoil contribution is consistently destabilising, whereas the surface tension term is always stabilising. The only term that changes sign between evaporation ($\hat{\eta}=1$) and condensation ($\hat{\eta}=-1$) is the thermocapillary (Marangoni) term. Since thermocapillary stresses drive fluid from cold to hot regions along the interface, this mechanism promotes wave growth in evaporating films (moving fluid from the trough to the crest) and suppresses wave growth in condensing films (moving fluid from the crest to the trough).

Among the various contributions, the inertial term exhibits the largest increase with film thickness $\overline{H}$. The surface tension and hydrostatic pressure terms scale as $\overline{H}^{3}$, the vapour recoil term scales as $\overline{H}^3 / (\overline{H} + \K)^3$, and the thermocapillary term scales as $\overline{H}^2 / (\overline{H} + \K)^2$. Moreover, the thermocapillary is the only term that depends linearly on $\K$ in the numerator. This occurs because, in the stress balance equation for the small-amplitude perturbation \eqref{eq:OS_main_bc1_2}, the thermocapillary effects are weighted by the mean free-surface temperature $\overline{\Theta}_s|_{\overline{H}}$ \eqref{eq:ss_rel_q_theta}, which is linear in $\K$ at the numerator.

The magnitude of the cut-off wavenumber $\bm{k}_c$, which defined the threshold between unstable ($\bm{k}<\bm{k}_c$) and stable ($\bm{k}>\bm{k}_c$) wavenumbers, is calculated by setting the imaginary part of $\omega$ in \eqref{eq:expansion_lw_3DOS} to zero and solving for $\lVert \bm{k} \rVert$, which gives the non-trivial solution reading:
\begin{equation}
\label{eq:cut-off_k}
    \lVert \bm{k}_c \rVert = \sqrt{\frac{1}{\Gamma}}\sqrt{\frac{\hat{\eta}\,\K\,\M}{(\overline{H}+\K)^2\,\Pr\,\overline{H}} + \frac{\hat{\eta}^2\,\Vr}{(\overline{H}+\K)^3}-\frac{\Ct}{3}+ \frac{2}{15}\,\overline{H}^3\,\hat{k}_{x1}^2\,\R^2}.
\end{equation}
In the case without phase change ($\Vr=0$) and thermocapillary effects ($\M=0$), the critical wavenumber depends on the balance of hydrostatic ($\Ct$) and inertial ($\R$) effects that underpin the H mode instability, also known as the \textit{Kapitza instability}. At the same time, purely spanwise perturbations remain linearly stable. In a phase-changing film, the instability in the spanwise direction is driven by vapour recoil and thermocapillary effects, with hydrostatic pressure providing stabilisation. 
\begin{figure}
    \centering
    \begin{tabular}{cc}  % Create a 3-column table
        \subfloat[]{\includegraphics[width=0.5\textwidth]{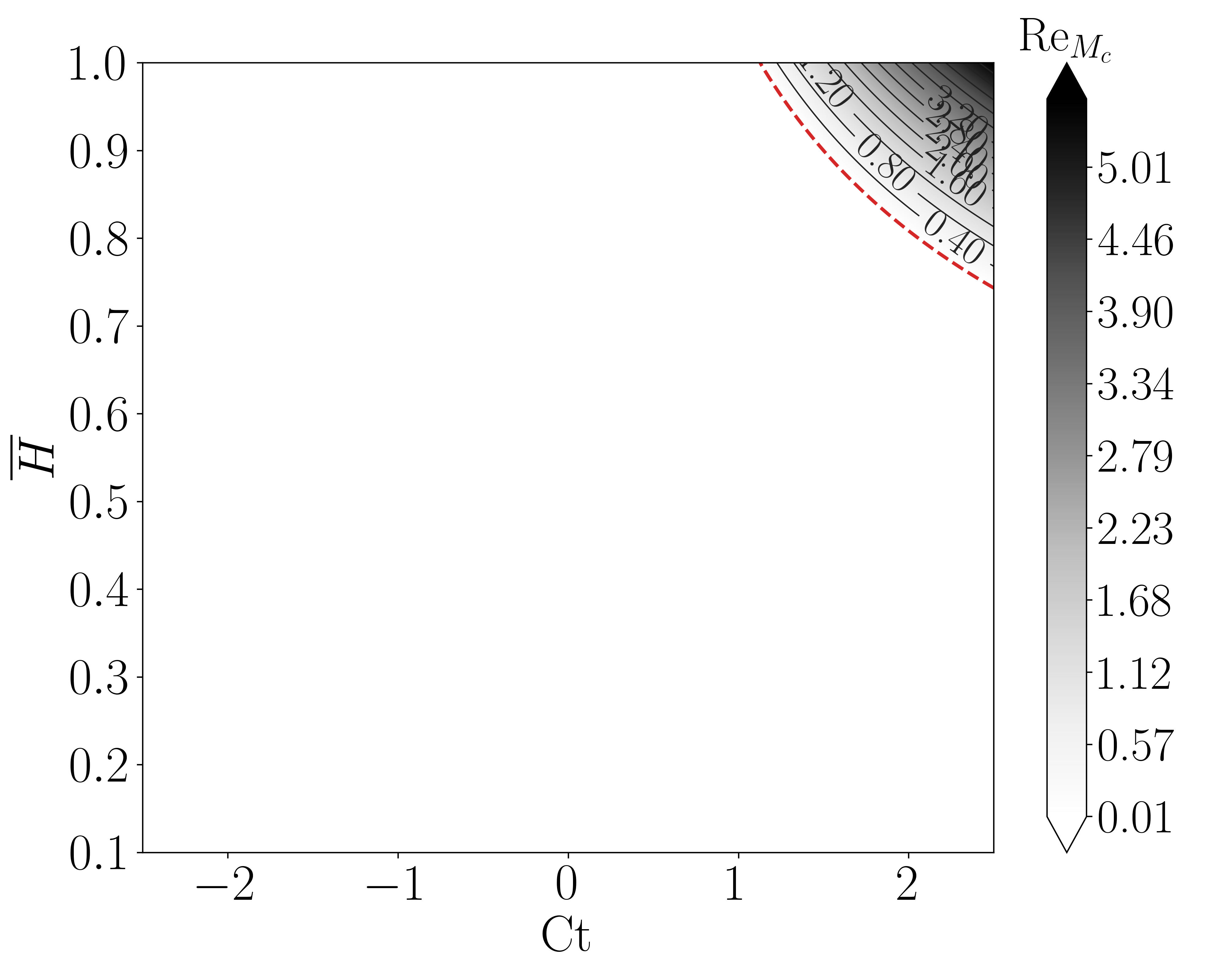}} &
        \subfloat[]{\includegraphics[width=0.5\textwidth]{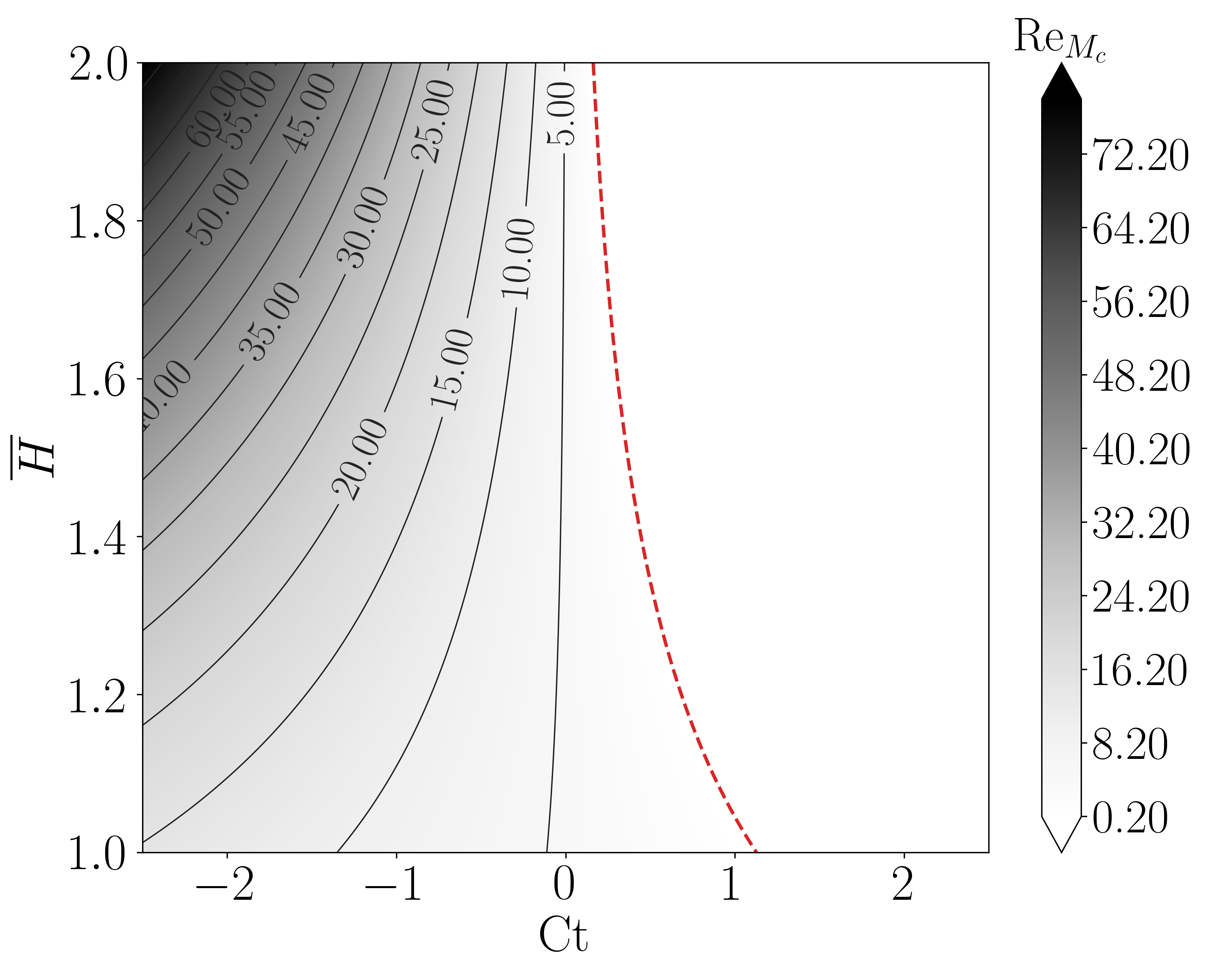}}
    \end{tabular}
    \caption{Critical Marangoni Reynolds number $\R_{M_c}$ for spanwise perturbations as a function of $\overline{H}$ and $\Ct$ for $\Vr=0.5$ in (a) evaporating ($\hat{\eta}=1$) and (b) condensing conditions ($\hat{\eta}=-1$).}
    \label{fig:critical_ReM_transversal}
\end{figure}

For 3D perturbations, the condition for $\R$ to have a stable liquid film for any wavenumber is obtained by imposing that the argument inside the square root in \eqref{eq:cut-off_k} is negative, which gives the condition reading:
\begin{equation}
    \R_{c-}<\R<\R_{c+}
\end{equation}
where $\R_{c-}>0$ and $\R_{c+}>0$ represent the minimum and the maximum critical Reynolds numbers, respectively, which are given by:
\begin{equation}
\label{eq:c_Re}
    \R_{c\pm} = \frac{\sqrt{5}}{4\overline{H}^3 \hat{k}_{x1}^2}\left(\sqrt{5}\cot{\beta}\pm \sqrt{5(\cot{\beta})^2-\frac{24\hat{\eta}\overline{H}^2\hat{k}_{x1}^2 \left(\hat{\eta}\,\overline{H}\Pr\Vr+\overline{H}\K \M+\K^2\M\right)}{\Pr(\overline{H}+\K)^3}}\right)
\end{equation}
where $\Ct$ was replaced with its definition \eqref{eq:adim_num_00}, to express it as a function of $\R$.

Setting $\Vr = 0$, assuming $\overline{H} = 1$, and considering two-dimensional streamwise perturbations, the above expression recovers the critical conditions for a falling liquid film, which represent a balance between hydrostatic and inertial forces \citet[Section~3.5]{kalliadasis2011falling}.

The vapour recoil and thermocapillary effects reinforce the destabilising influence of inertia, leading to a smaller critical Reynolds number $\R_{c+}$. Furthermore, the imbalance between phase-change effects and hydrostatic pressure gives rise to an additional critical Reynolds number $\R_{c-}$, below which the system becomes unstable again and is primarily governed by vapour recoil and thermocapillary mechanisms.

To have a positive $\R_{c-}$, the nondimensional group and the inclination angle $\beta$ should respect the following condition:
\begin{equation}
    \cot{\beta}\geq 2 \sqrt{\frac{6}{5}} \sqrt{\frac{\hat{\eta}\overline{H}^2\,\hat{k}_{x1}^2\, \left(\overline{H}\,\K\,\M+\hat{\eta}\overline{H}\,\Pr\,\Vr+\K^2\, \M\right)}{\Pr\,(\overline{H}+\K)^3}}.
\end{equation}

A critical minimal Reynolds number arises under both condensing and evaporating conditions, provided that the vapour recoil and Marangoni effects dominate over the hydrostatic contribution. The inclination angle $\beta$ increases as $\hat{k}_{x1}$ decreases, since larger spanwise components in the three-dimensional perturbations correspond to weaker inertial effects. As a consequence, the hydrostatic pressure needs to balance only the phase-change and thermocapillary effects, leading to a broader range of $\beta$ values for which a critical minimal Reynolds number cannot be attained.

Considering the case without hydrostatic pressure, such as a vertical condensing liquid film subject to streamwise perturbations, under these conditions, \eqref{eq:c_Re} still admits a critical Reynolds number, which is expressed as:
\begin{equation}
\label{eq:Re_c_cond}
    \R_c = \frac{1}{\overline{H}^3}\left(\sqrt{\frac{15}{2}} \sqrt{\frac{\overline{H}^2(\K\M(\overline{H}+\K)-\overline{H}\Pr  \Vr)}{\Pr(\overline{H}+\K)^3}}\right).
\end{equation}

In the absence of phase-change effects, the liquid film would be unstable for all values of $\R$. However, in the present case, the stabilising thermocapillary effects counteract the destabilising influence of vapour recoil. Thus, for \eqref{eq:Re_c_cond} to admit a real solution, the ratio $\M/\Pr$, defined as the Marangoni–Reynolds number in \eqref{eq:adim_thermal_nums}, must satisfy the following condition:
\begin{equation}
    \frac{\M}{\Pr}=\R_M > \R_{M_c},\qquad \text{with} \qquad \R_{M_c} = \frac{\overline{H}\,\Vr}{\K\,(\overline{H}+\K)}
\end{equation}
where $\R_{M_c}$ is the critical Marangoni–Reynolds number.
\begin{figure}
    \centering
    \begin{tabular}{cc}  % Create a 3-column table
        \subfloat[]{\includegraphics[width=0.5\textwidth]{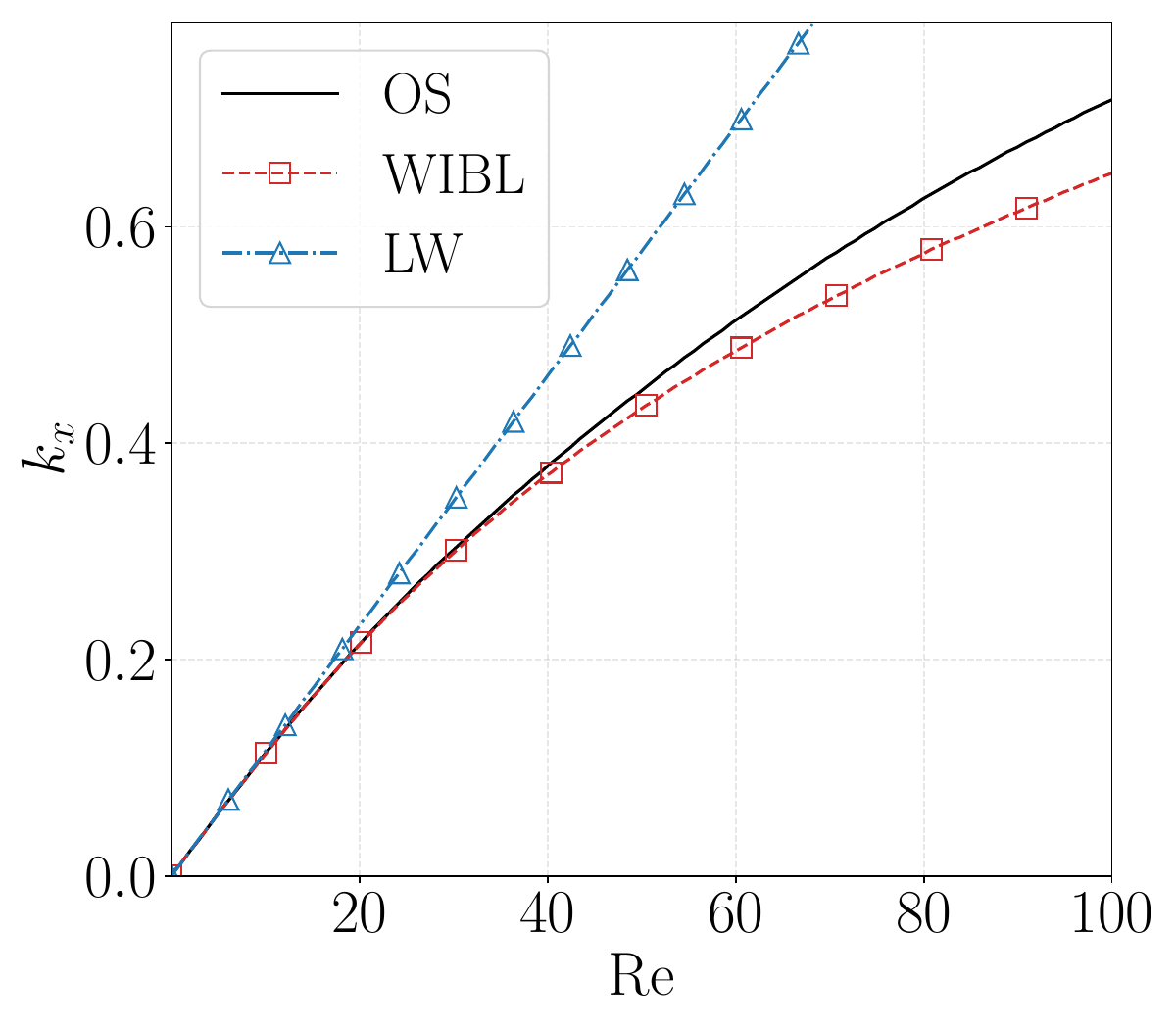}} &
        \subfloat[]{\includegraphics[width=0.5\textwidth]{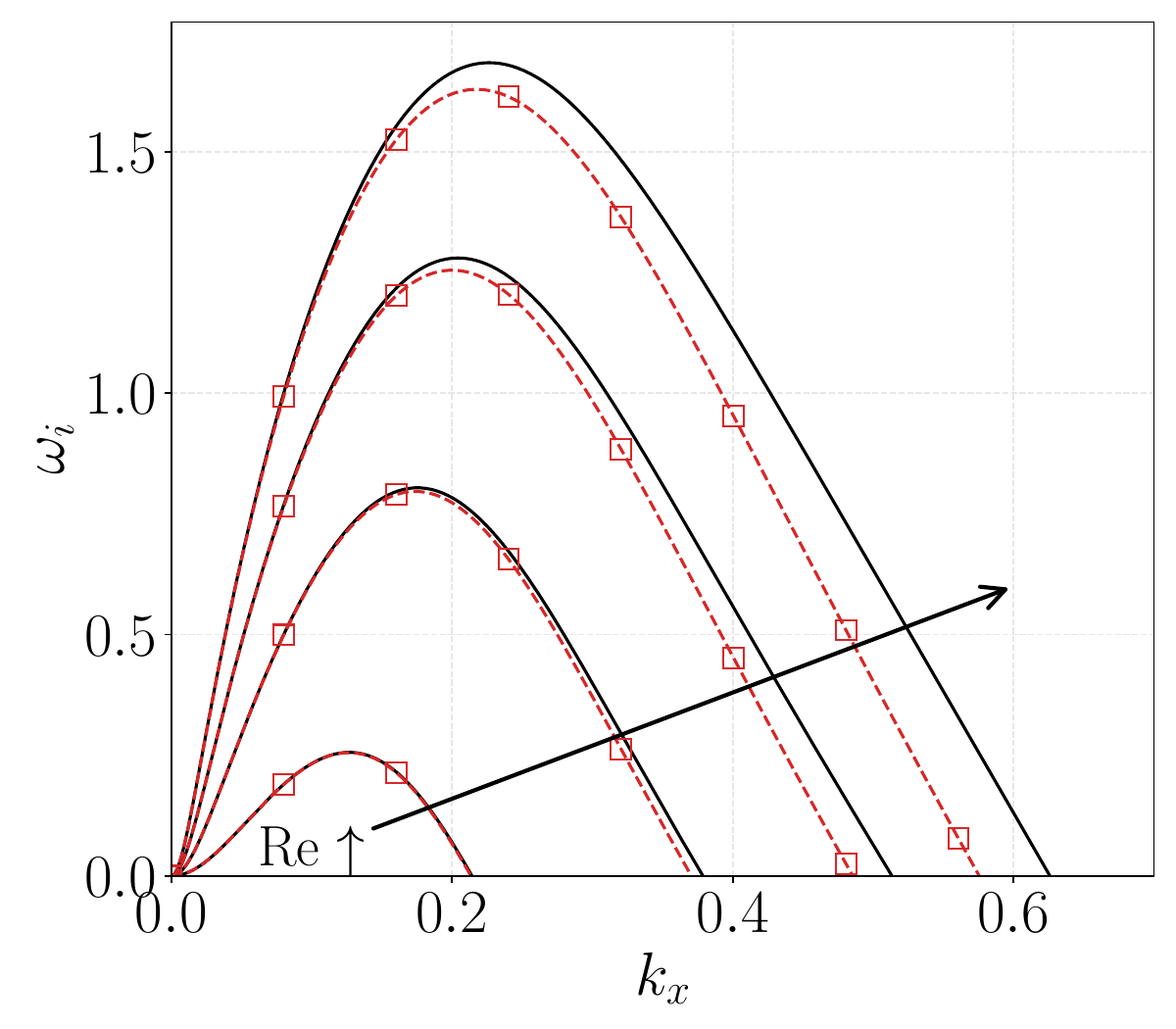}}
    \end{tabular}
    \caption{Linear stability properties of an isothermal vertically falling film in terms of (a) Neutral Curve in the plane ($k_x,\R$) and (b) growth rate $\omega_i$ as a function of $k_x$ for $\R=20,40,60,80$ given by the WIBL equations (red dashed line with squares) and the Orr-Sommerfeld eigenvalue problem solved with long-wave expansion (blue dash-dotted line) and numerically (black continuous line) for vertical film with 2D streamwise perturbations with $\Gamma=1000$ and  $\overline{H}=1$.}
    \label{fig:H_mode_alone}
\end{figure}

Considering now spanwise perturbations ($\hat{k}_{x1}=0$), the critical Marangoni Reynolds number $\R_{M_c}$ is obtained by setting the argument of the square root in \eqref{eq:cut-off_k} to zero, which gives:
\begin{equation}
\label{eq:critical_Ma_Re_span}
 \R_{M_c} = \frac{\overline{H}\,\left(\Ct\,(\overline{H}+\K)^3-3\hat{\eta}^2\Vr\right)}{3\hat{\eta}\,\K\,(\overline{H}+\K)}
\end{equation}
A critical condition exists in both the evaporating and condensing regimes, depending on whether the film is falling ($\Ct > 0$; $0 < \beta < \pi/2$) or hanging ($\Ct < 0$; $\pi/2 < \beta < \pi$). Moreover, a critical condition can also occur in the absence of the stabilising influence of hydrostatic pressure $\Ct = 0$ ($\beta = \pi/2$). For such a vertical falling film, the existence of a threshold is possible only in the condensing regime ($\hat{\eta}=-1$).

Figure~\ref{fig:critical_ReM_transversal} shows the range of $\R_{M_c}>0$ and their corresponding values (in grey), as given by~\eqref{eq:critical_Ma_Re_span}, plotted as functions of the liquid film thickness $\overline{H}$ and the inclination number $\Ct$ for $\Vr = 0.5$, considering (a) the evaporating and (b) the condensing regimes, with the red dashed line indicating the threshold between positive and negative values. In the evaporating regime, the admissible region is confined to the upper-right corner, corresponding to the early stages of evaporation when $\overline{H}$ remains above approximately 0.7, and the film is either mildly inclined or characterised by very low Reynolds numbers. As the film thins below 0.7, temperature variations at the free surface become sharper, and the vapour recoil effect tends to dominate. In contrast, in the condensing regime, the admissible region is broader, exhibits sharper variations, and extends primarily into the negative half-plane of $\Ct$, indicating its association with hanging-film configurations. Although in this setting the pressure exerts a destabilising influence, the stabilising thermocapillary effect compensates for both the destabilising pressure and vapour recoil effects.

Although the asymptotic long-wave approach provides closed-form expressions for the growth rate and the associated critical nondimensional parameters, its linear-stability predictions are accurate only at small wavenumbers and for low $\R$, as extensively demonstrated in the isothermal falling-film literature. Figure~\ref{fig:H_mode_alone} presents a comparison of (a) the cut-off wavenumber as a function of $\R$ and (b) the growth rate $\omega_i$ as a function of the streamwise wavenumber $k_x$ for several values of $\R$, considering a vertical film ($\beta=\pi/2$) with $\overline{H}=1$ and two-dimensional, streamwise H-mode perturbations (thermal and phase-change effects are neglected). The WIBL model accurately reproduces the neutral-stability curve up to moderately large Reynolds numbers, whereas the long-wave approximation deviates from the OS solution for $\R>20$. Beyond this range, WIBL tends to underpredict the threshold wavenumber, while the long-wave approximation overpredicts it. In terms of growth rates, the WIBL model captures both the location and the magnitude of the most unstable mode, although it slightly underpredicts the maximum growth rate for $\R>20$.
\begin{figure}
    \centering
    \begin{tabular}{cc}  % Create a 3-column table
        \subfloat[]{\includegraphics[width=0.5\textwidth]{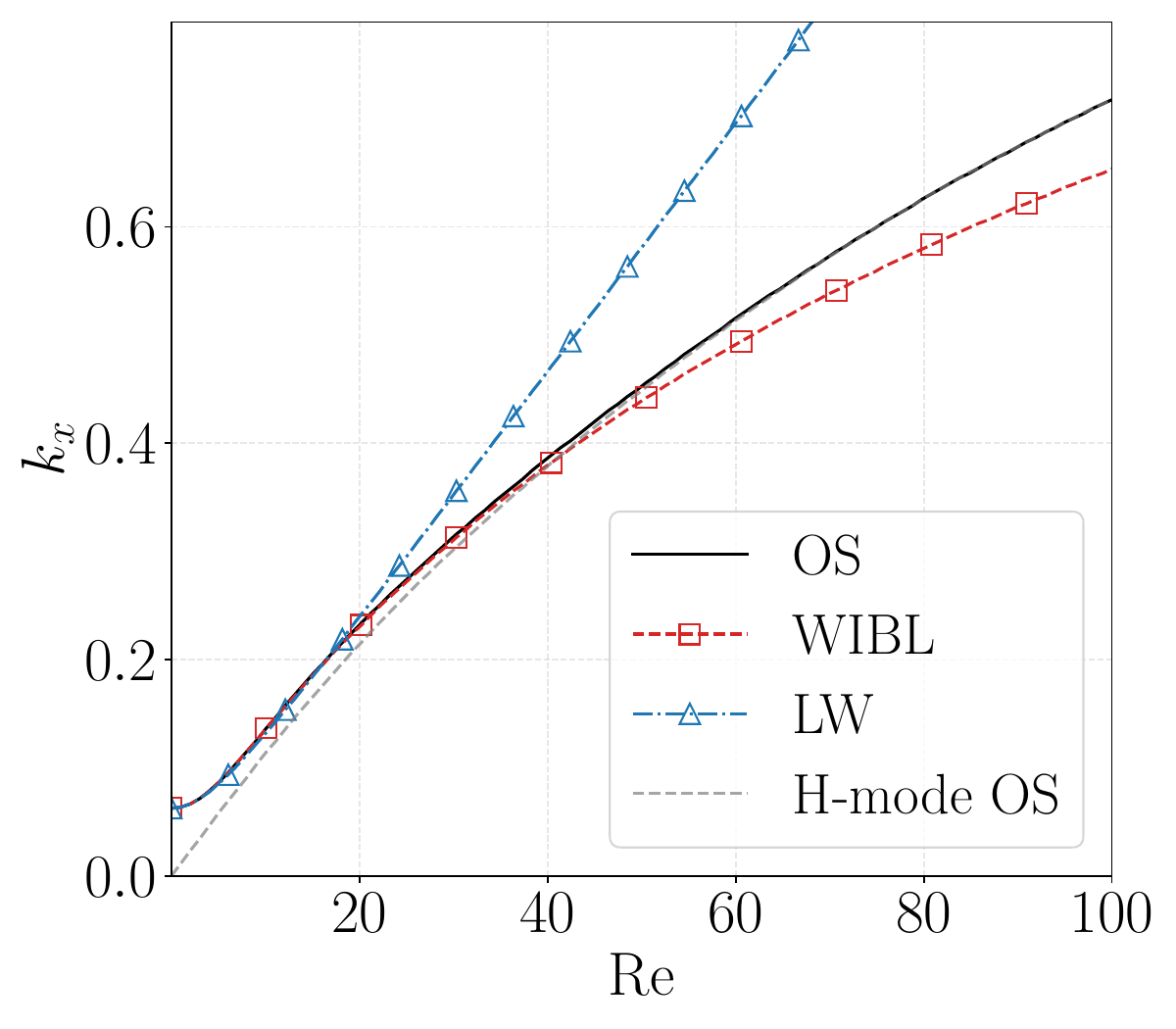}} &
        \subfloat[]{\includegraphics[width=0.5\textwidth]{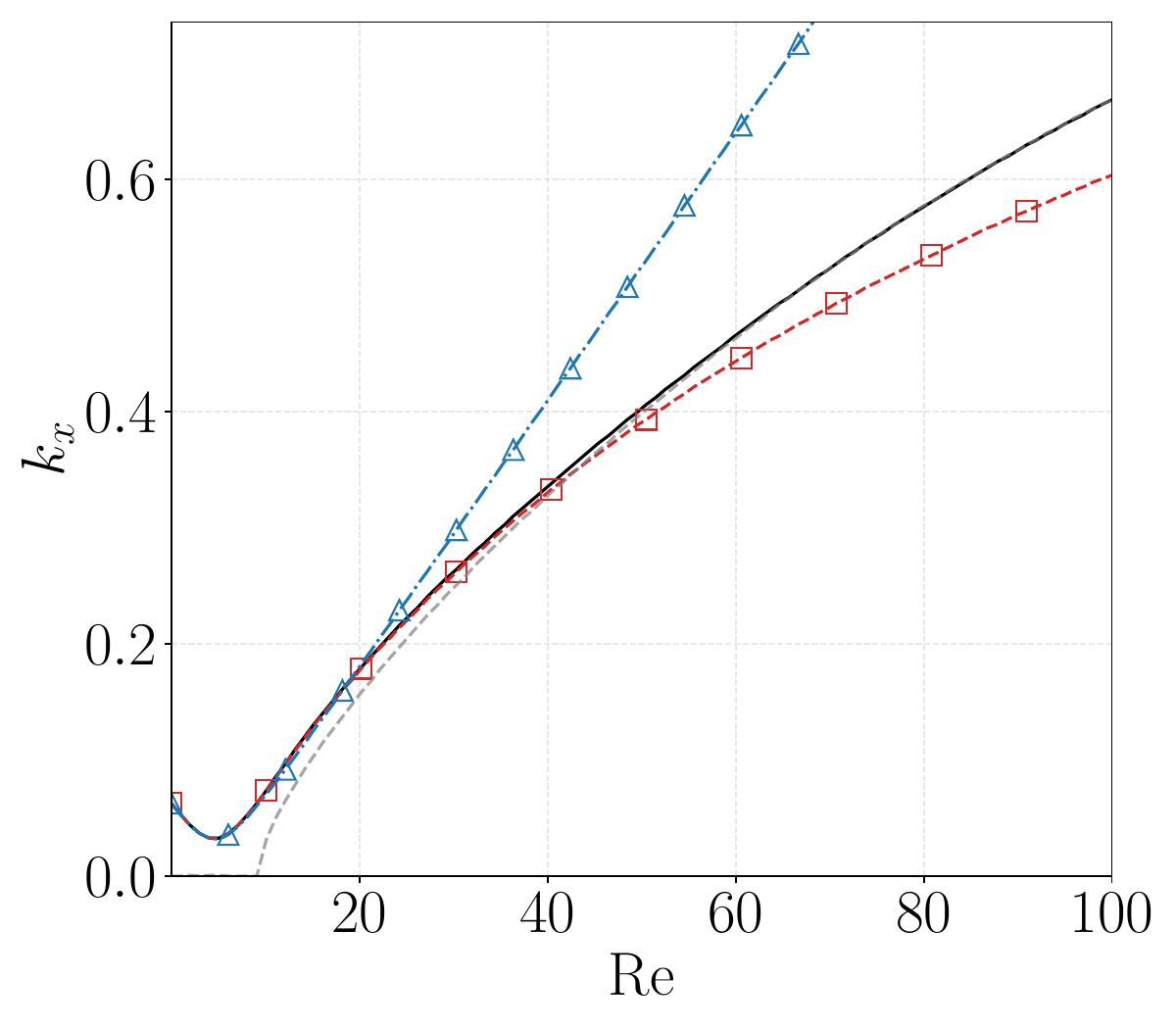}}
    \end{tabular}
    \caption{Neutral curves in ($\R,k_x)$ with $k_z=0$ obtained solving the linearized WIBL model (red dashed line with squares) and the Navier-Stokes equation solve numerically (continuous black line) and with the long wave expansion (dash-dotted line with triangles) for $\M=10$ and $\Vr=4$ and numerical solution of Navier-Stokes for $\M=\Vr=0$ (gray dashed line) for inclination angle $\beta$ equals to (a) 90\degree and (b) 15\degree.}
    \label{fig:multiple_modes}
\end{figure}

In terms of stability characteristics involving also M and E modes, figure~\ref{fig:multiple_modes} shows the neutral curve for $\M=10$ and $\Vr=4$ for an inclination angle $\beta$ of (a) $\pi/2$ and (c) $\pi/6$ for $\overline{H}=1$. The Marangoni and vapour recoil effects are most significant at small Reynolds numbers when inertia does not have a dominant role. As $\R$ increases, the H mode becomes dominant; consequently, the neutral curve obtained from the Orr-Sommerfeld (OS) eigenvalue problem tends to the solution obtained without $\M$ and $\Vr$. The simplified long-wave model and the asymptotic expansion agree well with the OS solution up to $\R=20$ for the long-wave model and up to $\R=40$ for the WIBL model. In both cases, the error between the WIBL prediction and the OS result remains below $0.05$. Moreover, the neutral curve predicted by the WIBL model systematically underestimates the instability threshold, as observed above for the H mode.

\begin{figure}
    \centering
    \begin{tabular}{cc}  % Create a 3-column table
        \subfloat[]{\includegraphics[width=0.5\textwidth]{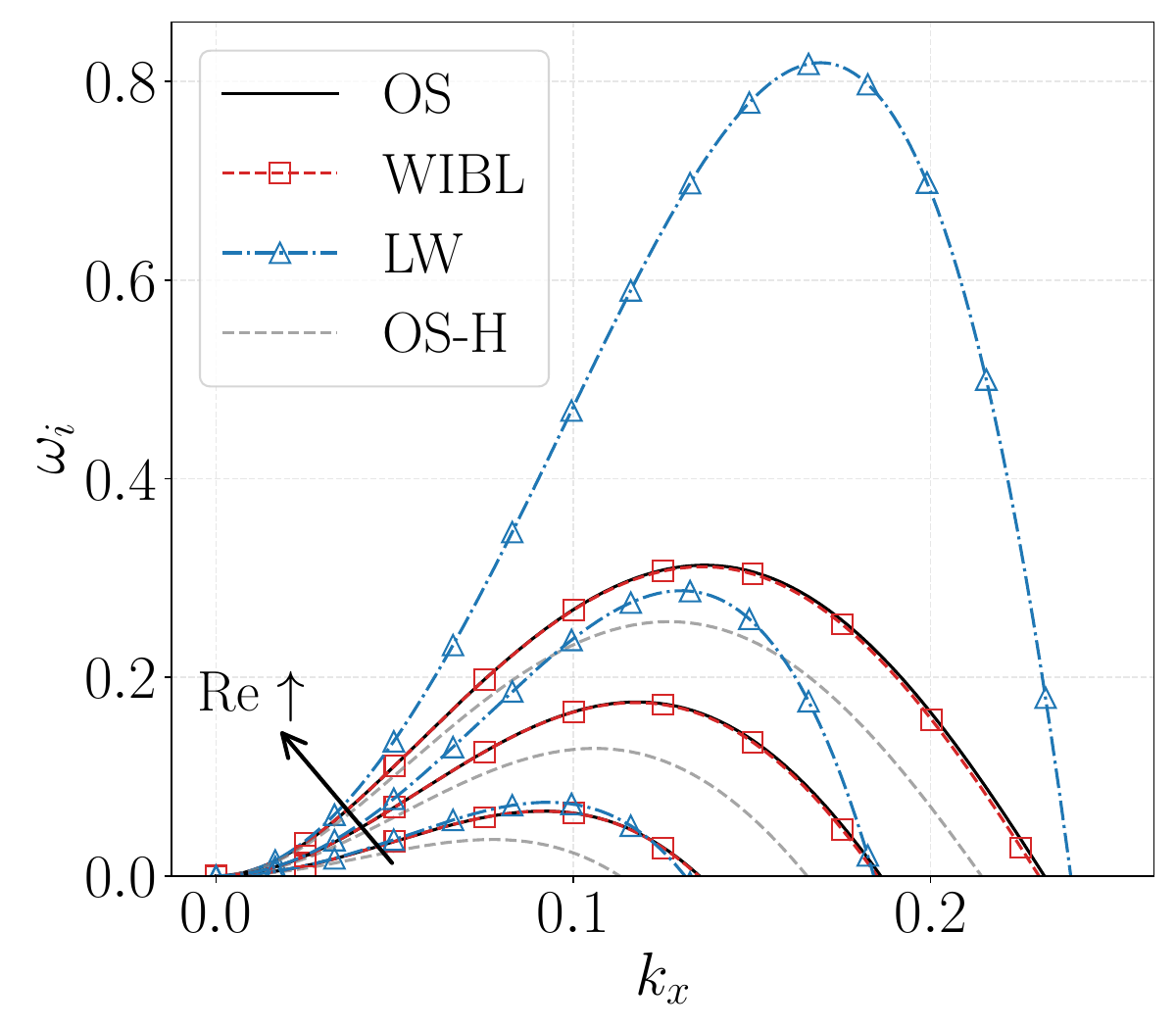}} &
        \subfloat[]{\includegraphics[width=0.5\textwidth]{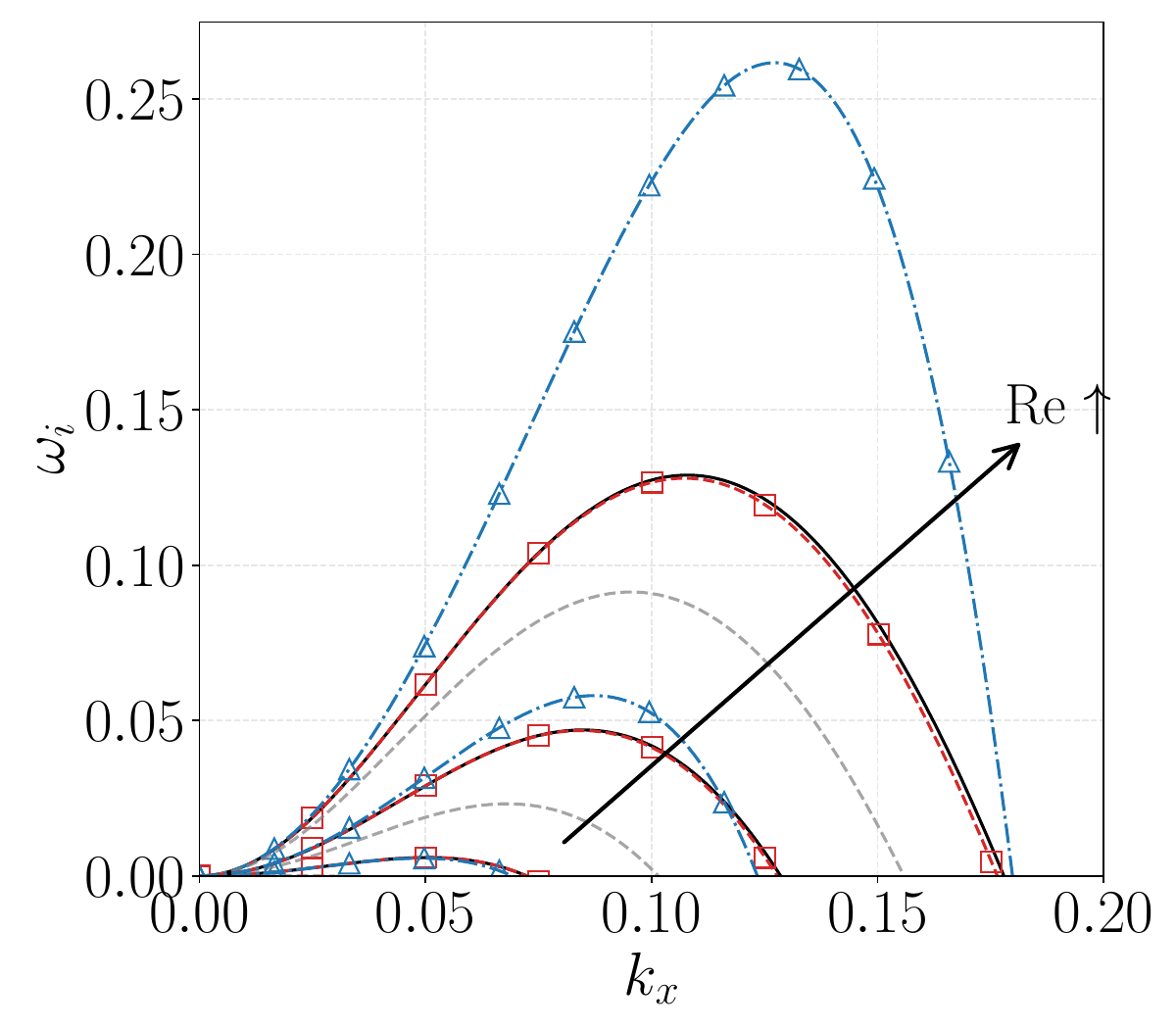}} \\
        \subfloat[]{\includegraphics[width=0.5\textwidth]{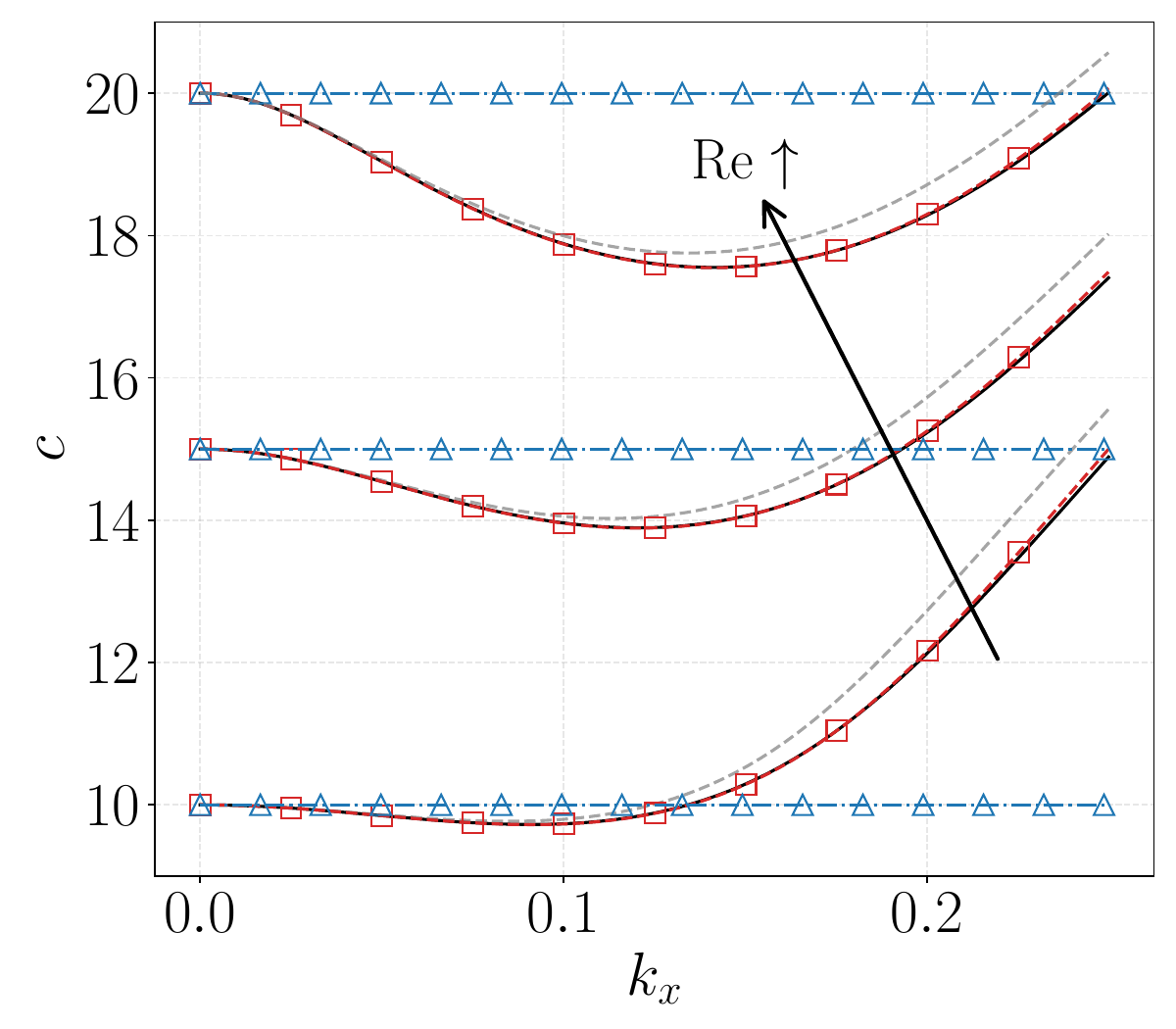}} &
        \subfloat[]{\includegraphics[width=0.5\textwidth]{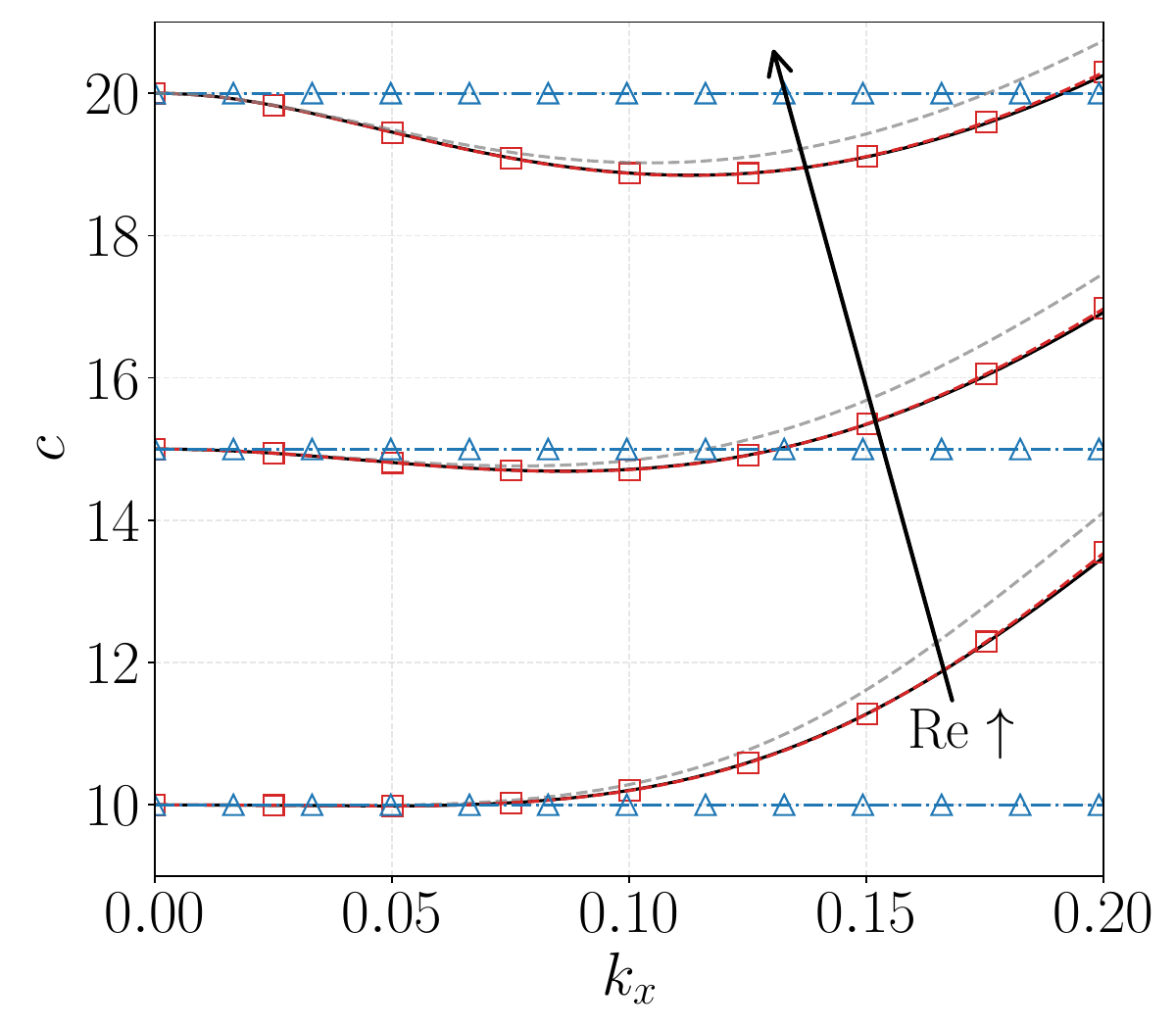}}
    \end{tabular}
    \caption{(a,b) Growth rate $\omega_i$ and (c,d) phase speed $c$ as functions of $k_x$ for $k_z=0$, with $\overline{H}=1$, $\Vr=4$, and $\M=10$ obtained using the WIBL model (red dashed lines with markers), the numerical (continuous black lines), and the long-wave (dash-dotted lines with triangles) solutions of the Navier-Stokes equations and the numerical solution of the Navier-Stokes equations for $\M=\Vr=0$ (gray dashed lines), for inclination angles $\beta$ equal to (a,c) $\pi/2$ and (b,d) $\pi/6$ and $\R=10$, 15, and 20.}
    \label{fig:growth_rates_comp}
\end{figure}

Although there is good agreement between the long-wave predictions and the numerical OS solutions for the neutral stability curves at low Reynolds numbers, noticeable differences arise in their predicted growth rates. Figure~\ref{fig:growth_rates_comp} presents the (a, b) growth rate $\omega_i$ and (c, d) phase speed $c$ for $10 \leq \R \leq 20$ and inclination angles $\beta$ of (a, c) $\pi/2$ and (b, d) $\pi/6$, for an evaporating liquid film with $\overline{H}=1$. The destabilising effects of vapour recoil and thermocapillarity reinforce the inertial instability, resulting in larger maximum values of $\omega_i$ for all $\R$. Furthermore, the maximum shifts slightly toward higher $k_x$ values. In most cases, the long-wave model overestimates the maximum $\omega_i$, whereas the WIBL prediction agrees very well with the OS solution. This discrepancy can influence the development of nonlinear waves and modal interactions in fully nonlinear simulations, as artificial amplification of small-amplitude perturbations can lead to different trajectories and stationary states. For $\beta = \pi/6$, the maximum $\omega_i$ decreases across all $\R$ as the stabilising hydrostatic effect becomes more significant, and again the WIBL model shows excellent agreement with the OS results.

Regarding the phase speed $c$, it initially decreases relative to its long-wave value as $ k_x$ increases, then rises again at larger $k_x$. Both vapour recoil and thermocapillary effects slightly reduce the wave speed compared with the purely hydrodynamic (H-mode) case. Across all simulated scenarios, the WIBL predictions show excellent agreement with the OS results, whereas the long-wave approximation remains accurate only in the limit $k \to 0$. As the Reynolds number increases, the minimum phase speed decreases, and the subsequent increase in $c$ with $k$ becomes less pronounced. The curves for $\beta = \pi/6$ display a similar trend, with smaller variation between the maximum and minimum values and slightly higher peak speeds. In this case as well, the WIBL predictions match the OS results almost perfectly.

\begin{figure}
    \centering
    \begin{tabular}{cc}  % Create a 3-column table
        \subfloat[]{\includegraphics[width=0.5\textwidth]{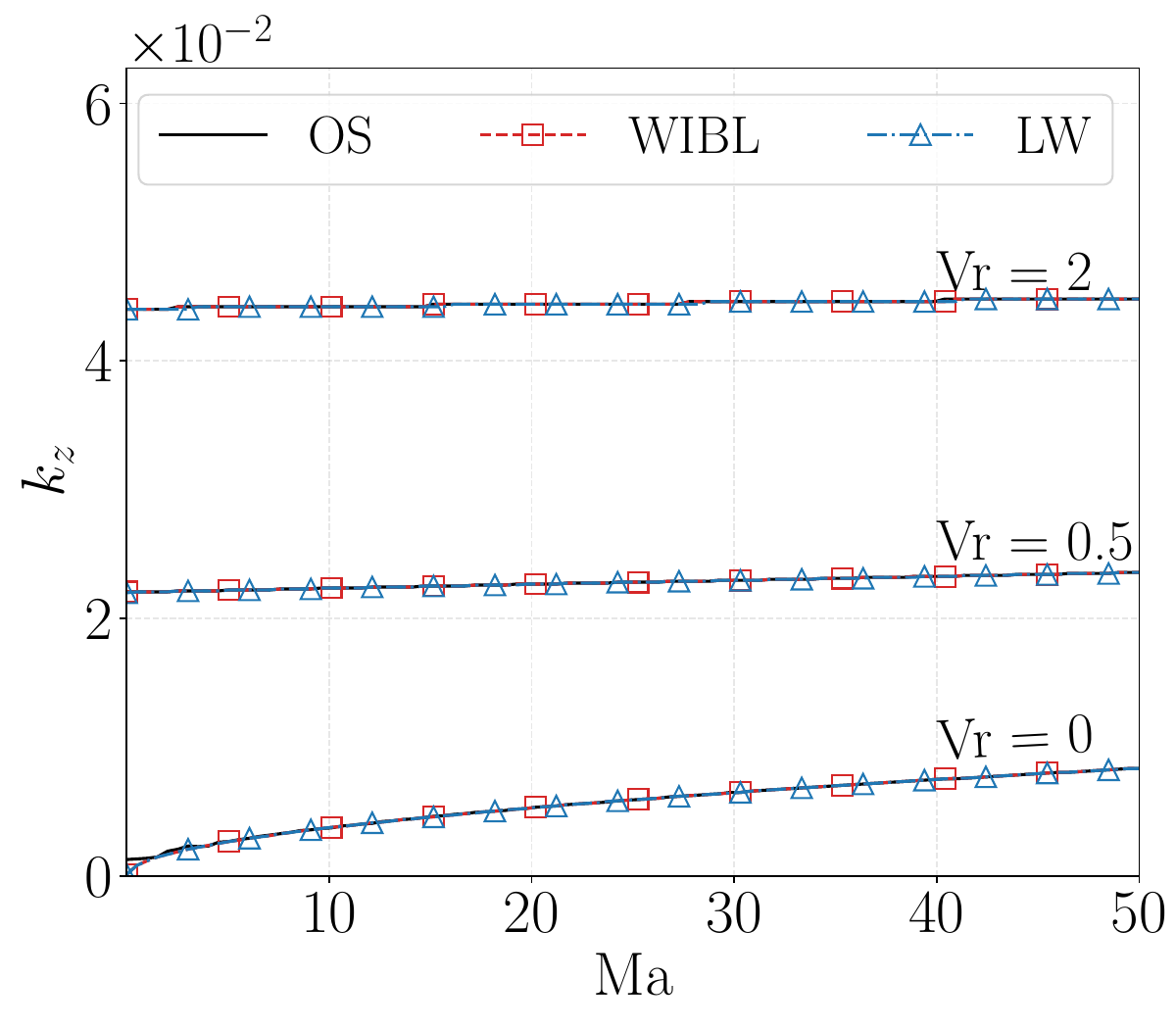}} &
        \subfloat[]{\includegraphics[width=0.5\textwidth]{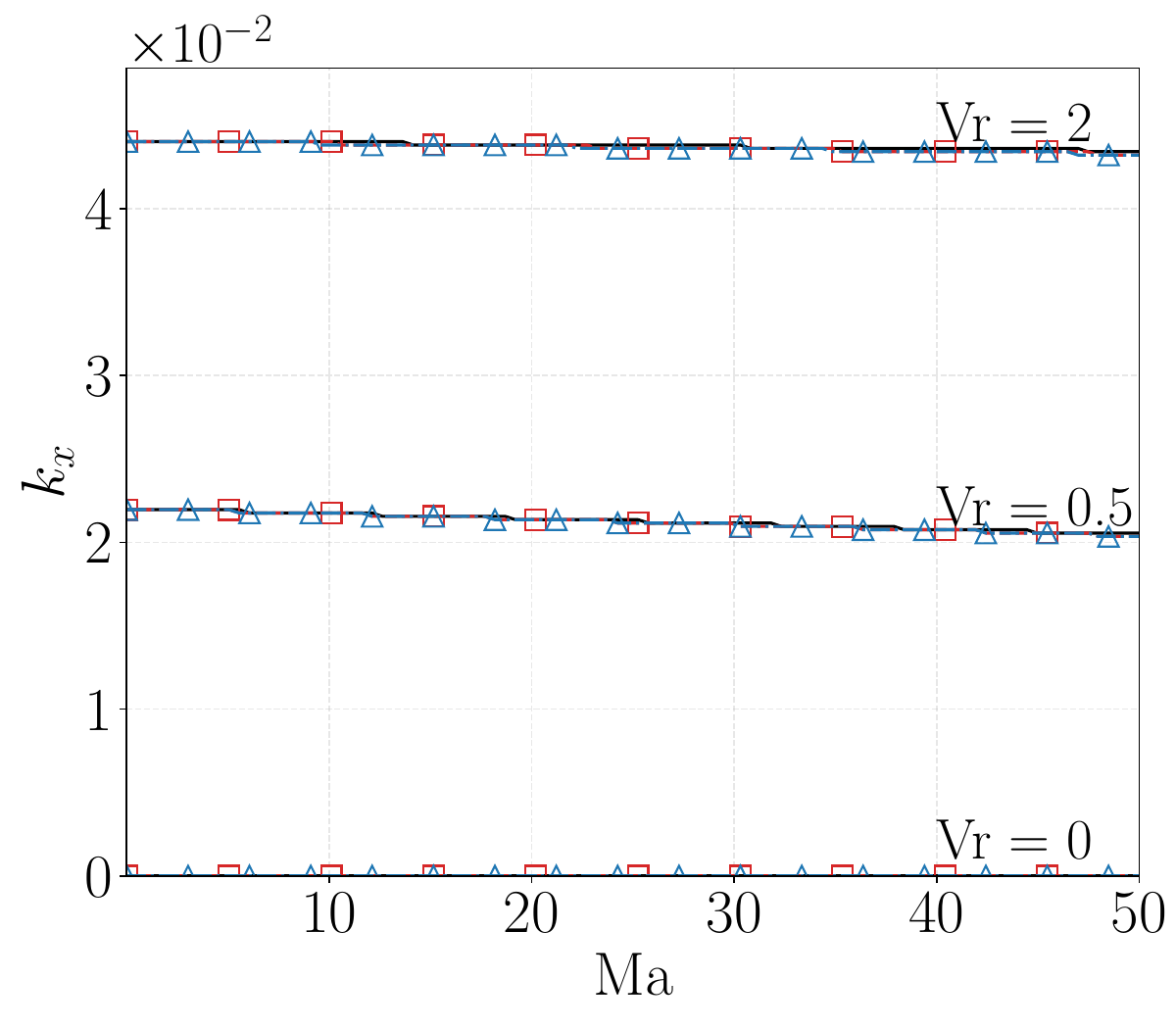}}\\
        \subfloat[]{\includegraphics[width=0.5\textwidth]{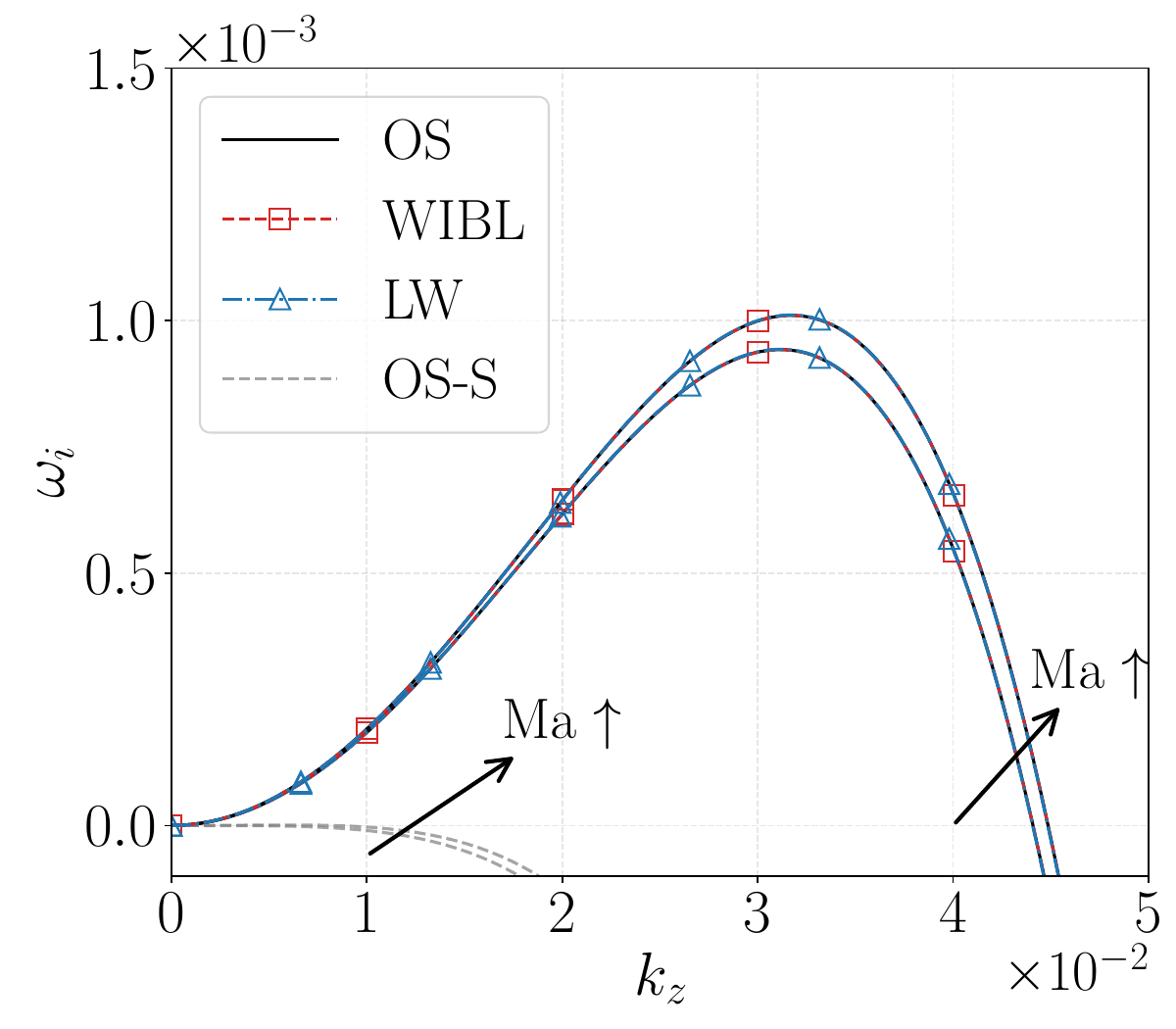}} &
        \subfloat[]{\includegraphics[width=0.5\textwidth]{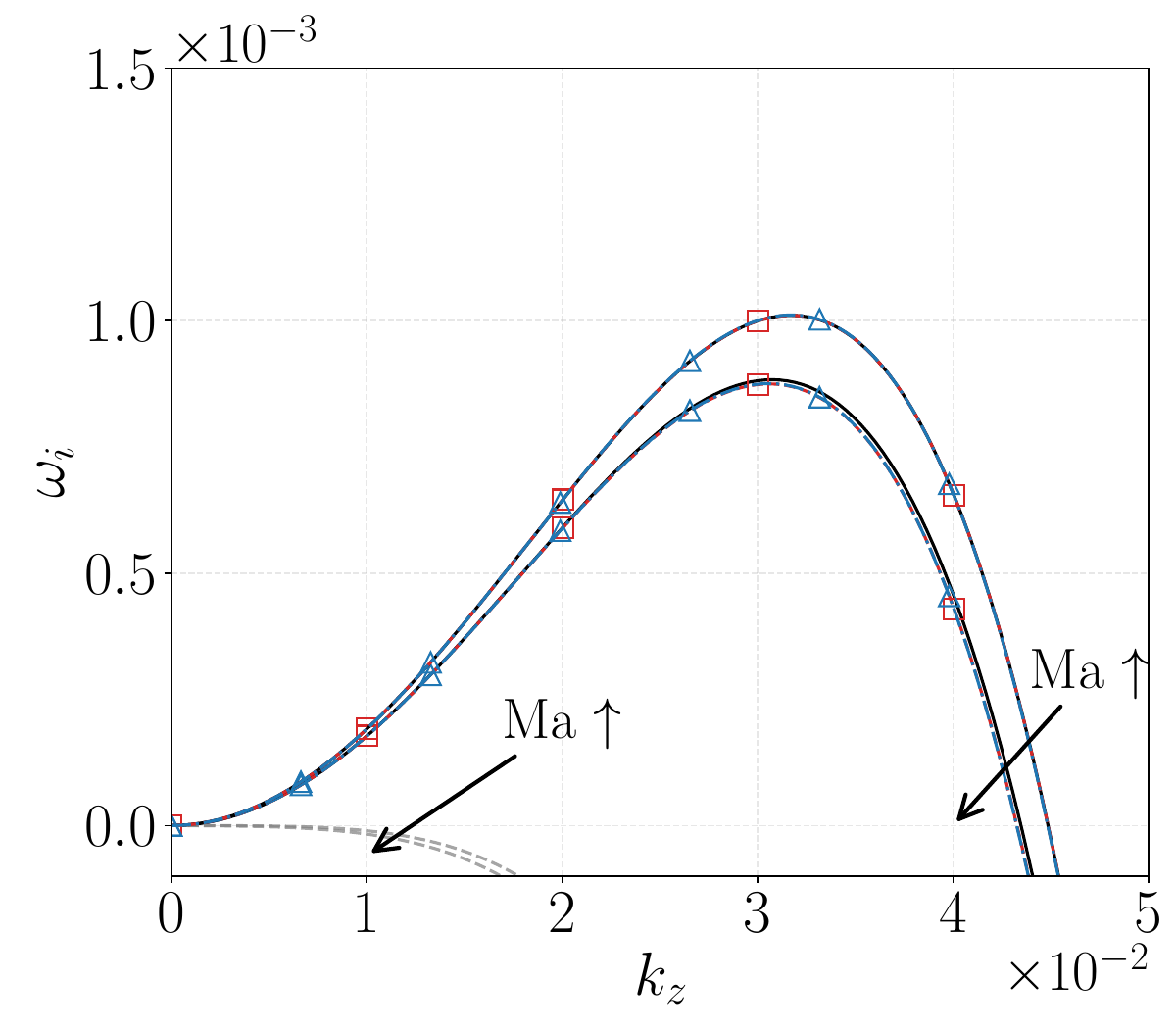}}
    \end{tabular}
    \caption{(a,b) Neutral stability curves in the $\M$-$k_z$ plane for $\R=0.1$, $\Pr=7$, $\K=0.01$, $\Gamma=1000$, and $\beta=90^\circ$, shown for different values of $\Vr$ with $\overline{H}=1$ and (c,d) growth rates $\omega_i$ as a function of $k_z$ for $\M=1$ and 50 with $\Vr=2$ considering (a,c) evaporation and (b,d) condensation. The predictions from the WIBL model (red dashed lines with markers), the numerical (continuous black lines), and the long-wave (dash-dotted lines with triangles) solution of Navier-Stokes are compared against the numerical solution of the Navier-Stokes equations for $\Vr=0$ (grey dashed lines).}
    \label{fig:neutral_curve_z_cond_evp}
\end{figure}

Considering spanwise perturbations, figure~\ref{fig:neutral_curve_z_cond_evp} shows (a,b) the neutral curves in the $(\M,k_z)$ plane and (c,d) the corresponding growth rates as functions of $k_z$, including solutions without vapour recoil (grey dashed lines), for $\overline{H}=1$ under (a,c) evaporating and (b,d) condensing conditions, and for various values of $\Vr$. Overall, within the examined parameter range, predictions from the Orr-Sommerfeld (OS) analysis, the WIBL model, and the long-wave (LW) solution are in excellent quantitative agreement. In both evaporation and condensation, the cutoff wavenumber $k_z$ is of order $O(10^{-2})$, with slightly larger values in the evaporating case, reflecting the destabilising influence of the Marangoni effect. During evaporation, the film becomes increasingly susceptible to transverse perturbations as $\M$ increases, since stronger Marangoni forcing destabilises the interface. Vapour recoil further enhances destabilisation, so that for $\Vr \neq 0$ the film can become unstable even at $\M=0$. In contrast, under condensing conditions, increasing $\M$ shifts the neutral curve toward lower wavenumbers, indicating that thermocapillary effects are stabilising.  

The LW solution exhibits only mild sensitivity to $\M$, as the Marangoni contribution in the long-wave scaling is multiplied by the small ratio $K/\Pr$, explaining the limited variation of the LW neutral curves. Similar trends are observed in the growth rates, where vapour recoil significantly increases the maximum $\omega_i$ relative to the thermocapillary-only case (grey line). In both evaporation and condensation, the spanwise growth rates remain three orders of magnitude smaller than the corresponding streamwise values, highlighting that, for vertical liquid films, streamwise perturbations dominate over spanwise perturbations.

\begin{figure}
    \centering
    \begin{tabular}{cc}  % Create a 3-column table
        \subfloat[]{\includegraphics[width=0.5\textwidth]{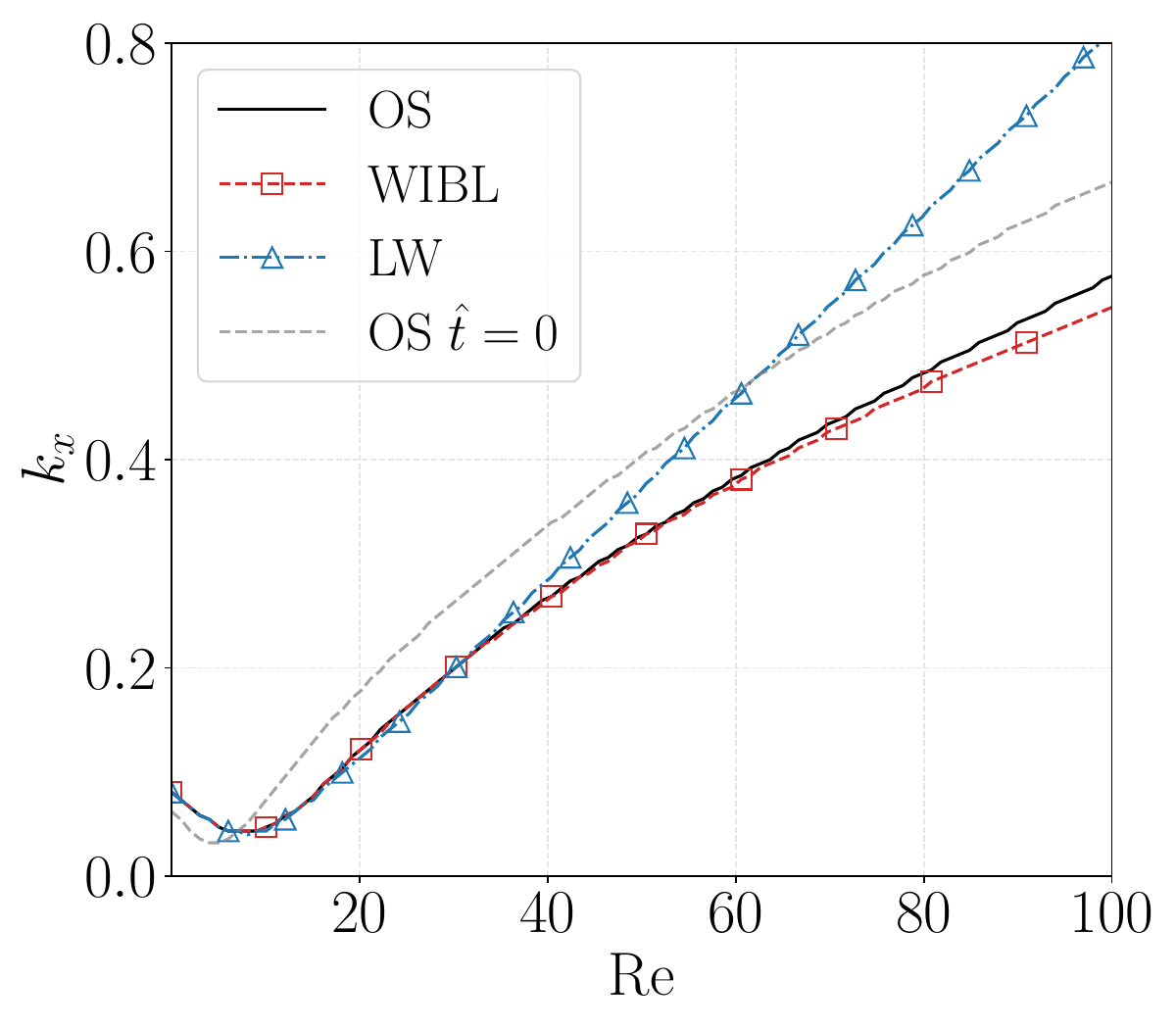}} &
        \subfloat[]{\includegraphics[width=0.5\textwidth]{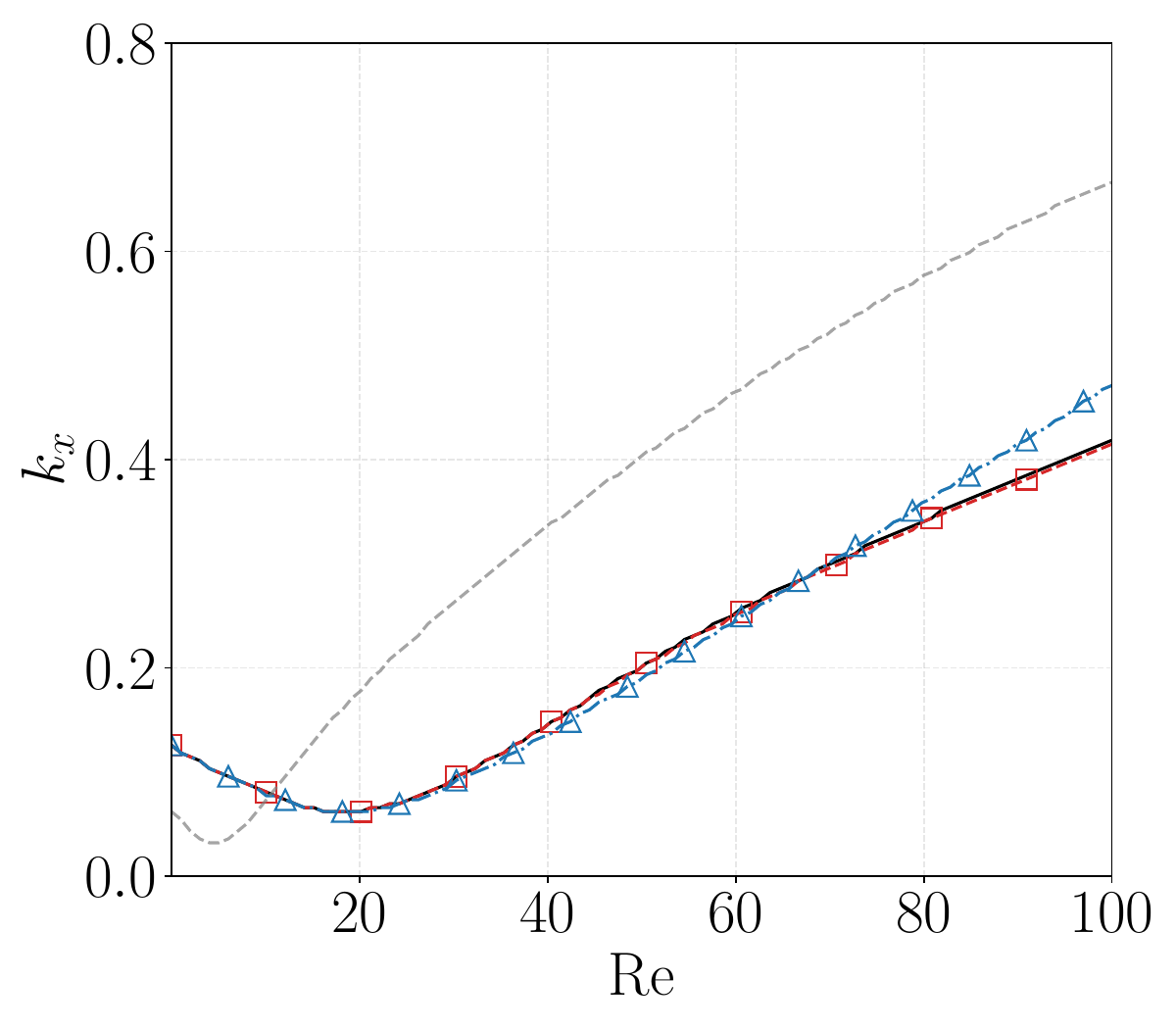}}\\
        \subfloat[]{\includegraphics[width=0.5\textwidth]{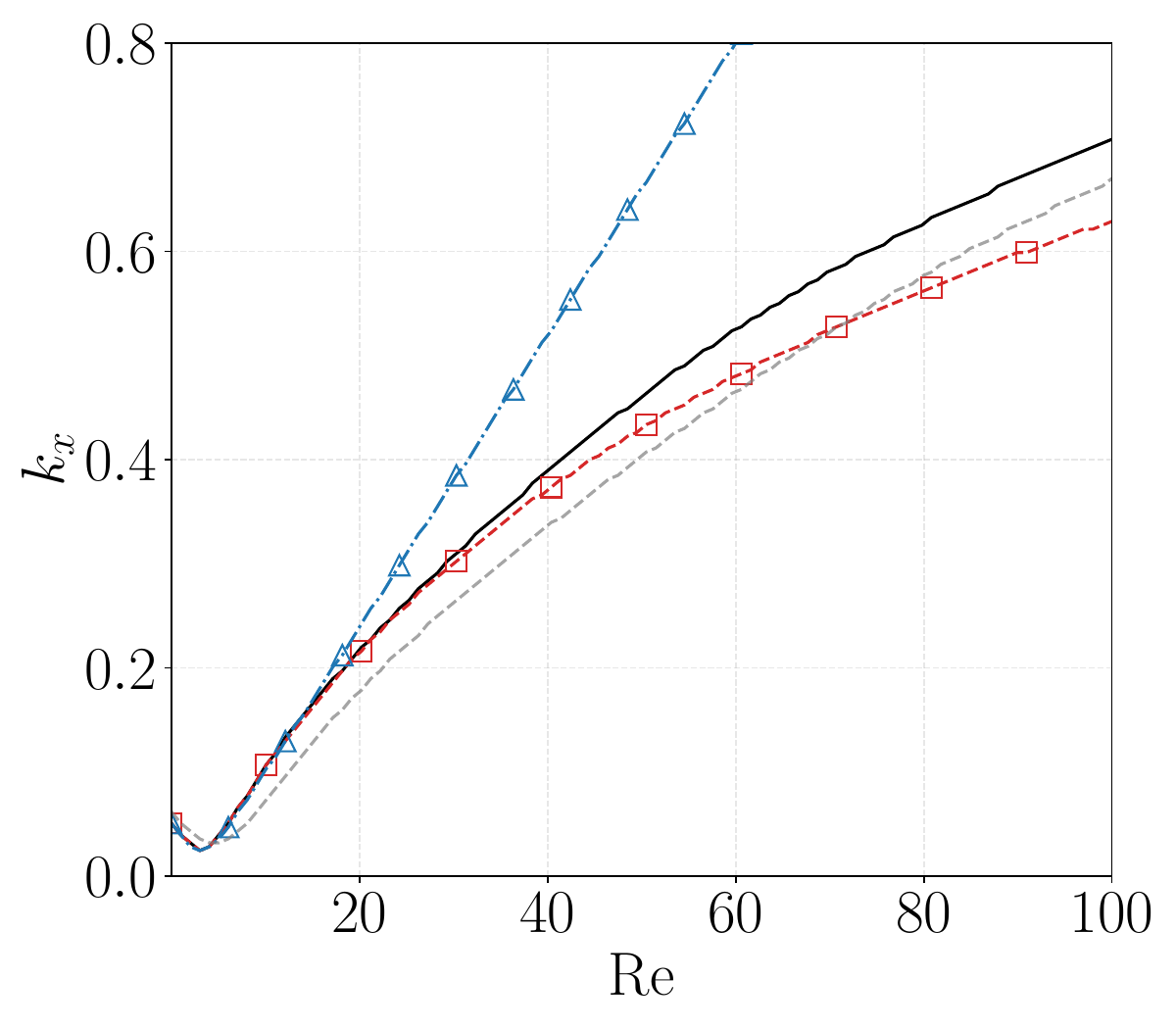}} &
        \subfloat[]{\includegraphics[width=0.5\textwidth]{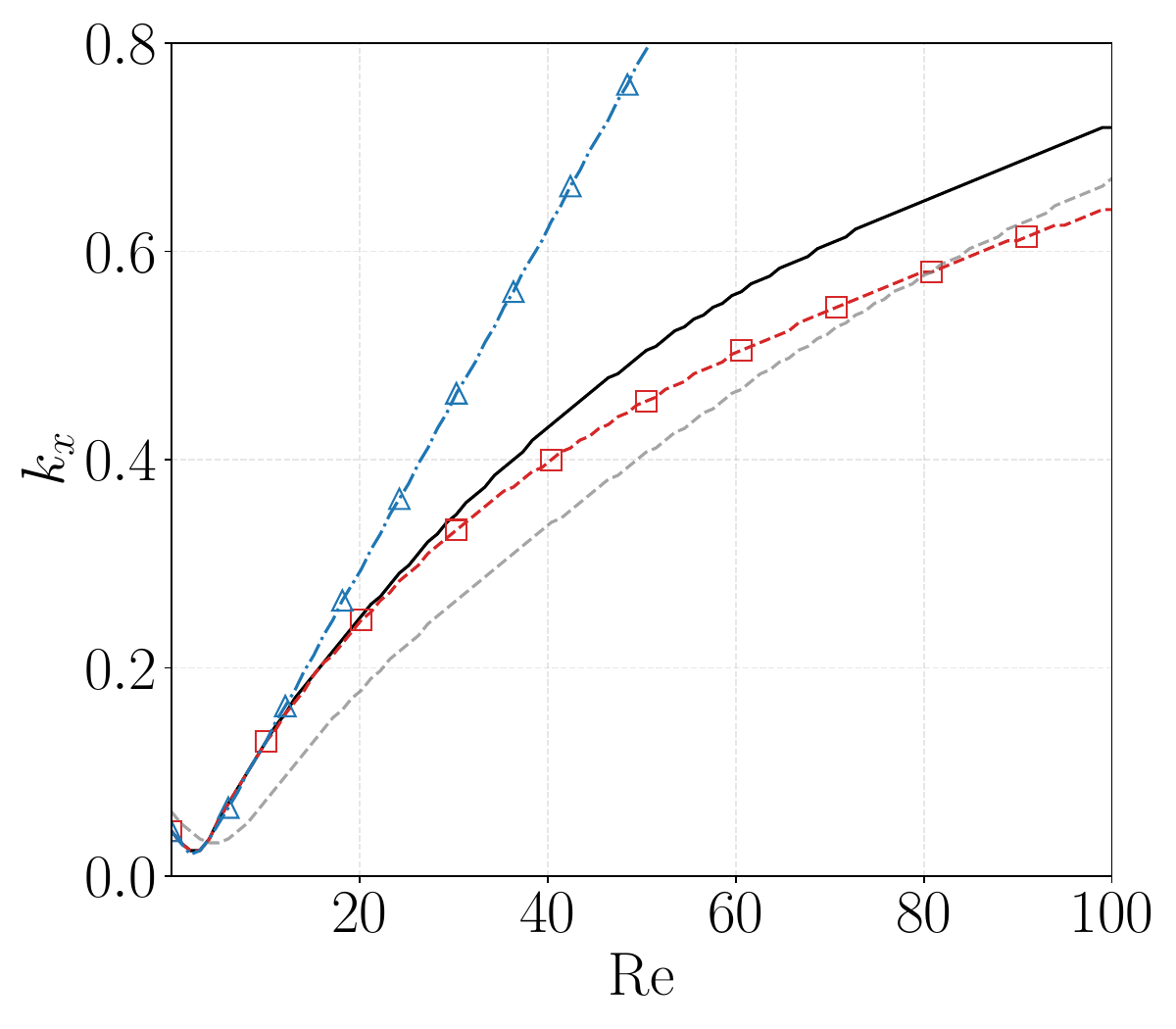}}
    \end{tabular}
    \caption{Neutral curve at different times for $\beta = \pi/12$, $\Pi = 0$, $\Gamma = 1000$, $\Pr = 7$, $\Vr = 4$, $\K = 0.01$, $\E = 0.1$, $\hat{\eta} = 1$, $\M = 10$ at time $\hat{t}$ equal to (a-c) $0.3\,\hat{t}_d$ and (b-d) $0.6\,\hat{t}_d$ for (a,b) evaporation and (c,d) condensation, obtained from the WIBL model (red dashed lines with markers), the numerical (continuous black lines and grey dotted line for $\hat{t}=0$), and the long-wave (dash-dotted lines with triangles) solutions of Navier-Stokes.}
    \label{fig:nc_var_time}
\end{figure}

The destabilising effects of vapour recoil and thermocapillarity are enhanced/reduced when the film thins/thickens during evaporation/condensation. Figure~\ref{fig:nc_var_time} presents the neutral curves in the $(k_x, \Rey)$ plane for streamwise perturbations for different instant of time with respect to the dry-out time $\hat{t}_d$, (a and c)  $\hat{t}=0.3 \hat{t}_D$ and (b and d) $\hat{t}=0.6 \hat{t}_d$, considering (a, b) evaporation and (c, d) condensation. For reference, the neutral curve at $\hat{t}=0$ is plotted as a grey dashed line. In the evaporation case, as the liquid film thins, the unstable wavenumber range shrinks due to the stronger influence of viscous effects. Concurrently, the minimum of the neutral curve shifts toward higher Reynolds numbers, and the maximum Reynolds number increases. This trend arises because, for thinner films, thermal effects become more pronounced, enhancing the destabilising influence of thermocapillarity and vapour recoil. Both LW and WIBL accurately capture this evolution. Moreover, as the film becomes thinner, their predictions deviate less from the OS solution over a broader range of Reynolds numbers, since inertial effects are weakened.

\begin{figure}
    \centering
    \begin{tabular}{cc}  % Create a 3-column table
        \subfloat[]{\includegraphics[width=0.5\textwidth]{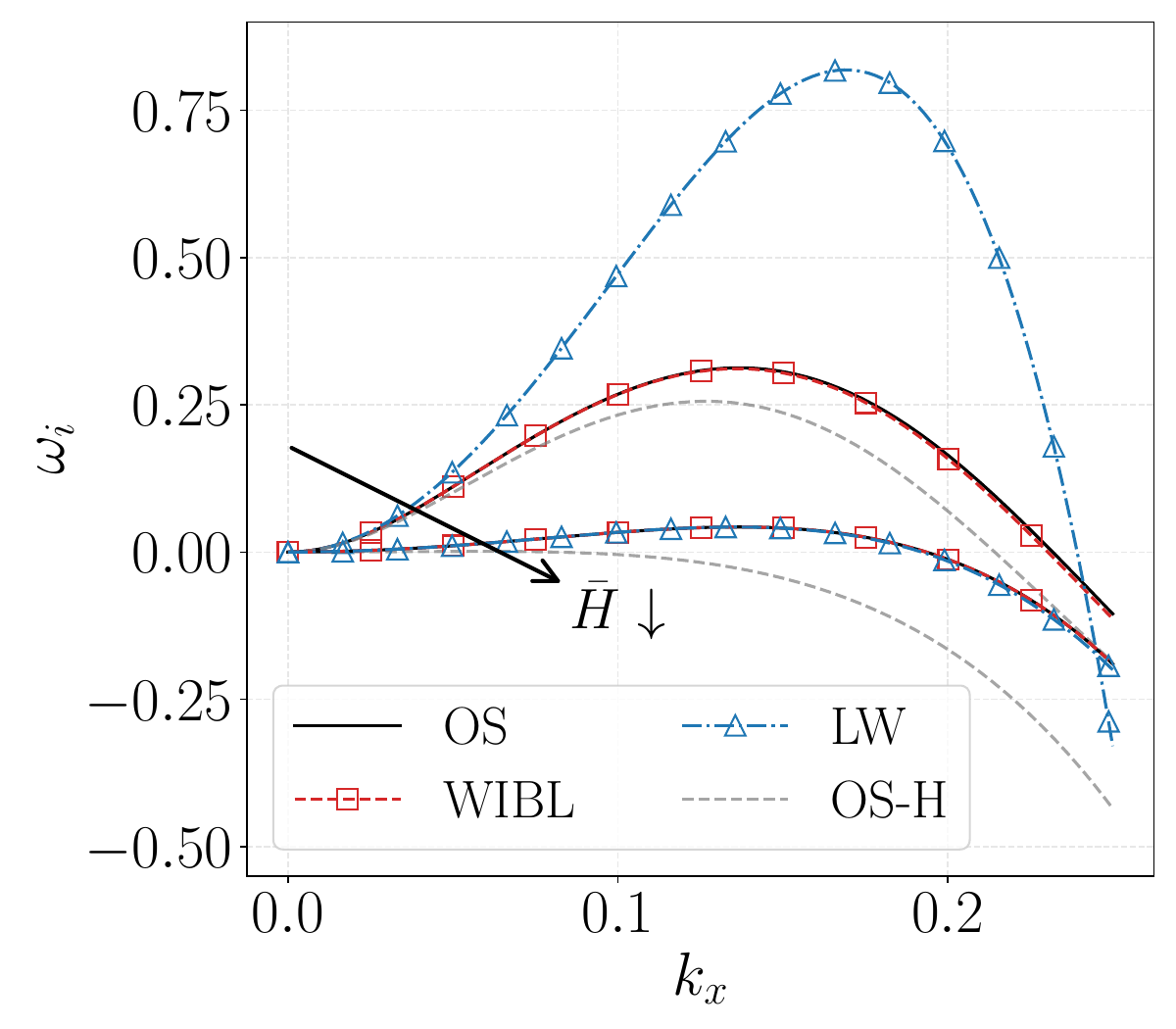}} &
        \subfloat[]{\includegraphics[width=0.5\textwidth]{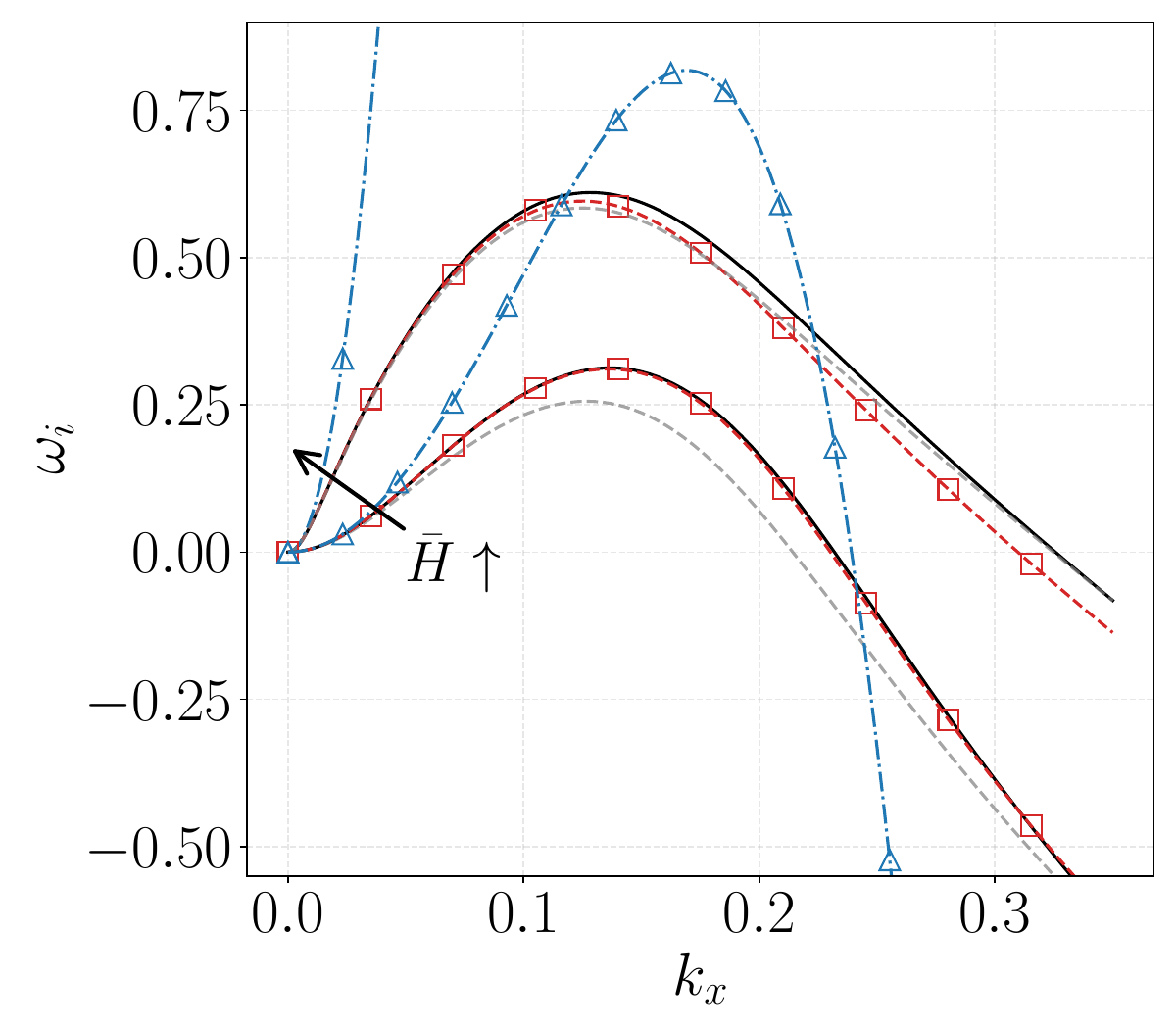}}\\
        \subfloat[]{\includegraphics[width=0.5\textwidth]{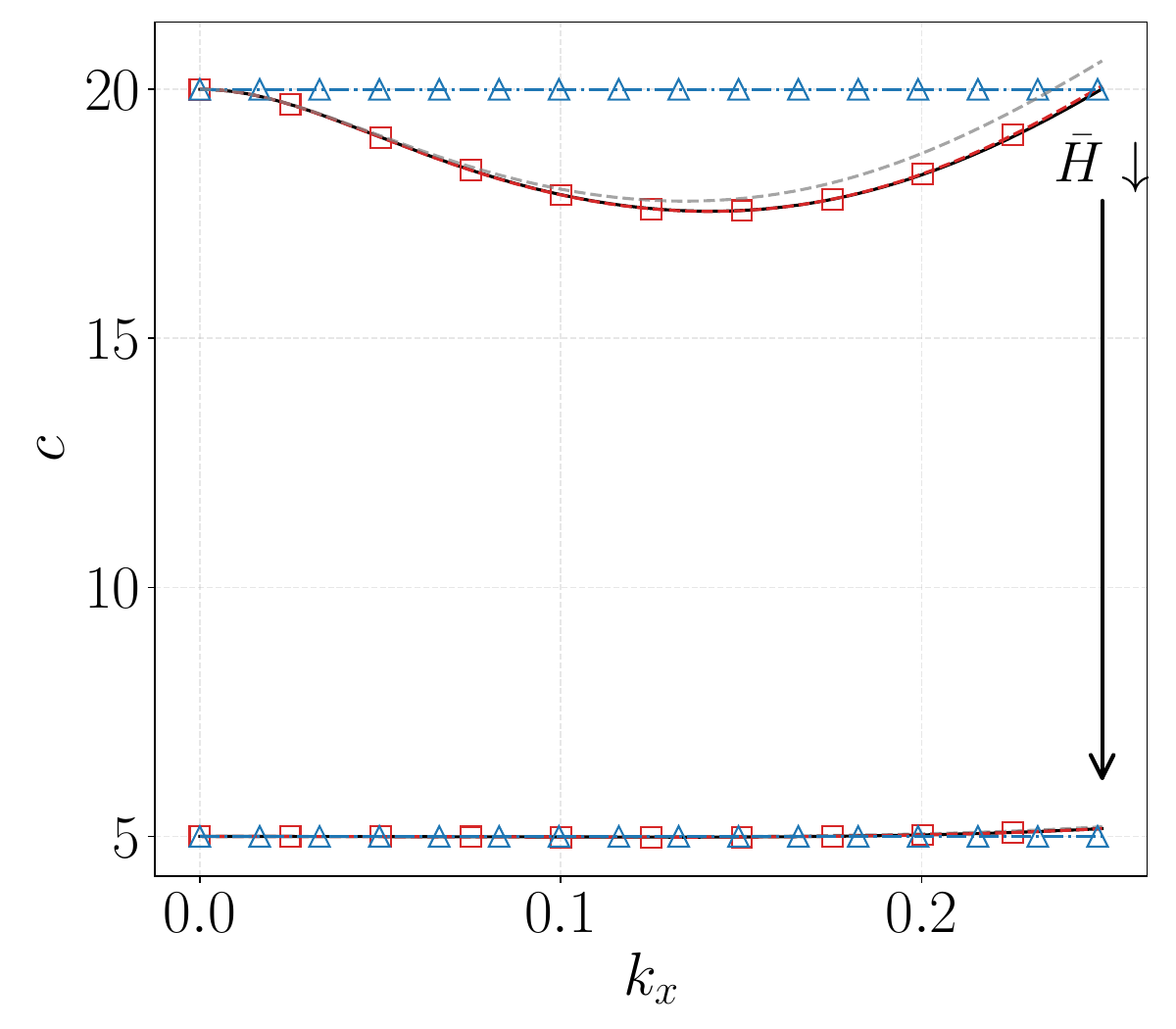}} &
        \subfloat[]{\includegraphics[width=0.5\textwidth]{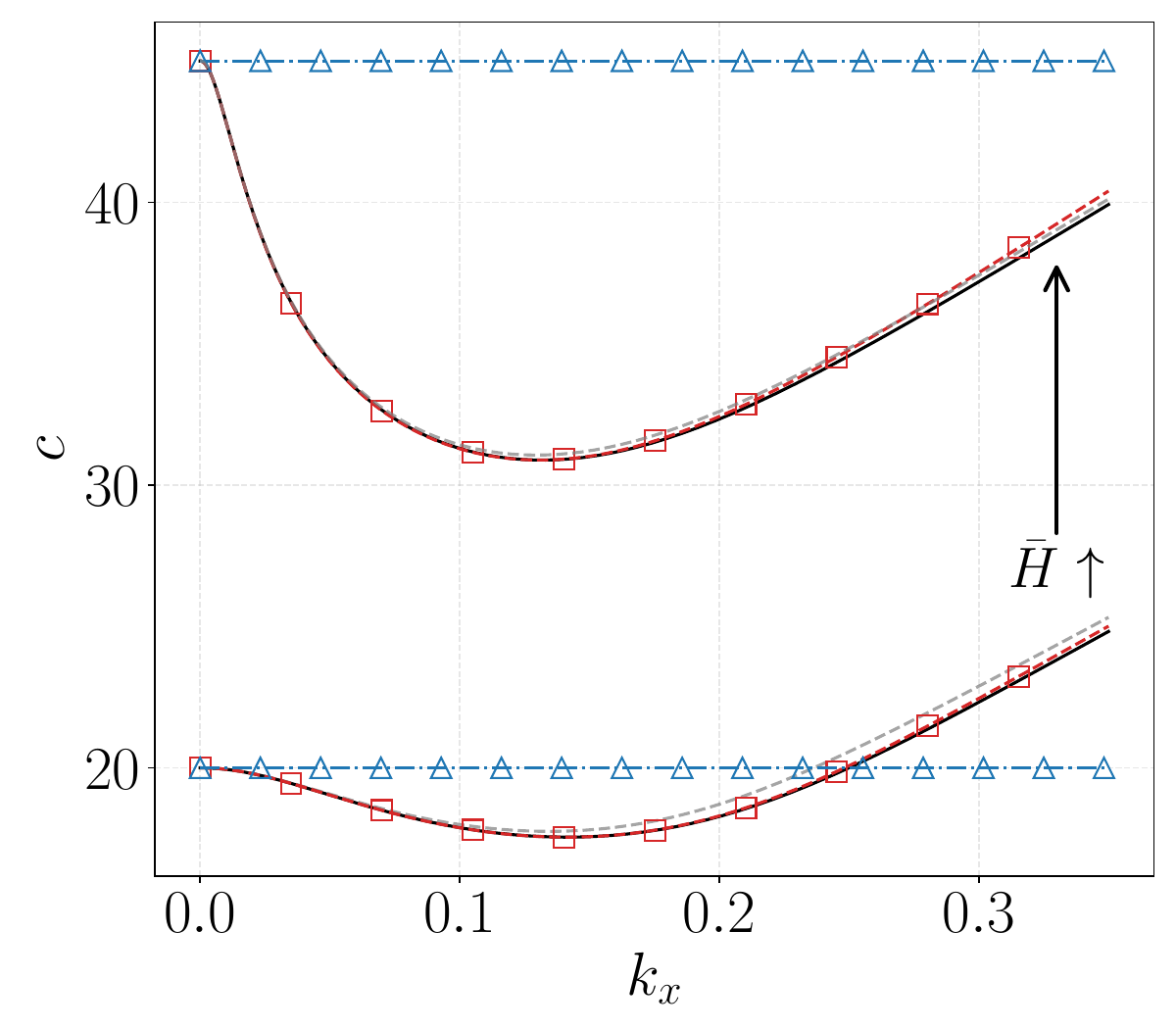}}
    \end{tabular}
    \caption{(a,b) growth rate $\omega_i$ and (c,d) phase speed $c$ for (a,c) evaporation and (b,c) condensation for different film thicknesses $\overline{H}$ (1 and 0.5 for evaporation) and (1 and 1.5 for condensation)  considering $\beta = \pi/12$, $\R=20$, $\Pi = 0$, $\Gamma = 1000$, $\Pr = 7$, $\Vr = 4$, $\K = 0.01$, $\E = 0.1$, $\M = 10$.}
    \label{fig:nc_var_time_omegai_c}
\end{figure}

The opposite trend occurs for condensation. As the film thickens, the unstable wavenumber region expands, though the variation is less pronounced than in the evaporation case. A thicker film strengthens inertial effects, thereby increasing the discrepancy between the WIBL and OS solutions and causing divergence at lower Reynolds numbers. In this case, the minimum of the neutral curve shifts slightly toward lower Reynolds numbers, while the maximum $k_x$ at $\R=0$ remains unchanged. This behaviour reflects the reduced influence of thermal effects at the free surface in thicker films, together with the stabilising role of the Marangoni effect. As shown by the analytical long-wave solution, the contributions of thermocapillarity and Marangoni effects scale approximately as $h^{-3}$ for small $k_x$.

Evaporation and condensation have a marked influence on the growth rates of the most unstable modes. Figure~\ref{fig:nc_var_time_omegai_c} shows (a,b) the growth rates and (c,d) the corresponding phase speeds for (a,c) evaporating and (b,d) condensing films of varying thickness $\overline{H}$, with the OS prediction for the H-mode indicated by the grey dashed line. As the film becomes thicker, both the growth rate and phase speed increase, and the curves converge toward the H-mode prediction, reflecting the dominance of the hydrodynamic mode in this regime. Conversely, as the film thins, both quantities decrease in magnitude, while discrepancies with respect to the isolated H-mode solution become more pronounced, owing to the increasing influence of the M- and E-modes.

Across all cases, the WIBL model captures these trends with excellent accuracy, closely matching the OS results. In contrast, the long-wave (LW) approximation increasingly deviates from the OS growth-rate predictions for $\overline{H} \geq 1$, and fails to reproduce the correct phase-speed behaviour except in the thinnest-film case $\overline{H}=0.5$, where the long-wave scaling inherent to the H- and E-modes remains valid.
\begin{figure}
    \centering
    \begin{tabular}{cc}  % Create a 3-column table
        \subfloat[]{\includegraphics[width=0.5\textwidth]{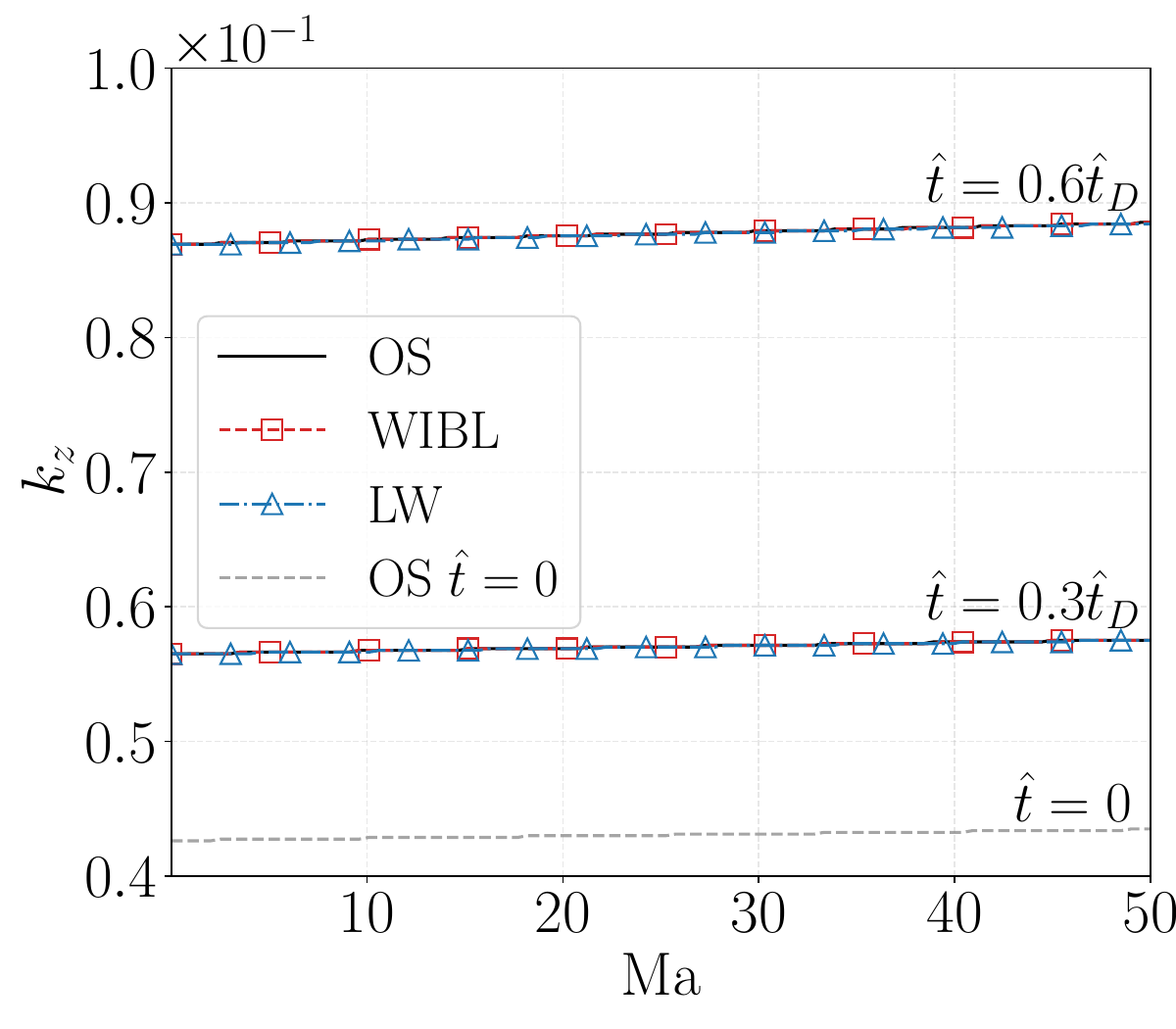}} &
        \subfloat[]{\includegraphics[width=0.5\textwidth]{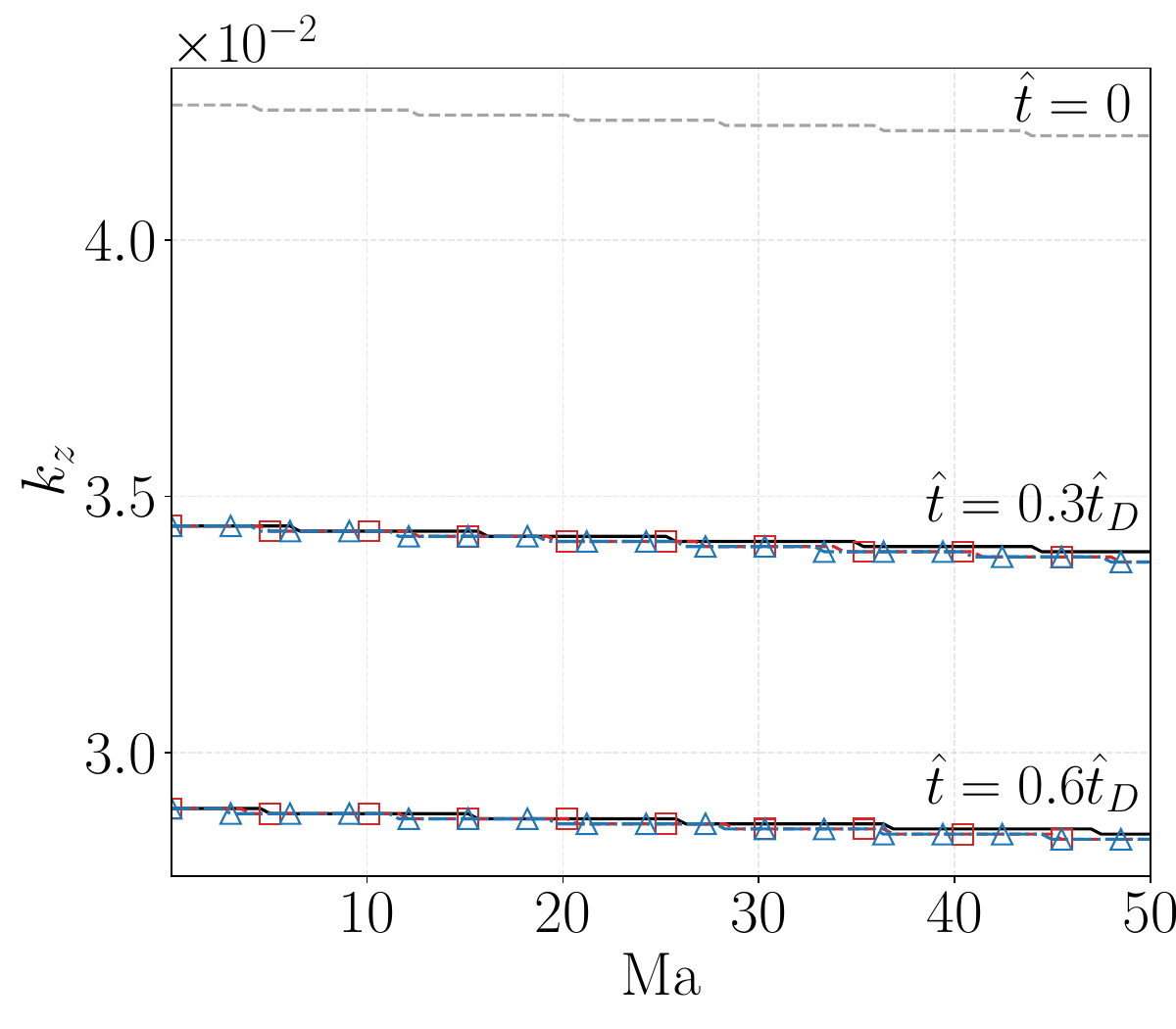}}
    \end{tabular}
    \caption{Neutral curve for evaporation at different times $\beta = \pi/12$, $\Pi = 0$, $\Gamma = 1000$, $\Pr = 7$, $\Vr = 2$, $\K = 0.01$, $\R = 0.1$ for (a) evaporation and (b) condensation}
    \label{fig:nc_var_time_kz}
\end{figure}

Moving to the spanwise perturbations, figure \ref{fig:nc_var_time_kz} presents the neutral curve in the $(k_z, \M)$ plane for $k_x=0$, considering the liquid film at different times during (a) evaporation and (b) condensation. As the film thins or thickens, the range of unstable wavenumbers expands or contracts, respectively. As with streamwise perturbations, thermal effects become increasingly important as the film thins, whereas their influence diminishes as the film thickens. A noteworthy observation is the difference in the rates of stabilisation and destabilisation. At the same time, the film becomes unstable much more rapidly under evaporation than under condensation. Specifically, in evaporation, the value of the neutral curve at $\hat{t}=0$ is less than half of its value at $\hat{t}=0.6$, whereas under condensation, the neutral curve at $\hat{t}=0$ is still less than twice its value at $\hat{t}=0.6$. This indicates that evaporation significantly enhances film instability, whereas condensation also improves stability for spanwise perturbations.

\begin{table}
  \centering
  \begin{tabular}{ccccccccccccccc}
    idx & Re & $\beta$ & $\Pi$ & $\Gamma$ & $\Pr$ & $\Vr$ & $\K$ & $\M$ & $\E$ & $\hat{t}$ & $\eta$ & $k_x^*$ & $k_z^*$ \\
    1 & 25 & 178.5 & 0 & 1000 & 7 & 4 & 0.01 & 10 & 0.808 & 0 & -1 & &\\
    2 & 44.65 & 170.727 & 0.041 & 649.69 & 2.00 & 0.76 & 0.73 & 9.20 & 0.522 & 3.18 & -1 & 8.03e-05 & 0.259 \\
    3 & 45.06 & 170.727 & 0.040 & 642.00 & 2.02 & 0.76 & 0.73 & 9.36 & 0.530 & 2.91 & -1 & 0        & 0.262 \\
    4 & 45.06 & 165.217 & 0.040 & 928.57 & 1.79 & 0.76 & 0.73 & 9.36 & 0.791 & 2.91 & -1 & 4.87e-05 & 0.172 \\
    5 & 45.06 & 174.082 & 0.053 & 642.00 & 2.16 & 0.33 & 0.63 & 0.80 & 0.371 & 2.91 & -1 & 3.49e-04 & 0.326 \\
    6 & 45.06 & 174.082 & 0.053 & 642.00 & 2.16 & 0.33 & 0.63 & 0.80 & 0.371 & 2.91 & -1 & 3.49e-04 & 0.326 \\
    7 & 45.06 & 175.270 & 0.094 & 5482.74& 2.44 & 5.47 & 0.84 & 0.81 & 0.464 & 2.63 & -1 & 2.95e-05 & 0.128 \\
    8 & 46.30 & 170.727 & 0.010 & 1053.96& 2.02 & 3.96 & 0.74 & 9.36 & 0.808 & 4.89 & -1 & 1.59e-05 & 0.207 \\
    9 & 59.01 & 170.727 & 0.056 & 1224.79& 2.02 & 0.76 & 0.58 & 1.80 & 0.433 & 2.80 & -1 & 3.31e-05 & 0.218 \\
    10 & 64.54 & 165.878 & 0.005 & 120.95 & 2.15 & 8.68 & 1.00 & 2.45 & 0.792 & 3.11 & -1 & 3.97e-05 & 0.542 \\
    11 & 72.40 & 173.341 & 0.058 & 1784.37& 1.45 & 4.02 & 0.53 & 2.76 & 0.493 & 2.70 & -1 & 7.88e-06 & 0.237 \\
    12 & 81.76 & 173.624 & 0.093 & 1347.20& 1.15 & 2.61 & 0.57 & 3.98 & 0.976 & 2.78 & -1 & 0        & 0.291 \\
    13 & 82.69 & 177.156 & 0.090 & 5686.76& 2.26 & 3.79 & 0.79 & 8.98 & 0.712 & 3.41 & -1 & 8.93e-07 & 0.217 \\
    14 & 85.73 & 173.624 & 0.026 & 1301.44& 2.14 & 0.70 & 0.69 & 2.00 & 0.961 & 2.78 & -1 & 5.12e-06 & 0.303 \\
    15 & 92.64 & 165.217 & 0.007 & 576.70 & 2.13 & 8.94 & 0.03 & 3.28 & 0.245 & 3.02 & -1 & 1.09e-05 & 0.313 \\
  \end{tabular}
\caption{Values of the nondimensional groups, including the inclination angle $\beta$, substrate temperature $\hat{\eta}$, and evaporation/condensation time $\hat{t}$, for which the maximum growth rate $\omega_i^*$ has $k_z^* \neq 0$, along with their associated index (idx), ordered by ascending $\R$.}
\label{tab:res_opt_kz}
\end{table}
\begin{figure}
    \centering
    \subfloat[]{\includegraphics[width=0.48\textwidth]{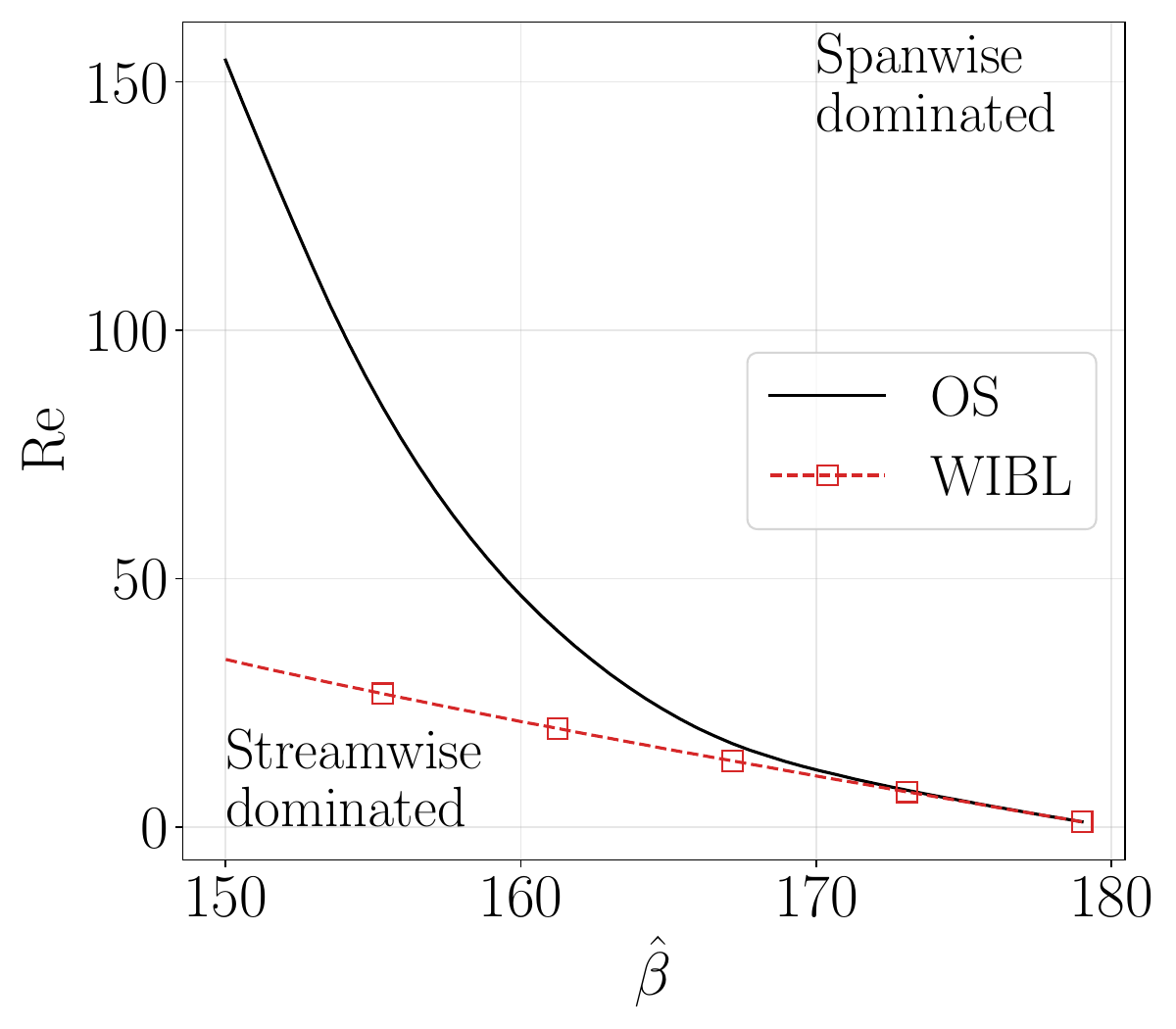}}
    \hfill
    \subfloat[]{\includegraphics[width=0.48\textwidth]{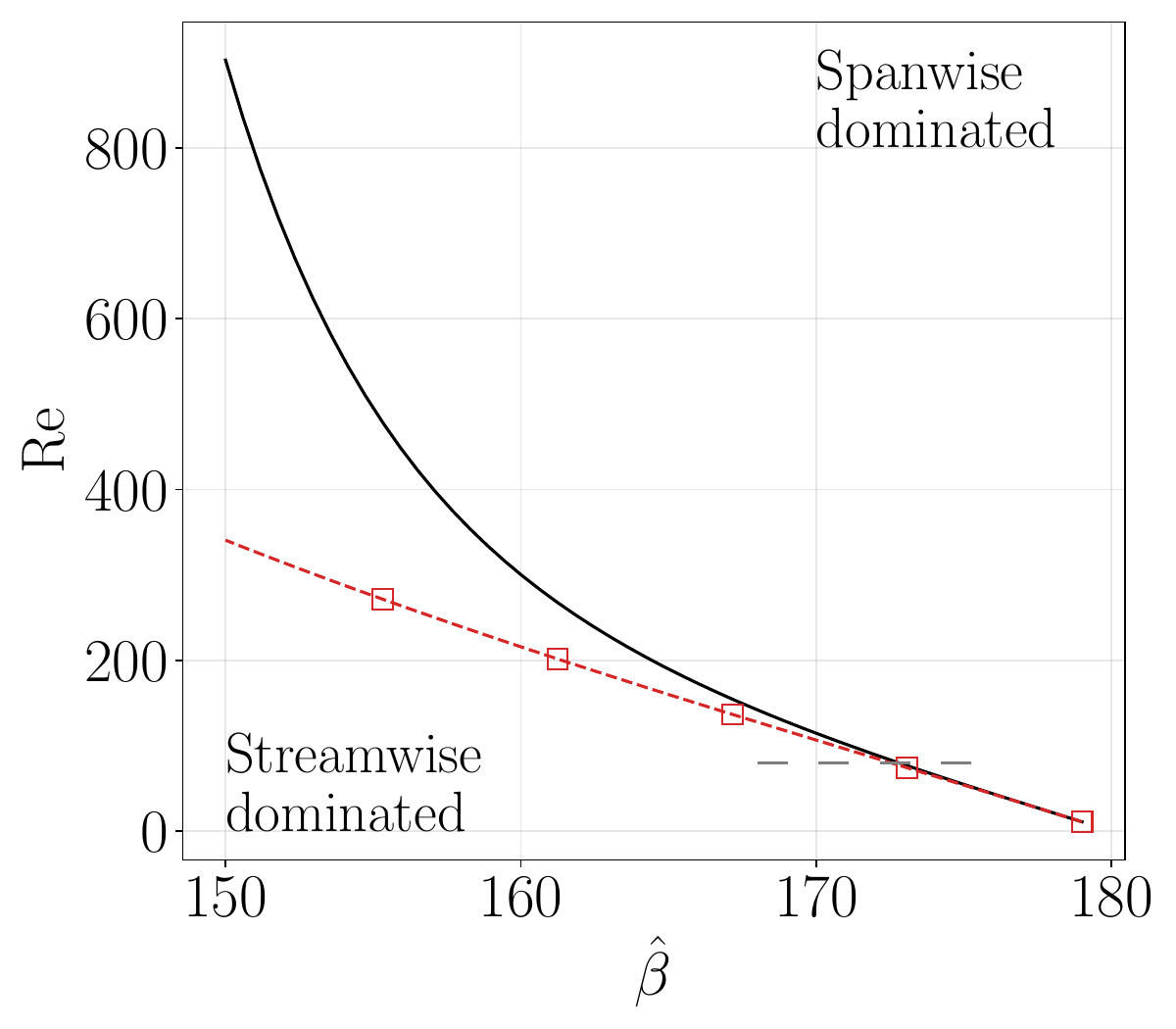}
    \label{fig:threshold_stream_span_max_growth_b}}
    \caption{Threshold between streamwise (below) and spanwise (above) dominated instability obtained with the WIBL model (red dashed line with squares) and by solving the Orr–Sommerfeld eigenvalue problem (black continuous line), shown in the $(\beta,\R)$ plane for cases (a)~8 and (b)~1.}
    \label{fig:threshold_stream_span_max_growth}
\end{figure}

Although the growth rates associated with spanwise perturbations are typically an order of magnitude smaller than those of streamwise perturbations in the cases examined above, this does not guarantee that spanwise modes are always less unstable. Since Squire’s theorem has not been established for liquid films with phase change, it is not certain that streamwise disturbances ($k_z = 0$) always yield the largest growth rate $\omega_i^*$. To investigate in which condition 3D perturbation might be the most unstable, the location of the maximum growth rate $\omega_i^*$ in the $(k_x, k_z)$ plane was analysed by solving an optimisation problem involving the nondimensional numbers and the flow conditions as described in section \ref{subsec:linear_stab_analysis}. 

Table~\ref{tab:res_opt_kz} reports the results of this optimisation, listing representative parameter combinations for which $\omega_i^*$ corresponds to a nonzero spanwise wavenumber ($k_z \neq 0$), as obtained from numerical solutions of the Orr-Sommerfeld problem. In all such cases, the most unstable mode occurs for nearly pure spanwise perturbations with $k_x \approx 0$, and for inclination angles between $160^\circ$ and $170^\circ$, corresponding to an almost horizontal hanging film.

This spanwise-dominated instability is not represented by the LW solution, which consistently links $\omega_i^*$ to a streamwise disturbance. By maximising the imaginary part of $\omega \approx \omega_0 \lVert \bm{k} \rVert + i \omega_1 \lVert \bm{k} \rVert^2$ with respect to $k_x$ and $k_z$ considering the solution \eqref{eq:growth_rate_o1_lw_OS}, $\omega_i^*$ is given by:
\begin{equation}
    \omega_i^* = -\frac{\Ct\overline{H}^3}{3}+\frac{\overline{H}^3\Vr}{(\overline{H}+\K)^3}+\frac{\overline{H}^9 \R^4}{36 \Gamma }+\frac{\hat{\eta}\overline{H}^2 \K\M}{\Pr(\overline{H}+\K)^2},
\end{equation} 
with associated maximum streamwise $k_x^*$ and spanwise $k_z^*$ wavenumber components reading:
\begin{equation}
 k_x^* = \frac{\overline{H}^3\R}{6\Gamma}, \qquad\qquad\qquad\qquad k_z^* = 0.
\end{equation} 

\begin{figure}
    \centering
    \begin{tabular}{cc}  % Create a 3-column table
        \subfloat[]{\includegraphics[width=0.5\textwidth]{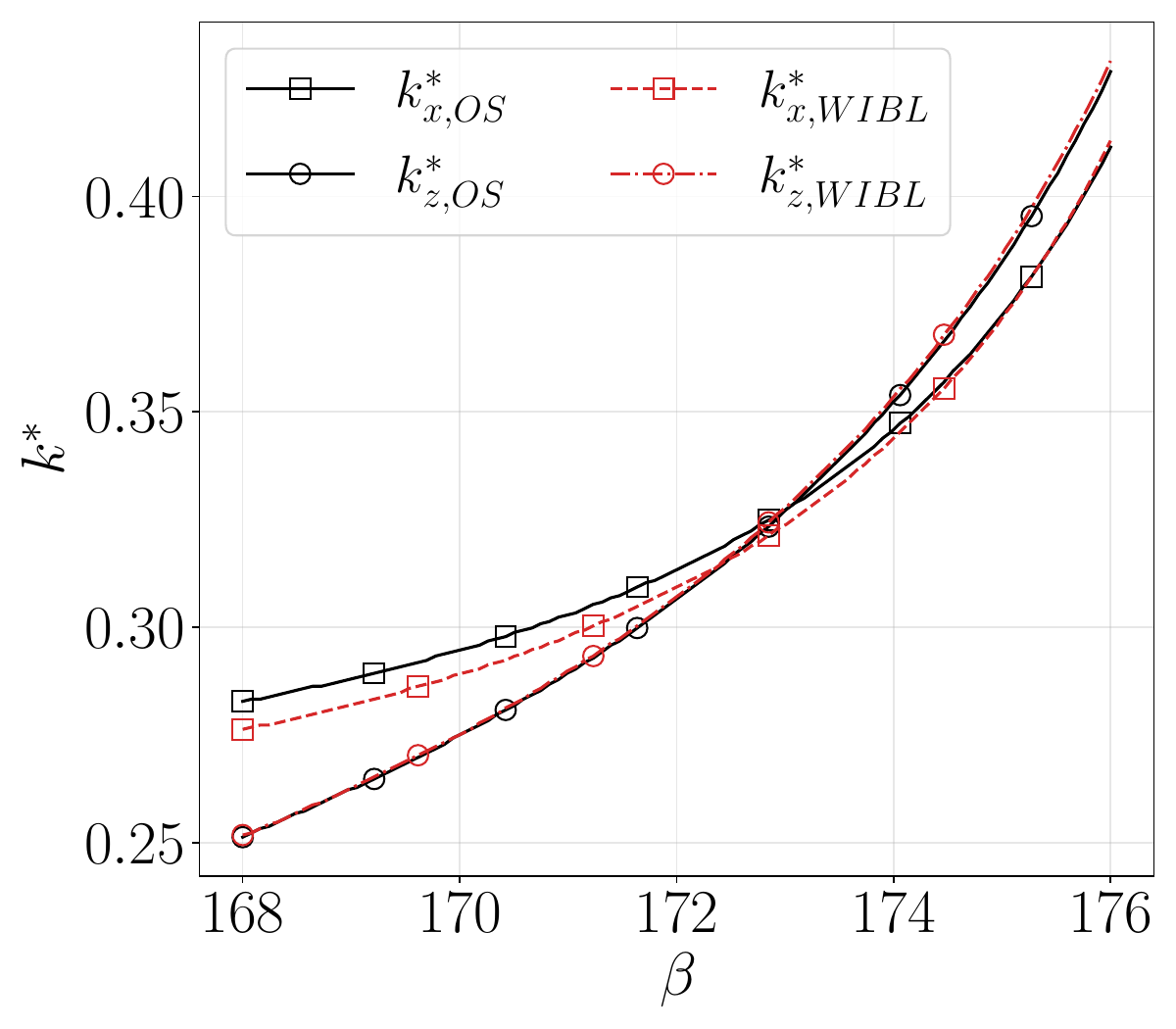}} &
        \subfloat[]{\includegraphics[width=0.5\textwidth]{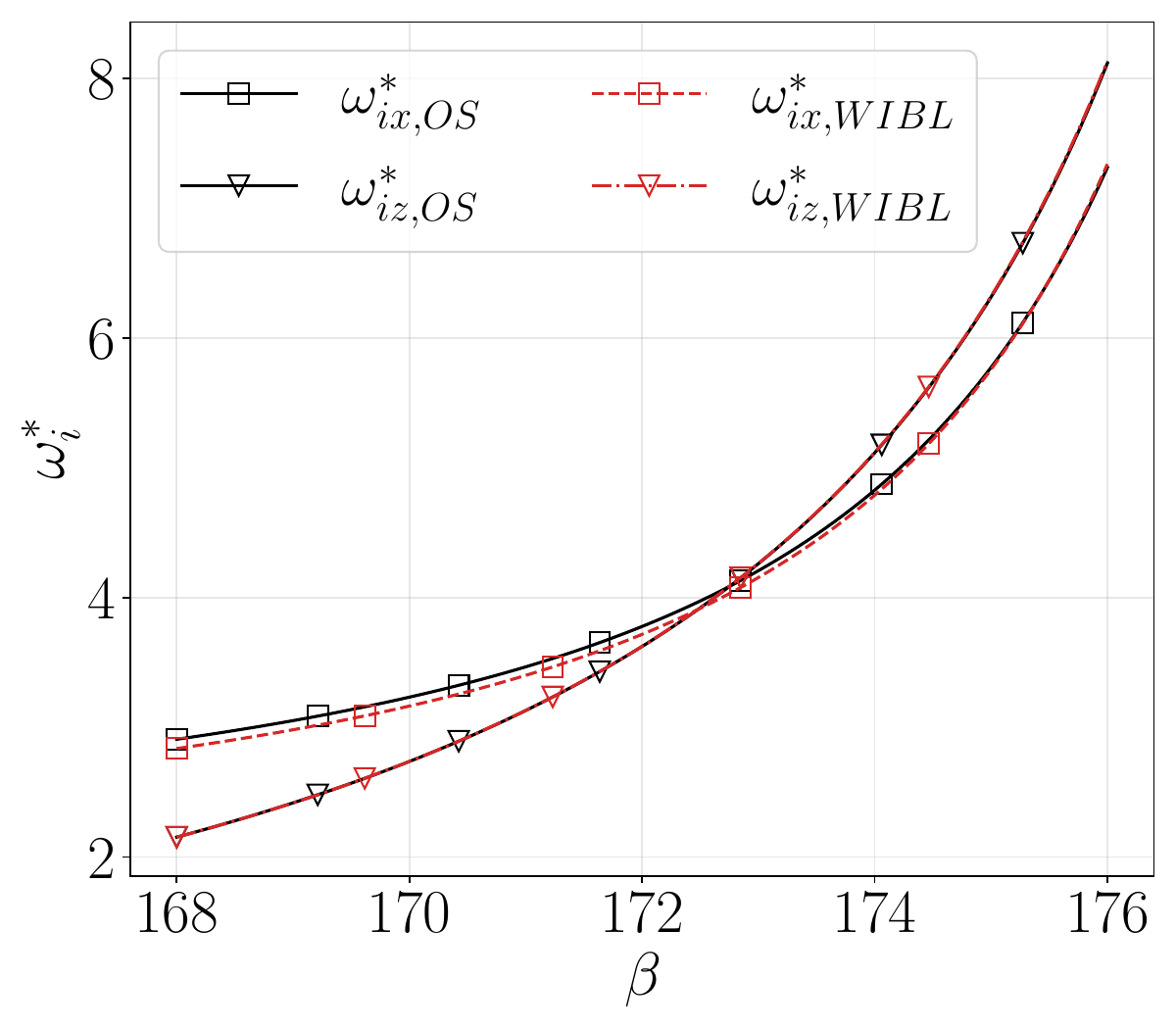}}\\
        \subfloat[]{\includegraphics[width=0.5\textwidth]{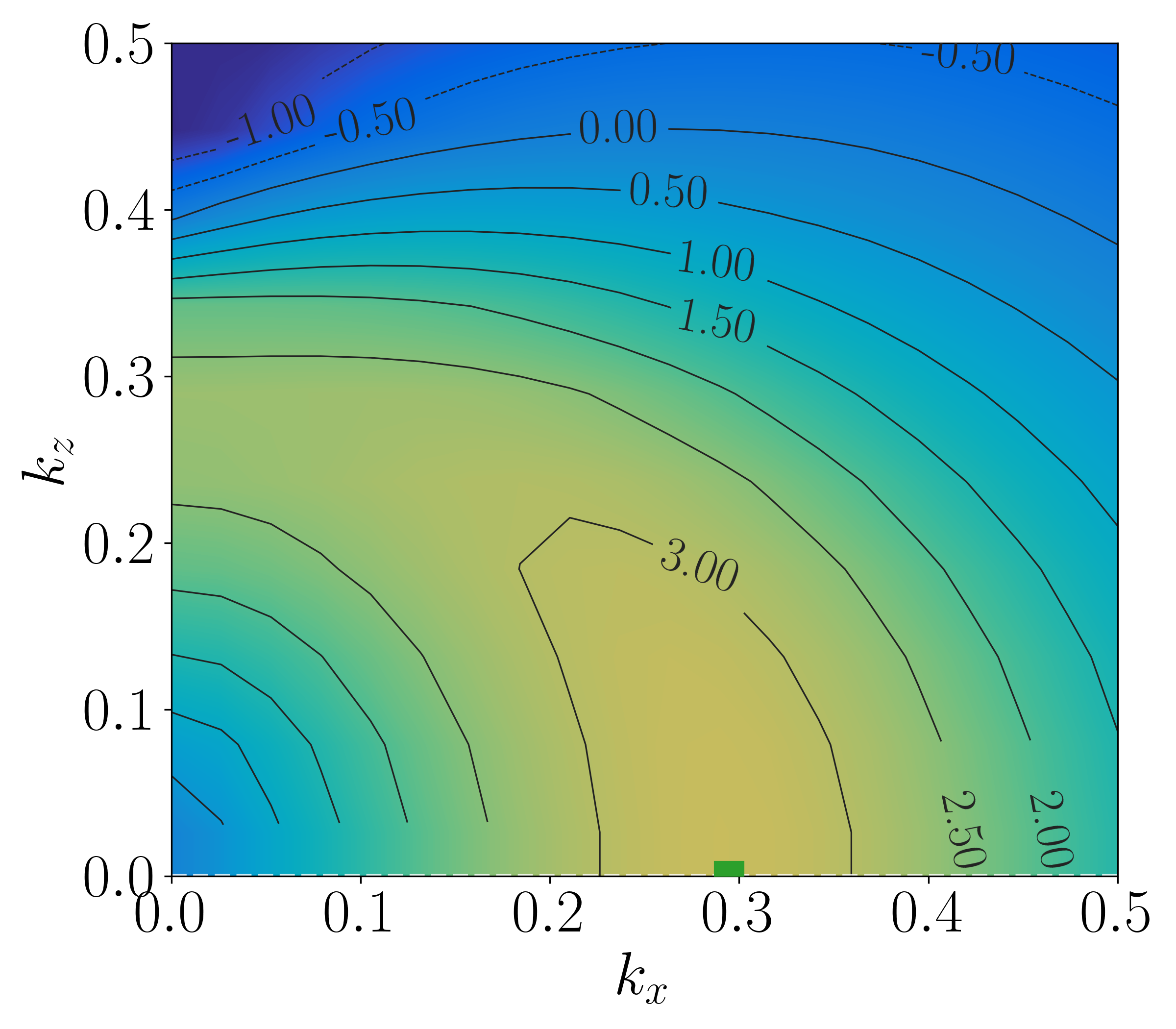}} &
        \subfloat[]{\includegraphics[width=0.5\textwidth]{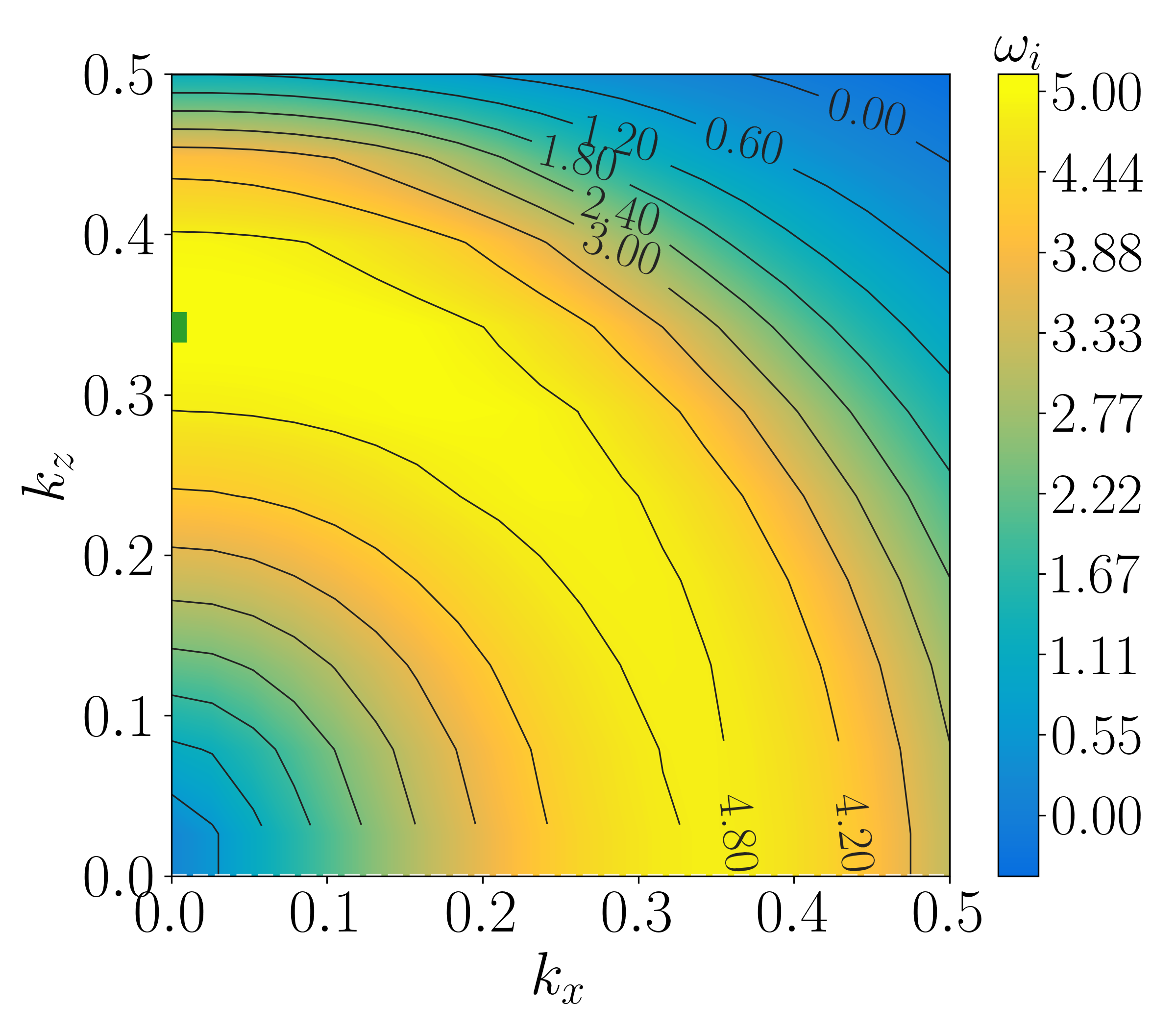}}
    \end{tabular}
    \caption{(a,b) maximum streamwise $\omega_{ix}^*$ and spanwise $\omega_{iz}^*$ growth rates with the associated wavenumbers $k_x^*$ and $k_z^*$ for cases 1 keeping the nondimensional number constants and varying $\beta$ and (c,d) growth rate $\omega_i$ (colours) in the wavenumbers space ($k_x,k_z$) for conditions 3 for inclination angle $\beta$ equals to (a) $170^{\circ}$ and (b) $174^{\circ}$ with the position of the maximum growth rate $\omega_i^*$ (green square).}
    \label{fig:omegai_vs_kx_kz}
\end{figure}

In contrast, the WIBL model captures the spanwise-dominated regime with reasonable accuracy. For fixed nondimensional parameters and flow conditions, I tracked the threshold separating streamwise- and spanwise-dominant instabilities in the $(\R,\beta)$ plane and compared predictions from the Orr-Sommerfeld (OS) equations with those from the WIBL formulation. Figure~\ref{fig:threshold_stream_span_max_growth} shows this threshold as given by the WIBL model (red dashed line with square markers) and by the OS computations (black solid line), along with the regions of streamwise-dominated (below) and spanwise-dominated (above) instability corresponding to parameter sets (a)~8 and (b)~1 in Table~\ref{tab:res_opt_kz}. The two curves agree closely for small~$\R$, where inertia is weak. The transition reflects the Rayleigh–Taylor mechanism overtaking the Kapitza (streamwise) mode, thereby switching the linear operator's most unstable direction.

This mode switch is evident when examining the evolution of the maximum growth rate $\omega_i^*$ in both directions. Figure \ref{fig:omegai_vs_kx_kz} reports (a) the streamwise and spanwise wavenumbers $k_x^*$ and $k_z^*$ associated with the largest growth rate $\omega_i^*$, and (b) the corresponding maximum growth rates $\omega_{ix}^*$ and $\omega_{iz}^*$ along x and z, respectively, as functions of the inclination angle $\beta$. Results are shown for the OS solution (black continuous lines with markers) and the WIBL model (dashed lines with markers), using parameter set 1 in Table~\ref{tab:res_opt_kz} for $\R=50$ and varying $\beta$ in the range $168^{\circ}$–$174^{\circ}$ (highlighted in Figure~\ref{fig:threshold_stream_span_max_growth_b} by the loosely dashed grey segment). The figure also displays the full dependence of $\omega_i(k_x,k_z)$ for $\beta=170^{\circ}$ and $174^{\circ}$, with the global maximum marked by a green square. As $\beta$ increases, the spanwise maximum growth rate exceeds the streamwise one, and the associated spanwise wavenumber exceeds the streamwise counterpart. This confirms that, for sufficiently steep inclinations, the spanwise mode dominates because Rayleigh–Taylor destabilisation outweighs inertial effects in the streamwise direction. The WIBL model reproduces this behaviour with good accuracy, capturing both the growth rates and their associated wavenumbers, in agreement with the OS results.

Increasing $\beta$ beyond the transition also induces a qualitative change in the topology of $\omega_i(k_x,k_z)$. In the streamwise-dominated regime, the growth rate decreases as perturbations acquire a stronger spanwise component. In contrast, in the spanwise-dominated regime, the growth rate increases with spanwise wavenumber, reaching its maximum at $k_x \neq 0$. This behaviour results from the combined amplification of the Rayleigh–Taylor and Kapitza instabilities: as the film approaches vertical, gravity becomes increasingly destabilising in the spanwise direction, leading to larger spanwise growth rates and reshaping the instability landscape.

% ---- Nonlinear dynamics ---- 
\subsection{Nonlinear test cases with substrate temperature varying in space and time}
\label{subsec:res_nonlinear_sims}
This subsection analyses the nonlinear response of the liquid film to unsteady and spatially non-uniform substrate temperature variations in two and three dimensions, with physical properties and nondimensional groups reported in table \ref{tab:temperature_conditions} and following the setup described in Subsection~\ref{subsec:nonlin_sim}. In two dimensions, WIBL predictions for the evolution of the free-surface profile and surface temperature are compared with full-model simulations. In three dimensions, the WIBL equations are solved on a square periodic domain to investigate the evolution of an initial hump under three thermal conditions: an isothermal substrate, a spatially uniform but time-dependent substrate temperature, and a substrate with combined spatial and temporal temperature variations. The film dynamics are characterised by the minimum film thickness $\hat{h}_{\min}$ and the spatially averaged surface mass flux $\langle \hat{J} \rangle_{\Omega}$.

\begin{figure}
    \centering
    \begin{tabular}{cc}  
        \subfloat[]{\includegraphics[width=0.5\textwidth]{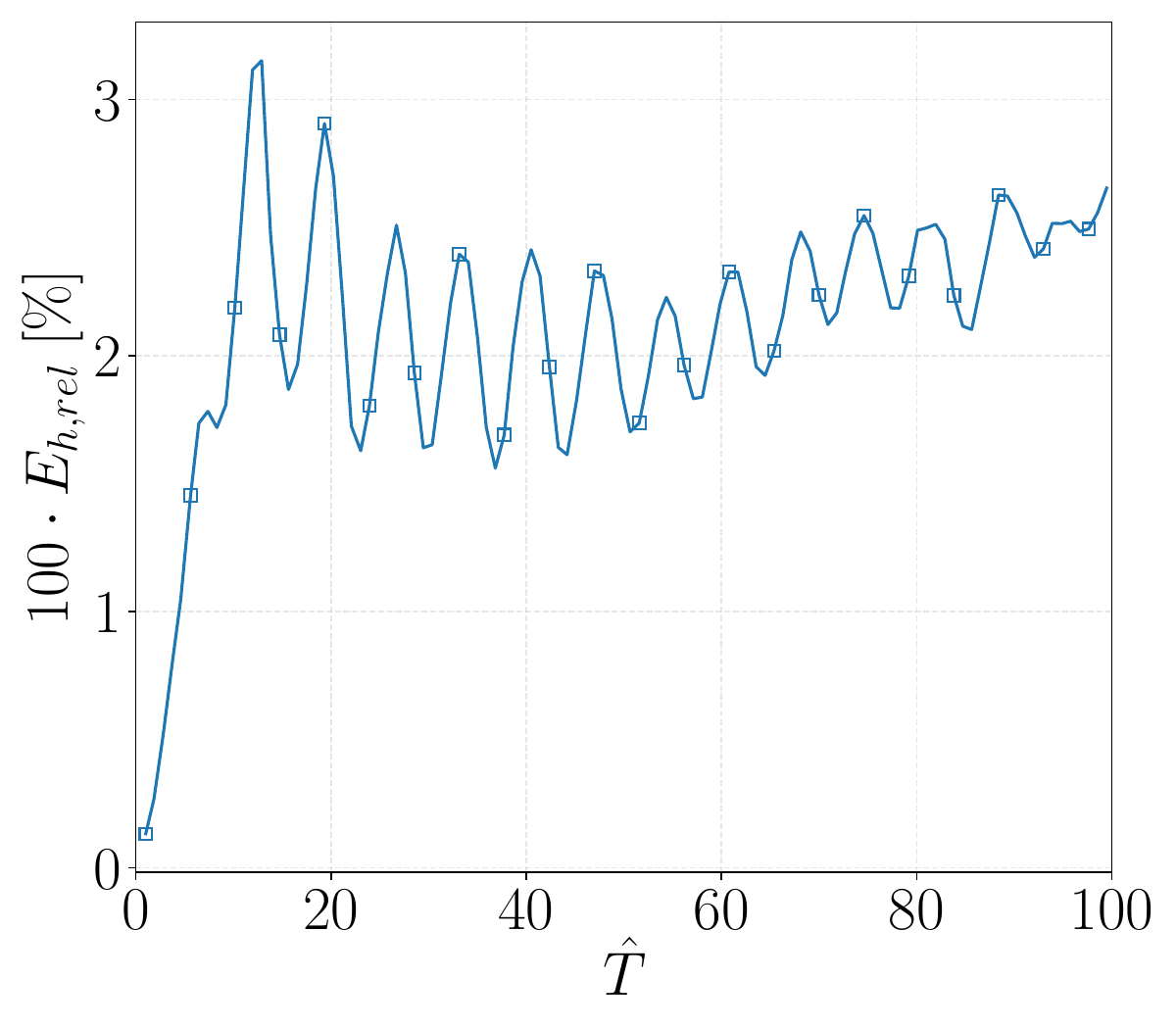}} &
        \subfloat[]{\includegraphics[width=0.5\textwidth]{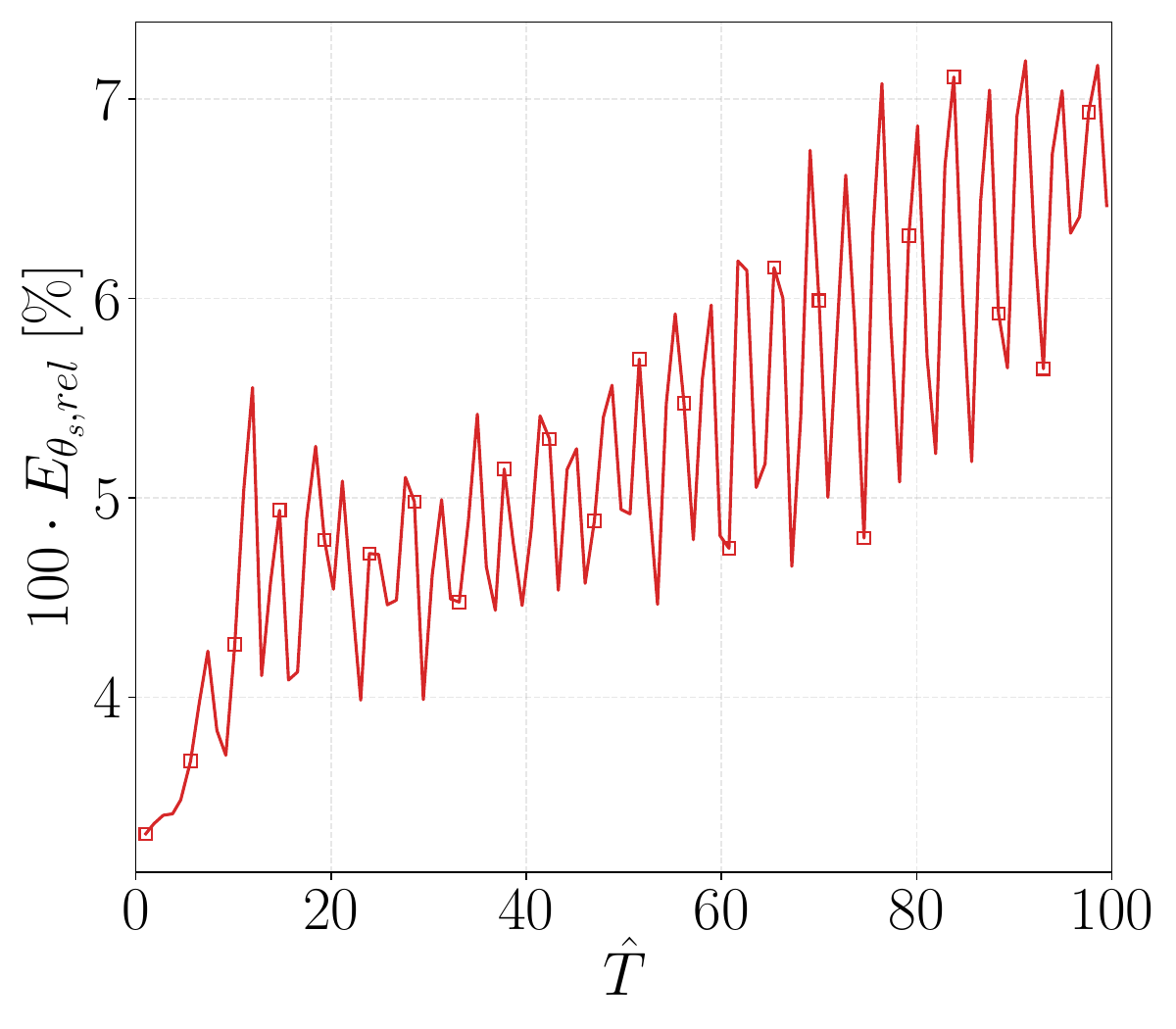}}
    \end{tabular}
    \caption{Error between the simulation of the full model equation and the WIBL equations expressed in percentage for (a) the film thicknesses and (b) the free-surface temperature for a 2D simulation with substrate temperature variations.}
    \label{fig:error_WIBL_COMSOL}
\end{figure}
\begin{figure}
  \centering
  % ===== Row 1 =====
  \begin{subfigure}{0.495\textwidth}
    \centering
    \includegraphics[width=\textwidth]{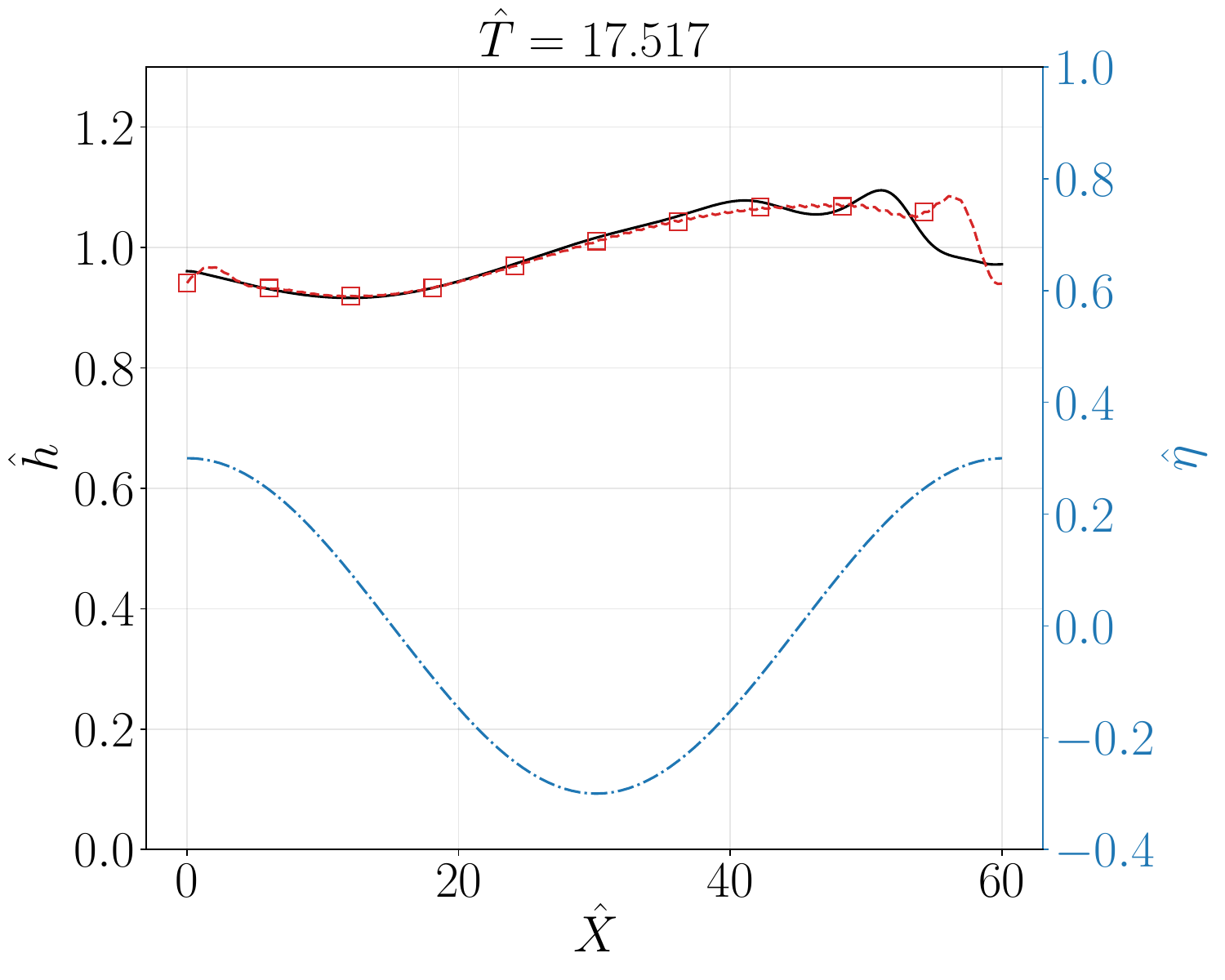}
    \caption{}
    \label{fig:caseA}
  \end{subfigure}
  \hfill
  \begin{subfigure}{0.495\textwidth}
    \centering
    \includegraphics[width=\textwidth]{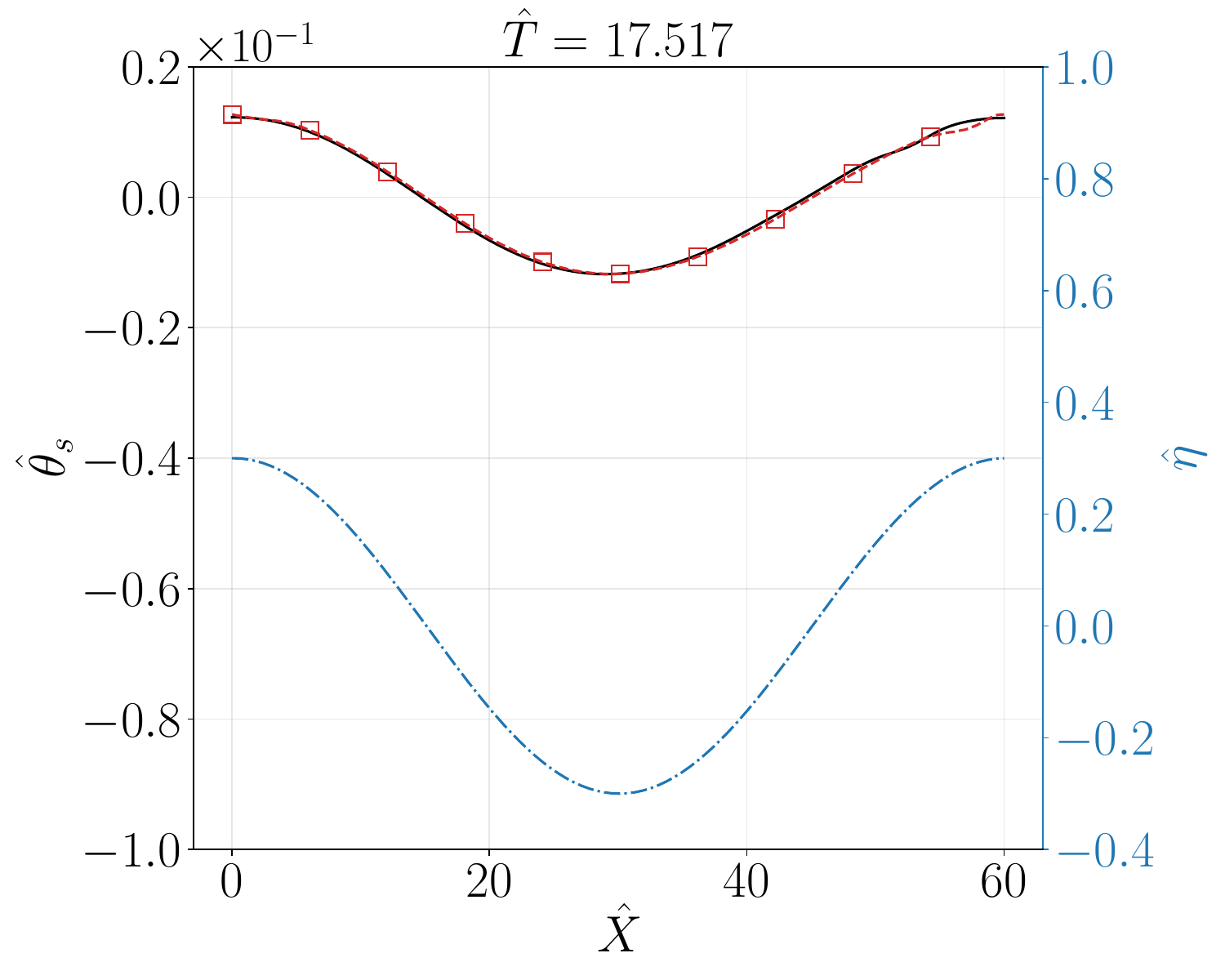}
    \caption{}
    \label{fig:caseB}
  \end{subfigure}
  % ===== Row 2 =====
  \vskip\baselineskip  % vertical space between rows
  
  \begin{subfigure}{0.495\textwidth}
    \centering
    \includegraphics[width=\textwidth]{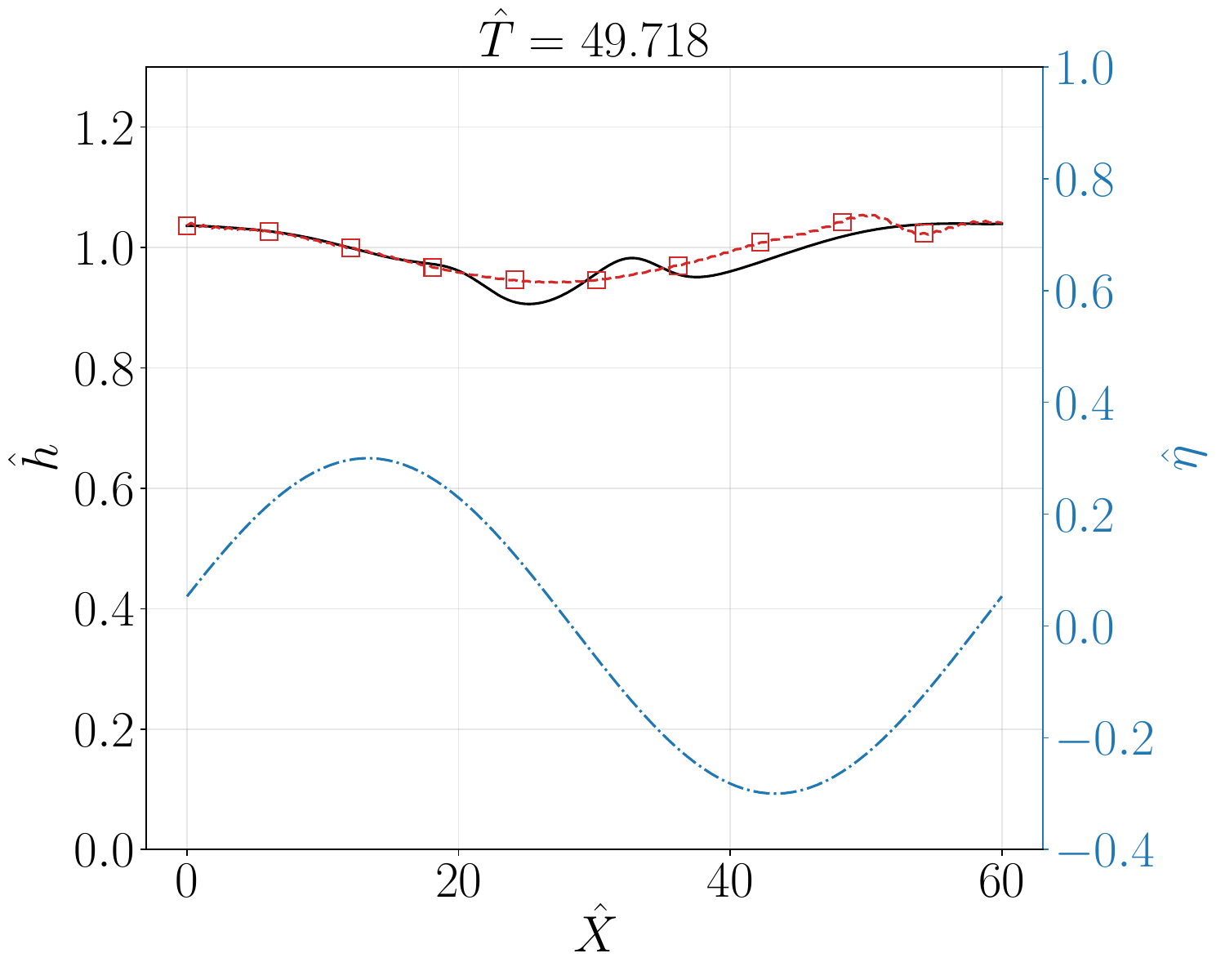}
    \caption{}
    \label{fig:caseC}
  \end{subfigure}
  \hfill
  \begin{subfigure}{0.495\textwidth}
    \centering
    \includegraphics[width=\textwidth]{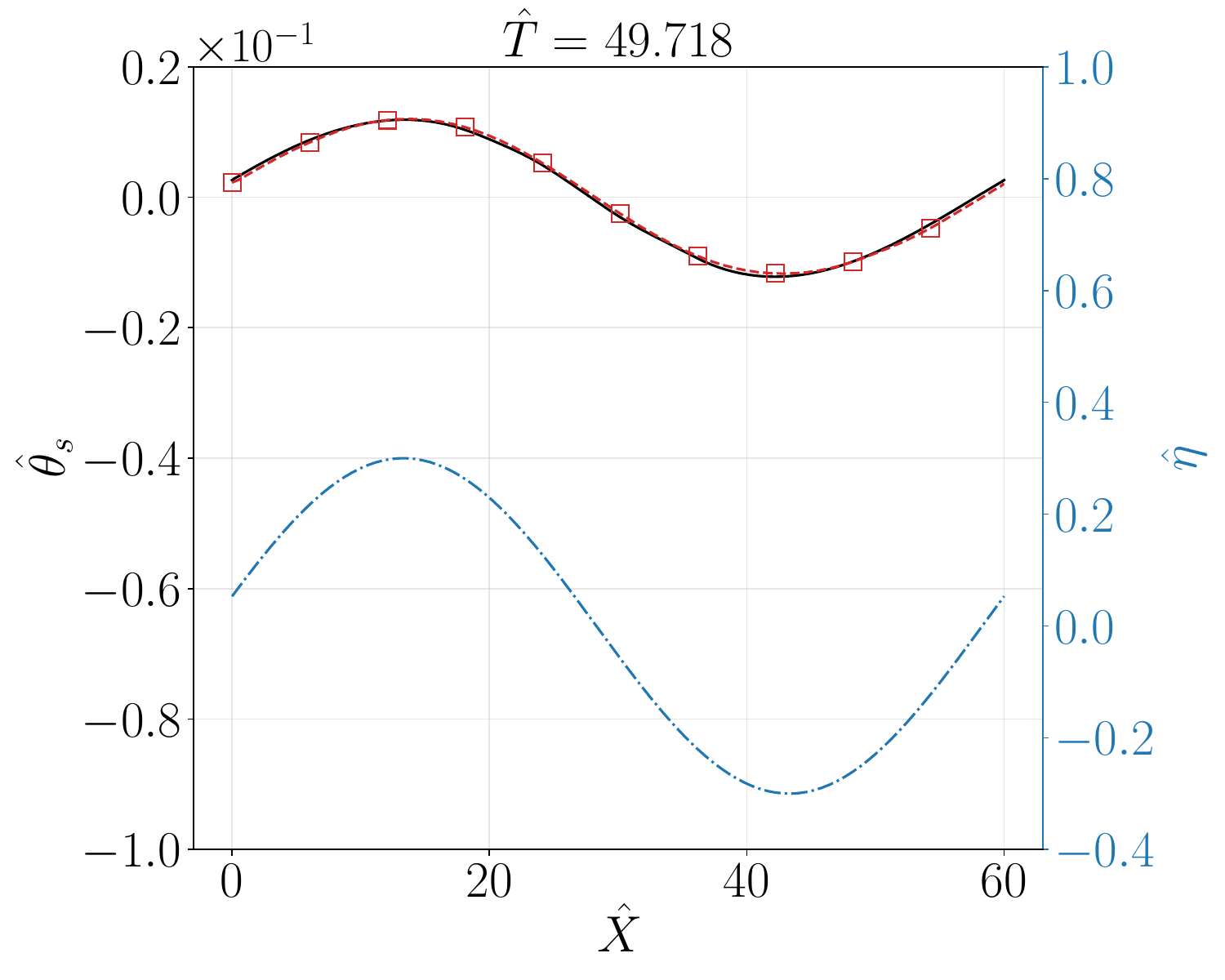}
    \caption{}
    \label{fig:caseD}
  \end{subfigure}
  % ===== Row 3 =====
  \vskip\baselineskip  % vertical space between rows

  \begin{subfigure}{0.495\textwidth}
    \centering
    \includegraphics[width=\textwidth]{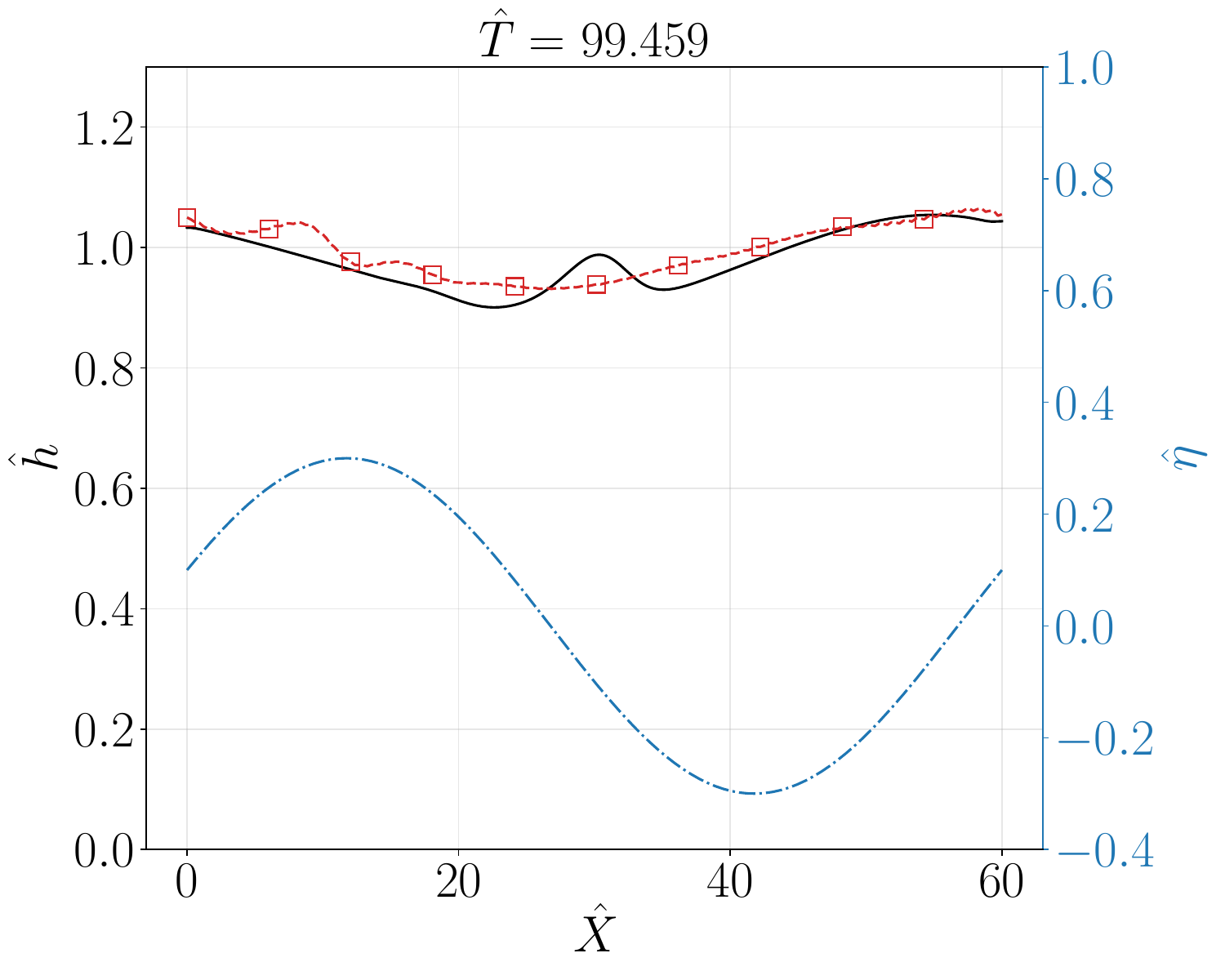}
    \caption{}
    \label{fig:caseC}
  \end{subfigure}
  \hfill
  \begin{subfigure}{0.495\textwidth}
    \centering
    \includegraphics[width=\textwidth]{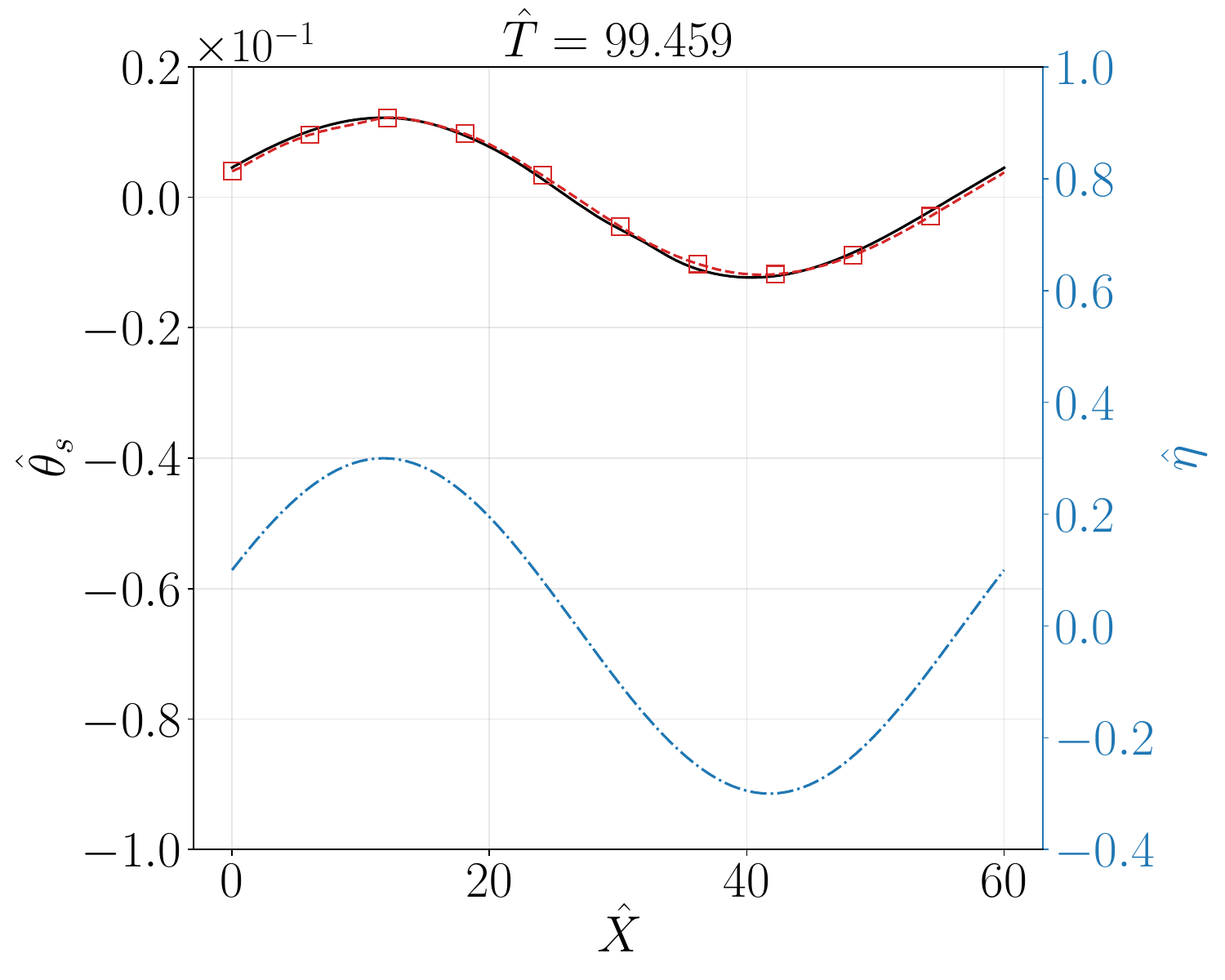}
    \caption{}
    \label{fig:caseD}
  \end{subfigure}
  
    \caption{Evolution of the free-surface thickness (left) and temperature (right) in the nonlinear test case, obtained by solving the WIBL equations (dashed red line with squares) and the full governing equations (solid black line), with harmonic temperature variation of the substrate shown (green dash-dotted line, right $y$-axis) for different time steps.}
    \label{fig:comp_COMSOL_2D}
\end{figure}

\begin{figure}
    \centering
    \includegraphics[width=\textwidth]{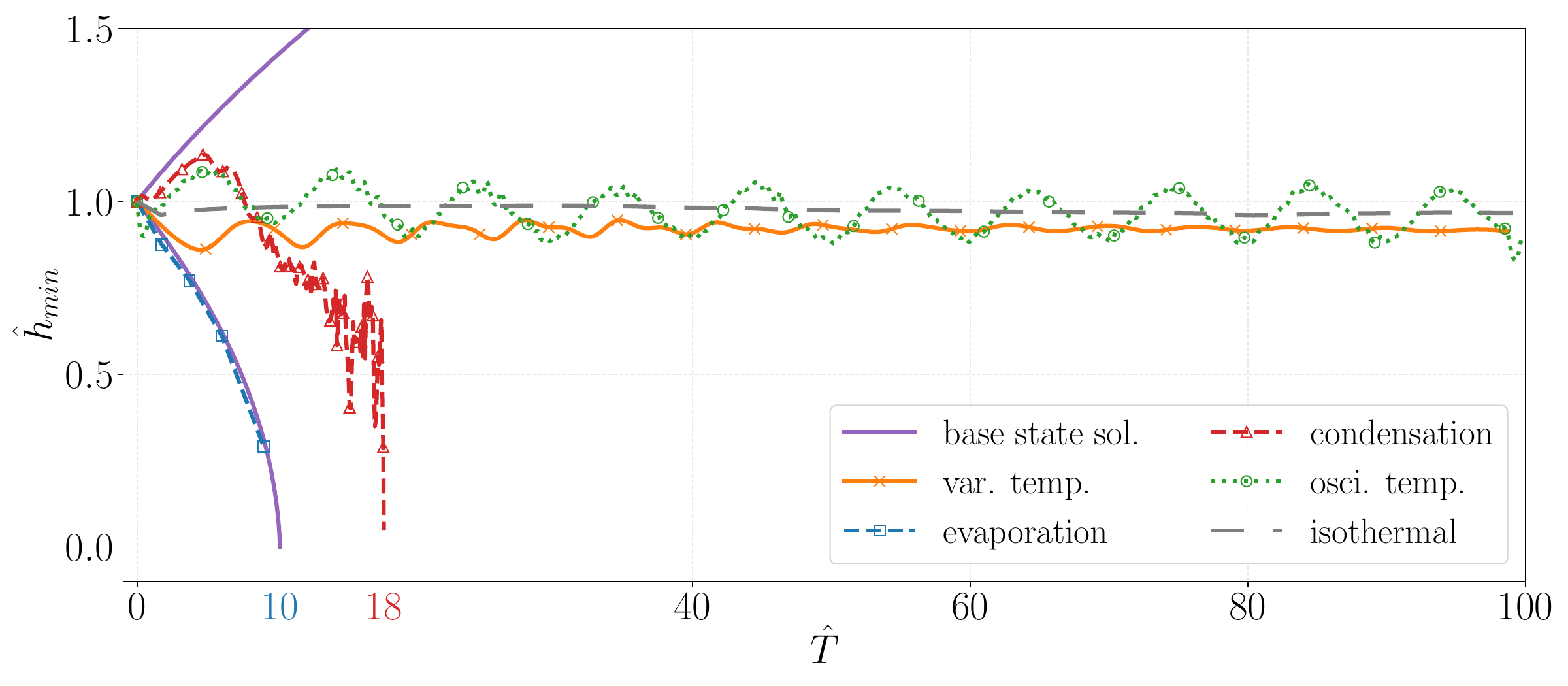}
    \includegraphics[width=\textwidth]{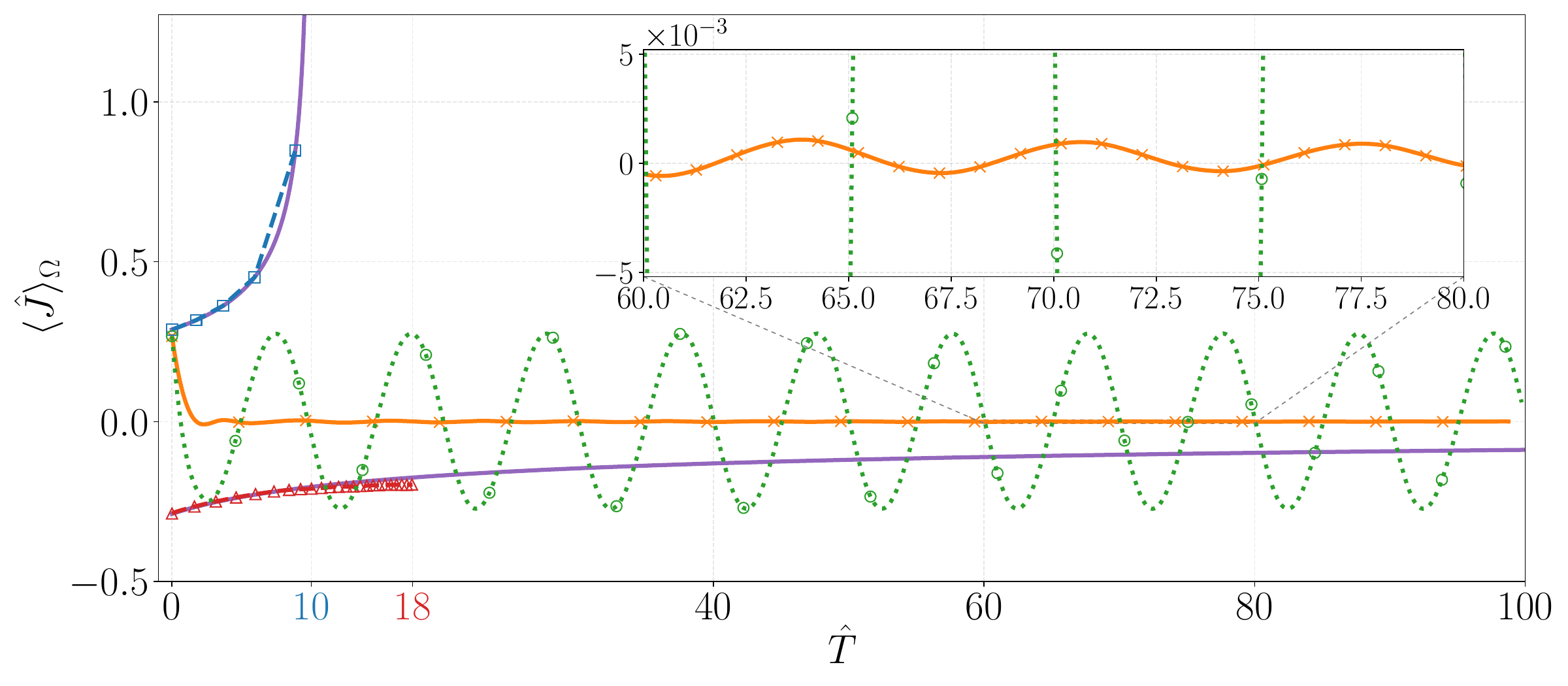}
    \caption{Evolution of (top) the minimum film thickness $\hat{h}_{\min}$ and (bottom) the spatially averaged free-surface mass flux $\langle\hat{J}\rangle_{\Omega}$, considering the flat-film weakly evaporating solution (purple solid line), the isothermal case (gray dashed line), the steady-state substrate temperature with evaporation $\hat{\eta}=0.3$ (blue dashed line with squares) and condensation $\hat{\eta}=-0.3$ (red dashed line with triangles), as well as a substrate temperature oscillating in time (green dotted line with circles) and varying in space and time (orange solid line with crosses).}
    \label{fig:evol_J_avg_h_min}
\end{figure}
\begin{figure}
    \centering
    \includegraphics[width=\textwidth]{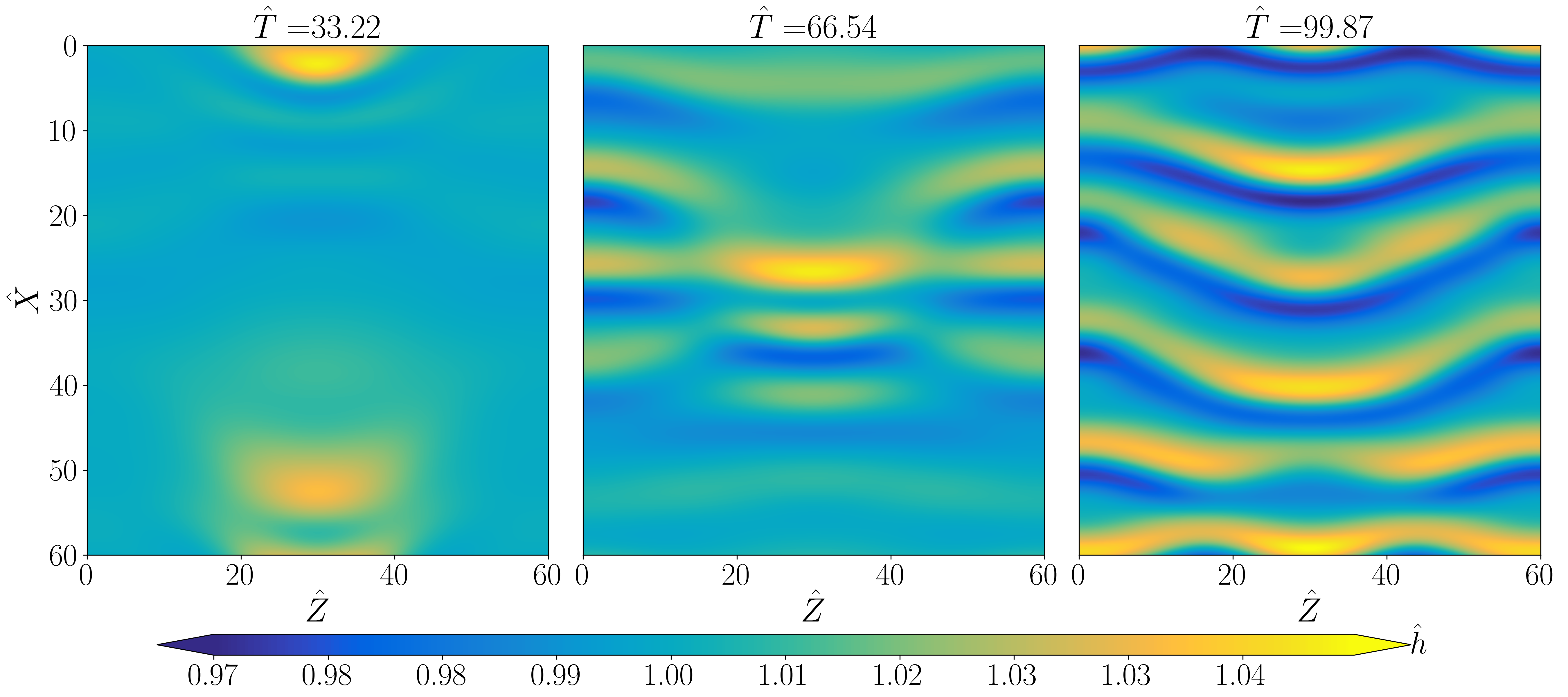}
    \caption{Liquid film thickness $\hat{h}$ (colour map) for the isothermal case at different times.}
    \label{fig:3D_isothermal}
\end{figure}

\begin{figure}
  \centering
  \begin{subfigure}{\textwidth}
    \centering
    \includegraphics[width=\textwidth]{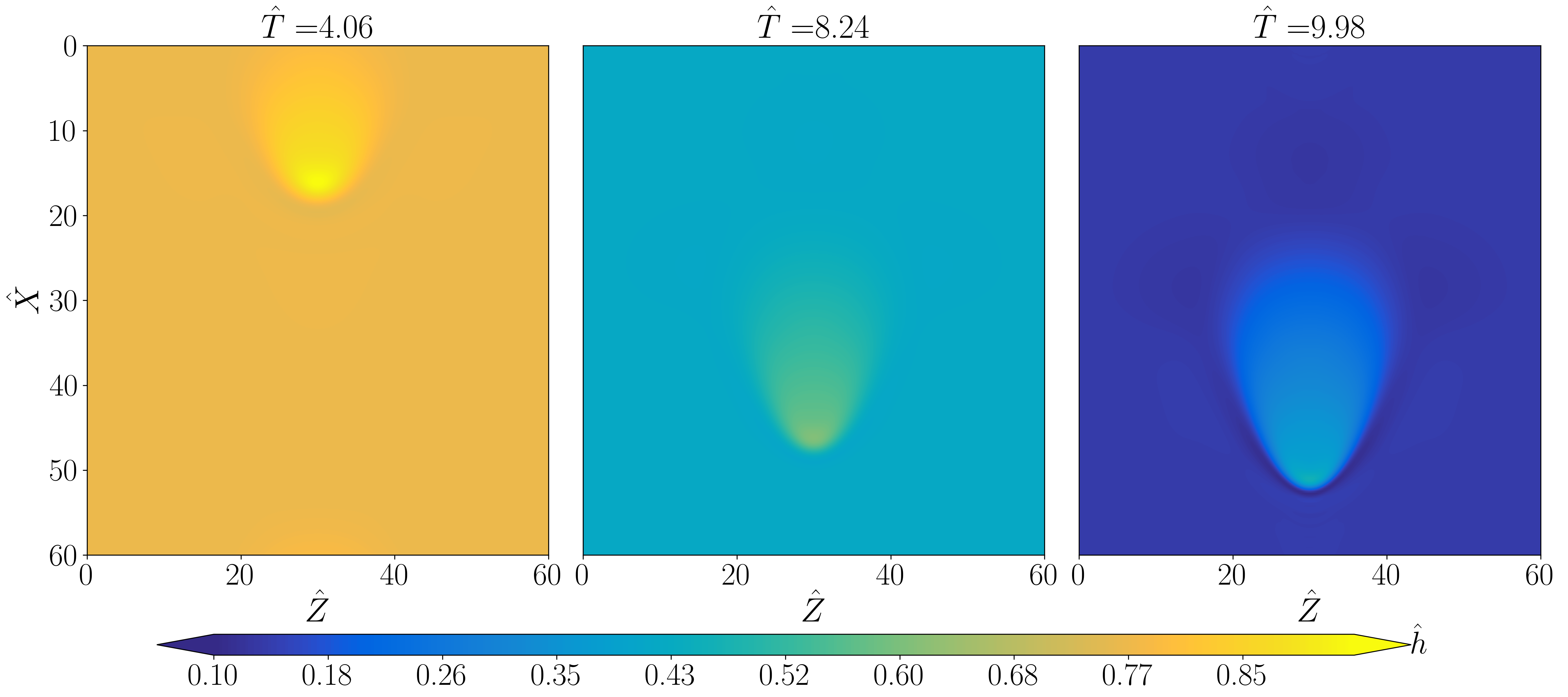}
    \caption{Constant substrate temperature $\hat{\eta}=0.3$ (evaporation).}
    \label{fig:3D_const_temp_evap}
  \end{subfigure}\\[6pt]
  \begin{subfigure}{\textwidth}
    \centering
    \includegraphics[width=\textwidth]{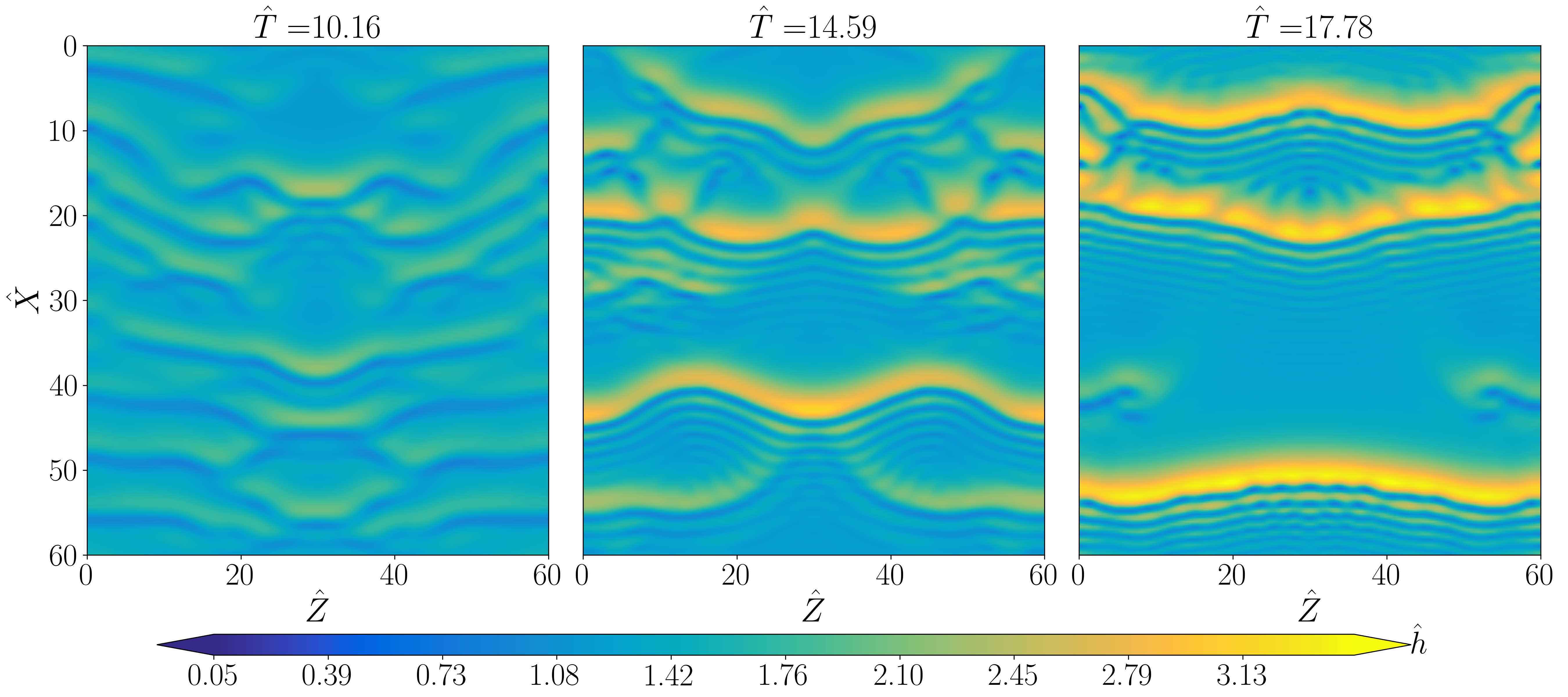}
    \caption{Constant substrate temperature $\hat{\eta}=-0.3$ (condensation).}
    \label{fig:3D_const_temp_cond}
  \end{subfigure}\\[6pt]
  \begin{subfigure}{\textwidth}
    \centering
    \includegraphics[width=\textwidth]{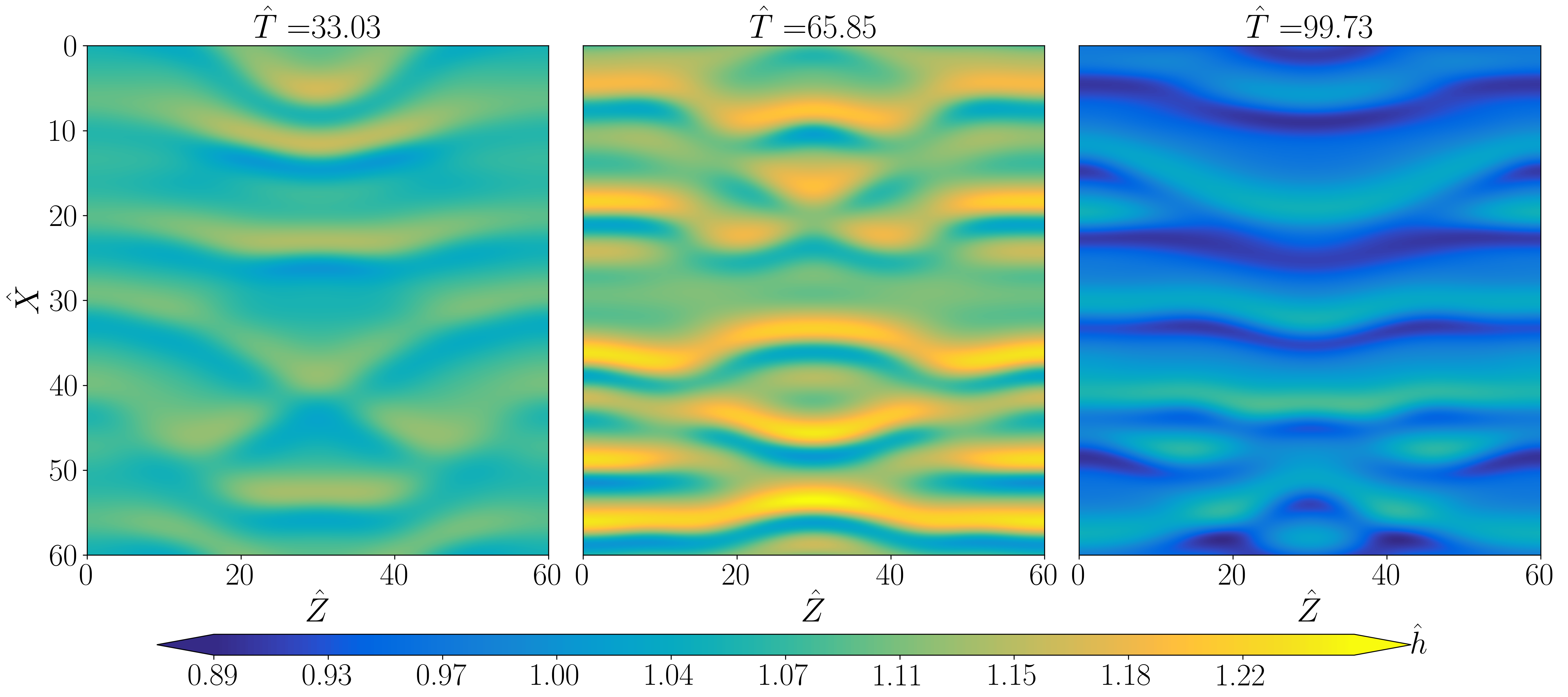}
    \caption{Oscillating substrate temperature in time.}
    \label{fig:3D_osci_temp}
  \end{subfigure}
  \caption{Evolution of the three-dimensional liquid film thickness $\hat{h}$ (colour map) for non-isothermal cases, shown at three different time instances.}
  \label{fig:3D_variable_temp}
\end{figure}

\begin{figure}
    \centering
    \begin{tabular}{cc}
        \subfloat[]{\includegraphics[width=0.5\textwidth]{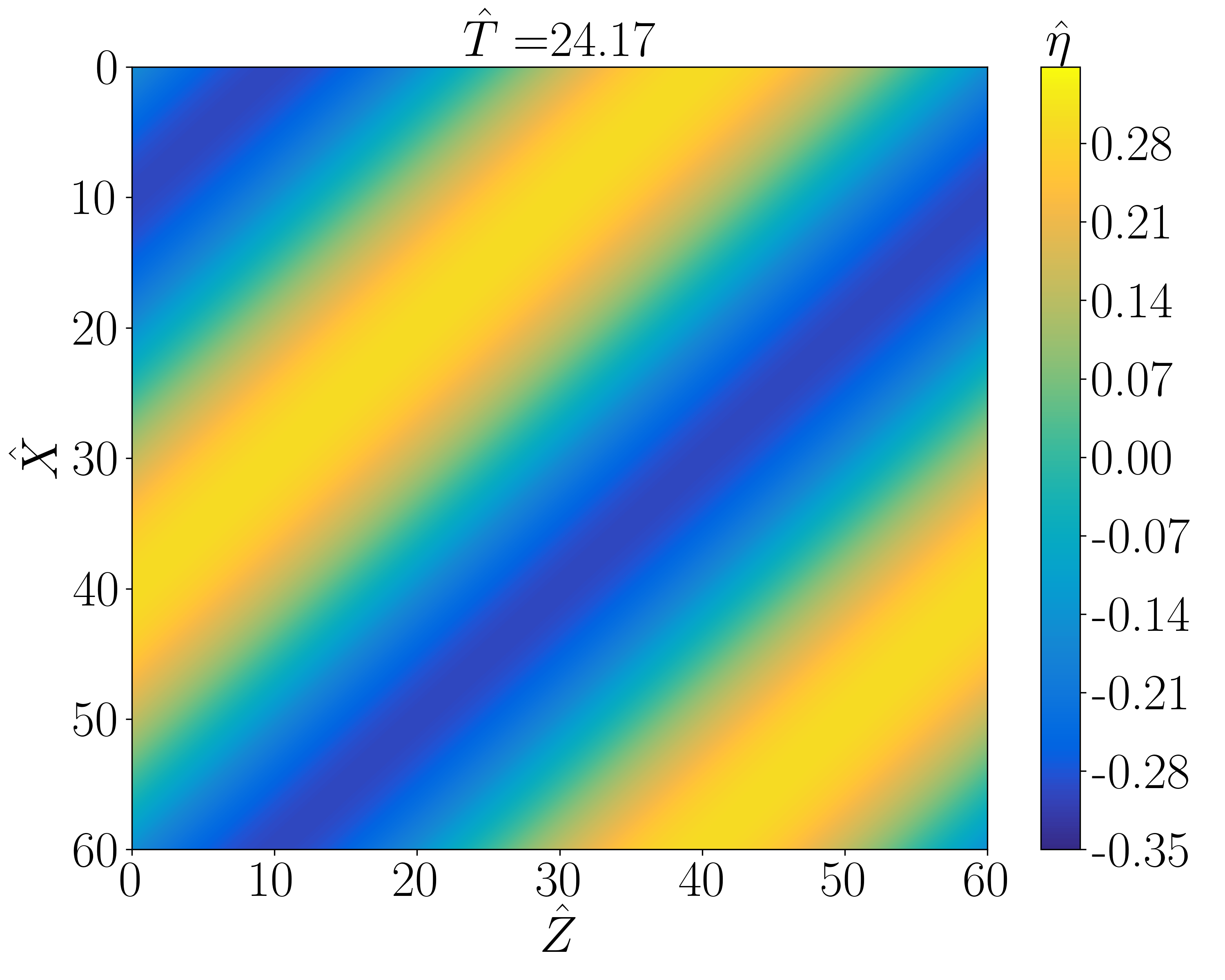}} &
        \subfloat[]{\includegraphics[width=0.5\textwidth]{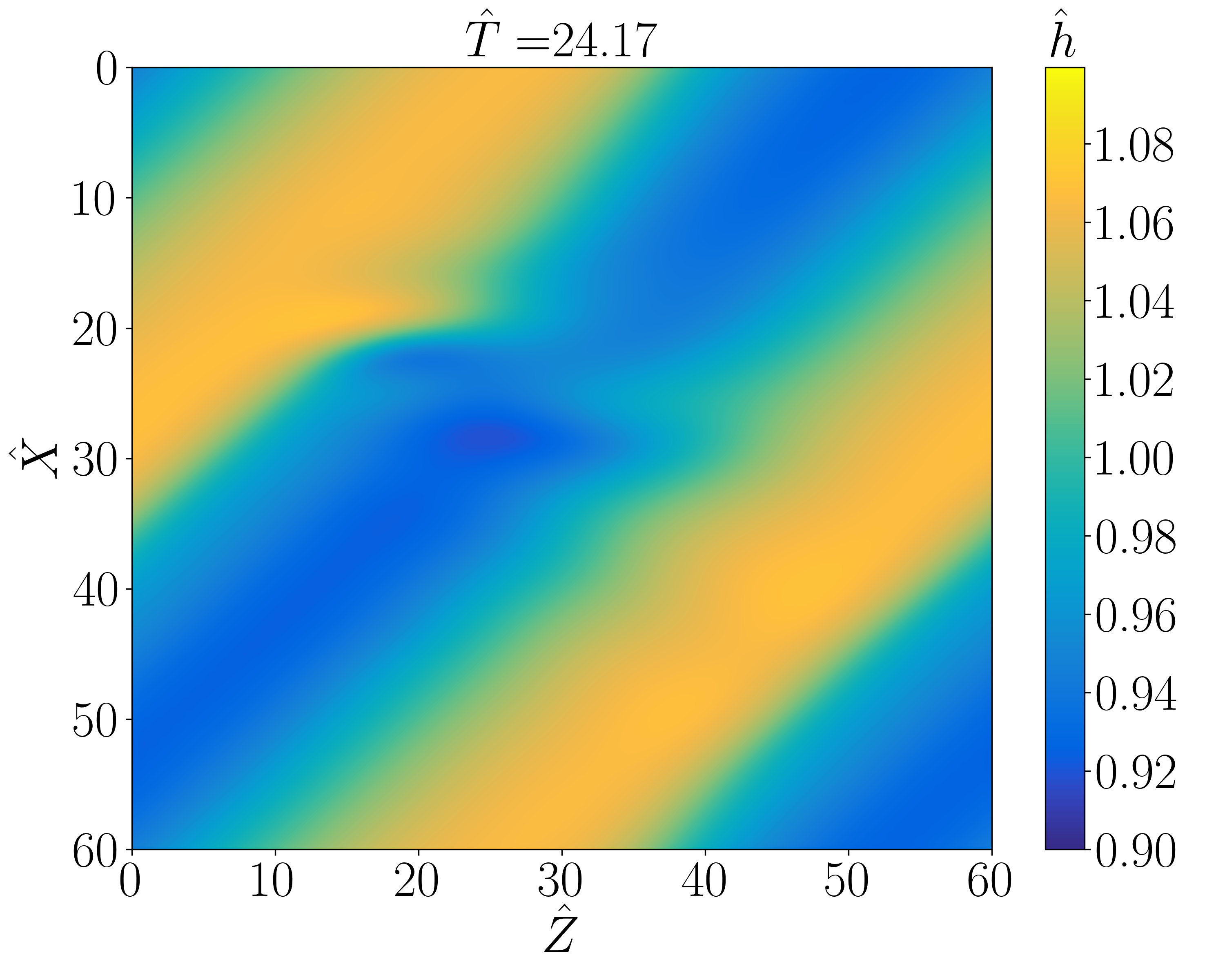}}\\
        \subfloat[]{\includegraphics[width=0.5\textwidth]{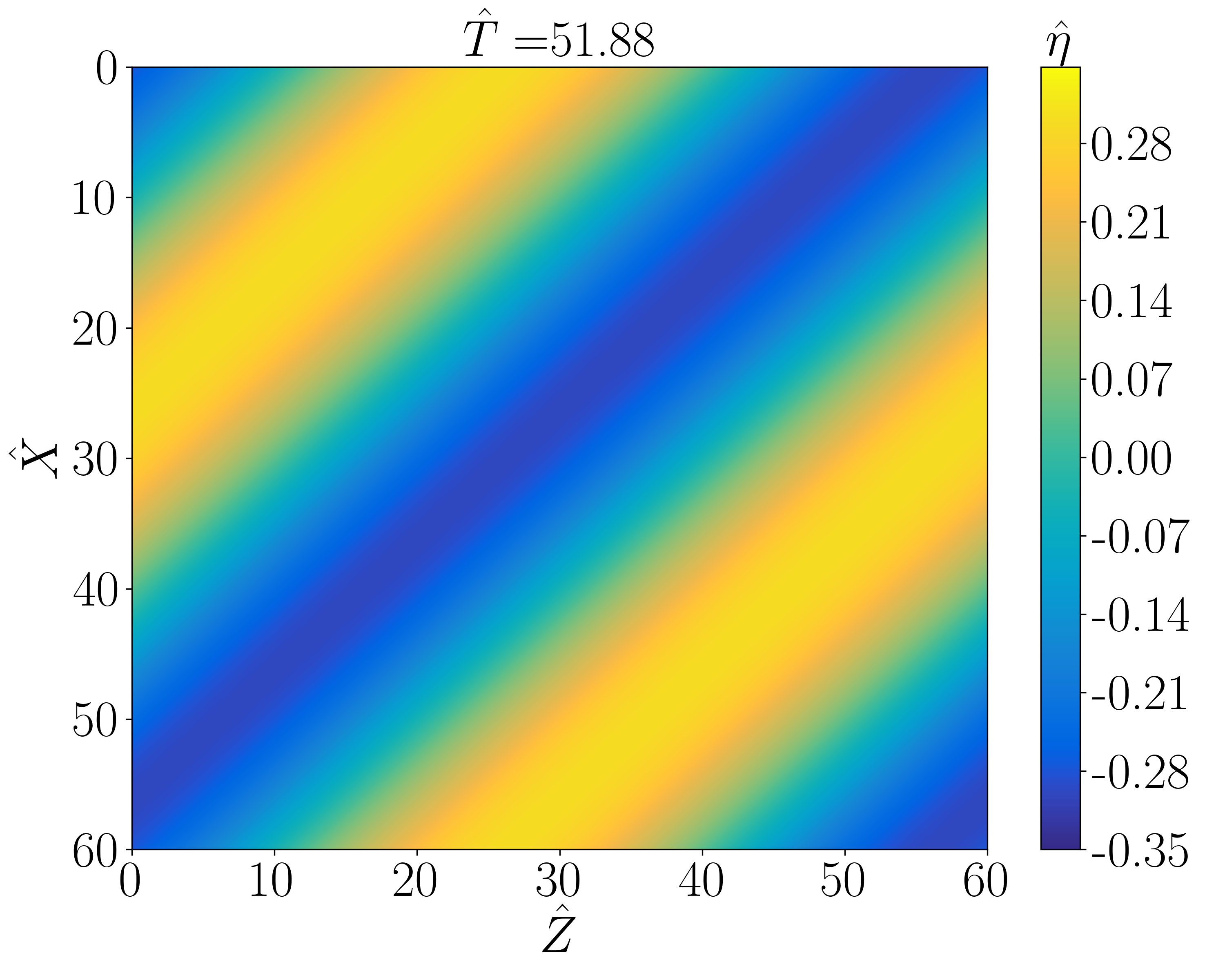}} &
        \subfloat[]{\includegraphics[width=0.5\textwidth]{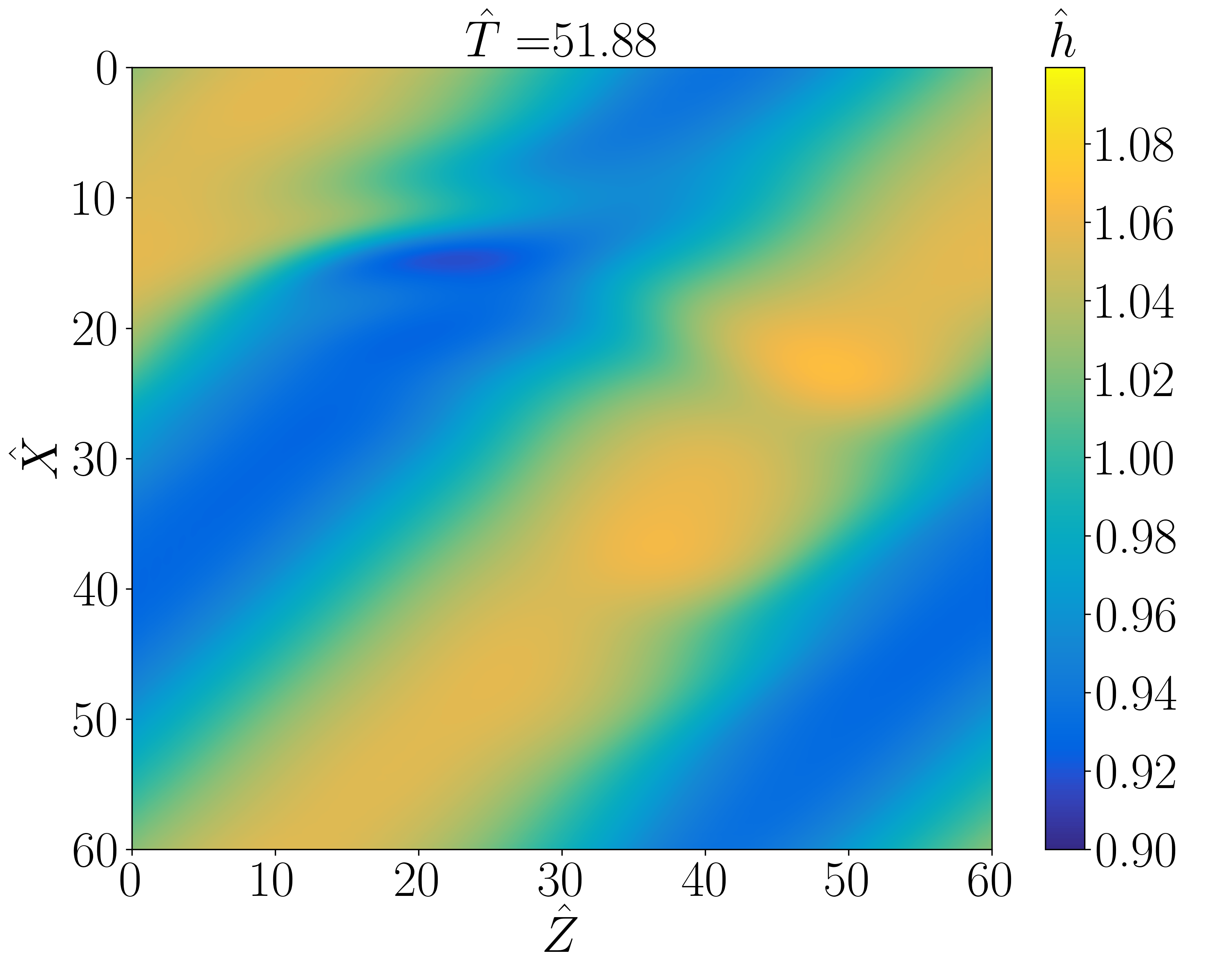}}\\
        \subfloat[]{\includegraphics[width=0.5\textwidth]{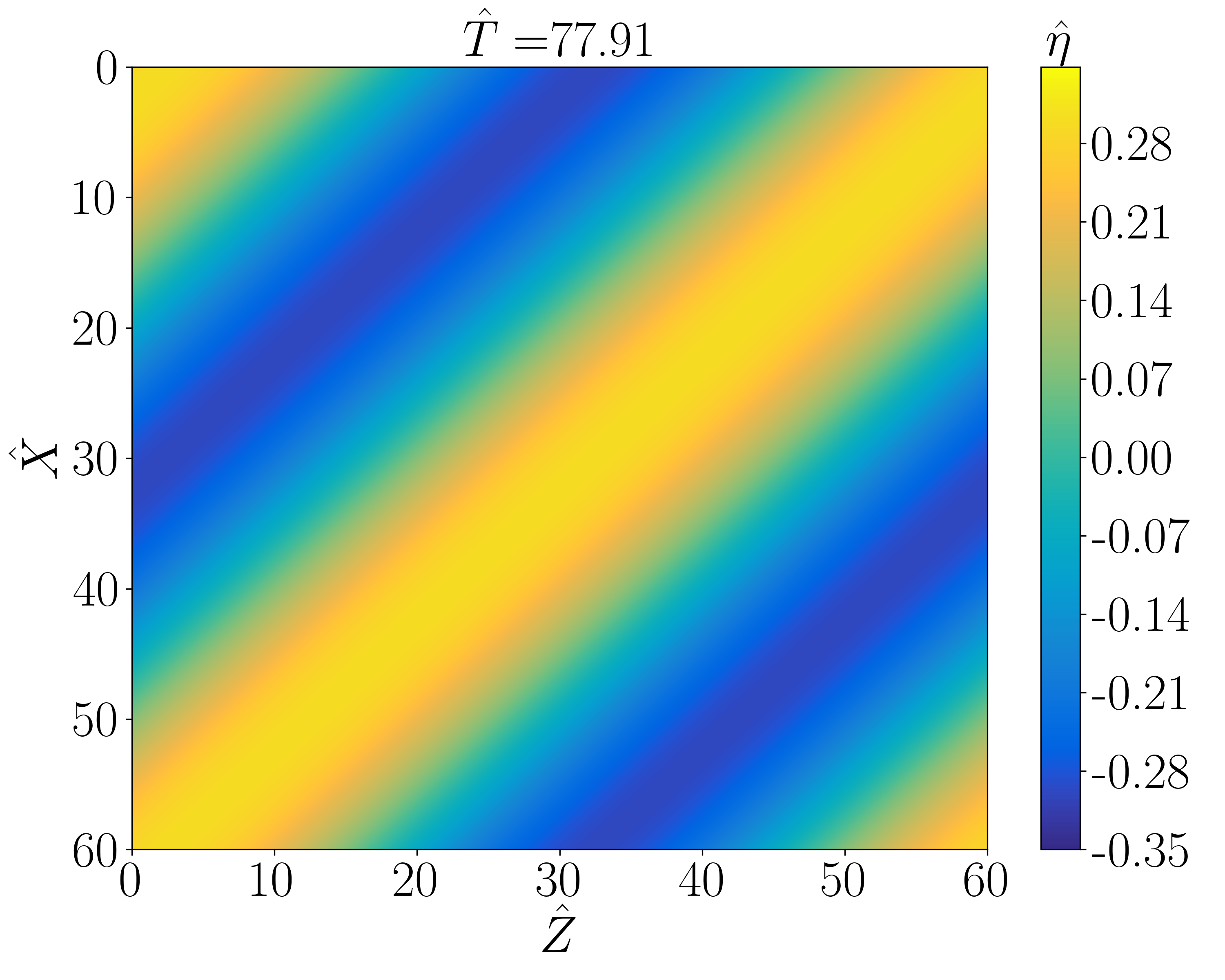}} &
        \subfloat[]{\includegraphics[width=0.5\textwidth]{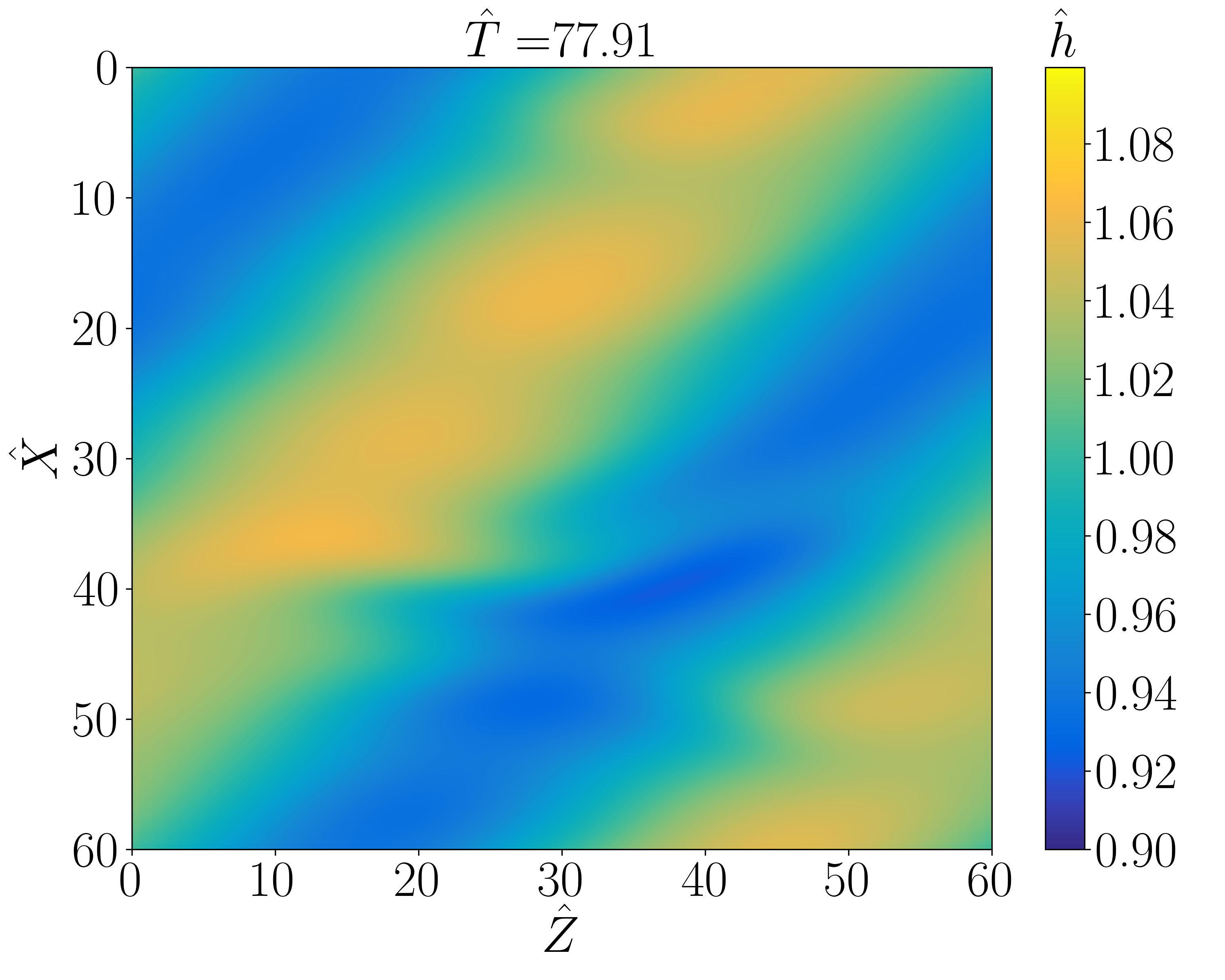}}
    \end{tabular}
    \caption{Comparison between substrate temperature (left column) and free-surface thickness (right column) in the travelling-wave test case at three time instants. Left axes: nondimensional substrate temperature (green dash-dotted, right $y$-axis in the original plot); right axes: free-surface thickness.}
    \label{fig:comp_3D_travelling_Waves}
\end{figure}

For the two-dimensional simulations, figure~\ref{fig:error_WIBL_COMSOL} reports the relative error between the WIBL and COMSOL predictions, expressed as a percentage using \eqref{eq:error_WIBL_COMSOL}, for (a) the film thickness $\hat{h}$ and (b) the free-surface temperature $\hat{\theta}_s$. The WIBL model reproduces both quantities with reasonable accuracy, with errors remaining below $3\%$ for thickness and $7\%$ for temperature.

The thickness error oscillates throughout the simulation, with its amplitude gradually decreasing. A sharp spike appears at startup due to the sudden change in substrate temperature, even though the temporal forcing is exponentially smoothed. After this transient, the error drops below $2\%$, before increasing slightly once $\hat{T}>70$, where it remains in the range $2$–$2.5\%$. These oscillations arise from small-scale surface structures created by intense local mass fluxes. Because such short-wavelength features lie outside the assumptions of the boundary-layer formulation, the reduced model captures them only approximately.

The free-surface temperature exhibits similar oscillations but with a shorter period. Following an initial spike around $\hat{T}\approx 15$, the error falls below $5\%$ and then gradually increases for $\hat{T}>40$, reaching about $7\%$ by the end of the simulation. The oscillation amplitude increases slowly over time, whereas the frequency remains nearly constant. These variations are likely driven by small-scale interfacial features that modulate heat transport. Additionally, the leading-order temperature approximation $\hat{\theta}^{(0)}$, based on thermal relaxation eigenfunctions, cannot fully reproduce the rapid temperature variations introduced by the imposed substrate heating.

Further insight is provided by examining the time evolution of thickness and free-surface temperature. Figure~\ref{fig:comp_COMSOL_2D} compares the trajectories predicted by WIBL (red dashed line with squares) with those from COMSOL (solid black line), together with the imposed substrate temperature (blue dashed line, right axis). The WIBL model captures the free-surface temperature with only minor amplitude discrepancies. The evolution of film thickness is also well reproduced overall. However, short-wavelength structures generated in the full model appear with a slight phase shift and weaker damping in the WIBL solution, consistent with the validity limit of the long-wave approximation. The agreement at this moderate Reynolds number confirms that the reduced model can accurately describe nonlinear dynamics when its limitations are acknowledged.

These benefits come at a fraction of the computational cost: the WIBL simulation required under one minute of CPU time, compared with over ten hours for the corresponding COMSOL computation. Because a full three-dimensional COMSOL simulation would need several weeks with the available resources, a direct 3D comparison was not feasible. 

The three-dimensional dynamics under variable substrate heating were therefore analysed exclusively with the WIBL model. Figure~\ref{fig:evol_J_avg_h_min} shows the evolution of (a) the minimum film thickness $\hat{h}_{\min}$ and (b) the spatially averaged mass flux $\langle \hat{J} \rangle_{\Omega}$ defined in \eqref{eq:int_variable_comp_3D}, for an initial-hump configuration. The curves correspond to the isothermal case (grey dashed line), to constant heating and cooling ($\hat{\eta}=0.3$ and $\hat{\eta}=-0.3$), to the weakly evaporating/condensing base-state solutions (solid purple), and to temporally or spatio-temporally varying substrate temperatures (green dotted and orange solid).

In the isothermal case, $\hat{h}_{\min}$ oscillates around unity with a long period before settling near $\hat{h}\approx 0.95$, indicating that the film approaches a travelling-wave regime with only mild variation in wave amplitude. For the weakly evaporating flat-film solutions, the minimum thickness at which dry-out occurs is $\hat{t}_d=10$. Nonlinear simulations with constant heating exhibit the same trend, drying out early in both evaporation and condensation. In the evaporating case, the predicted dry-out time $\hat{t}_d\approx 10$ closely matches the base-state estimate, confirming that nonlinear wave development does not significantly alter the global evaporation dynamics. As the film thins, thermal effects dominate over advection, and the system evolves toward the weakly evaporating solution. In the condensing case, the film dries out at $\hat{T}\approx 18$. After initially following the increasing trend of the flat-film solution and peaking around $\hat{T}\approx 5$, $\hat{h}_{\min}$ undergoes oscillations with a steadily decreasing envelope. This behaviour is associated with the onset of Kapitza-type instabilities. As the film thickens, inertial effects strengthen and destabilise the interface, leading to wave growth and, eventually, blow-up of the WIBL solution. Because steep gradients and rupture lie beyond the long-wave regime, these results must be interpreted with caution.

When the substrate temperature varies in time or in both space and time, the film does not undergo dry-out. For temporal oscillations, $\hat{h}_{\min}$ responds nearly harmonically, oscillating between $\hat{h}\approx 1.05$ and $\hat{h}\approx 0.90$. In the travelling-temperature case, multiple harmonics appear, with slowly decaying amplitude, before the thickness settles near $\hat{h}\approx 0.90$. In both scenarios, the long-time minimum thickness approaches the values of the isothermal configuration.

The behaviour of the spatially averaged mass flux mirrors that of the minimum thickness. Under constant heating, the mean mass flux closely follows the base-state prediction, with only a small deviation near dry-out in the condensation case. Under temporal forcing, it oscillates at a single dominant frequency with small amplitude. When the substrate temperature varies in space and time, the mean mass flux rapidly decays to values of order $5\times10^{-3}$ and remains near zero thereafter, reflecting the zero-mean periodic forcing.

The full spatiotemporal evolution further illustrates these trends. Figure~\ref{fig:3D_isothermal} shows that under isothermal conditions, the initial hump develops into three-dimensional waves that gradually align in the streamwise direction. In the absence of vapour recoil or thermocapillarity, hydrodynamic effects dominate, favouring streamwise alignment.

Figure~\ref{fig:3D_variable_temp} summarises the dynamics for constant heating, constant cooling, and temporal oscillations. In the evaporating case, the falling hump creates a shallow trough that steadily thins by evaporation and eventually dries out; the phase speed decreases as the film thins, producing a steep rear-facing peak. In the condensing case, enhanced spreading leads to rapid development of streamwise-dominated waves, with capillary ripples preceding a large hump; as the crest steepens under gravity, dry-out occurs. During temporal oscillations, the film again spreads in the spanwise direction, but the overall pattern resembles that of the isothermal case.

Finally, figure~\ref{fig:comp_3D_travelling_Waves} contrasts the isothermal configuration with a substrate temperature varying in both space and time. In the isothermal case, wave amplitudes remain mild, with structures largely aligned streamwise. Under travelling heating, vapour-recoil and evaporation effects distort the wave pattern and gradually lock it to the imposed temperature wave: crests form above hot regions, troughs above cool ones. Alternating heating and cooling zones delay the formation of dry spots relative to the constant-temperature case.

% ---------------- Conclusion ----------------
\section{Conclusion}
Phase-changing liquid films are used to enhance heat transfer in a wide range of thermal management systems. Yet their analysis and control are often hindered by the high computational cost of solving the full set of governing equations. To circumvent this problem, I developed a non-isothermal Weighted Integral Boundary-Layer (WIBL) model for three-dimensional evaporating and condensing films flowing over an inclined surface, also accounting for spatial and temporal variations in the substrate temperature. The derivation is based on second-order long-wave equations, assuming weak evaporation/condensation and strong surface tension conditions. The first-order velocity and temperature corrections, in the film parameter $\varepsilon$, are computed analytically, and suitable gauge conditions are enforced to obtain a closed system of three nonlinear hyperbolic PDEs for the film thickness, the streamwise and spanwise flow rates, and the free-surface temperature. This reduced formulation enables fast simulations while retaining a high degree of fidelity to the full governing equations.

Model performance was assessed against the full system's linear and nonlinear dynamics. In the linear regime, the WIBL formulation significantly improves upon classical Benney-type models, accurately predicting maximum growth rates and cutoff wavenumbers across a broad range of wavenumbers and Reynolds numbers. Although discrepancies appear for $\R>40$, the model remains robust; it systematically underestimates short-wave amplification, thereby avoiding unphysical instabilities in nonlinear computations. The long-wave Orr-Sommerfeld analysis reveals a critical Reynolds number for vertical condensing films, set by the interplay of thermocapillary and vapour-recoil effects. For hanging films, this study identified a threshold that separates streamwise- and spanwise-dominated instabilities, reflecting a mode switch between Kapitza and Rayleigh–Taylor mechanisms as the inclination angle increases. This transition, absent in simpler long-wave models, is reproduced by the WIBL formulation with reasonable accuracy for small $\R$ and large inclination angles ($\beta>170^\circ$).

Nonlinear simulations with sharp substrate-temperature variations show that the WIBL model maintains quantitative accuracy within approximately $3\%\div 7\%$ of the full governing equations in two dimensions, while reducing computational time by several orders of magnitude. The free-surface temperature is captured with high accuracy, although some short-wave features remain under-resolved due to the limitations of the long-wave assumption. Three-dimensional simulations reveal that condensing films undergo dry-out driven by H-mode amplification, while condensation enhances lateral momentum redistribution, producing hump-like streamwise structures with capillary ripples. Substrate-temperature oscillations accelerate the onset of three-dimensional waves. Overall, spatial and temporal variations in material properties strongly influence film dynamics, enabling control of interfacial mass flux and delaying dry-out.

The reduced-order model developed in this work captures the essential dynamics of phase-changing films at a fraction of the computational cost of the full governing equations. This makes the WIBL framework well-suited to extensive parametric exploration, nonlinear stability analysis, and control-oriented studies. In particular, it provides a promising foundation for designing optimal substrate-temperature control strategies, for example, to maximise heat transfer over a prescribed time horizon without altering the substrate topography. Nonetheless, some limitations persist: the long-wave approximation deteriorates when strong short-wave structures arise or when the film approaches dry-out, and the assumptions of continuum vapour transport and simplified interfacial kinetics become restrictive when the vapour layer is thick or far from equilibrium. In addition, to bridge the gap to practical applications, particularly under high heat-flux conditions, it will be necessary to account for variations in material properties as a function of film temperature.

Several extensions could naturally overcome these limitations. The velocity and temperature profiles used in the model formulation can be adjusted to account for changes in the flow structure as the film approaches dry-out conditions via an asymptotic analysis. Furthermore, under intense evaporation conditions, the simplified linear relation for the mass flux can be replaced by the solution of the Boltzmann-equation Moment Method Model, which requires solving the continuity, momentum, and energy equations across the Knudsen layer in the vicinity of the free surface.

% ---------------------------
%   Acknowledgements and Funding        
% ---------------------------
\section*{Acknowledgements}
The author gratefully thanks Prof. Richard Kerswell for helpful discussions and Prof. Benoit Scheid for his valuable comments and suggestions.

\section*{Funding and Declaration of Interest}
F.Pino is supported by a postdoctoral fellowship from the Fondation Wiener-Anspach. The authors report no conflict of interest.

\newpage
% ---------------------------
%          Appendix        
% ---------------------------
\appendix

% ------- Boundary Layer equations ------- 
\section{Derivation of the Second-order Boundary Layer equations}
\label{appx:boundary_layer_equations}
The boundary layer equations, consistent, are obtained by assuming \eqref{eq:ass_ROMS_nondimnum} and introducing the slow spatial and temporal scales $\hat{X}$ and $\hat{T}$ and the slowly varying vertical velocity $\hat{V}$ \eqref{eq:ROM_stretching} into the governing equations \eqref{eq:nondim_gov_eq} and the boundary conditions \eqref{eq:bc_noslip}, \eqref{eq:kin_cond_adim}, \eqref{eq:bc_thermal} and \eqref{eq:bc_dyn},  retaining terms up to $O(\varepsilon^2)$, gives the set of simplified governing equations, reading: 
\begin{subequations}
\label{eq:bl_governing_equations}
\begin{gather}
    \partial_{X}\hat{u} + \partial_{y}\hat{V} + \partial_Z\hat{w} = 0,\label{eq:bl_cont}\\
    \varepsilon(\partial_T\hat{u} + \hat{u}\partial_{X}\hat{u} + \hat{V}\partial_{y}\hat{u} + \hat{w}\partial_Z\hat{u}) = -\varepsilon\partial_X\hat{p} + \partial_{yy}\hat{u} + \varepsilon^2(\partial_{XX}\hat{u} + \partial_{ZZ}\hat{u}) +\R,\label{eq:bl_mom_x}\\
    \varepsilon^2(\partial_T\hat{V} + \hat{u}\partial_X\hat{V} + \hat{V}\partial_y\hat{V} + \hat{w}\partial_Z\hat{V}) = -\partial_y\hat{p} +\varepsilon\partial_{yy}\hat{V} - \Ct, \label{eq:bl_mom_y}\\
    \varepsilon(\partial_T\hat{w} + \hat{u}\partial_{X}\hat{w} + \hat{V}\partial_{y}\hat{w} + \hat{w}\partial_Z\hat{w}) = -\varepsilon\partial_Z\hat{p} + \partial_{yy}\hat{w} + \varepsilon^2(\partial_{XX}\hat{w} + \partial_{ZZ}\hat{w}), \label{eq:bl_mom_z}\\
    \varepsilon\Pr(\partial_T\hat{\theta} + \hat{u}\partial_X\hat{\theta} + \hat{V}\partial_y\hat{\theta} + \hat{w}\partial_Z\hat{\theta}) = \partial_{yy}\hat{\theta} + \varepsilon^2(\partial_{XX}\hat{\theta} + \partial_{ZZ}\hat{\theta}),\label{eq:bl_ene}
\end{gather}
\end{subequations}
with boundary conditions at the substrate ($\hat{y}=0$) reading:
\begin{equation}
\label{eq:bc_bl_equations_wall}
    \hat{u}=0,\qquad\qquad \hat{w}=0,\qquad\qquad \hat{V}=0,\qquad\qquad \hat{\theta}=\hat{\eta},
\end{equation}
and at the free-surface ($\hat{y}=\hat{h}$), the boundary conditions reading:
\begin{subequations}
\label{eq:bc_fs_bl}
\begin{gather}
    \hat{V} = \partial_{T}\hat{h} + \hat{u}\partial_{X}\hat{h} + \hat{w}\partial_{Z}\hat{h} + \overline{\E}\,\hat{J},\label{eq:bc_cont_bl}\\
    \hat{p}= 2\varepsilon(\partial_y\hat{V} - \partial_X\hat{h}\partial_y\hat{u}-\partial_Z\hat{h}\partial_y\hat{w}) - 3\varepsilon^{-1}\overline{\Gamma}(\partial_{XX}\hat{h} + \partial_{ZZ}\hat{h}),\label{eq:bc_p_bl}\\
     \partial_y\hat{u} = -2\varepsilon\frac{\M}{\Pr}\partial_X\hat{\theta}_s + \varepsilon^2\big(\partial_Z\hat{h}(\partial_Z\hat{u}+\partial_X\hat{w}) + 2\partial_X\hat{h}(2\partial_X\hat{u} + \partial_Z\hat{w}) - \partial_X\hat{V}\big),\label{eq:bc_bl_shear_x}\\
     \partial_y\hat{w} = -2\varepsilon\frac{\M}{\Pr}\partial_Z\hat{\theta}_s + \varepsilon^2\big(\partial_X\hat{h}(\partial_Z\hat{u}+\partial_X\hat{w}) + 2\partial_Z\hat{h}(2\partial_Z\hat{u} + \partial_X\hat{w}) - \partial_Z\hat{V}\big),\label{eq:bc_bl_shear_z}\\
    \hat{J}\Big(1 + \frac{\varepsilon^2}{2}\big((\partial_X\hat{h})^2 + (\partial_Z\hat{h})^2\big)\Big) + \varepsilon^2\overline{\Pi}\hat{J}^3 = -\partial_y\hat{\theta} + \varepsilon^2(\partial_X\hat{\theta}\,\partial_X\hat{h} + \partial_Z\hat{\theta}\,\partial_Z\hat{h}),\label{eq:bc_bl_ener}
\end{gather}
\end{subequations}
where $\hat{J}$ is expressed by the constitutive relation \eqref{eq:J_cond_FS_adim}.

Integration \eqref{eq:bl_mom_y} in $\hat{y}$, retaining terms up to $O(\varepsilon)$, and making use of the dynamic boundary condition \eqref{eq:bc_p_bl}, I obtain the pressure field $\hat{p}$, which reads:
\begin{equation}
\label{eq:press_bl}
    \hat{p} = \Ct(\hat{h}-\hat{y}) + \varepsilon\Big(\partial_y\hat{V} + (\partial_y\hat{V})|_{\hat{y}=\hat{h}}\Big) - 3\,\varepsilon^{-1}\,\overline{\Gamma}\,(\partial_{XX}\hat{h} + \partial_{ZZ}\hat{h}). 
\end{equation}

Since the pressure gradients $\partial_X\hat{p}$ and $\partial_Z\hat{p}$ appear at order $O(\varepsilon)$ in the simplified streamwise~\eqref{eq:bl_mom_x} and spanwise~\eqref{eq:bl_mom_z} momentum equations, to be consistent at $O(\varepsilon^2)$, I have retained only terms up to $O(\varepsilon)$ in the derivation of $\hat{p}$. The shear stress terms $\partial_y \hat{u}$ and $\partial_y \hat{w}$ have been neglected as the are at most $O(\varepsilon)$ quantities according to the boundary conditions~\eqref{eq:bc_bl_shear_x} and~\eqref{eq:bc_bl_shear_z}.

Replacing \eqref{eq:press_bl} in \eqref{eq:bl_mom_x} and \eqref{eq:bl_mom_z} and making use of the continuity equation \eqref{eq:bl_cont}, gives:
\begin{subequations}
\label{eq:mom_eq_bl_mod}
\begin{equation}
\label{eq:bl_mom_x_mod}
\begin{gathered}
    \varepsilon(\partial_T\hat{u} + \hat{u}\partial_{X}\hat{u} + \hat{V}\partial_{y}\hat{u} + \hat{w}\partial_Z\hat{u}) = \partial_{yy}\hat{u} + 3\overline{\Gamma}(\partial_{XXX}\hat{h} + \partial_{XZZ}\hat{h}) + \R+\\ - \varepsilon\,\Ct\,\partial_X\hat{h}  +\varepsilon^2\,\Big(2\partial_{XX}\hat{u} + \partial_{XZ}\hat{w} + \partial_{ZZ}\hat{u} - \partial_X\Big((\partial_y\hat{V})|_{\hat{y}=\hat{h}}\Big)\Big),
\end{gathered}
\end{equation}
\begin{equation}
\label{eq:bl_mom_z_mod}
\begin{gathered}
    \varepsilon(\partial_T\hat{w} + \hat{u}\partial_{X}\hat{w} + \hat{V}\partial_{y}\hat{w} + \hat{w}\partial_Z\hat{w}) = \partial_{yy}\hat{w} + 3\,\overline{\Gamma}\,(\partial_{XXZ}\hat{h} + \partial_{ZZZ}\hat{h}) +\\ - \varepsilon\,\Ct\,\partial_Z\hat{h}  +\varepsilon^2\Big(\partial_{XX}\hat{w} + \partial_{XZ}\hat{u} + 2\partial_{ZZ}\hat{w} - \partial_Z\Big((\partial_y\hat{V})|_{\hat{y}=\hat{h}}\Big)\Big).
\end{gathered}
\end{equation}
\end{subequations}

The set of equations~\eqref{eq:bl_cont}, \eqref{eq:bl_ene} and \eqref{eq:mom_eq_bl_mod}, together with the boundary conditions \eqref{eq:bc_bl_equations_wall}, \eqref{eq:bc_cont_bl}, \eqref{eq:bc_bl_shear_x}, \eqref{eq:bc_bl_shear_z}, \eqref{eq:bc_bl_ener}, and the constitutive relation~\eqref{eq:J_cond_FS_adim}, form the set of boundary layer equations and boundary conditions consistent at $O(\varepsilon^2)$. By neglecting phase-change effects ($\overline{E} = \Vr = \overline{\Pi} = 0$), the boundary layer equations reduce to those for a non-isothermal falling liquid film~\citep[Section~4.1]{kalliadasis2011falling}.

% ------------- Description terms WIBL model
\section{Details for the WIBL integral model derivation}

% ---- Equations for first order correction terms
\subsection{Governing equation for $O(\varepsilon)$ correction terms.}
\label{appx_subsec_gov_eqs_oesp}
In this subsection, I present the governing equation for the $O(\varepsilon)$ correction terms, $\hat u^{(1)}, \hat w^{(1)}$ and $\hat \theta^{(1)}$, used in the derivation of the WIBL model. By substituting the approximated fields \eqref{eq:vel_decomp} and \eqref{eq:V_bl} into the boundary-layer equations \eqref{eq:bl_governing_equations}, and using the continuity equation \eqref{eq:bl_cont} together with the boundary conditions 
\eqref{eq:bc_bl_equations_wall} and \eqref{eq:bc_fs_bl}, 
while retaining the inertial terms at $O(\varepsilon)$ 
and all other terms at $O(\varepsilon^2)$, and considering $\hat{J}=O(1)$, I obtain the following set of equations governing the correction terms at $O(\varepsilon)$, reading:
\begin{subequations}
\label{eq:first_order_system}
\begin{align}
\partial_{yy}\hat{u}^{(1)} 
&= \varepsilon\big[(\partial_T 
    + (\overline{U} + \hat{u}^{(0)})\partial_X 
    + \hat{V}\partial_y 
    + (\overline{U} + \hat{w}^{(0)})\partial_Z)
    (\overline{U} + \hat{u}^{(0)})\big]\nonumber \\
&\quad - \varepsilon^2\big[2\partial_{XX}(\overline{U} + \hat{u}^{(0)}) 
    + \partial_{ZZ}(\overline{U} + \hat{u}^{(0)}) 
    + \partial_{XZ}(\overline{U} + \hat{w}^{(0)}) 
    - \partial_X((\partial_y\hat{V})|_{\hat{y}=\hat{h}})\big] \\
&\quad - \partial_{yy}(\overline{U} + \hat{u}^{(0)}) 
    - \R 
    + \varepsilon\big(\Ct\,\partial_X\hat{h} 
    + 3\Vr\,\hat{J}\partial_X\hat{J}\big)
    - 3\overline{\Gamma}(\partial_{XXX}\hat{h} + \partial_{ZZX}\hat{h}),\nonumber
\\[0.5em]
\partial_{yy}\hat{w}^{(1)} 
&= \varepsilon(\partial_T 
    + (\overline{U} + \hat{u}^{(0)})\partial_X 
    + \hat{V}\partial_y 
    + (\overline{U} + \hat{w}^{(0)})\partial_Z)
    (\overline{U} + \hat{w}^{(0)})\nonumber\\
&\quad - \varepsilon^2\big[\partial_{XX}(\overline{U} + \hat{w}^{(0)}) 
    + 2\partial_{ZZ}(\overline{U} + \hat{w}^{(0)}) 
    + \partial_{XZ}(\overline{U} + \hat{u}^{(0)}) 
    - \partial_Z((\partial_y\hat{V})|_{\hat{y}=\hat{h}})\big] \\
&\quad - \partial_{yy}(\overline{U} + \hat{w}^{(0)}) 
    + \varepsilon\big(\Ct\,\partial_Z\hat{h} 
    + 3\Vr\,\hat{J}\partial_Z\hat{J}\big)
    - 3\overline{\Gamma}(\partial_{XXZ}\hat{h} + \partial_{ZZZ}\hat{h}),\nonumber
\\[0.5em]
\partial_{yy}\hat{\theta}^{(1)} 
&= \varepsilon\,\Pr(\partial_T 
    + (\overline{U} + \hat{u}^{(0)})\partial_X 
    + \hat{V}\partial_y 
    + (\overline{U} + \hat{w}^{(0)})\partial_Z)
    (\overline{\Theta} + \hat{\theta}^{(0)})\\
&\quad - (\partial_{XX} + \partial_{ZZ})(\overline{\Theta} + \hat{\theta}^{(0)}) 
    - \partial_{yy}\hat{\theta}^{(0)}.\nonumber
\end{align}
\end{subequations}

The boundary conditions at the substrate $\bar{y}=0$ reading:
\begin{equation}
\label{eq:bc_wall_WIBL_der}
    \hat{u}^{(1)} = 0\qquad\qquad \hat{w}^{(1)} = 0\qquad\qquad\hat{\theta}^{(1)} = 0,
\end{equation}
and at the free-surface $\bar{y}=1$ reading:
\begin{subequations}\label{eq:first_order_gradients}
\begin{align}
\partial_y\hat{u}^{(1)} 
=& \varepsilon^2\Big[
\partial_Z\hat{h}\Big(\partial_Z(\overline{U} + \hat{u}^{(0)}) 
+ \partial_X(\overline{U} + \hat{w}^{(0)})\Big) + \\
&\quad+ 2\partial_X\hat{h}\Big(2\partial_X(\overline{U} + \hat{u}^{(0)}) 
+ \partial_Z(\overline{U} + \hat{w}^{(0)})\Big)
\Big]  - \varepsilon^2\,\partial_X\hat{V}
- 2\varepsilon\,\frac{\M}{\Pr}\,\partial_X\hat{\theta}_s,\nonumber
\\[0.5em]
\partial_y\hat{w}^{(1)} 
=& \varepsilon^2\Big[
2\partial_Z\hat{h}\Big(\partial_X(\overline{U} + \hat{u}^{(0)}) 
+ 2\partial_Z(\overline{U} + \hat{w}^{(0)})\Big)
+ \\&\quad + \partial_X\hat{h}\Big(\partial_Z(\overline{U} + \hat{u}^{(0)}) 
+ \partial_X(\overline{U} + \hat{w}^{(0)})\Big)
\Big] 
 - \varepsilon^2\,\partial_Z\hat{V}
- 2\varepsilon\,\frac{\M}{\Pr}\,\partial_Z\hat{\theta}_s,\nonumber
\\[0.5em]
\partial_y\hat{\theta}^{(1)} 
=& -\hat{J}\Big[1 + \frac{\varepsilon^2}{2}\Big((\partial_X\hat{h})^2 + (\partial_Z\hat{h})^2\Big)\Big] - \varepsilon^2\overline{\Pi}\hat{J}^3 - \partial_y(\overline{\Theta} + \hat{\theta}^{(0)})+ \\ &\quad+ \varepsilon^2\Big(
\partial_X\hat{h}\,\partial_X(\overline{\Theta} + \hat{\theta}^{(0)}) 
+ \partial_Z\hat{h}\,\partial_Z(\overline{\Theta} + \hat{\theta}^{(0)})
\Big).
\end{align}
\end{subequations}

The inertial terms at $O(\varepsilon^2)$, which are neglected in \eqref{eq:first_order_system}, read: 
\begin{subequations}
\begin{equation}
    (\partial_T + (\overline{U} + \hat{u}^{(0)})\,\partial_X + \hat{V}\,\partial_y + (\overline{W} + \hat{w}^{(0)})\,\partial_W)\,\hat{u}^{(1)},
\end{equation}
\begin{equation}
    (\partial_T + (\overline{U} + \hat{u}^{(0)})\,\partial_X + \hat{V}\,\partial_y + (\overline{W} + \hat{w}^{(0)})\,\partial_W)\,\hat{w}^{(1)},
\end{equation}
\begin{equation}
    \Pr\,(\partial_T + (\overline{U} + \hat{u}^{(0)})\,\partial_X + \hat{V}\,\partial_y + (\overline{W} + \hat{w}^{(0)})\,\partial_W)\,\hat{\theta}^{(1)}.
\end{equation}
\end{subequations}

Neglecting these second-order corrections is reasonable, as these terms are generally small compared with the dominant advective contributions. Moreover, including such terms in the derivation of the reduced-order model would yield significantly more complex formulations with a narrower range of applicability, as shown by \citet{scheid2006wave}.

% ---- Description terms appearing in WIBL model
\subsection{Description terms appearing in WIBL integral model}
\label{sec:des_terms_WIBL}
In this subsection, I discuss the physical interpretation of the terms appearing in the WIBL model introduced in Subsection~\ref{subsec:WIBL_2nd_order}. Each term is denoted as $F_{i,j}^{(k)}$, where the subscript $i$ identifies the equation they belong to (streamwise $x$, spanwise $z$ momentum or free-surface temperature $\hat{\theta}$), $j$ denotes the physical effect, and the superscript $k$ represents the order in $\varepsilon$. A special emphasis is placed on their relations with the boundary-layer equations presented in Appendix~\ref{appx:boundary_layer_equations}. The explicit expressions for the terms in the streamwise \eqref{eq:int_qx}, and spanwise \eqref{eq:int_qx} integral equations read:
\begin{equation}
\begin{aligned}
    F_{x,in}^{(1)}  \coloneqq& -\frac{23\,\overline{E}\,\hat{J}\,\hat{q}_x}{16\,\hat{h}}
    + \frac{9\,\hat{q}_z\,\hat{q}_x\,\partial_Z\hat{h}}{7\,\hat{h}^2}
    + \frac{9\,\hat{q}_x^2\,\partial_X\hat{h}}{7\,\hat{h}^2}
    - \frac{9\hat{q}_z\partial_Z\hat{q}_x}{7\hat{h}} - \frac{17 \hat{q}_x\partial_X\hat{q}_x}{7\hat{h}} 
    - \frac{8\hat{q}_x\partial_Z\hat{q}_z}{7\hat{h}} \\
    F_{x,shb}^{(2)} \coloneqq& -\frac{23\,\partial_X\hat{h}\,\partial_Z\hat{q}_z}{16\,\hat{h}}
    - \frac{23\,\partial_Z\hat{h}\,\partial_X\hat{q}_z}{16\,\hat{h}} 
    + \frac{21\,\hat{q}_z\,\partial_Z\hat{h}\,\partial_X\hat{h}}{8\hat{h}^2} - \frac{23\,\hat{q}_z\,\partial_{XZ}\hat{h}}{16\,\hat{h}} + \partial_{XZ}\hat{q}_z, \\
    F_{x,shfs}^{(1)} \coloneqq& -\frac{5\M\,\partial_X\hat{\theta}_s}{2\Pr}, \;
    F_{x,shfs}^{(2)} \coloneqq \frac{15\partial_Z\hat{h}\, \partial_Z\hat{q}_x}{8\hat{h}}
    + \frac{15\partial_X\hat{h}\partial_X\hat{q}_x}{4\hat{h}}
    - \frac{15\hat{q}_x(\partial_Z\hat{h})^2}{8\hat{h}^2} 
    - \frac{15\hat{q}_x\partial_{XX}\hat{h}}{8\hat{h}}
    \\ &- \frac{15\hat{q}_x(\partial_X\hat{h})^2}{4\hat{h}^2}
    + \frac{15\partial_X\hat{h}\partial_Z\hat{q}_z}{8\hat{h}} - \frac{15\hat{q}_z\partial_X\hat{h}\partial_Z\hat{h}}{8\hat{h}^2}
    - \frac{15\hat{q}_z\partial_{XZ}\hat{h}}{8 \hat{h}}
    + \frac{5}{4}\partial_{XX}\hat{q}_x + \frac{5}{4}\partial_{XZ}\hat{q}_z, \\
    F_{x,el}^{(0)} \coloneqq& -\frac{5\hat{q}_x}{2\hat{h}^2},\qquad F_{x,el}^{(2)} \coloneqq -\frac{23\partial_{Z}\hat{h}\partial_{Z}\hat{q}_x}{8\hat{h}}
    - \frac{23\partial_X\hat{h}\partial_X\hat{q}_x}{4 \hat{h}}
    + \frac{21\hat{q}_x(\partial_Z\hat{h})^2}{8\hat{h}^2} + \frac{21\hat{q}_x(\partial_X\hat{h})^2}{4\hat{h}^2}\\  &
    - \frac{23\hat{q}_x\partial_{ZZ}\hat{h}}{16\hat{h}}
    - \frac{23\hat{q}_x\partial_{XX}\hat{h}}{8 \hat{h}} + \partial_{ZZ}\hat{q}_x + 2\partial_{XX}\hat{q}_x, \qquad
    F_{x,g}^{(0)} \coloneqq \frac{5\R\hat{h}}{6},\\
    F_{x,p}^{(0)} \coloneqq& \frac{5}{2}\,\overline{\Gamma}\,\hat{h}\,(\partial_{XZZ}\hat{h}+\partial_{XXX}\hat{h}), \qquad\qquad
    F_{x,p}^{(1)} \coloneqq -\frac{5}{6}\,\Ct\,\hat{h}\partial_X\hat{h} - \frac{5}{2}\,\Vr\,\hat{h}\,\hat{J}\,\partial_X\hat{J}, \\
    F_{x,p}^{(2)} \coloneqq& - \frac{\hat{q}_z\partial_{XZ}\hat{h}}{4\hat{h}} -\frac{5\partial_X\hat{h}\partial_X\hat{q}_x}{2 \hat{h}}
    + \frac{5\hat{q}_x(\partial_X\hat{h})^2}{2\hat{h}^2}
    - \frac{5\hat{q}_x\,\partial_{XX}\hat{h}}{4\hat{h}} - \frac{5 \partial_X\hat{h}\,\partial_Z\hat{q}_z}{4\hat{h}}
    - \frac{5 \partial_Z\hat{h}\,\partial_X\hat{q}_z}{4\hat{h}}\\  &
    + \frac{5\hat{q}_z\,\partial_Z\hat{h}\,\partial_X\hat{h}}{2\hat{h}^2}- \frac{5\hat{q}_z\,\partial_{XZ}\hat{h}}{4\hat{h}}
    + \frac{5}{4} \partial_{XX}\hat{q}_x
    + \frac{5}{4}\partial_{XZ}\hat{q}_z, \\
F_{z,in}^{(1)} \coloneqq& -\frac{23\,\overline{E}\,\hat{J}\,\hat{q}_z}{16 \hat{h}}+\frac{9\hat{q}_x\,\hat{q}_z\,\partial_X\hat{h}}{7 \hat{h}^2}+\frac{9\hat{q}_z^2\partial_Z\hat{h}}{7\hat{h}^2}-\frac{9\hat{q}_x \partial_X\hat{q}_z}{7\hat{h}}-\frac{17\,\hat{q}_z\,\partial_Z\hat{q}_z}{7\hat{h}}-\frac{8\,\hat{q}_z\,\partial_X\hat{q}_x}{7\hat{h}} \\
    F_{z,shb}^{(2)} \coloneqq& -\frac{23 \partial_X\hat{h}\,\partial_Z\hat{q}_x}{16\hat{h}}-\frac{23\partial_Z\hat{h}\,\partial_X\hat{q}_x}{16 \hat{h}}+\frac{21\hat{q}_x\,\partial_Z\hat{h}\,\partial_X\hat{h}}{8\hat{h}^2}-\frac{23\hat{q}_x\,\partial_{XZ}\hat{h}}{16 \hat{h}}+\partial_{XZ}\hat{q}_x,\\
    F_{z,shfs}^{(0)} \coloneqq&-\frac{5\M\partial_Z\hat{\theta}_s}{2\Pr},\; F_{z,shfs}^{(2)} \coloneqq \frac{15\partial_Z\hat{h}\,\partial_X\hat{q}_x}{8 \hat{h}}-\frac{15\hat{q}_x\,\partial_X\hat{h}\,\partial_Z\hat{h}}{8\hat{h}^2}-\frac{15\hat{q}_x\,\partial_{XZ}\hat{h}}{8 \hat{h}}+\frac{15\partial_Z\hat{h}\,\partial_Z\hat{q}_z}{4\hat{h}}\\  &+\frac{15\partial_X\hat{h}\,\partial_X\hat{q}_z}{8\hat{h}}-\frac{15\hat{q}_z\,(\partial_Z\hat{h})^2}{4 \hat{h}^2}-\frac{15\hat{q}_z\,\partial_{ZZ}\hat{h}}{8 \hat{h}}-\frac{15\hat{q}_z\,(\partial_X\hat{h})^2}{8 \hat{h}^2}+\frac{5}{4} \partial_{XZ}\hat{q}_x+\frac{5}{4}\partial_{ZZ}\hat{q}_z,\\
    F_{z,ely}^{(0)} \coloneqq& -\frac{5\hat{q}_z}{2\hat{h}^2},\qquad F_{z,elxz}^{(2)} \coloneqq -\frac{23\partial_Z\hat{h}\,\partial_Z\hat{q}_z}{4\hat{h}}-\frac{23\partial_X\hat{h}\,\partial_X\hat{q}_z}{8 \hat{h}}+\frac{21\hat{q}_z\,(\partial_Z\hat{h})^2}{4 \hat{h}^2}\\&+\frac{21\hat{q}_z\,(\partial_X\hat{h})^2}{8 \hat{h}^2}-\frac{23\hat{q}_z\,\partial_{ZZ}\hat{h}}{8 \hat{h}}-\frac{23\hat{q}_z\,\partial_{XX}\hat{h}}{16 \hat{h}}+2 \partial_{ZZ}\hat{q}_z+\partial_{XX}\hat{q}_z,\\
    F_{z,p}^{(0)} \coloneqq& \frac{5}{2}\,\overline{\Gamma}\,\hat{h}\,(\partial_{XXZ}\hat{h}+\partial_{ZZZ}\hat{h})\qquad\qquad
    F_{z,p}^{(1)} \coloneqq -\frac{5}{6}\,\Ct\,\hat{h}\,\partial_Z\hat{h} -\frac{5}{2}\,\Vr\,\hat{h}\,\hat{J}\,\partial_Z\hat{J},\\
    F_{z,p}^{(2)} \coloneqq& -\frac{5\partial_X\hat{h}\,\partial_Z\hat{q}_x}{4 \hat{h}}-\frac{5\partial_Z\hat{h}\,\partial_X\hat{q}_x}{4 \hat{h}}+\frac{5\partial_Z\hat{h}\,\partial_X\hat{h}\hat{q}_x}{2\hat{h}^2}-\frac{5\hat{q}_x\,\partial_{XZ}\hat{h}}{4\hat{h}}-\frac{5\partial_Z\hat{h}\,\partial_Z\hat{q}_z}{2\hat{h}}\\  &+\frac{5\hat{q}_z\,(\partial_Z\hat{h})^2}{2 \hat{h}^2}-\frac{5\hat{q}_z\,\partial_{ZZ}\hat{h}}{4 \hat{h}}+\frac{5}{4}\partial_{XZ}\hat{q}_x+\frac{5}{4}\partial_{ZZ}\hat{q}_z,
\end{aligned}
\end{equation}
where the terms $F_{x,\mathrm{in}}^{(1)}$ and $F_{z,\mathrm{in}}^{(1)}$ denote the inertial contributions arising from the integration of the left-hand sides of~\eqref{eq:bl_mom_x_mod} and~\eqref{eq:bl_mom_z_mod}, respectively. Using the continuity equation~\eqref{eq:bl_cont} to replace $\partial_T \hat{h}$ in the derivation introduces a convection term that is linearly dependent on the evaporation mass flux at the free surface $\hat{J}$ weighted by the flow rate. The terms $F_{x,\mathrm{shb}}^{(2)}$ and $F_{z,\mathrm{shb}}^{(2)}$ represent bulk shear stresses associated with $\partial_{XZ}\hat{w}$ and $\partial_{XZ}\hat{u}$, while $F_{x,\mathrm{shfs}}$ and $F_{z,\mathrm{shfs}}$ account for the shear stress effects at the free surface at $O(1)$ and $O(\varepsilon)$, as specified by the boundary conditions~\eqref{eq:bc_bl_shear_x} and~\eqref{eq:bc_bl_shear_z}. The terms $F_{x,\mathrm{el}}$ and $F_{z,\mathrm{el}}$ correspond to streamwise viscosity diffusion effects in the streamwise and spanwise directions, associated with $\partial_{yy}\hat{u}$, $\partial_{XX}\hat{u}$ and $\partial_{ZZ}\hat{u}$ in the streamwise integral momentum equation, and with $\partial_{yy}\hat{w}$, $\partial_{XX}\hat{w}$ and $\partial_{ZZ}\hat{w}$ in the spanwise integral momentum equation. The contribution $F_{x,\mathrm{g}}^{(0)}$ arises from gravitational acceleration acting in the streamwise direction. Finally, the terms $F_{x,\mathrm{p}}$ and $F_{z,\mathrm{p}}$ correspond to pressure contributions, which include the surface tension associated to $\overline{\Gamma}$ at $O(1)$, the hydrostatic pressure proportional to $\Ct$ and vapour recoil $\Vr$ effects at $O(\varepsilon)$, and viscous effects at $O(\varepsilon^2)$. The mass flux at the free surface $\hat{J}$ associated with phase change contributes to both the streamwise and spanwise integral momentum equations as a local term in $F_{,\mathrm{in}}$ and as a non-local term through the vapour recoil pressure in $F_{,p}$. As a consequence, in the evaporating regime ($\hat{J} > 0$), phase change reduces the film’s momentum and, therefore, its kinetic energy. As a result, the wave amplitude and phase speed both decrease. In contrast, under condensing conditions, the liquid gains kinetic energy.

Moving to the equation for the free-surface temperature evolution \eqref{eq:int_temp}, the explicit expressions for the terms appearing in this equation are given by:
\begin{equation}
\begin{aligned}
    F_{\theta,conv}^{(1)} \coloneqq& \Big(3\Pr\,\hat{h}\,(14\,\overline{\E}\, \hat{J}\,(\hat{\theta}_s\,(7\hat{h}-2\K)+2\K\,\hat{\eta})-38\hat{q}_x\, \hat{\theta}_s\,\partial_X\hat{h}-38\hat{q}_z\,\hat{\theta}_s\,\partial_Z\hat{h}\\&\quad -2\hat{h}\,(7\K\,\partial_T\hat{\eta}+19(-\hat{\theta}_s(\partial_X\hat{q}_x+\partial_Z\hat{q}_z)+\hat{q}_x \partial_X\hat{\theta}_s+\hat{q}_z\,\partial_Z\hat{\theta}_s))+11\K\, \hat{\eta}\,\partial_X\hat{q}_x\\&\quad - 11\K\,\hat{\theta}_s\,\partial_X\hat{q}_x -11\K\,\hat{q}_x\,\partial_X\hat{\eta}-164\K\,\hat{q}_x\,\partial_X\hat{\theta}_sd+11\K\,\hat{\eta}\,\partial_Z\hat{q}_z\\&\quad -11 \K\,\hat{\theta}_s\,\partial_Z\hat{q}_z -11\K\,\hat{q}_z\,\partial_Z\hat{\eta}-164\K\,\hat{q}_z\,\partial_Z\hat{\theta}_s)\Big)/(14\,\Pr\,\hat{h}^2\,(7\hat{h}+27\,\K))\\
    F_{\theta,diff}^{(0)} \coloneqq& (840\K\,(\hat{\eta}-\hat{\theta}_s)-840\, \hat{h}\,\hat{\theta}_s)/(14\K^2\,\Pr\hat{h}^2\,(7\hat{h}+27\K)),\\ 
    F_{\theta,diff}^{(2)} \coloneqq& \Big(14\,(\hat{h}\,(3\hat{h}(14\partial_X\hat{h}\,\partial_X\hat{\theta}_s+\K\,(\partial_{ZZ}\hat{\eta}+\partial_{XX}\hat{\eta}+9\partial_{ZZ}\hat{\theta}_s+9\partial_{XX}\hat{\theta}_s))\\&\quad+6\partial_Z\hat{h}\,(\partial_Z\hat{\theta}_s\,(7\hat{h}-2\K)+2\K\,\partial_Z\hat{\eta})+12\K\,\partial_X\hat{h}\,(\partial_X\hat{\eta}-\partial_X\hat{\theta}_s)\\&\quad +7 \hat{h}^2\,(\partial_{ZZ}\hat{\theta}_s+\partial_{XX}\hat{\theta}_s))+6\K\,((\partial_Z\hat{h})^2+(\partial_X\hat{h})^2+\hat{h}\,(\partial_{ZZ}\hat{h}+\partial_{XX}\hat{h}))\,\hat{\eta}\\&\quad-3\hat{\theta}_s\,(2\hat{h}\,(\K\,(\partial_{ZZ}\hat{h}+\partial_{XX}\hat{h})+3(\partial_X\hat{h})^2)+2 (\partial_Z\hat{h})^2\,(3\hat{h}+\K)+2\K\,(\partial_X\hat{h})^2\\&\quad-7(\partial_{ZZ}\hat{h}+\partial_{XX}\hat{h})\,\hat{h}^2))\Big)/(14\,\Pr\,\hat{h}^2\,(7\hat{h}+27\K)),\\
    F_{\theta,fs}^{(2)} \coloneqq & (420\hat{h}\,(-\K\,\hat{J}\,((\partial_Z\hat{h})^2+(\partial_X\hat{h})^2)+2)+2\K\,(\partial_Z\hat{h}\,\partial_Z\hat{\theta}_s+ \partial_X\hat{h}\,\partial_X\hat{\theta}_s)\\&\quad+2\hat{\theta}_s(((\partial_Z\hat{h})^2+(\partial_X\hat{h})^2)+1)-2\K\overline{\Pi}\, \hat{J}^3))/(14\,\K^2\,\Pr\,\hat{h}^2\,(7\hat{h}+27\K)),
\end{aligned}
\end{equation}
where $F_{\theta,\mathrm{conv}}^{(1)}$ arises from convective effects and is associated with the left-hand side of equation~\eqref{eq:bl_ene}. As for the integral momentum equations, there are some terms which depend on the mass flux at the free surface $\hat{J}$. In addition, the convective effects are also dependent on the spatial and temporal variation of the substrate temperature $\hat{\eta}$. Moving to right-hand side of the thermal equation, the terms $F_{\theta,\mathrm{diff}}$ account for diffusive contributions, corresponding to $\partial_{yy}$ at order $O(1)$ and to $\partial_{XX}$ and $\partial_{ZZ}$ at order $O(\varepsilon^2)$. It is worth noticing that the temperature variation at the substrate has a direct effect on the thermal diffusion across the liquid film via the term $\hat{\eta}-\hat{\theta}_s$. Finally, the term $F_{\theta,\mathrm{fs}}^{(2)}$ represents the contribution from the free-surface boundary condition~\eqref{eq:bc_bl_ener}.

% Description of the pseudospectral numerical method used to implement the WIBL equations
\section{Solution WIBL equations via Fourier pseudo-spectral method}
\label{appx:Fourier_Spectral_method}
In this section, I describe the numerical method used to solve the WIBL equations \eqref{eq:cont_eq_ROMS} and \eqref{eqs:WIBL_equations} for the 2D and 3D nonlinear test cases presented in subsection~\ref{subsec:nonlin_sim}. In what follows, I focus on solving the 3D equations, as the 2D case can be viewed as a simplification of the method described here.

The 3D equations are solved on a square domain with periodic boundary conditions,$\Omega = \{ (\hat{X}, \hat{Z}) \in \mathbb{R}^2 \mid 0 < \hat{X} < L, \ 0 < \hat{Z} < L \}$. The first step of the numerical solution consists of discretising $\Omega$ with a uniform grid of $N \times N$ points along the $\hat{X}$ and $\hat{Z}$ directions, denoted by $\hat{X}_i$ and $\hat{Z}_j$. The resulting discrete 2D domain is given by:
\begin{equation}
\Omega_d=\{(\hat{X}_i = i\,\Delta\hat{X}, \hat{Z}_i = j\,\Delta\hat{Z} \mid i=0,\dots,N-1; \ j=0,\dots,N-1\},\qquad
\Delta\hat{X}=\Delta\hat{Z}=\frac{L}{N}.
\end{equation}

The solution of the dependent variable at the grip points is sought by approximating these quantities with Fourier modes. The film thickness $\hat{h}$, the flow rates $\hat{q}_x$ and $\hat{q}_z$ and the free-surface temperature $\hat{\theta}_s$ are approximated by a truncated Fourier series with $N$ harmonics (with $N\in\mathbb{N}^+$, even) along the streamwise $\hat{X}$ and spanwise direction $\hat{Z}$, reading:
\begin{subequations}
\label{eq:approx_fourier_re}
    \begin{align}
        \hat{h}(\hat{X},\hat{Z},\hat{T}) \approx&  
\sum_{m=-N/2}^{N/2-1}\sum_{n=-N/2}^{N/2-1} H_{(m,n)}(\hat{T})\,exp\Big(i\Big(\frac{2\pi\,m}{L}\hat{X} + \frac{2\pi\,n}{L}\hat{Z}\Big)\Big)\\
        \hat{q}_x(\hat{X},\hat{Z},\hat{T}) \approx&  
\sum_{m=-N/2}^{N/2-1}\sum_{n=-N/2}^{N/2-1} Q_{x(m,n)}(\hat{T})\,exp\Big(i\Big(\frac{2\pi\,m}{L}\hat{X} + \frac{2\pi\,n}{L}\hat{Z}\Big)\Big)\\
        \hat{q}_z(\hat{X},\hat{Z},\hat{T}) \approx&  
\sum_{m=-N/2}^{N/2-1}\sum_{n=-N/2}^{N/2-1} Q_{z(m,n)}(\hat{T})\,exp\Big(i\Big(\frac{2\pi\,m}{L}\hat{X} + \frac{2\pi\,n}{L}\hat{Z}\Big)\Big)\\
        \hat{\theta}_s(\hat{X},\hat{Z},\hat{T}) \approx&  
\sum_{m=-N/2}^{N/2-1}\sum_{n=-N/2}^{N/2-1} \Theta_{s(m,n)}(\hat{T})\,exp\Big(i\Big(\frac{2\pi\,m}{L}\hat{X} + \frac{2\pi\,n}{L}\hat{Z}\Big)\Big)
    \end{align}
\end{subequations}
where $m,n\in\mathbb{Z}$ and $H_{(m,n)}$, $Q_{x(m,n)}$, $Q_{z(m,n)}$ and $ \Theta_{s(m,n)}$ are the complex Fourier amplitudes, collected in the matrices $\mathsfbi{H}$, $\mathsfbi{Q}_{x}$, $\mathsfbi{Q}_{z}$ and $\bm{\Theta}_{s}$.

Substituting \eqref{eq:approx_fourier_re} into the WIBL equations yields the residual $\bm{r}(\hat{X},\hat{Z},\hat{T})$. In the residual, the unknowns are represented by Fourier amplitude matrices, with spatial derivatives replaced by algebraic expressions derived from differentiating the harmonics.

A closed system of ODEs for the Fourier amplitudes is obtained by imposing that the projection of $\bm{r}(\hat{X},\hat{Z},\hat{T})$ onto a basis of Dirac delta functions $\delta$ located at the grid point $\Omega_d$ is zero, which gives:
\begin{equation}
\big\langle \mathbf{r}(\hat{X},\hat{Z},\hat{T}),\delta(\hat{X}-\hat{X}_i,\hat{Z}-\hat{Z}_i)\big\rangle = \int_0^L\int_0^L\, \mathbf{r}\delta(\hat{X}-\hat{X}_i,\hat{Z}-\hat{Z}_i)\;d\hat{X}d\hat{Z} = \mathbf{r}(\hat{X}_i,\hat{Z}_i,\hat{T},) =  \mathbf{0}.
\end{equation}

By doing so, I obtain an $ N$-dimensional system of ordinary differential equations for the amplitude matrices, which is then solved in time with the explicit Runge-Kutta method of order 5(4), using the Python \textit{scipy.integrate} library. The solution for the 2D case follows the same steps, but it solely considers harmonics in the streamwise direction in \eqref{eq:approx_fourier_re}

From an implementation standpoint, the transforms between physical and Fourier space are carried out using the discrete Fourier transform (DFT) via the \texttt{scipy.fft.fft2} function. Nonlinear terms are computed in physical space before each integration step and then transformed to Fourier space using the FFT. To mitigate aliasing errors introduced by the truncated Fourier series \eqref{eq:approx_fourier_re}, the transformed quantities are filtered using the standard $2/3$-rule by zeroing the highest one-third of Fourier modes \citep[p.~294]{kalliadasis2011falling}.

\bibliographystyle{jfm}
\bibliography{jfm}

\section*{Supplementary Material}
\addcontentsline{toc}{section}{Supplementary Material}
\FloatBarrier
This section presents the supplementary material for the article, consisting of a description of the linearised full-model equations and a mesh-convergence study used to determine an appropriate number of grid points or approximating polynomials for the various numerical simulations.

\subsection*{Linearized Navier-Stokes equations}
\addcontentsline{toc}{subsection}{Linearized Navier-Stokes equations}
\label{subsec:linearized_NS}
The dependent variables are decomposed into a base-state solution and a small perturbation, namely
\begin{equation}
\label{eq:linear_decomposition}
\begin{gathered}
    \hat{u} = \overline{U}(\hat{y}) + \iota\,\tilde{u}(x,\hat{y},t), \qquad
    \hat{v} = \overline{V}(\hat{y}) + \iota\,\tilde{v}(x,\hat{y},t), \qquad
    \hat{w} = \overline{W}(\hat{y}) + \iota\,\tilde{w}(x,\hat{y},t),\\
    \hat{\theta} = \overline{\Theta}(\hat{y}) + \iota\,\tilde{\theta}(x,\hat{y},t), \qquad
    \hat{p} = \overline{P}(\hat{y}) + \iota\,\tilde{p}(x,\hat{y},t),\\
    \hat{J} = \overline{J} + \iota\,\tilde{J}(x,t), \qquad
    \hat{h} = \overline{H} + \iota\,\tilde{h}(x,t),
\end{gathered}
\end{equation}
where the overbar denotes base-state quantities associated with the weakly evaporating/condensing flat-film solution introduced in Subsection~\ref{sec:flat_film}, $\iota \ll 1$ is the perturbation amplitude, and tilded variables are $O(1)$ perturbations.

Substituting \eqref{eq:linear_decomposition} into the nondimensional governing equations presented in Subsection~\ref{subsec:gov_eqs_adim}, and retaining terms up to $O(\iota)$, yields the linearised system reading:
\begin{subequations}
    \begin{equation}
    \label{eq:lin_cont}
       \partial_x\tilde{u}+\partial_y\tilde{v}+\partial_z\tilde{w}=0
    \end{equation}
    \begin{equation}
    \label{eq:lin_x_mom}
\partial_t\tilde{u}  + \Bar{U}\partial_x\tilde{u} + D\Bar{U}\tilde{v}= -\partial_x\tilde{p} + \partial_{xx}\tilde{u} + \partial_{yy}\tilde{u} + \partial_{zz}\tilde{u}
    \end{equation}
    \begin{equation}
    \label{eq:lin_y_mom}
\partial_t\tilde{v}  + \partial_y\tilde{p} + \bar{U}\partial_x\tilde{v} = -\partial_y\tilde{p} + \partial_{xx}\tilde{v} + \partial_{yy}\tilde{v} + \partial_{zz}\tilde{v}
    \end{equation}
    \begin{equation}
    \label{eq:lin_z_mom}
 \partial_t\tilde{w} + \bar{U}\partial_x\tilde{w}= -\partial_z\tilde{p} + \partial_{xx}\tilde{w} + \partial_{yy}\tilde{w} + \partial_{zz}\tilde{w}
    \end{equation}
    \begin{equation}
    \label{eq:lin_temp}
\Pr(\partial_t\tilde{\theta} + \Bar{U} \partial_x\tilde{\theta} + D\Bar{\Theta}\tilde{v}) = \partial_{yy}\tilde{\theta } + \partial_{xx}\tilde{\theta} + \partial_{zz}\tilde{\theta}
    \end{equation}
\end{subequations}
where $D(\bullet)=\partial_{y}(\bullet)$ is the wall-normal differential operator. 

The boundary conditions at the solid substrate $(\hat{y}=0)$, read:
\begin{equation}
\label{eq:bc_wall_lin}
    \tilde{u} = 0,\qquad\qquad \tilde{v} = 0,\qquad\qquad\tilde{w} = 0,\qquad\qquad\tilde{\theta}=0.
\end{equation}

The values of the dependent variables at the perturbed free surface $\hat{h}$ are obtained by performing a first-order Taylor expansion about the flat base-state film thickness $\overline{H}$, yielding:
\begin{equation}
\label{eq:approx_bc_values_ns_lin}
    \hat{u}|_{\hat{h}} \approx \Bar{u}|_{\overline{H}} + \iota\tilde{u}|_{\overline{H}} + \iota(D\bar{u})|_{\overline{H}}\tilde{h}.
\end{equation}

Introducing the approximation \eqref{eq:approx_bc_values_ns_lin} in the boundary conditions and retaining terms up to $O(\iota)$ give the set of boundary conditions at the free-surface ($\hat{y}=\hat{h}$), which read:
\begin{subequations}
    \begin{equation}
    \label{eq:cont_lin_fs}
    \E\,\tilde{J}+\Bar{U}|_{\overline{H}}\,\partial_x\tilde{h} + \partial_t\tilde{h} = \tilde{v}
    \end{equation}
    \begin{equation}
    \label{eq:linear_normal_stress_balance}
\tilde{p} = 3\Vr\bar{J}\,\tilde{J}+(\partial_{xx}\tilde{h}+\partial_{zz}\tilde{h})\Big(\frac{2\M\,\Bar{\Theta}|_{\overline{H}}}{\Pr}-3\Gamma\Big)+2\partial_y\tilde{v}-\tilde{h}\,( D\Bar{P})|_{\overline{H}}
    \end{equation}
    \begin{equation}
    \label{eq:lin_normal_stress_x}
\frac{2\M}{\Pr}(\partial_x\tilde{h}\,(D\Bar{\Theta})|_{\overline{H}}+\partial_x\tilde{\theta})+ \tilde{h}\,(D^2\Bar{U})|_{\overline{H}}+\partial_y\tilde{u}+\partial_x\tilde{v}=0
    \end{equation}
    \begin{equation}
    \label{eq:lin_normal_stress_z}
        0 = \partial_z\tilde{v} + \partial_y\tilde{w} + \frac{2\M}{\Pr}(\partial_z\tilde{\theta} + (D\Bar{\Theta})|_{\overline{H}}\,\partial_z\tilde{\theta})
    \end{equation}
    \begin{equation}
    \label{eq:lin_temp_bc_fs}
(3\Pi\bar{J}^2+1)\tilde{J}+\partial_y\tilde{\theta}+\tilde{h}\,(D^2\Bar{\Theta})|_{\overline{H}}=0
    \end{equation}
    \begin{equation}
    \label{eq:const_rel_lin}
        \K\,\tilde{J}=\tilde{\theta} + \tilde{h}\,(D\Bar{\Theta})|_{\overline{H}}
    \end{equation}
\end{subequations}

The streamwise $\tilde{u}$ and spanwise $\tilde{w}$ velocities can be eliminated by rewriting the equations and the boundary conditions in terms of the wall normal velocity $\tilde{v}$. This recast costing in taking the divergence of the linearised momentum conservation equations ($\partial_x\eqref{eq:lin_x_mom}+\partial_y\eqref{eq:lin_y_mom}+\partial_z\eqref{eq:lin_z_mom}$), and making use of the continuity equation \eqref{eq:lin_cont}, yields:
\begin{equation}
\label{eq:111}
    \nabla^2\tilde{p} = -2D\Bar{U}\partial_x\tilde{v},
\end{equation}
then, by applying the Laplacian to the wall-normal linearised momentum conservation equation \eqref{eq:lin_y_mom} and using \eqref{eq:111} to remove the pressure, gives:
\begin{equation}
\label{eq:new_gov_eq_lin}
    \nabla^2(\partial_t\tilde{v} - \nabla^2\tilde{v}) + (-D^2\Bar{U} + \Bar{U}\nabla^2)\partial_{x}\tilde{v} = 0.
\end{equation}

Moving to the boundary conditions at the free-surface, I differentiate the wall-normal linearised momentum conservation equation \eqref{eq:lin_y_mom} with respect to $\hat{y}$ and evaluate it at the unperturbed free surface $\overline{H}$, noting that $D\Bar{U}|_{\overline{H}} = 0$, this yields:
\begin{equation}
\label{eq:222}
    -\partial_{yy}\tilde{p}|_{\overline{H}} = \partial_{ty}\tilde{v} + \Bar{U}|_{\overline{H}}\partial_{xy}\tilde{v} - \partial_{y}\nabla^2\tilde{v}|_{\overline{H}}
\end{equation}

Evaluating \eqref{eq:111} at $\overline{H}$ and expanding the Laplacian operator as $\nabla^2 = D^2 + \nabla^2_{xz}$, where $ \nabla_{xz}^2 = \partial_{xx} + \partial_{zz}$ is the two dimensional Laplacian operator, gives:
\begin{equation}
\label{eq:223}
    D^2\tilde{p} = -\nabla_{xz}^2\tilde{p}|_{\overline{H}} - 2(D\bar{U}\partial_x\tilde{v})|_{\overline{H}}.
\end{equation}

Equating \eqref{eq:222} with \eqref{eq:223}, and noting that $D\bar{U}|_{\overline{H}}=0$, gives:
\begin{equation}
\label{eq:333}
    \nabla_{xz}^2\tilde{p}|_{\overline{H}} = \partial_{ty}\tilde{v} + \Bar{U}|_{\overline{H}}\partial_{xy}\tilde{v} + \Bar{V}\partial_{yy}\tilde{v} - \partial_{y}\nabla^2\tilde{v}|_{\overline{H}}.
\end{equation}

Moving to the normal stress balance, I apply the two dimensional Laplacian operator to the linearized normal stress balance $\partial_x\eqref{eq:lin_normal_stress_x}+\partial_z\eqref{eq:lin_normal_stress_z}$, noting that $\nabla_{xz}^2(\tilde{p}|_1)=\nabla_{xz}^2\tilde{p}|_1$ and equation with \eqref{eq:333} gives:
\begin{equation}
\begin{gathered}
\label{eq:383}
   - \partial_{ty}\tilde{v} - \Bar{U}\partial_{xy}\tilde{v} -3\Vr\Bar{J}\nabla_{xz}^2\tilde{J} + \nabla_{xz}^2\nabla_{xz}^2\tilde{h}\Big(\frac{2\M\,D\Theta|_{\overline{H}}}{\Pr}-3\Gamma\Big) + \\+ \Ct\nabla_{xz}^2\tilde{h} + 3\nabla_{xz}^2\partial_{xxt}\tilde{v} + \partial_{yyy}\tilde{v} = 0
\end{gathered}
\end{equation}

Moving to the normal stress balance, taking  2D divergence of the tangential stress balance in vector form $\partial_x\eqref{eq:lin_normal_stress_x}+\partial_z\eqref{eq:lin_normal_stress_z}$ and using the continuity condition gives:
\begin{equation}
    \partial_x\tilde{h} = -\frac{2\M}{\Pr}(D\Bar{\Theta}\nabla_{xz}^2\tilde{h} + \nabla_{xz}^2\tilde{\theta}) - (\nabla_{xz}^2 - \partial_{yy})\tilde{v} = 0
\end{equation}

Equations \eqref{eq:new_gov_eq_lin} and \eqref{eq:lin_temp} with the boundary condition \eqref{eq:bc_wall_lin} at the wall and \eqref{eq:cont_lin_fs}, \eqref{eq:lin_temp_bc_fs}, \eqref{eq:333} and \eqref{eq:383} with the constitutive relation \eqref{eq:const_rel_lin} represent the reformulated linear problem in terms of the wall-normal velocity $\tilde{v}$, the temperature $\tilde{\theta}$, the liquid film thickness $\tilde{h}$ and the mass flux at the free-surface $\tilde{J}$. These equations represent the base for the derivation of the generalised eigenvalue problem (known as the Orr-Sommerfeld eigenvalue problem) in subsection \ref{subsec:linear_stab_analysis}.

% ---------- Mesh convergence study
\subsection*{Mesh convergence study}
\addcontentsline{toc}{subsection}{Mesh convergence study}
\label{subsec:mesh_conv_sutd}
In this section, I present the mesh-convergence study used to determine the optimal number of Chebyshev functions for solving the Orr--Sommerfeld eigenvalue problem, the appropriate number of grid points for the 2D and 3D WIBL simulations, and the suitable mesh configurations for solving the full set of equations in COMSOL.

The error associated to the different meshes for the WIBL and COMSOL implementations is expressed by the following formulas:
\begin{subequations}
\label{eq:error_mesh_WIBL}
    \begin{align}
        E_{h,N}(\hat{T})
        &=
        \frac{
        \left( \displaystyle\int_0^L 
        \| \hat{h}_{N}(\hat{X},\hat{T}) - \hat{h}_{fine}(\hat{X},\hat{T}) \|_2 \, \mathrm{d}\hat{X} \right)^{1/2}
        }{
        \left( \displaystyle\int_0^L 
        \| \hat{h}_{fine}(\hat{X},\hat{T}) \|_2 \, \mathrm{d}\hat{X} \right)^{1/2}
        },\\
        E_{\theta_s,N}(\hat{T})
        &=
        \frac{
        \left( \displaystyle\int_0^L 
        \| \hat{\theta}_{s,N}(\hat{X},\hat{T}) - \hat{\theta}_{s,fine}(\hat{X},\hat{T}) \|_2 \, \mathrm{d}\hat{X} \right)^{1/2}
        }{
        \left( \displaystyle\int_0^L 
        \| \hat{\theta}_{s,fine}(\hat{X},\hat{T}) \|_2 \, \mathrm{d}\hat{X} \right)^{1/2}
        },
    \end{align}
\end{subequations}
where $N$ indicates the number of grid points for the WIBL implementation or the mesh index for the COMSOL implementaiton, and \textit{fine} stands for the mesh with the largest number of elements in the two cases.

% ---------- Mesh convergence study Orr-Sommerfeld
\subsubsection*{Orr-Sommerfeld eigenvalue problem}
\addcontentsline{toc}{subsubsection}{Orr-Sommerfeld eigenvalue problem}
To determine a suitable number of Chebyshev polynomials for the numerical solution of the Orr-Sommerfeld eigenvalue problem, I evaluated the relative error of the most unstable eigenvalue $\omega$ (i.e.\ the eigenvalue with the largest $\omega_i$) for various numbers of Chebyshev polynomials $N$, using the case with $N=80$ as reference. This relative error $\Delta \omega_N$ is expressed as:
\begin{equation}
    \Delta \omega_N = \frac{|\omega_N - \omega_{80}|}{|\omega_{80}|}.
\end{equation}

Table~\ref{tab:mesh_ind_OS} reports the values of the most unstable eigenvalue for both streamwise and spanwise perturbations, together with the corresponding relative error $\Delta \omega_N$. For streamwise perturbations, the error falls below $10^{-6}$ once $N > 20$. In contrast, for spanwise perturbations, the error remains on the order of $10^{-2}$ across all tested resolutions. Nevertheless, for streamwise perturbations, the magnitude of the most unstable eigenvalue varies only marginally with increasing $N$.
\begin{table}
\centering
\begin{tabular}{ccccccc}
   & \multicolumn{3}{c}{$k_x = 0.2, \quad k_z=0, \quad \R =50$} & \multicolumn{3}{c}{$k_x = 0, \quad k_z=0.05, \quad \R=0.1$} \\
N  & $\omega_r$  & $\omega_i$ & $\Delta \omega_N$ & $\omega_r$  & $\omega_i$  & $\Delta \omega_N$ \\
10 & 7.77288003  & 0.70126896 & 1.27042e-05       & 0           & 0.003182    & 0.0633836163      \\
20 & 7.7727815   & 0.70127054 & 7.8071e-06        & 0           & 0.003182    & 0.0633840691      \\
50 & 7.77277885  & 0.70127186 & 8.0709e-06        & 2e-06       & 0.003188    & 0.065102486       \\
80 & 7.7727988   & 0.70121211 & -                 & 5.5e-05     & 0.002999    & -                
\end{tabular}
\caption{Values of $\omega_i$ and $\omega_r$ of the most-unstable eigenvalue (largest $\omega_i$) for $\beta = 15^{\deg}$, $\Pi = 0$, $\Gamma = 1000$, $Pr = 7$, $Vr = 4$, $\K = 0.01$, $\overline{H} = 0$, $\E = 0.2$, $\eta = 1$, $\M = 10$, $Re = 50$, for four different numbers of Chebyshev polynomials N with their relative accuracy $\Delta \omega_N$ with respect to the N = 80 case}
\label{tab:mesh_ind_OS}
\end{table}

\begin{figure}
    \centering
    \begin{tabular}{ccc}  % Create a 3-column table
        \subfloat[]{\includegraphics[width=0.33\textwidth]{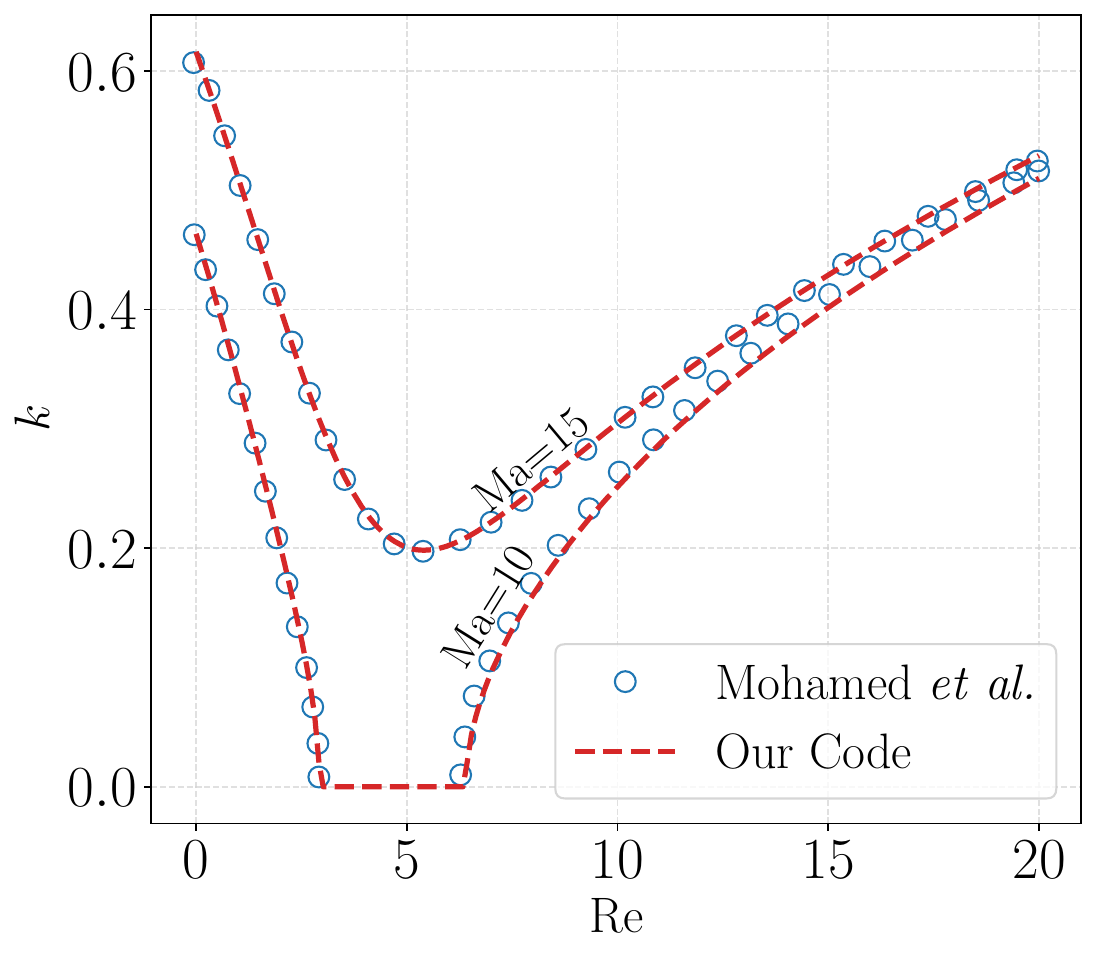}} &
        \subfloat[]{\includegraphics[width=0.33\textwidth]{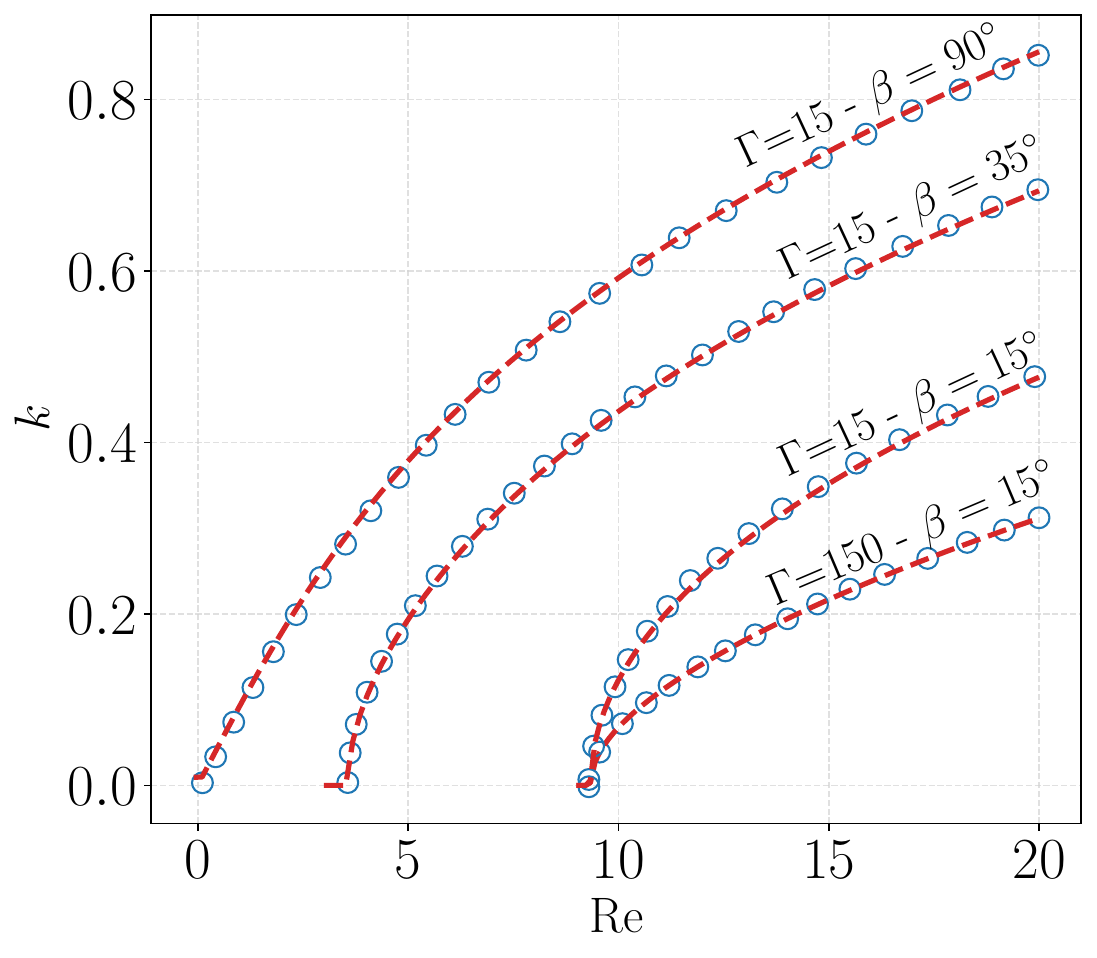}} &
        \subfloat[]{\includegraphics[width=0.33\textwidth]{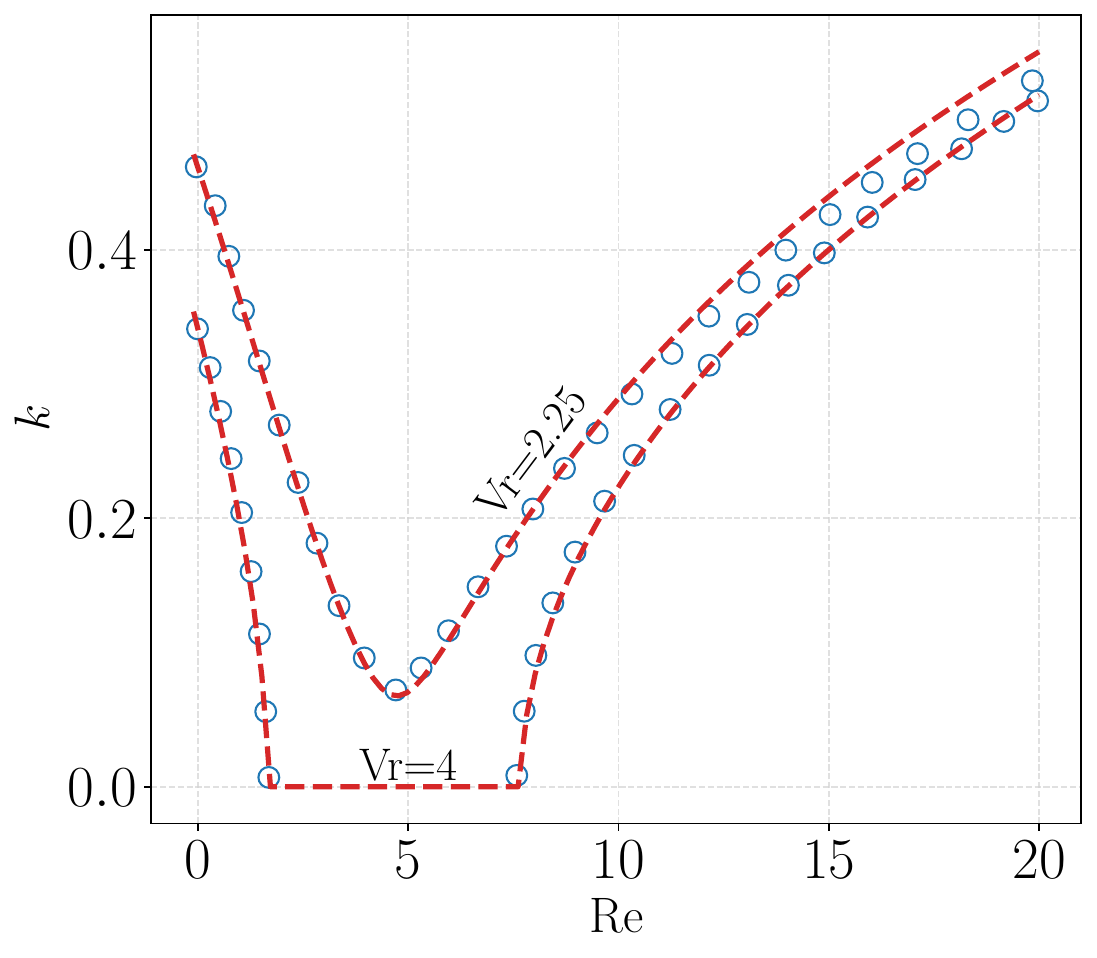}}
    \end{tabular}
    \caption{Comparison of neutral from \citet{mohamed2020linear} against those obtained with our code for (a) M-mode, (b) H-mode and (c) E-mode.}
    \label{fig:validation}
\end{figure}

I then solved the OS problem using 20 Chebyshev polynomials and compared the resulting neutral curves with those reported by \citet{mohamed2020linear} for streamwise perturbations in the evaporation case. Figure~\ref{fig:validation} shows the neutral curves for the H, M, and E modes considered independently (with the other nondimensional numbers set to zero) for different value of surface tension $\Gamma$, inclination angle $\beta$, Marangoni number $\M$ and vapour recoil number $\Vr$, obtained with my solver (dashed lines) alongside the results of \citet{mohamed2020linear} (blue empty dots). As shown, the agreement is excellent.

% ------- Mesh convergence study WIBL
\subsubsection*{WIBL simulations}
\begin{figure}
  \begin{subfigure}[b]{0.49\textwidth}
    \includegraphics[width=\textwidth]{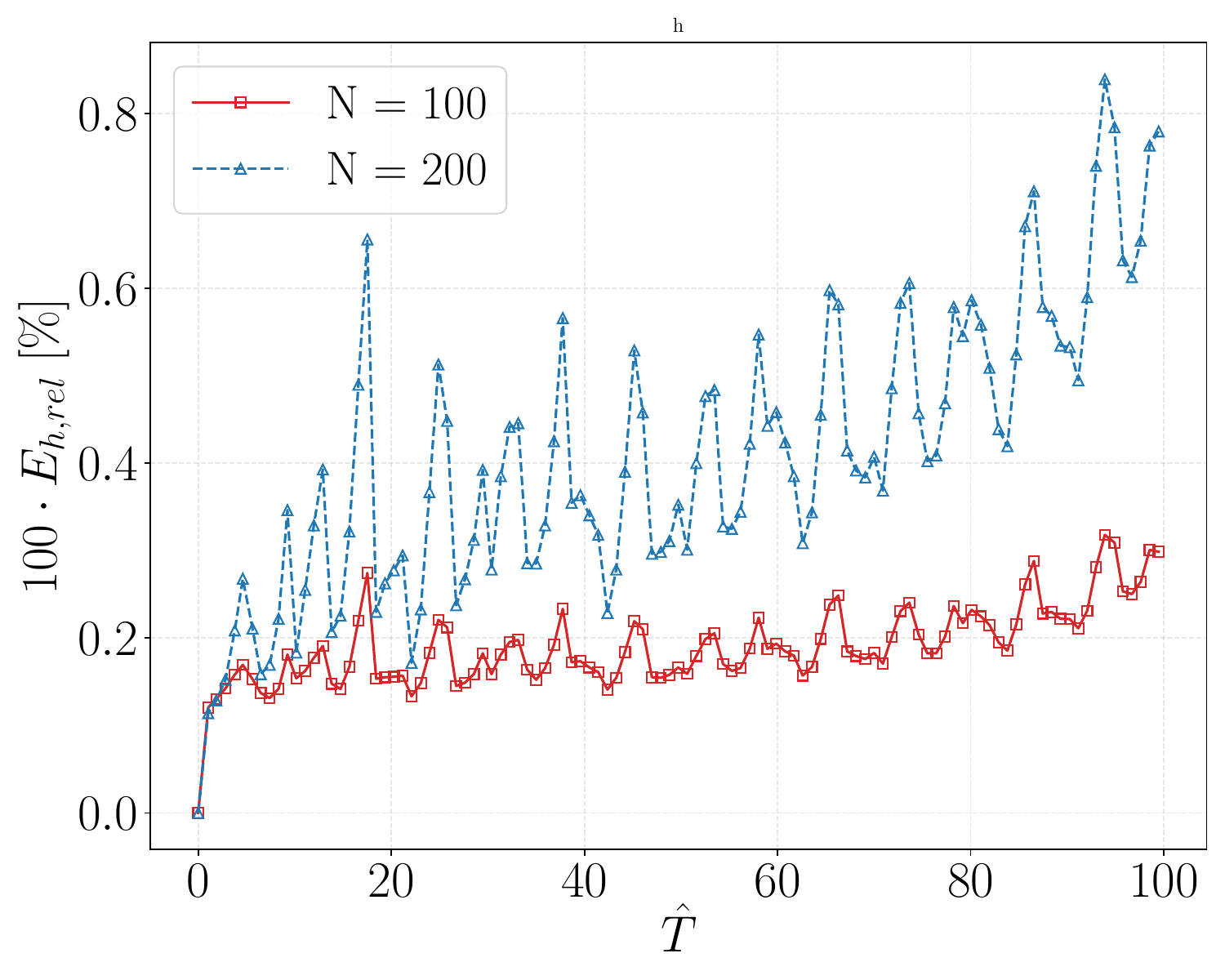}
    \caption{}
  \end{subfigure}
  \hfill
  \begin{subfigure}[b]{0.49\textwidth}
    \includegraphics[width=\textwidth]{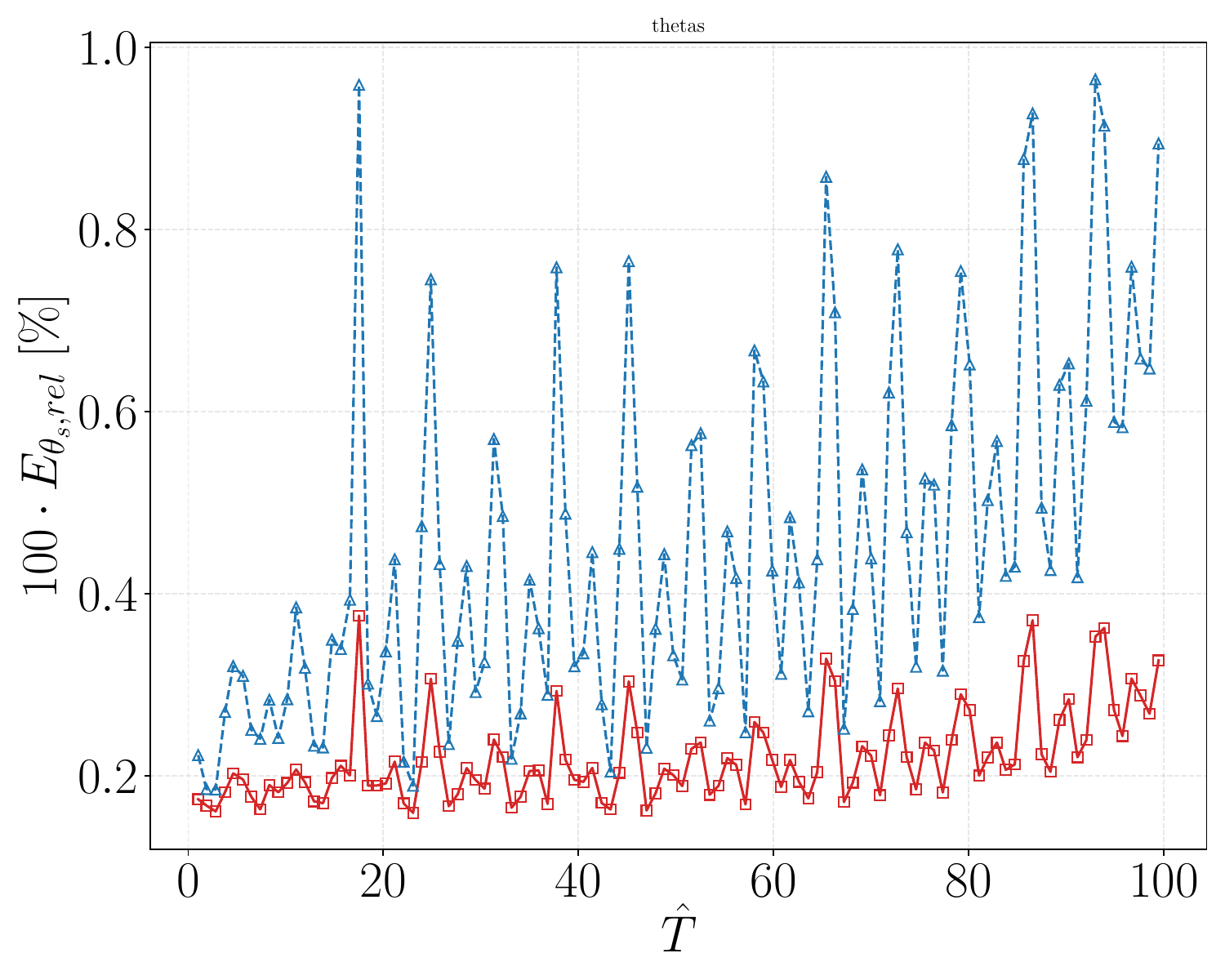}
    \caption{}
\end{subfigure}
\caption{Evolution of the error for the 2D WIBL simulation, relative to the reference case with 500 grid points, expressed as a percentage in terms of (a) the film thickness $\hat{h}$ and (b) the free-surface temperature $\hat{\theta}_s$. Results are shown for grids with 100 points (red continuous line with squares) and 200 points (dashed line with triangles).}
\label{fig:error_mesh_WIBL_2D}
\end{figure}

\addcontentsline{toc}{subsubsection}{WIBL simulations}
To determine a suitable number of grid points for the WIBL simulations in 2D and 3D, I define three grids with 100, 200, and 500 points and compute the errors in liquid film thickness and free-surface temperature relative to the 500-point case at different time instants. The error function read:

Figure~\ref{fig:error_mesh_WIBL_2D} shows the evolution of the error defined in~\eqref{eq:error_mesh_WIBL}, expressed as a percentage, for (a) the liquid film thickness and (b) the free-surface temperature, for simulations using $N=100$ (red continuous line with squares) and $N=200$ (dashed line with triangles) grid points. In both cases, the relative error remains below 1\% when compared with the reference solution obtained using 500 grid points.

% ---------- Mesh convergence study
\subsection*{COMSOL Simulations}
\addcontentsline{toc}{subsection}{COMSOL Simulations}

\begin{figure}
    \centering
    \includegraphics[width=0.55\textwidth]{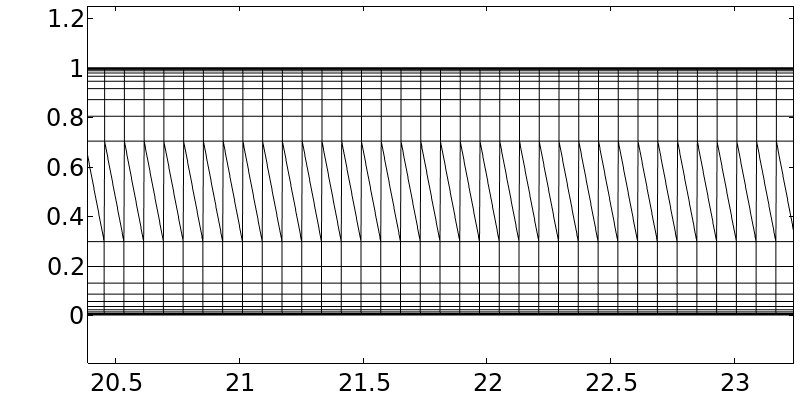}
    \caption{Mesh structure used in the COMSOL simulations.}
    \label{fig:COMSOL_mesh}
\end{figure}

For the mesh convergence study of the COMSOL simulation setup, I consider a boundary-layer mesh near the free surface, with the solid boundary and the free surface connected through triangular elements, as illustrated in figure~\ref{fig:COMSOL_mesh}. This mesh structure captures the development of the thermal boundary layers while providing sufficient flexibility to deform as free-surface waves evolve.

\begin{figure}
  \begin{subfigure}[b]{0.49\textwidth}
    \includegraphics[width=\textwidth]{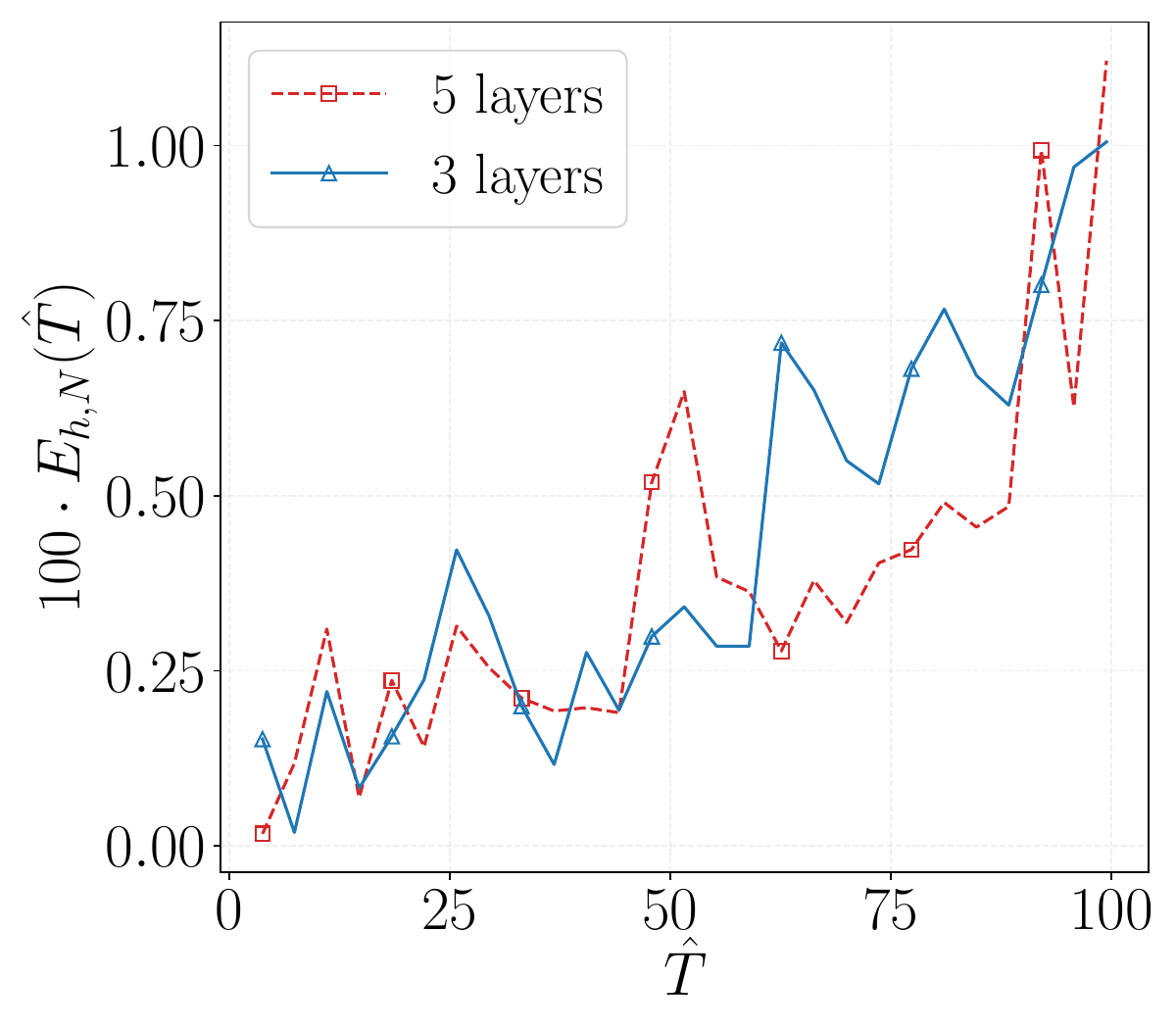}
    \caption{}
  \end{subfigure}
  \hfill
  \begin{subfigure}[b]{0.49\textwidth}
    \includegraphics[width=\textwidth]{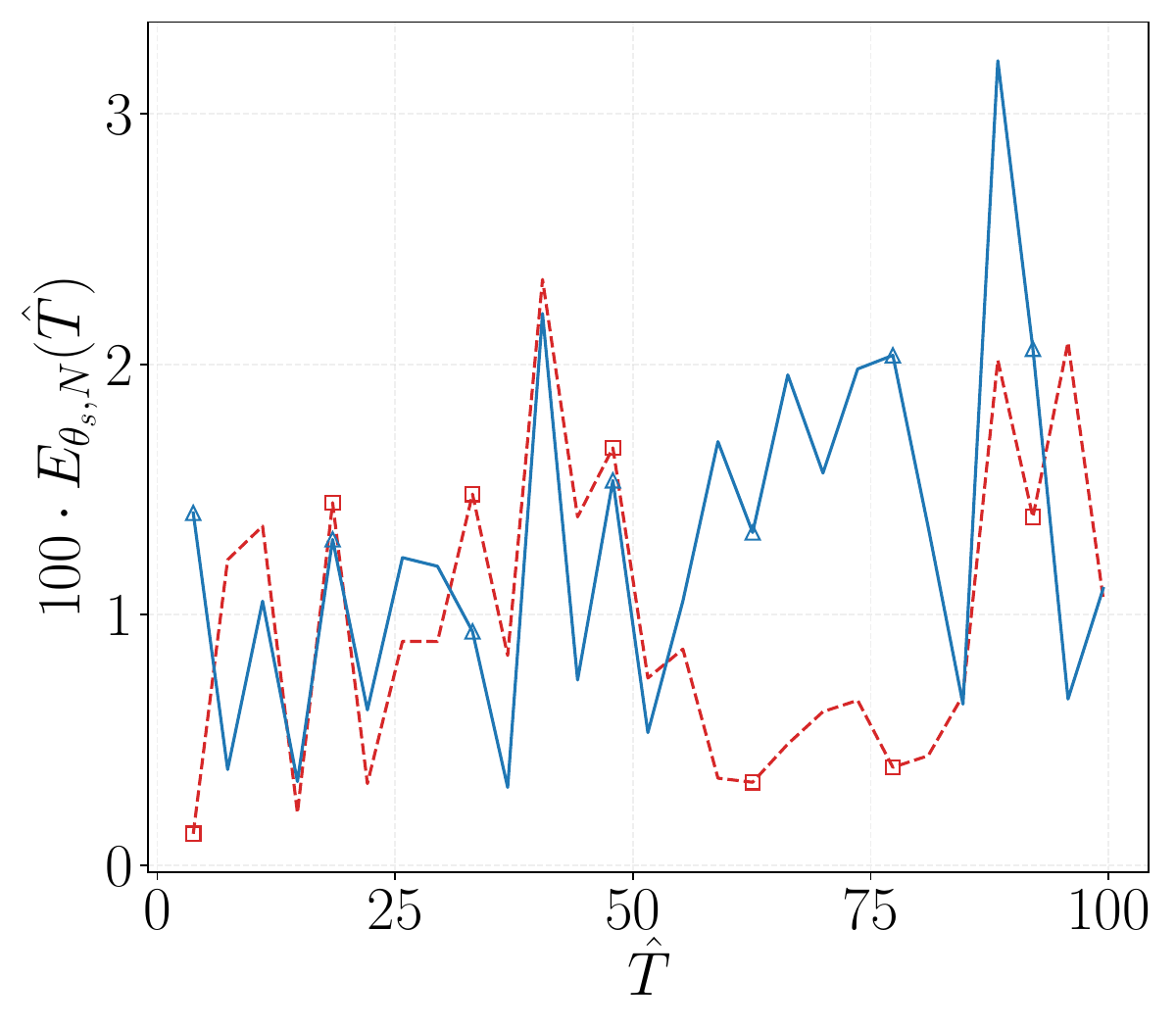}
    \caption{}
\end{subfigure}
\caption{Evolution of the error for the COMSOL simulation, relative to the case with 10 boundary layers, expressed as a percentage in terms of (a) the film thickness $\hat{h}$ and (b) the free-surface temperature $\hat{\theta}_s$. Results are shown for a boundary layer with 3 (continuous red line with squares) and 5 points (dashed blue with triangles) layers.}
\label{fig:error_mesh_COMSOL}
\end{figure}

Three different meshes with \(N = 3\), \(5\), and \(10\) layers were tested. Figure~\ref{fig:error_mesh_COMSOL} reports the error defined in \eqref{eq:error_mesh_WIBL}, computed relative to the reference mesh with \(N = 10\) layers, for both the free-surface thickness and the temperature throughout the simulation. For the thickness field, the error remains below \(1\%\) compared to the 10-layer mesh, while for the temperature, it remains below \(2\%\), except for a peak of approximately \(3\%\) near the end of the simulation. These results demonstrate that the solution is well converged and that a 10-layer mesh provides a satisfactory level of accuracy.

\end{document}